# SETI OBSERVATIONS OF EXOPLANETS WITH THE ALLEN TELESCOPE ARRAY

G. R. Harp, Jon Richards, Jill C. Tarter, John Dreher, Jane Jordan, Seth Shostak, Ken Smolek, Tom Kilsdonk, Bethany R. Wilcox, M. K. R. Wimberly, John Ross, W. C. Barott, R. F. Ackermann, Samantha Blair

SETI Institute, Mountain View, CA 94043, USA

## ABSTRACT

We report radio SETI observations on a large number of known exoplanets and other nearby star systems using the Allen Telescope Array (ATA). Observations were made over about 19000 hours from May 2009 to Dec 2015. This search focused on narrow-band radio signals from a set totaling 9293 stars, including 2015 exoplanet stars and Kepler objects of interest and an additional 65 whose planets may be close to their Habitable Zone. The ATA observations were made using multiple synthesized beams and an anticoincidence filter to help identify terrestrial radio interference. Stars were observed over frequencies from 1- 9 GHz in multiple bands that avoid strong terrestrial communication frequencies. Data were processed in near-real time for narrow-band (0.7- 100 Hz) continuous and pulsed signals, with transmitter/receiver relative accelerations from -0.3 to 0.3 m/s$^2$. A total of 1.9 x 10$^8$ unique signals requiring immediate follow-up were detected in observations covering more than 8 x 10$^6$ star-MHz. We detected no persistent signals from extraterrestrial technology exceeding our frequency-dependent sensitivity threshold of 180 - 310 × 10$^{-26}$ W m$^{-2}$.

## 1 INTRODUCTION

The first discovery of a planet orbiting a main sequence star (Mayo & Queloz 1995) has had a major impact on the search for extraterrestrial intelligence (SETI) for the last 20 years. Prior to 1995, we had no observational information about the probability that any star has planets or which stars do. By December 2015, about 2000 exoplanets had been identified. Many of these were initially discovered by the Kepler spacecraft which has also contributed thousands of Kepler objects of interest (KOIs), many of which are likely to become confirmed exoplanets with further observations (Borucki et al. 2010; Han et al. 2014). From almost the moment of the first exoplanet discovery, many SETI programs have been performing observations of exoplanets and KOIs. Special attention is given to those planets close to the "habitable zone" of their star where the HZ is roughly defined as the range of planetary orbital radii where liquid water may be present on the surface of a planet with an atmosphere. The motivation for studying exoplanets is simple since life as we know it originated on a planet, and life as we know it thrives anywhere there is liquid water.

Cocconi and Morrison ( 1959) established the basic rationale for searching for interstellar radio transmissions generated by technological civilizations. The radio band from ~1- 10 GHz, called the terrestrial microwave window (Oliver & Billingham 1971), is a particularly attractive observation band with low atmospheric radio absorption and minimal galactic background radiation. Radio observations began in 1960 (Drake 1961a) and have continued, often sporadically, at multiple locations around the globe.

Since 2007, the SETI Institute (SI) has used the Allen Telescope array in Northern California, actively performing SETI observations (Tarter et al. 2011) for approximately 12 hours each day. SI's main instrument, called "SETI on the ATA" or SonATA, is primarily a targeted search system which for many years has focused on stars with known exoplanets or objects of interest identified by the Kepler space telescope. The basic operation of SonATA involves pointing the telescope at three stellar targets simultaneously for typically 30 minutes at a time, while searching for narrowband (artificial) signals coming from the direction of those stars. When signals are identified and are not immediately revealed to be radio frequency interference (RFI), they are followed-up in near real time and tracked until they disappear or are positively identified as not actually arriving from the direction of any of the stars under investigation.

SI's observations complement observations performed as part of programs at many other observatories including Arecibo, the Green Bank Telescope (GBT), Low Frequency Array (LOFAR), the Very Large Array (VLA) and others (Penny 2011; Korpela et al. 2011) and those planned to be performed on the Square Kilometer Array (SKA) in the future (Siemion et al. 2014). The present work extends from SETI Institute's earlier campaign Project Phoenix (Backus 1996, Tarter 1996, Backus 1997, Backus 1998, Cullers 2000, Backus 2001, Backus et al. 2004), which used large single dish telescopes to explore the radio spectrum one star at a time over a frequency range from 1.2 - 3 GHz. Here we have used the interferometer capabilities of the ATA to observe 2 - 3 stars at a time with SonATA's automated system in frequency ranges from 1- 9 GHz.

This paper describes the first substantial SETI campaign that uses an interferometer with multiple phased array beams, and by example shows that interferometers can be dramatically more effective than single dishes for SETI observations. Another element that sets this work apart is SonATA's unique near-real time follow-up of interesting signals with automated logic for real time signal classification. This allows us to keep a minute by minute up-to-date catalog of time-variable, terrestrial interference which serves as a highly effective classification tool to avoid false positives. Our near-real time system is sensitive to transient signals with lifetimes on the scale of hours, as compared with other searches that rely on post-observation data reduction requiring signal persistence for days, months or years at a time.

As it is dedicated to SETI observations 12 hours every day, the ATA can effectively perform the targeted work described here, under the supervision of the nearly-autonomous SonATA system control. Since the ATA's commission in 2007, roughly nineteen thousand hours have been dedicated to SETI observing, a record that has not been duplicated at any other mid- to large scale telescope in the world.

## 1.1 TECHNOLOGICAL VS. ASTROPHYSICAL SIGNALS

A signal transmitted by an extraterrestrial civilization has to be detected against the combined background noise from the cosmos, the receiving system, and most importantly from our own terrestrial signals. The terrestrial microwave window (Oliver & Billingham 1971) represents a broad minimum in cosmic and atmospheric background noise at microwave frequencies, so transmissions in this range are more discernable. Astrophysical sources are broadband radio emitters when compared to many types of communications signals. The narrowest astrophysical line emission sources are saturated maser lines having a width of about 300 Hz (Grimm et al. 1987). Narrowband signals with linewidths smaller than this are mostly likely engineered, and are the object of our SETI observations.

The observed linewidth of an extrasolar signal has a lower bound. This is supported by theoretical studies (Ekers et al. 2002; Drake & Helou 1977) considering multiple sources of phase decoherence. Scintillation in the interstellar medium (ISM) broadens an infinitesimally narrow extrasolar signal limiting coherence times to a maximum of $10^4$ seconds for transmitters 1000 light years (LY) away. Scintillations in our solar system's interplanetary medium (IPM) are worse, limiting extrasolar signals to $10^3$ - $10^2$ seconds depending on the direction of arrival compared to the sun position, not counting similar effects on the transmitter side. For this reason, during SETI observations at the ATA, we maintain a 60$^{\circ}$ solar avoidance angle. Between 1- 2 GHz, the Earth's ionosphere limits signal coherence to 1000 - 100 seconds depending on solar activity. These factors informed our choice of main observatory clock, which is based on a rubidium standard with coherence time of ~100 sec, disciplined by GPS to avoid long term drifts. Likewise in our observations we choose coherent integration times of approximately 100 sec for a coherent spectral resolution of ~0.01 Hz.

To detect a narrowband extraterrestrial signal, we must also consider the rate of change in frequency or "drift" of any signal. An extraterrestrial transmitter might be on the surface of a rotating planet or an orbiting spacecraft, and our own receiver participates in the diurnal rotation and orbital motions of the Earth. Thus there will be a relative acceleration between the transmitter and receiver that makes the signal drift in frequency. For example, a transmitter on an Earth-sized planet with an eight-hour day has maximal acceleration is 0.3 m $s^{-2}$ and the received frequency would drift by about one part in $10^9$ per second (1 Hz $s^{-1}$ at 1 GHz). The rotation of the terrestrial receiver will impose another acceleration of 0.03 m $s^{-2}$ or frequency drift of about one part in $10^{10}$ per second. The signal detection algorithms employed by SonATA specifically accommodate positive and negative drifts. Since the drift rate is often proportional to frequency, higher frequency observations use wider spectral channel widths than lower frequency observations to minimize channel crossing during an observation.

Regularly pulsed carrier waves are another identifiable artificial signal type. Excluding pulsars, the minimum variability timescale of fluctuating astronomical sources is on the order of tens of minutes. Therefore a signal with pulse period of less than a two minutes and a bandwidth near the inverse of its duration would be clearly artificial and an energy-effective beacon. The present observations search for pulses with repetition rates between 0.03 and 0.7 Hz.

In summary, this study focuses on narrowband, slowly drifting, continuous or slowly pulsed signals that are unlike any known astrophysical source. Unfortunately, human-made signals frequently contain components of this type. We have developed an arsenal of mitigation techniques for terrestrial interference, described in the section § 3 on interference mitigation.

# 2 OBSERVATIONS AND SIGNAL PROCESSING

## 2.1 SOURCE SELECTION

The observations reported here were made during a 6-year campaign to observe stars with exoplanets. As described below, the ATA supports three simultaneous beams with high sensitivity that are usually all pointed within the large field of view enabled by the small apertures of the dishes. This field of view (FOV) depends on observing frequency (3.5º full-width at half maximum (FWHM) at 1 GHz, 0.4º FWHM at 9 GHz). When selecting three sources for observation, a star orbited by a known exoplanet or a Kepler Object of Interest (KOI) is chosen at field center where the first beam is placed. Then two other targets are chosen within the FOV taken from catalogs that include, in order of selection, exoplanets/KOI, HabCat (Turnbull & Tarter 2003a, 2003b) stars, and Tycho stars (Høg, E. et al. 2000). Our main catalog[1] contains confirmed Kepler exoplanets and KOI as well as all other known exoplanet stars discovered by other means (typically radial velocity and gravitational lensing measurements). At higher frequencies where the FOV is reduced in size, it is not always possible to find two exoplanet/KOI stars in the FOV, at which point stars are chosen from the HabCat catalog (Turnbull & Tarter 2003a, 2003b) containing stars with properties thought to be favorable for the development of life. Failing that, stars are then chosen from the Tycho-2 catalog containing 2.5 million bright stars until all three beams are assigned to stars.

Throughout the campaign special attention was given to so-called “habitable zone” (a.k.a. HZ) planets. The HZ targets have been selected from variously defined catalogs by different authors since the beginning of these observations, but is intended to be the zone where liquid water could exist on the surface of a planet with an atmosphere. This subset of exoplanet/KOI stars included 65 targets compiled from the Arecibo HZ catalog (Arecibo Planetary Habitability Laboratory at University of Puerto Rico 2015) and stars identified in the Kepler catalog as potential HZ stars (Borucki et al. 2011, table 5).

## 2.2 THE TELESCOPE AND SENSITIVITY

---

[1] The exoplanets catalog was prepared in-house at the SETI Institute and was periodically updated after major data releases, with a most recent update in spring 2015.

The Allen Telescope Array (ATA) is a LNSD array (large number, small diameter or large number of small dishes) consisting of 42 dishes with 6.1 meter diameter placed within an area approximately 300 by 150 meters on the ground. It is described in (Welch et al. 2009). A 6.1 m dish has a half-power beam width of 3.5° divided by the observing frequency in GHz. This is the maximum possible field of view of the array. Each dish is instrumented with a wideband feed (0.5 – 11.2 GHz) and low noise amplifier (LNA). The resulting analog radio frequency (RF) voltages are upconverted and sent over buried optical fibers to the array control building. There signal from each antenna is down-converted by four independent intermediate frequency (IF) systems, filtered and digitized to a 100 MHz bandwidth.

Digitized signals were processed with three dual-polarization beamformers (Barott et al. 2011). The beamformers contain a digital filter that limits the usable bandwidth to about 70 MHz. Each beamformer synthesizes a beam with spatial resolution corresponding to the maximum extent of the array. At 1.4 GHz the synthesized beam is about 3 by 6 arcminutes with a field of view of 2.5°. The three beamformers can simultaneously observe 3 different point sources at different positions in the field of view.

The antennas have a frequency-dependent system temperature (Tsys, typically ranging over 40-120 K at 1.4 GHz). Since the array was in development during these observations, some antennas, feeds and LNAs were always being upgraded. Observations typically used 27±4 antennas.

The minimum detectable flux density $S_{\min}$ for an observation using a single polarization, with spectral resolution $b = 0.7$ Hz, and a user-defined detection threshold $SNR$ (units of mean power per bin), is given by

$$S_{\min} = \frac{SNR}{\sqrt{bt_{obs}}}\left(\frac{2\,k_B T_{sys}}{A_{eff}}\right) \qquad \left(\frac{W}{m^2 Hz}\right) \tag{1}$$

where $k_B$ is Boltzmann’s constant and $A_{eff}$ is the effective collecting area of the array[2]. Initially, $t_{obs}$ was set to 192 s with $SNR = 9$, and as the computational capacity grew, our system could tolerate a greater number of noise-induced false positives so $t_{obs}$ was decreased to 93 s with SNR = 6.5 which results in the same value for $S_{\min}$.

[2] SonATA observations for continuous narrowband signals are carried out independently on each polarization, whereas observations for narrowband pulses are carried out on the union of the two polarizations.

Table 1

Signal detection limit and sensitivity threshold $S_{\text{min}}$ as a function of observation frequency. System temperatures were determined via observations of sources with known flux. Assumes 25 antennas were used for the observations.

| Frequency (GHz) | Tsys (K) | 1σ Narrowband Detection threshold ($10^{-26}$ W $m^{-2}$) | $S_{\text{min}}$ ($10^{-26}$ W $m^{-2}$) |
|---|---|---|---|
| 1.43 | 80 | 20 | 181 |
| 3.04 | 120 | 30 | 271 |
| 6.667 | 95 | 24 | 215 |
| 8.4 | 137 | 34 | 310 |

The resulting minimum detectable flux densities are 180 - 310 $10^{-26}$ W $m^{-2}$ as shown in Table 1. These limits correspond approximately to the strength of a signal from a narrowband transmitter that has an effective isotropic radiated power equivalent to the effective isotropic radiated power of the Arecibo planetary radar ($2 \times 10^{13}$ W), if that transmitter was at a distance of 100 LY. In other words, the present ATA system could detect the Arecibo transmitter at that distance, assuming a lining up in both space and time.

## 2.3 SIGNAL PROCESSING

SonATA is the evolutionary product of a full-custom hardware system that began observations in 1992 with the NASA High Resolution Microwave Survey and later Project Phoenix at Parkes, Green Bank, and Arecibo observatories. Both campaigns involved constant human supervision. Over time, custom hardware was replaced by rack-mounted PCs with accelerators, and in 2004 the system was moved to Hat Creek, reconfigured to work with the ATA (then under construction) and used in a sequence of different search strategies for the purpose of increasing autonomous control and conducting preliminary observations as the Prelude Project, following the array commissioning in 2007. Installation of enterprise servers and switches in 2010 enabled the software incarnation of SonATA as used for the observations described in this paper. SonATA continues to evolve in capability and control. The description of the signal processing that follows describes the manner of observation that prevailed during most of the reported observing window.

At the start of each daily observing session, the SonATA software automatically performed a series of calibrations for the beamformers (the equivalent of focusing the beams). A strong radio source such as Cas A was used to calibrate the line delays, and then a point source (quasar) was used for the frequency-dependent phase correction. The point source was re-observed at ten frequencies distributed with increasing separation across a given 300 MHz observing band chosen for a single observing session in order to measure and fit the phase calibration as a function of frequency. At 1400 MHz, the calibration phases are stable to within ~10° phase over at least 12 hours. SETI observations usually occur at night, but can be performed at any time.

After each beamformer was calibrated, SonATA automatically selected the targets and observing frequencies (within the chosen band) for the first observation. The selection is based on the primary catalog type, in this case exoplanets, the local sidereal time, and the observation history of the available targets. Once the target for the first beam was selected, the software attempted to find other targets from the exoplanets catalog for the other beams, subject to the constraints that targets must be separated by at least three half-power beam widths on the sky and lie within the telescope FOV. After each observation, the software would determine whether to follow-up interesting candidate signals, to choose a new frequency band, or to switch to new targets.

At various times during the survey we updated the hardware and software of our SETI signal processing system, but the basic processing scheme has remained the same. Data processing occurs in a two stage pipeline: 1) data collection, spectrum analysis, and normalization, and 2) signal detection and interference mitigation. To emphasize the combination of data collection and analysis, we refer to the two stages jointly as an “activity”, and each activity is given a unique identifier.

In the first stage of the pipeline, two nested digital polyphase filters produced spectra with a resolution of ~1 kHz. These data were then processed in two ways for data collection and normalization. The data were accumulated for 1.5 seconds and a third polyphase filter transformed the data to a resolution of ~0.7 Hz and stored it for signal detection. The 1 kHz data were also accumulated to form a “baseline” spectrum to be used in scaling the fine resolution data. Each 1.5 second baseline spectrum contributed to a running average combining the previous and current spectra with a 0.9 and 0.1 weighting respectively. Each observation started with ~15 seconds of baseline accumulation before the fine resolution data collection commenced. Successive fine resolution spectra overlapped in time by 50% to reduce the loss of sensitivity for signals not aligned with a ~1.5 second spectrum time window. Spectral data were scaled in units of the mean power using baseline spectra to facilitate statistical analysis based on unit normal data, and stored for signal detection during the data collection cycle of the next observation.

The second stage of the pipeline analyzed the (128 or 256) fine spectra for continuous wave (CW) and pulse signals. For CW detection, the data for all channels in all spectra was truncated at 3 times the mean noise. If pulses have the same mean power as a CW signal, then the pulse power is expected to be much higher than the noise data when the pulse is present. Therefore, only those data points (time, frequency, power) where the power exceeded a large threshold (typically 9 times the mean noise) were stored for subsequent analysis by the pulse detector system, thus producing a sparse data array with the non-truncated power. The performance of the signal processing systems is described below.

For the preponderance of observations described here, SonATA processed 20 - 40 MHz of bandwidth from each of three dual-polarization beams. SonATA accepts digital 100 MHz output from the beamformers in the form of IP packets over a 10 GbE network. A network switch distributes each beam to two “channelizers,” one for each polarization. Each channelizer uses a polyphase filter bank to create 128 frequency channels, each 0.8192 MHz wide, from the input; the channels are oversampled by 4/3. The 1024 point polyphase filter produces channels with ripple of less than .2 dB and almost 70 dB of adjacent-channel rejection. Of the 128 channels created, the SonATA channelizer outputs 49 channels via UDP for processing by the detectors. The center channel, which represents DC, is discarded. Of the remaining channels, 24 - 48 channels (approximately 20 - 40 MHz) are actually processed by detectors in SonATA. The final down-select adds the flexibility to skip channels with known RFI.

Software detector modules (DXs) each process two 0.8192 MHz channels, accepting channel data streams from a pair of channelizers to perform dual-polarization detections. Each DX participates in both stages of the activity: data collection and signal detection. In data collection, the first step consists of “subchannelizing” the channel data with a polyphase filter bank to produce 2048 subchannels each 533.333 Hz wide, of which 1536 are used (due to the channel oversampling). Subchannels are also oversampled by 4/3 using a filter similar to the channelizer filter and have similar response (0.2 dB ripple, 70 dB out-of-band rejection). The detector Fourier transforms each of the subchannels into “bins” of 0.694 Hz, and creates two datasets that are stored in memory as previously described: a truncated two-dimensional array of power spectrum vs. time for CW detection (waterfalls, *c.f.* Figure 1), and an untruncated, sparse matrix of all bins which exceed a specified power threshold for the detection of pulse trains (not shown). The truncation of the power data for CW detection at three times the mean noise minimizes the effect of very short, strong signals when integrating over straight-line signals in waterfalls.

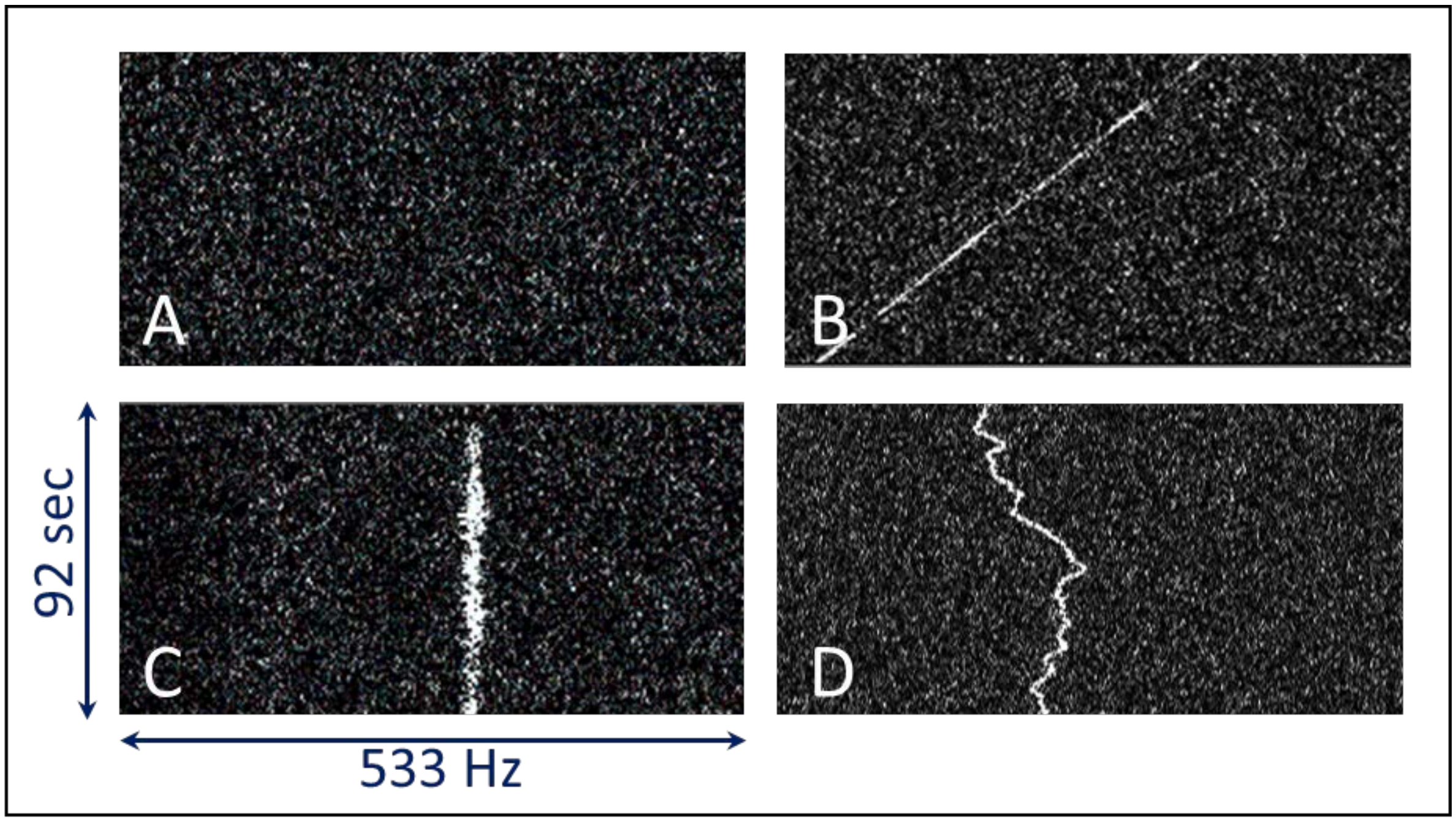

**Figure 1:** Four waterfall (power spectrum vs. time) plots showing (A) noise, (B) an example of the type of signal targeted in this survey – a drifting continuous wave, (C) RFI: a non-drifting continuous wave from an Earth-based transmitter, and (D) RFI: what we call a "squiggle," which may result from temperature variations of an unregulated oscillator (*a priori* we cannot rule out the possibility that this is part of a communication signal). None of the signals shown above passed our direction of origin tests for a true ET signal.

Signal detection for activity N is performed while data collection is being done for activity N+1 at a frequency higher than observation N. CW signals are found by an efficient algorithm that sums the truncated power along all possible straight-line paths in the CW data with a frequency drift rate $df/dt \leq \pm 1,\ 2, \text{or } 4$ bins per spectrum depending on observing frequency (corresponding to acceleration magnitudes less than $0.3\,\text{m s}^{-2}$). Path sums that exceed the statistical threshold (Table 1, column 4) are deemed signals. The CW algorithm, the Doubling Accumulation Drift Detector, recursively uses partial sums to achieve $2mn\log_2(m)$ performance for $n$ frequency bins and $m$ spectra (Cullers et al. 1985). Triplets are sets of 3 or more pulses that occur along a line in the sparse frequency-time plane, with nearly equal time spacing between pulses. These triplets are later combined to represent the entire signal pulse train.

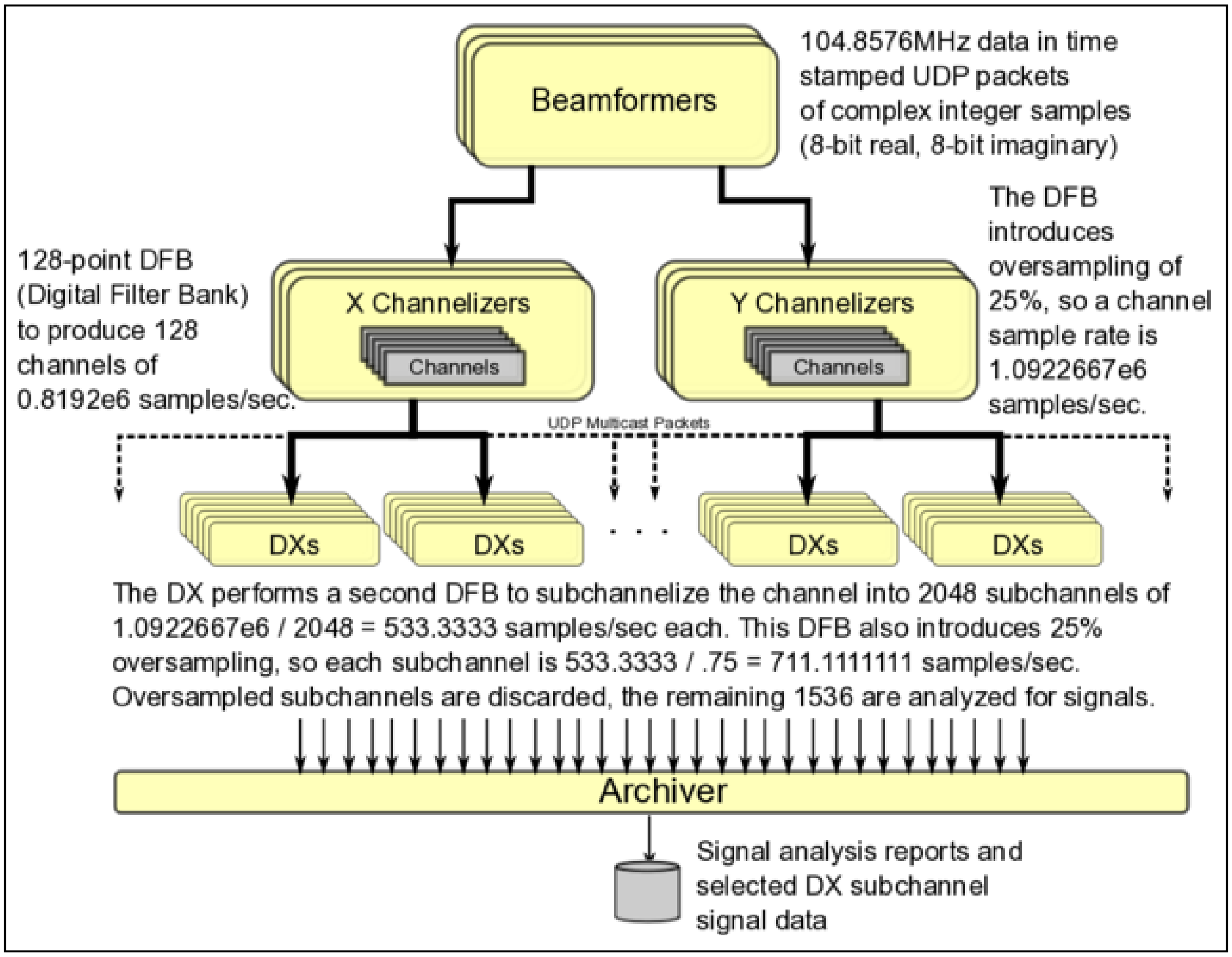

**Figure 2:** Schematic of the hardware and software components of the SonATA signal processing system.

The SonATA system (Figure 2) is quite physically compact; it consists of one 20-port Fujitsu XG2000 10 GbE switch, three Dell C6624P 10/1 GbE switches (one for each beam), one Dell C2100 server to host the control system, and six Dell C6100 servers to perform the channelization and signal detection. Each C6100 consists of four processing sleds; one sled acts as the channelizer, while the other three sleds serve as detector hosts, with eight DXs running on each host. Currently, the total configuration requires 2 rack units.

The top level SonATA software managed the observations, provided control of the observations, and performed high-level analysis of results. At the start of each observing session, it allocated the antennas, tuned the local oscillators, set the digitizer input levels, and calibrated the beamformers. Then SonATA selected the stars to observe and subsequently received the signal reports from the detectors, performed interference mitigation, decided which signals needed immediate follow-up observations and performed archiving.

## 3 INTERFERENCE MITIGATION

The most serious challenge facing any SETI project is distinguishing between strong terrestrial signals entering the sidelobes of the antennas and the potentially weak extraterrestrial signals being sought in the telescope FOV. While the terrestrial signals are generally due to licensed transmitters properly operating in assigned frequency bands, from the point of view of SETI observations, they are considered Radio Frequency Interference (RFI). Other signals generated at the observatory, by clocks and digital signal processing hardware, also pose a problem. The variability of the interference environment is a main driver for processing the data in near-real time. Over many years we developed a layered mitigation strategy to avoid ambiguous, untraceable results. Signals must be persistent long enough (>~ 1 hour) to pass direction of origin testing before human intervention is sought.

It is a consequence of modern technologies that some frequency bands are continuously occupied by strong signals and are unavailable for SETI or radio astronomy use, such as the GPS navigation service band centered at 1575.4 MHz. At intervals over the course of observations we have primed our RFI database using signals observed with a single ATA antenna pointed at the zenith for about 12 hours. These RFI scan observations stepped through all frequencies to characterize the persistent strong interference. Signals detected with the broad beam of a single dish and seen in more than one observation (pointed at a different position on the sky) were clearly persistent RFI. Based on those data we defined a preliminary set of “permanent RFI” bands. These RFI bands were refined and accumulated over time as RFI generally increases with time in this period of rapidly advancing technology. The scheduler software avoided observing those frequencies, even though these communications bands may represent likely places to find transponded replies from nearby civilizations that have previously detected Earth’s leakage radiation. By the end of the epoch of observations reported here, our mask of signals strong enough to be detected by a single dish covered 73 MHz of the terrestrial microwave window from 1 - 10 GHz, as listed in Table 2.

Table 2

RFI Masks

| Center Freq (MHz) | Width (MHz) | Min (MHz) | Max (MHz) |
|---|---|---|---|
| 1542.613 | 44.045 | 1520.519 | 1564.636 |
| 1575.285 | 8.192 | 1571.189 | 1579.381 |
| 1584.706 | 0.819 | 1584.296 | 1585.116 |
| 1599.548 | 12.480 | 1593.307 | 1605.788 |
| 1681.153 | 1.638 | 1680.334 | 1681.972 |
| 1684.840 | 0.819 | 1684.430 | 1685.249 |
| 1689.461 | 5.146 | 1686.887 | 1692.034 |
| Total | 73.140 | | |

Figure 3 plots the permanent masks of Table 2 as black vertical lines in a graph covering the frequency range from 0 - 10 GHz. For comparison, the International Astronomical Union radio astronomy protected frequency bands are displayed in green. Finally, we generate a new plot of congested bands by analysis of our database of about 2 x $10^9$ signals, which includes all candidate signals as well as those immediately identified as RFI. The signals are binned by frequency into 2 MHz bins and the probability of the detection of a signal is calculated. A threshold of a minimum of 10 signals per observation of the 2 MHz band is set, and frequencies where the average number of signals per observation exceed this threshold are presented as yellow vertical lines in the graph. The 10 signal per observation threshold is arbitrary and can be adjusted to guide future observations with more capable systems. Because of the relatively high density of RFI signals in the yellow bins, it is suggested that future SETI campaigns at the ATA should avoid these frequency ranges because it will be harder to establish the extraterrestrial origin of signals in those bands. Instead, limited observing time should focus on the remaining 99.2% of the 1 – 10 GHz spectrum show within the white regions of Figure 3 where little human-generated interference is observed at Hat Creek.

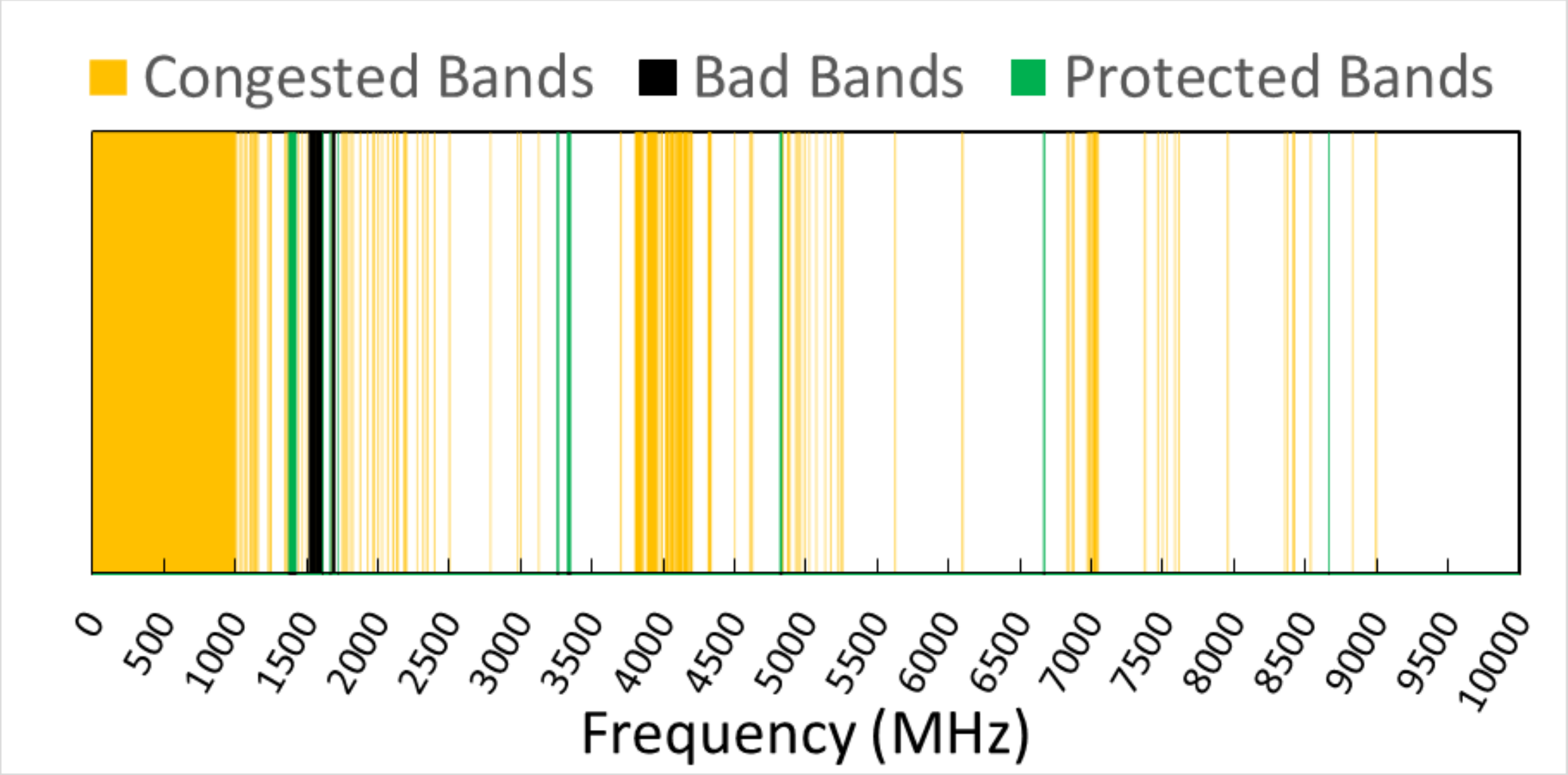

**Figure 3:** The frequency positions of radio-astronomical protected bands (green), the permanent RFI mask used for these observations (black), and regions of RFI congestion derived from these observations (yellow).

To date, our efforts have not attempted to identify the specific sources of detected RFI signals. Many radio services transmit intermittently on timescales of days or weeks. To handle such interferers and avoid classification as RFI in a band that may be clear much of the time, newly detected signals are compared to only those signals appearing in the past week in our RFI database. All detected signals are stored in the database and any signal identified as RFI by the methods described below is classified as such in the database.

Because the signal processing room is housed in an unshielded structure inherited from previous generations of radio astronomy projects at the Hat Creek Radio Astronomy Observatory, some signals from the observatory equipment inevitably leak in the RF/IF chain. Many of these signals are harmonics or intermodulation products of the digital equipment. They are very easy to identify because they are all locked to the observatory frequency standard and at a resolution of ~1 Hz have a very stable frequency; they are identified by "zero frequency drift." The maximum frequency drift rate for these signals was set to seven millihertz per second, a drift of less than one frequency bin per observation. In addition, note that satellites in geosynchronous orbit generally have drifts lower than this threshold.

## 3.1 MULTI-BEAM INTERFERENCE REJECTION

The new RFI mitigation technique that has been enabled by the ATA is the use of simultaneous, multiple synthesized beams. Each beam observes a different star system at the same frequency and at the same time. Thus each beam has two "off-target" beams for comparison. Since RFI mainly enters through the antenna sidelobes, it is often detectable at a similar strength in more than one beam. Signals detected in multiple beams at similar strength are classified as RFI. In order to make sure we don't miss a strong ET signal in one beam that might be detected in the other beams, each beam is modified to put an offset-null on the other position(s) observed (Barott et al. 2011). Theoretically, beams with such offset nulls have zero sensitivity in the direction of the null but calibration errors reduce the depth of the nulls to typically -7 dB relative to the unphased sensitivity of the array, which is about -7 dB relative to the beam main lobe, for an expected cross correlation between beams of -14 dB.

If a detected signal was not in the recent RFI database, had a non-zero frequency drift rate, and was not seen in the other beams, it was classified as a candidate ET signal. For all candidates, SonATA stored the voltage data centered on the signal for that observation and subsequent follow-up observations. This archival 'raw' data, available for subsequent processing, has a full bandwidth 10.5 or 8.5 kHz depending on channel width. All other raw data are discarded at the end of each activity.

SonATA then automatically conducted a series of follow-up observations (see logic diagram Figure 4) of any candidate ET signals that remained at the end of the analysis stage of an activity, starting with a re-observation of the star system (target1-on). Our detection thresholds are set such that approximately one in a million waterfall plots show apparent signals caused by noise, alone. Many RFI signals, possibly from aircraft or low Earth orbit satellites, only persist for a few minutes. So a re-observation is the fastest way to eliminate these noise and RFI events. If a candidate signal did not persist or failed on one of the other tests below, it was added to the recent RFI database and normal observations resumed.

This persistence requirement makes our search insensitive to short bursts of illumination from an ET transmitter that might be characteristic of a sequential-target-list strategy of transmission. As pointed out in the "SETI 2020" workshops (Ekers, et al. 2002), an omnidirectional search instrument with a significant ring buffer is the instrument of choice for such transients. As the prodigious compute power required for such a search strategy becomes affordable, we intend to initiate transient searches.

When a signal is found, it is first tested for presence in the recent RFI database. This comparison uses a simple matching filter that any signal with a database entry within 100 Hz is classified as RFI (and then the RFI database is updated with new signal parameters and time).

Signals with no counterpart in the RFI database are then checked for zero drift and presence in more than one beam (ref. Figure 4). Any signal that gets this far is (temporarily) classified as a candidate signal. The signal rejection in follow-up observations is then dominated by the direction of origin tests offered with the multi-beam system, so the chances of survival of subsequent tests are largely determined by the probabilities mentioned below.

If a signal was still detected in the on-target candidate beam, SonATA automatically moved all the beams to different locations or off-target for any candidates reaching this stage in the processing. If the signal was not seen in any off-target beam, the signal remained a candidate and the on/off observations continued for up to five cycles. At the ATA to date, only one signal has ever survived these tests.  The flowchart below summarizes the RFI mitigation process.

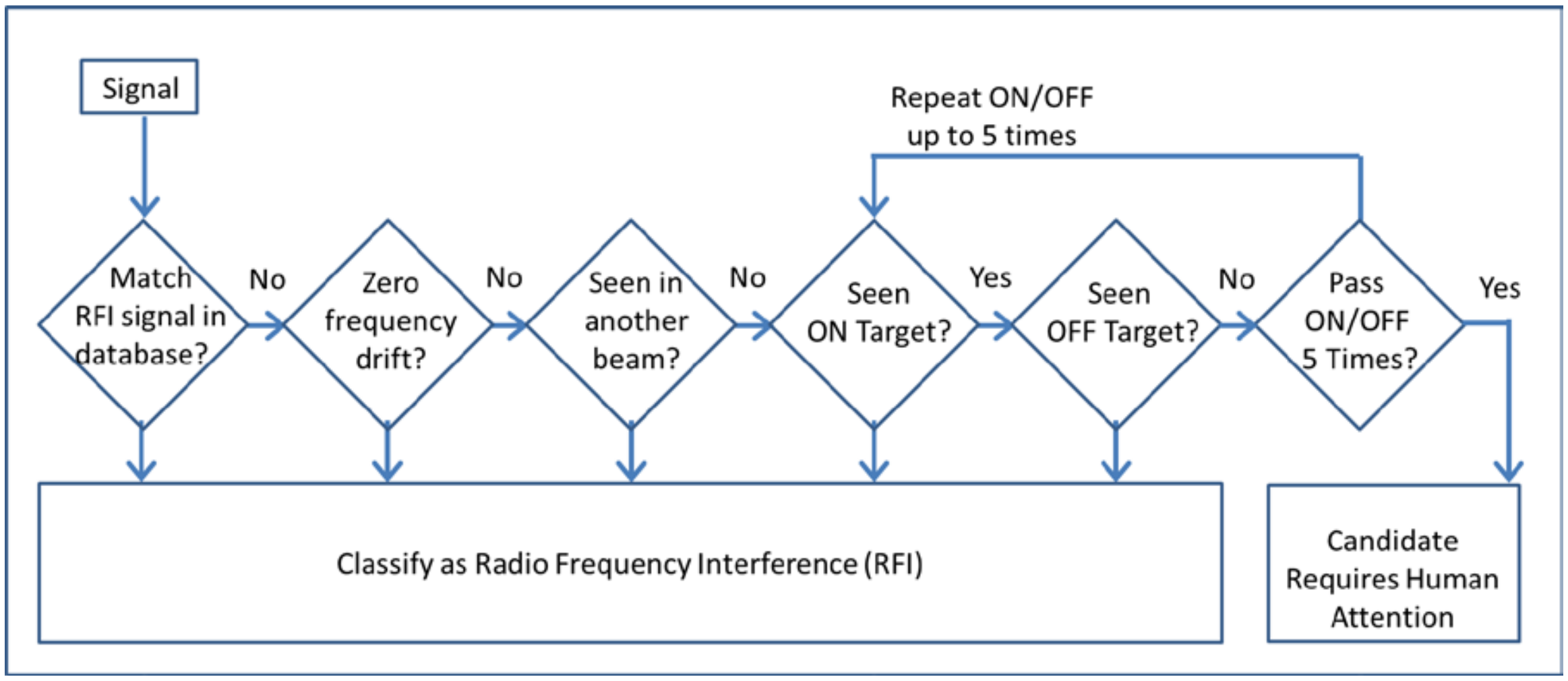

**Figure 4:** Schematic of signal classification logic used for SETI observations.

# 4 RESULTS

## 4.1 OBSERVATIONS OF EXOPLANETS

Observations reported here are from May 2009 through December, 2015, during a total of about 19,000 hours of observing, completing a total of 210,000 separate observations in 20 - 40 MHz blocks on 9293 stars covering a total of 7.3 million star·MHz at selected frequencies between 1 - 9 GHz. The unit of SETI observations star·MHz is a standard unit in the field representing a search over 1 MHz bandwidth on one target star. These results are summarized in Table 3. The detailed distribution of star·MHz versus frequency is shown in Figure 5.

Table 3

Summary of SETI observations, including all re-observations of candidate signals

| Catalog | HZ Exoplanets | Exoplanet not HZ | HabCat | Tycho (backup) |
|---|---|---|---|---|
| Number of Stars | 65 | 1959 | 2822 | 7459 |
| Star·MHz | 1,100,000 | 4,000,000 | 950,000 | 2,000,000 |
| <MHz/star> | 8000[a] | 2040 | 337 | 268 |

[a] Because of the high expected value of the HZ targets, many were observed over the full frequency range multiple times  The ratio in this case is greater than the maximum frequency range of observation, so this number is truncated to the latter value.

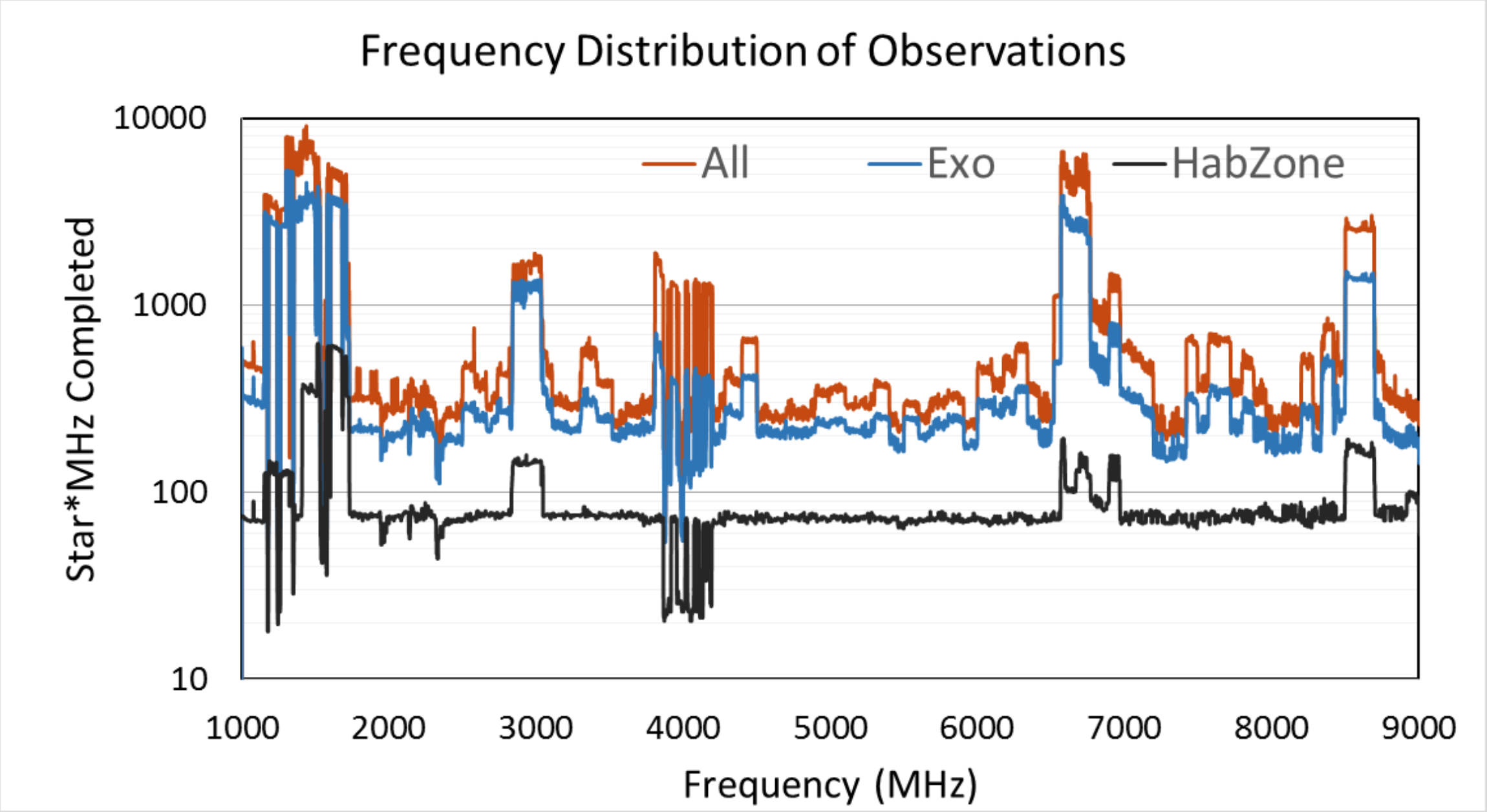

**Figure 5:** Plot of the total frequency coverage for all stars (orange) and the subset of exoplanet stars (blue) and the subset of HZ stars targeted in this campaign. The ordinal units star·MHz indicates the number of stars observed for each 1 MHz bin.

The variable nature of the coverage as a function of frequency is the result of changing priorities over the long period of observations. Initially, special attention was given to one range at L-band (1300 - 1710 MHz) for which 40% of the stars were fully covered (excepting permanent RFI bands mentioned above). This range corresponds to a slightly expanded version of the so-called water hole frequency range.  Another focus was placed on the range 6656 - 6676 MHz centered on the prominent methanol maser line.[3] In this band, 30% of the stars were completed. Other frequency ranges were also given priority over time as evidenced in the graph. More recent observations have attempted to obtain complete frequency coverage on specific stars, especially 65 HZ stars.

[3] A frequency at which the ATA has been previously well-characterized.

Table 6 lists the various targets that at one time or another were listed in the HZ catalog, along with the percentage of the 1 - 9 GHz available bandwidth over which each was measured. A table listing all the other observation targets in this campaign is found in Table 7 of the online version of this paper.

## 4.2 ANALYSIS OF SIGNAL CLASSIFICATION PERFORMANCE

The logic for signal classification outlined in Figure 4 can be characterized by the principle that a credible ETI signal comes from a point source moving at sidereal rate on the sky and will persist for long enough to allow direction of origin estimation. The logic tree that we have imposed has yielded no detections of a credible ETI signal. Our definition of RFI thus includes all observed signals to date, with little contribution from noise-alone events, and we shall show that this assumption is sufficient to describe the data. The ATA is fundamentally an imaging instrument, and the best way to understand multi-beam detection is in the image domain.

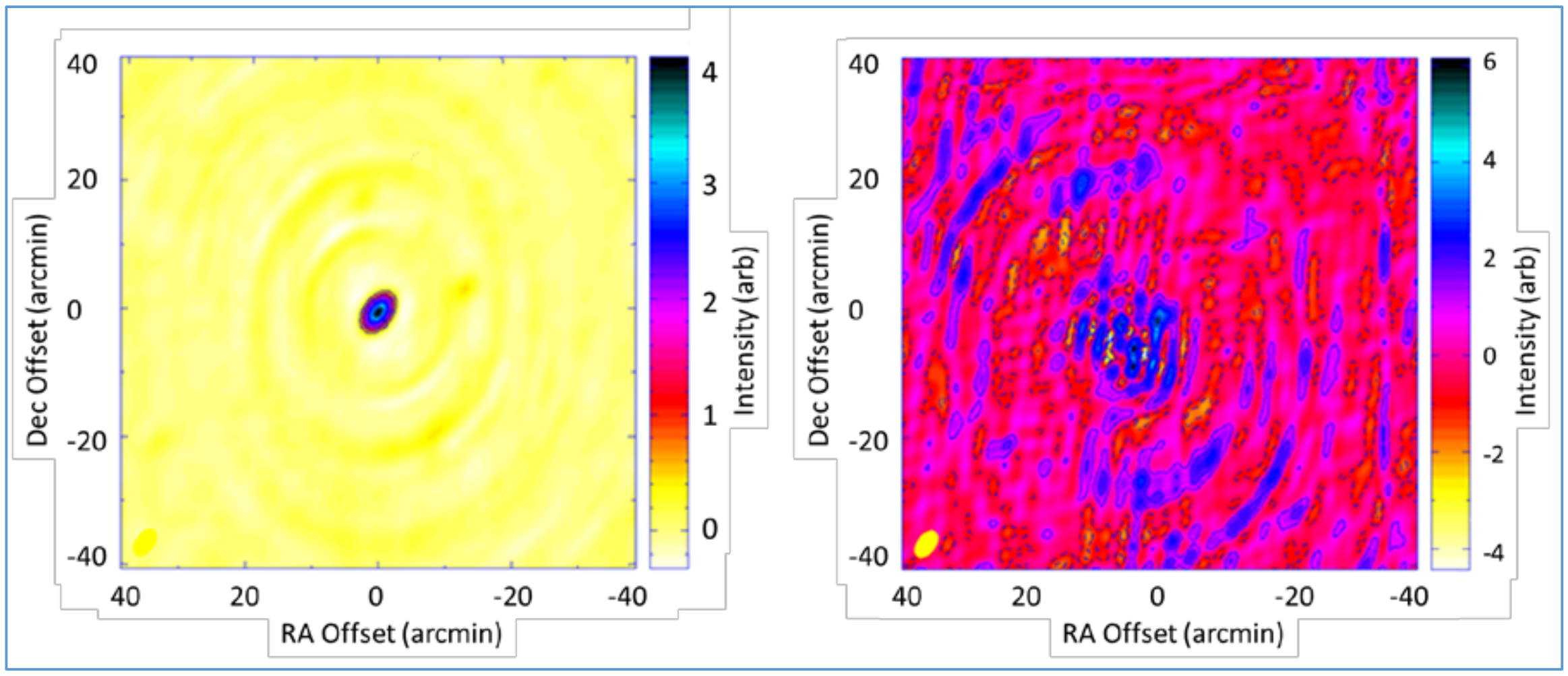

**Figure 6:** Comparison of two images taken at slightly different frequencies centered on a phase calibrator (blazar 0716+714). On the left there is a clear image of the blazar at image center. The right hand image was taken simultaneously, but because of strong radio interference in the chosen frequency, the image is dominated by non-imaging intensity.

Figure 6 displays two images taken with the ATA while pointed at an unresolved source (blazar). The left hand image shows a point at image center with a shape dictated by the ATA synthetic beam. On the right we see a comparable image at a different frequency that is spoiled by strong interference coming in sidelobes of the antenna. Since the blazar is within the field of view, it images to a point. But the large angle sidelobes of the antennas introduce different (random) phases into the RFI signals arriving at each antenna (Harp et al. 2011), so the RFI signal does not image down to a point in the FOV or generally anywhere on the sky. The RFI intensity is spread across the image with regions of high and low intensity unpredictably dispersed across the image.

The three beams in our SonATA system can be thought of as random but well-separated pixels (or rather, synthetic beams) selected from an image like those of Figure 6. We eliminate the vast majority of noise hits in the data by using only signals that passed at least one on-target observation after discovery. We model the detected signals as a single population of RFI sources having a finite probability $p$ of being observed in any randomly chosen beam with a corresponding probability $(1-p)$ of not being observed. Success with this single population model will indicate whether the detected sources actually include a true population of ETI signals, or noise-alone events. With these assumptions we can compute the probability that an RFI signal initially observed in one beam will survive an on-target observation (signal must be seen in the same beam but not in other two) as $p(1-p)^{N_b-1}$ where the number of beams $N_b = 3$ in this work. Similarly, the probability of an RFI signal surviving an off-target observation (signal is not seen in any beam) is $(1-p)^{N_b}$.

In Figure 7 we compare the observed signal survival probabilities from this work with the simple model of the previous paragraph. This fit has one free parameter, $p$ and the best fit to the data yield a value of $p = 0.225$. This is a reasonably good fit (coefficient of determination $R^2 = 0.996$) to the observed survival, which validates the proposed model. From this result we learn that RFI (as well as noise-alone events) are more quickly excluded by an on-target observation which eliminates 77.5% of false positives than by an off-target observation which eliminates only 22.5% of false positives. This validates the design of the SonATA search strategy and we consider what this means for the design of future searches below.

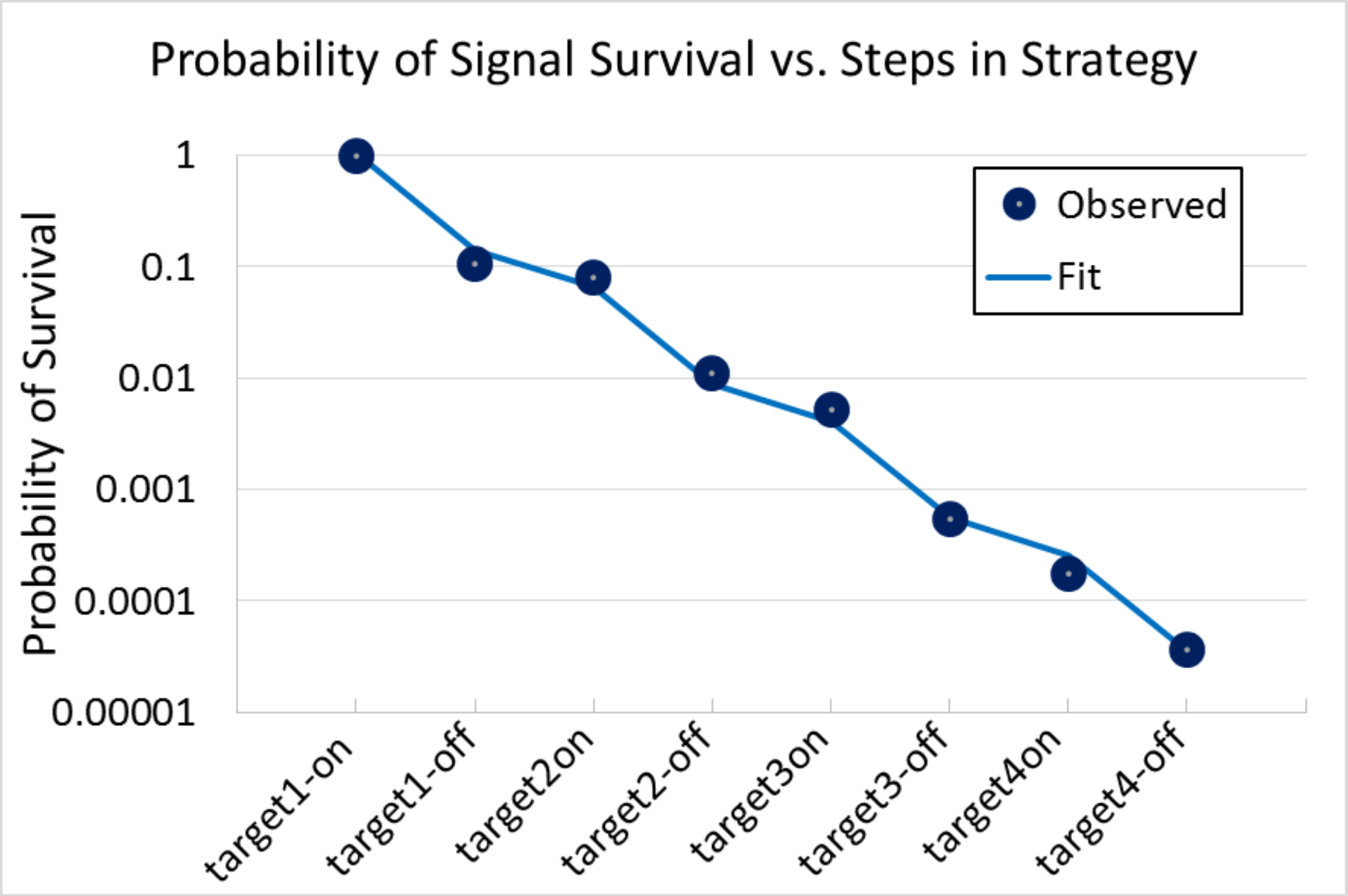

**Figure 7:** The observed probability of survival for signals detected in this study as a function of step in the logic sequence as outlined in Figure 4 and described in the text.

Besides the direction of origin classification scheme described above, a few other tests were applied to classify signals as RFI, as described in Figure 4. As discussed earlier, the SonATA system makes records[4] of all signals above threshold in a database along with their classification. During observations, this database is queried for each detected signal to determine whether a signal with similar frequency was seen before (within last 7 days) on a different target, which is strong evidence that the signal originates on Earth. This test immediately eliminates 61% of incoming signals. 4% of signals had a drift rate that was too high to be accurately detected within the search parameters of our system, and were neglected. While these signals are not necessarily RFI, the system parameters are adjusted to trade-off between maximum detectable drift and the minimum frequency channel width which directly impacts the detector sensitivity to the narrowest signals. Beyond the permanent RFI masks, some bands were so highly congested that too many candidates were detected to be classified within the near-real time constraints of the system. For this reason 3% of signals were dropped by necessity.

[4] The following is a subset of the information recorded for each signal: Beam RA Dec, Telescope RA Dec, Target Name, Beam Number (1-3), Type (CW, Pulse), RF Frequency at the start of the observation, Drift Rate (Hz/s), Signal Width (Hz), Integrated Power, Polarization (X, Y, both), Ultimate Classification (RFI, Unknown), Reason for classification, Probability of False Alarm, Signal to Noise Ratio, and where applicable Pulse Period and Number of Pulses.

About 1.3% of signals have very small drift in the coordinate frame of the observatory (zero drift). The vast majority of such signals are generated at the observatory, by ground based transmitters or by geosynchronous satellites. The excessive number of zero drift signals are dropped from the analysis since the likelihood that they are human generated is much higher than for other drift rates.

Table 4

Summary of classifications of observed signals

| Classification | Fraction Classified |
|---|---|
| Found in Recent RFI Database | 61% |
| Drift too high | 4% |
| Too many candidates (system overload) | 3% |
| Zero Drift | 1.3% |
| Signal drifted out of subband | 0.01% |
| Passed on as Candidate | 31% |

Because the frequencies are subjected to detection in overlapping blocks or subbands, there were a small number of cases (0.01%) where a signal drifted out of the subband under investigation during follow-up. The SonATA system does not have provision for following signals across subbands, so such signals were dropped from analysis and given a label in the database indicating they were unresolved.

Finally, the status of about 31% of all detected signals was not immediately resolved using the classifications of Table 4. These signals were passed on to the next stage of processing as candidates for on/off follow-up tests (see Figure 4).

# 5 DISCUSSION

## 5.1 ACCELERATING THE SEARCH WITH MORE BEAMS

In this section we discuss two outcomes of this research: 1) The lessons learned, especially those useful for the development of future searches, and 2) the contributions of this paper to our understanding of the prevalence of technological civilizations in the galaxy.

One result that is clear from Table 4 is that radio frequency interference seen in one direction on the sky is often seen in other directions. The recent RFI database therefore makes speedy assessments of 61% of observed signals. A similar conclusion is drawn from Figure 6. Using a direction of origin sieve, candidate signals are classified as RFI exponentially fast with the number of observations.

Figure 6 also shows that on-target observations generally exclude more RFI than target off observations. This is elucidated by our model fitting parameters where the probability of detecting RFI in any single beam is $p = 0.225$ and the probability of any RFI not being detected in a given beam is $(1-p)=0.775$. Hence the chance of survival of an on-target observation with 3 beams is $p(1-p)^2=0.14$ whereas the chance of surviving an off-target observation is $(1-p)^3=0.47$. From these results we draw the following conclusions for the design of efficient future searches: 1) Ideally, follow-up observations should always put one of the beams on the source where the signal was first detected, and 2) search efficiency can be greatly enhanced by the use of many more beams that the three beams used here.

We may predict the performance of the ATA outfitted with more beamformers as in Figure 7. Here we simulate an on-target observation where one beam is pointed in the direction where the signal was observed and $(N_b-1)$ beams are pointed in other directions. It is seen that with the addition of more off beams, the signal survival rate decreases exponentially. A relatively large number of beams would make direction of origin testing more effective, requiring 14 beams to reduce the survival rate to 1% and 50 beams to reduce the survival rate to below $10^{-6}$.

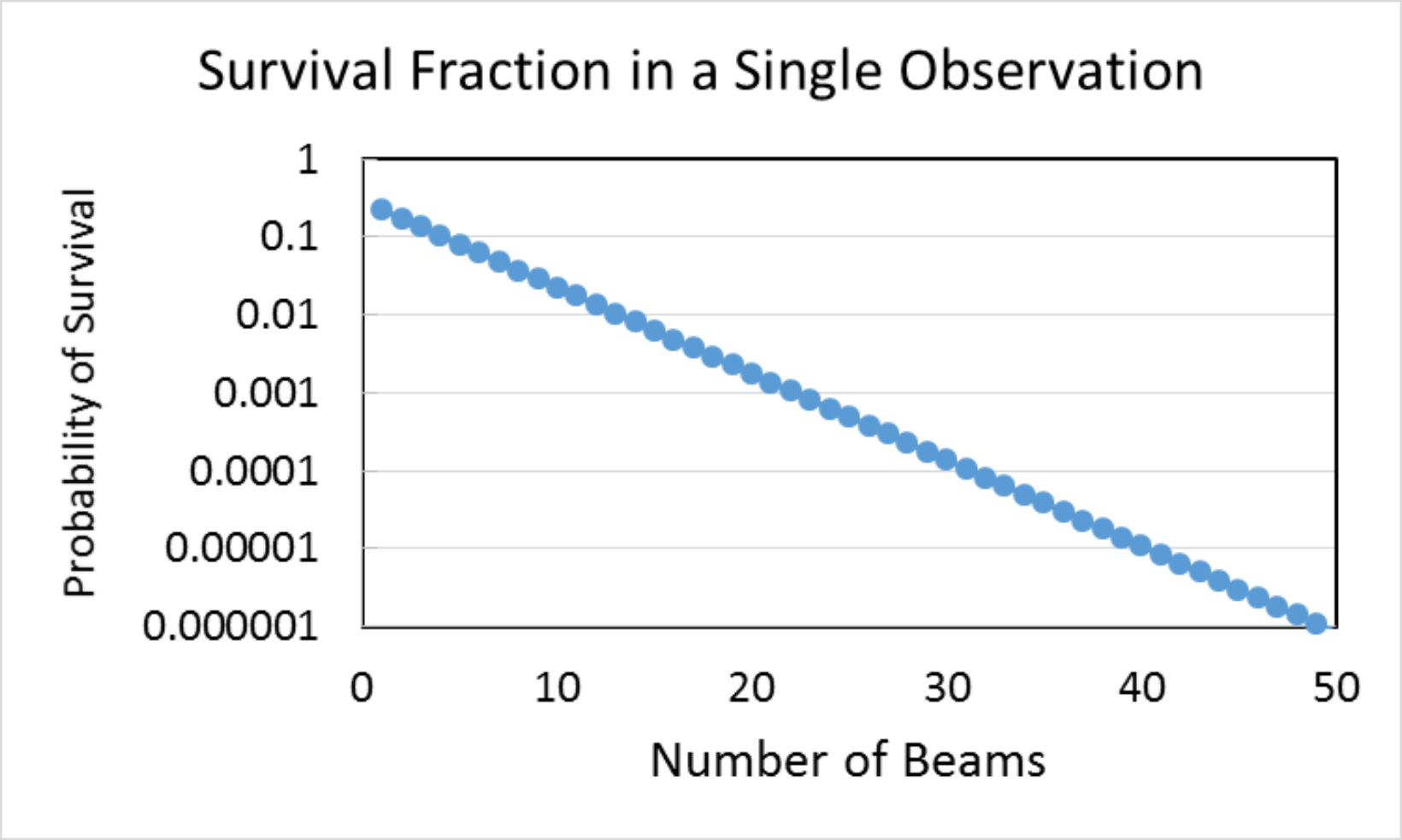

Figure 8: Simulation of the survival probability of an RFI signal versus number of beams in a direction of origin test.

This discussion relates to one of the weaknesses of the current SETI campaign and indeed all other SETI campaigns undertaken to date: Short duration transient signals will always be rejected by the classification system even if they are coming from a fixed direction on the celestial sphere, because they do not survive for the multiple follow-up observations required to have high confidence of their direction of origin. But if we could reduce the survival rate in a single observation sufficiently, then it would be possible to state with high confidence that a signal is arriving from the direction where it is first seen with only a single observation.

For example, an image with 1250 non-overlapping beams can be generated in a single ATA observation as in Figure 5. Images like these maximize the sampled solid angle of the sky in a single observation. If narrowband images could be generated at the frequencies where candidate signals are found, it would be possible to make a reliable decision about the signal's direction of origin with only a single observation. This method has been used successfully in (Harp, G. R. et al. 2015). Although a single image is not sufficient to prove extraterrestrial origin, it is enough to suggest that a transient signal is worthy of substantial follow-up to determine if the signal eventually repeats (thus lending it the opportunity to be confirmed as having an ET origin).

THIS METHOD COULD LEAD TO A FALSE POSITIVE FOR AN ORBITING SATELLITE THAT HAPPENS TO APPEAR IN THE TELESCOPE FOV. HOWEVER, BY AVOIDING FREQUENCIES WITH KNOWN INTERFERENCE (FIGURE 3) AND ALSO AVOIDING POINTING TOWARD THE FIXED PROJECTION OF GEOSTATIONARY ORBIT AS VIEWED AT THE OBSERVATORY, EXPERIENCE INDICATES THAT INSTANCES OF SATELLITES APPEARING IN THE FOV ARE VERY RARE.5.2 PROBABILITY LIMIT ON THE EXISTENCE OF EXTRATERRESTRIAL TRANSMITTERS

Here we pursue a simple approach to understanding the meaning of this research for our knowledge of extraterrestrial intelligence in the galaxy. As usual, this paper uses technology as a stand-in for intelligence, since the latter cannot be directly observed. Our model assumes the following:

1. All transmission frequencies between 1 - 9 GHz are equally probable.
2. All pointing directions centered on stars are equally likely to present a detectable signal.[5]
3. Prior to 1995 there was no significant prior knowledge affecting our best estimates of the probability for any pointing direction to harbor a detectable signal.

We pursue a Bayesian analysis to put a lower limit on the posterior probability (or belief) that in observations over the complete terrestrial microwave window any random pointing will result in a detectable signal from an extraterrestrial transmitter.

We use two datasets to constrain or model, the previously published targeted survey by the SETI Institute called Phoenix (1995 - 2004), and the data from the current work. The combined observations from Project Phoenix are summarized in Table 5. One could compare these results of approximately 1.2 $10^6$ star·MHz with the present work which covered 7.6 $10^6$ star·MHz (not counting re-observations of the same pointing and frequency). Because all pointings and frequencies are assumed to be equally likely for discovery of ET, we divide the observing coverage in each case by the full frequency range of the terrestrial microwave window (1-10 GHz). This allows us to state that in Phoenix and the present work, the equivalent number of stars observed over the full terrestrial window is $N_{obs} =$ 133 and 845 stars respectively, all of which gave a null result. We discuss the limitations of this model below.

[5] This assumption is problematic if transmitters are linked to stars, and those stars are not close by. This will be considered later.

Table 5

Summary of Phoenix Observations (1995 – 2004) compared with present work (2009-2015).

| Year | Observatories | Frequency Range | Number of Targets | $\text{Star}\cdot\text{MHz}$ | Threshold Signal Level ($10^{-26}$ W $m^{-2}$) |
|---|---|---|---|---|---|
| 1995[a] | ATNF, Parkes, MOPRA | 1200-1750 | 206 | 113,300 | 100 |
| 1995[a] | Parkes, ATNF, MOPRA | 1750-3000 | 105 | 131,000 | 100 |
| 1996-1998[b] | NRAO 140', Woodbury | 1200-3000 | 195 | 351,000 | 100 |
| 1998-2004[c] | Arecibo, Lovell | 1200-1750 | 290 | 160,000 | 16 |
| 1998-2004[c] | Arecibo, Lovell | 1750-3000 | 371 | 464,000 | 16 |
| Phoenix Total | | | | 1,200,000 | |
| This Work | ATA | 1000-9000 | Varies | 7,600,000 | 180-310 |

[a] Backus 1996, Tarter 1996, Backus 1997, Backus 1998.

[b] Cullers 2000.

[c] Backus, Project Phoenix Team 2001.

To pursue Bayesian inference it is necessary to specify a prior likelihood distribution for the desired quantity. In this case, we desire to constrain the probability $p_p$ that any future observation will result in a signal passing the criteria of this work. By assumption 3 and prior to Phoenix, any value of $p_p$ from 0 to 1 is equally likely, hence the prior likelihood distribution $\pi(p_p)=1$ (uniform distribution).

Our research question can be stated as follows. What is the posterior likelihood $\pi(p_p\,|\,\text{obs})$ for a given value of $p_p$ in light of our observations? We set up Bayes relation

$$\pi(p_p\,|\,\text{obs})=\frac{\pi(\text{obs}\,|\,p_p)\,\pi(p_p)}{\int_{p_p}\pi(\text{obs}\,|\,p_p)\,\pi(p_p)dp_p}\ . \tag{2}$$

But the likelihood of $N_{obs}$ null observations in a row is simply

$$\pi(\text{obs} \mid p_p) = \left(1 - p_p\right)^{N_{obs}} \propto \pi(p_p \mid \text{obs}) \quad (3)$$

where the final proportionality results after substitution for $\pi(p_p)$.

The results of this analysis are summarized in Figure 8a for two values of $N_{obs}$ corresponding to the posterior likelihood for the Phoenix campaign ($N_{obs}$ = 133) and for the combination of the present work with the Phoenix campaign ($N_{obs}$ = 978). Unsurprisingly, the most probable value for $p_p = 0$, but this is not the main result of the analysis.

We extract more information from our Bayesian inference by integrating the posterior likelihood $\pi(obs \mid p_p)$ from zero to a given maximum value $p_{\max}$ and display (in Figure 8b) the probability that the true value of $p_p$ satisfies $0 \le p_p \le p_{\max}$.

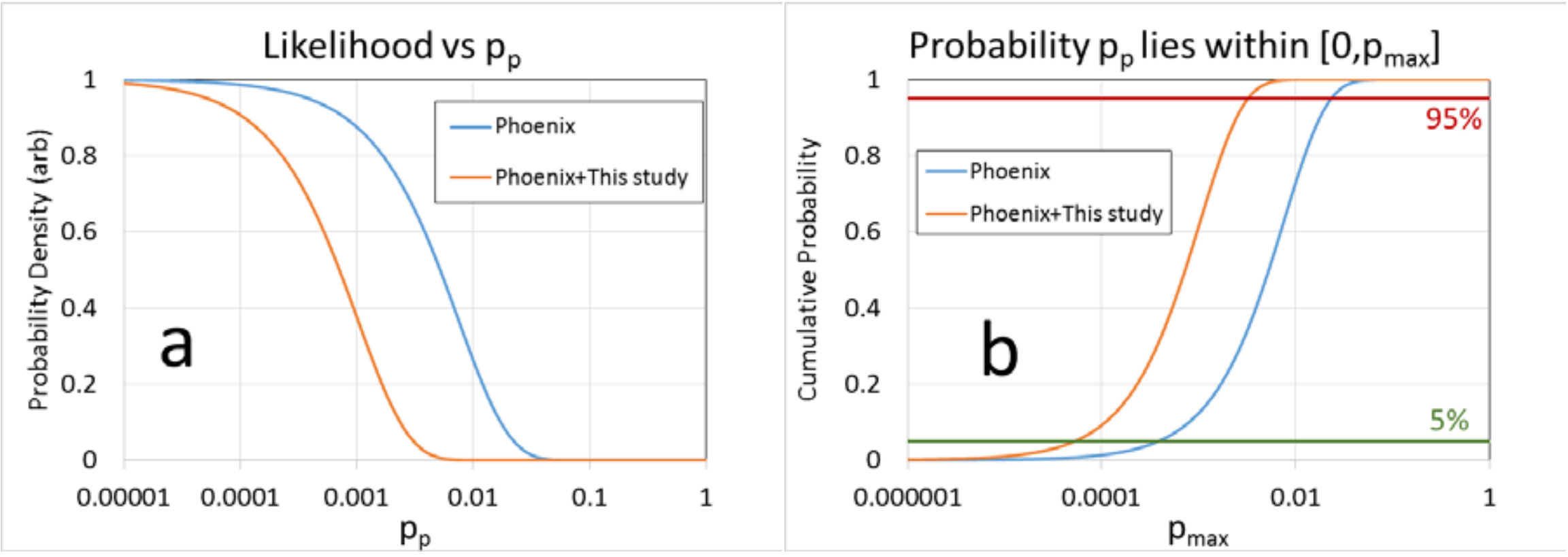

Figure 9: a) The posterior likelihood $\pi(p_p \mid \text{obs})$ that pointing in any direction will reveal an ETI transmitter after the Phoenix campaign, after the current campaign and b) The integrated probability from 0 to $p_{\max}$ of the likelihood functions $\pi(p_p \mid \text{obs})$ in a).

From Figure 8b we can compare the posterior knowledge of $p_p$ before and after the present work. Horizontal lines are drawn for cumulative probabilities at 5% and 95%. This gives a more illuminating picture of the results, showing that there is 90% probability $p_p$ is between these two lines, that is, the chances of finding a transmitting star are expected to be finite. We characterize these chances by observing the crossover point where there is a 50% probability $p_p$ is lower and 50% probability $p_p$ is higher which is $p_{p,50\%} = 3.5 \times 10^{-2}$ after completion of the Phoenix project and $p_{p,50\%} = 1.5 \times 10^{-3}$ after this work. The way we interpret these probabilities is to say that it is not reasonable to rule out, based on data alone, the chance that 1 in 29 stars are transmitting after Phoenix and the updated chance of 1 in 680 after this work.

It may surprise the reader that the probability for a transmitting civilization on a given pointing is so large. This is because we have used the entire terrestrial microwave window (1-10 GHz) as the full range of frequencies that are available to ET. No other ETI search prior to this one has used 9000 MHz in the denominator for calculating $N_{obs}$. For example, in previous descriptions of the Phoenix work, only the frequency range 1200-3000 MHz was considered, for which the same number of 1.2 $10^6$ star-MHz would correspond to a "complete" set of observations on $N'_{obs} = 667$ resulting in $p_{p,50\%} = 1.7 \times 10^{-3}$ for the old frequency range. This highlights the importance of using the same figure of merit to compare across different surveys.

The 1 in 588 number can be compared with an estimate by the founder of observational SETI Frank Drake who speculates the chances of finding a transmitting civilization to be 1 in $10^7$ (David 2015) based on reasonable estimates using the Drake equation (Drake 1961b). Our results speak to the very large parameter space over which ET might transmit in the terrestrial microwave window. A survey with $N_{obs} \approx 1 \times 10^7$ pointings is required before we can meaningfully test Drake's estimate using the frequency range 1 - 10 GHz. A survey that seriously tests Drake's estimate will take decades of more searching.

Clearly, a new paradigm is necessary to break through to this large number of observations at some point in the future. A hint about how to perform this task can be found in our suggestion that ATA should be backed up by a SETI correlator supplying the maximum of 1250 simultaneous distinct beams and hundreds to thousands of times the throughput of our SonATA system.

We hasten to criticize this simple model on several accounts. Firstly, this model does not take into account the sensitivity of the observations or the distance to the stars, which are closely related. To account for this prior information, it would be necessary to specify an *a priori* distribution for transmitters versus their output power, for which little is known. Possibly a future analysis could consider these elements testing different scenarios but this is beyond the scope of this paper, which focuses on observational results.

Similarly, some of the target pointings were near the galactic plane where our beam might cover many stars, whereas other pointings were selected near the galactic North Pole where only one or a few stars may be covered. However, all targets had one star at the pointing center so they are comparable to this extent.

Another criticism is that we do not attempt to include the results from all previous observations performed by our colleagues elsewhere in the world. Furthermore, it is not true that we have no other prior knowledge of the distribution of $p_p$. Indeed a large sky survey of stars over a limited frequency range is ongoing at Arecibo (Korpela 2011; Werthimer et al. 2000) as well as other observatories worldwide. Again, incorporating all the results from all campaigns to date is a worthy goal for a future paper but beyond the scope of this one.

We justify the simple model and comparisons described here as they accomplish multiple goals including 1) giving insight to the increase in our knowledge of the probability of detection represented by the observations summarized here, 2) illustrating the kind of useful information that can be derived from these observations and 3) showing that current observations as of 2015 are far from adequately constraining our knowledge of the probability of transmitting civilizations to meaningful limits based on reasonable though speculative models of that probability.

# 6 CONCLUSIONS

We summarize and report on 19,000 hours of SETI observations made with the Allen Telescope array from 2009 - 2015. Many of these observations have focused on stars with exoplanets or Kepler objects of interest. Special focus was placed on stars with planets in or near the habitable zone of their star.

We described the almost fully automated, near-real time observing system called SonATA. With a frequency-dependent sensitivity between 180 - 310 $10^{-26}$ W $m^{-2}$, over the observed frequency range from 1-9 GHz. Comprising 9293 targets, this campaign covered $7.3\times10^{6}$ star·MHz of observation bandwidth.

A total of $2.0\times10^{8}$ candidate signal detections were made. Almost all of these detections were positively identified as terrestrial interference using some or another form of direction of origin classification (i.e. showing the signals did not originate from a single sky pointing). Our system uses multiple interferometer phased array beams, which is novel. Such multi-beam testing proves to be an effective method for eliminating terrestrial interference from our SETI searches. No ET signal candidate survived all of our stringent tests, hence we place a new constraint on the number of transmitting civilizations: there is a 50% posterior probability that less than 1 in 1500 pointings (or stars) would be detectable in this campaign.

Relating to the design of future systems, we conclude that a large number (of order 50 to decrease false positive rate to $10^{-6}$) simultaneous synthetic beams can enable searches that are sensitive to transient signals (lasting no more than a few minutes). Furthermore, an interferometer similar to the ATA could maximize the number of effective beams (hence observation efficiency) by employing an imaging correlator, and this approach is recommended for future SETI surveys of large solid angles on the sky.

## ACKNOWLEDGEMENTS

The authors would like to acknowledge the generous support of the Paul G. Allen Family Foundation, who has provided major support for design, construction, and operations of the ATA. Contributions from Nathan Myhrvold, Xilinx Corporation, Sun Microsystems, and many private donors have been instrumental in supporting the ATA. The ATA has been supported by the US Naval Observatory, in addition to National Science Foundation grants AST-050690, AST-0838268, and AST-0909245. Since 2011, the ATA has been operated and maintained by SRI International. Sun Microsystems and Xilinx Corporation contributed hardware for the interim Prelude system, and Dell Inc., Intel Corp., and Google donated the powerful servers and switches that enabled a transition to SonATA. We gratefully thank Franklin Antonio for funding the development and installation of new wideband feed/receiver systems as well as support of the survey of habitable zone planets, the compilation of all results and their publication. We further acknowledge Dave Messerschmitt for discussions on broad band signal dispersion and detection.

Finally, we gratefully acknowledge the comments of an anonymous reviewer which were very helpful in preparation of this manuscript.

## REFERENCES

Arecibo Planetary Habitability Laboratory at University of Puerto Rico. 2015, http://phl.upr.edu/projects/habitable-exoplanets-catalog/about.

Backus, 1996, BAAS, 27, 1336.

Backus et al. 1997, in Cosmovici, Batalli, Bowyer, and Werthimer, Astronomical and biochemical origins and the search for life in the universe," IAU Colloquia. 161, 661.

Backus 1998, Acta Astronaut., 42, 651.

Backus, Project Phoenix Team 2001, BAAS 33, 900.

Backus & Project Phoenix Team. 2004, Am. Astron. Soc. Meet., 36, 805.

Barott, Milgrome, Wright, et al. 2011, Radio Sci., 46, RS1016.

Borucki, Koch, Basri, et al. 2010, Science, 327, 977–980.

Borucki, Koch, Basri, et al. Ap. J. 2011, 19, 106.

Cocconi & Morrison. 1959, Nature, 184, 844–846.

Cullers, Linscott & Oliver. 1985, Commun. ACM, 28, 1151–1163.

Cullers 2000, Proc. ASP, 213, 451.

David. 2015, http://www.space.com/28665-seti-astronomer-frank-drake-interview.html.

Drake & Helou. 1977, NAIC Rep., 76.

Drake. 1961a, Phys. Today, April 1961, 40.

Drake 1961b in Natl. Acad. Sci. Conf. Extraterr. Life., Greenbank, West Virginia.

Ekers, Cullers, Billingham & Scheffer. SETI Press, 2002.

Grimm, Gulkis, Stevens & Tarter. 1987, Mon. Not. Roy. Astron. Soc., 225, 491–498.

Han, Sep, Wright & Feng. 2014, arXiv, 1409.7709, 1–20.

Harp, Ackermann, Nadler, et al. 2011, IEEE Antennas Propagation, 59, 2004–2021.

Harp, Richards, Shostak, et al. 2015, *Astrophys. J.*, 825, 155.

Høg, Fabricius, Makarov, et al. 2000, Astron. Astrophys., 355, L27.

Korpela, Anderson, Bankay, et al. 2011,Proc. SPIE 8152, 815212.

Mayor & Queloz. 1995, Nature, 378, 355–359.

Oliver & Billingham. 1971 *NASA/ASEE Summer Fac. Fellowsh. Progr. (NASA-CR-114445)*, 1.

Penny. 2011, Astron. Geophys., 52, 1.21–1.24.

Siemion, Benford, Cheng-Jin, et al. 2014, arXiv:1412.4867, 1-15.

Tarter 1996, SPIE, 2704, 24.

Tarter, Ackermann, Barott, et al. 2011, Acta Astronaut., 68, 340–346.

Turnbull & Tarter. 2003a, Ap. J. Supp., 145, 181–198.

Turnbull & Tarter. 2003b, App. J. Suppl., 149, 423–436.

Welch, Backer, Blitz, et al. 2009, Proc. IEEE, 97, 1438 – 1447.

Werthimer, Bowyer, Cobb, Lebofsky & Lampton. 2000, ASP Conf. Ser. Bioastronomy 99 A New Era Search Life, 213.

Table 6

List of all targets having exoplanets near their Habitable Zone and frequencies of observation.

| RA +Dec (Decimal Hr/Deg) | **Alias** |
|---|---|
| HZ | |
| 01.73 -15.94 | TauCeti |
| 04.90 -16.23 | GJ180b,c |
| 11.49 -01.45 | EPIC 201367065 d |
| 11.50 +07.59 | EPIC 201912552 b |
| 18.78 +41.95 | Kepler 438 b |
| 18.87 +48.83 | KOI1430.03 |
| 18.88 +45.35 | KOI701.03 |
| 18.88 +45.35 | Kepler-62 e, f |
| 18.92 +39.90 | KOI817.01 |
| 18.95 +48.81 | KOI1423.01 |
| 18.96 +49.31 | KOI351.01 |
| 19.02 +38.95 | KOI806.01 |
| 19.02 +39.28 | Kepler 442 b |
| 19.02 +41.45 | Kepler 440 b |
| 19.06 +38.38 | KOI401.01 |
| 19.07 +39.28 | KOI812.03 |
| 19.07 +48.43 | KOI536.01 |
| 19.09 +37.43 | KOI1026.01 |
| 19.10 +49.44 | KOI1422.01 |
| 19.11 +46.78 | KOI326.01 |
| 19.12 +41.99 | KOI416.01 |
| 19.14 +41.57 | KOI847.01 |
| 19.14 +44.88 | KOI1261.01 |
| 19.15 +51.25 | KOI1503.01 |
| 19.16 +43.83 | KOI518.01 |
| 19.18 +42.34 | KOI70.03 |
| 19.19 +43.90 | KOI902.01 |
| 19.20 +47.97 | KOI211.01 |
| 19.22 +42.26 | KOI1375.01 |
| 19.23 +51.08 | KOI438.02 |
| 19.24 +49.97 | Kepler-443 |
| 19.26 +51.21 | KOI1478.01 |
| 19.28 +47.88 | KOI87.01, Kepler-22 b |
| 19.30 +41.81 | KOI854.01 |
| 19.35 +48.36 | KOI2770.01 |
| 19.38 +44.87 | KOI374.01 |
| 19.41 +40.36 | KOI1564.01 |
| 19.44 +38.04 | KOI1099.01 |
| 19.44 +42.37 | KOI865.01 |
| 19.46 +46.43 | KOI947.01 |
| 19.49 +48.51 | KOI1429.01 |
| 19.54 +40.93 | KOI555.01 |
| 19.55 +41.61 | KOI415.01 |
| 19.57 +47.84 | KOI1298.01 |
| 19.60 +45.14 | KOI465.01 |
| 19.62 +43.63 | KOI1486 |
| 19.64 +43.40 | KOI1472.01 |
| 19.66 +42.71 | KOI1355.01 |
| 19.67 +39.27 | KOI2531.01 |
| 19.69 +42.48 | KOI1361 |
| 19.69 +46.27 | KOI711.01 |
| 19.69 +42.48 | Kepler-61 b |
| 19.72 +39.18 | KOI1328.01 |
| 19.73 +41.33 | KOI51.01 |
| 19.73 +51.26 | KOI622.01 |
| 19.78 +43.50 | KOI1527.01 |
| 19.79 +48.11 | KOI174.01 |
| 19.80 +40.87 | KOI448.02 |
| 19.80 +47.49 | KOI1159.01 |
| 19.81 +41.91 | KOI157.05 |
| 19.83 +46.96 | KOI1596.02 |
| 19.85 +43.26 | KOI683.01 |
| 19.86 +46.97 | KOI2493.01 |
| 19.91 +43.95 | KOI571.01 |
| 19.94 +41.87 | KOI372.01 |

Table 7

List of all targets where at least 100 MHz bandwidth was observed, not including HZ targets of Table 6.

| RA +Dec (Decimal Hr/Deg) | Alias | Percentage 1-9 GHz | BW (MHz) |
|---|---|---|---|
| Exoplanet/KOI | | | |
| 19.12 +49.32 | KOI1.01 | 100 | 9000 |
| 19.08 +50.04 | KOI20.01 | 100 | 9000 |
| 19.92 +44.00 | KOI317.01 | 100 | 9000 |
| 19.04 +50.14 | KOI7.01 | 100 | 9000 |
| 19.12 +49.06 | KOI241.01 | 100 | 9000 |
| 09.94 -24.10 | HIP48739 | 100 | 9000 |
| 19.48 +47.97 | KOI2.01 | 100 | 9000 |
| 10.98 -31.14 | GJ3634b | 100 | 9000 |
| 19.58 +45.11 | KOI464.01 | 100 | 9000 |
| 08.61 -30.04 | HIP42214 | 100 | 9000 |

Table 7

List of all targets where at least 100 MHz bandwidth was observed, not including HabZone targets of Table 6.

| RA +Dec (Decimal Hr/Deg) | Alias | Percentage 1-9 GHz | BW (MHz) |
|---|---|---|---|
| Exoplanet/KOI | | | |
| 19.12 +49.32 | KOI1.01 | 100 | 9000 |
| 19.08 +50.04 | KOI20.01 | 100 | 9000 |
| 19.92 +44.00 | KOI317.01 | 100 | 9000 |
| 19.04 +50.14 | KOI7.01 | 100 | 9000 |
| 19.12 +49.06 | KOI241.01 | 100 | 9000 |
| 09.94 -24.10 | HIP48739 | 100 | 9000 |
| 19.48 +47.97 | KOI2.01 | 100 | 9000 |
| 10.98 -31.14 | GJ3634b | 100 | 9000 |
| 19.58 +45.11 | KOI464.01 | 100 | 9000 |
| 08.61 -30.04 | HIP42214 | 100 | 9000 |
| 09.58 -12.13 | HIP47007 | 100 | 9000 |
| 19.04 +38.40 | KOI377.01 | 100 | 9000 |
| 18.88 +45.14 | KOI42.01 | 100 | 9000 |
| 19.85 +48.08 | KOI3.01 | 100 | 9000 |
| 07.19 +30.25 | tauGem | 100 | 9000 |
| 19.68 +49.56 | KOI353.01 | 100 | 9000 |
| 19.63 +49.77 | KOI167.01 | 100 | 9000 |
| 18.68 +43.84 | KOI2594.01 | 100 | 9000 |
| 09.34 +33.88 | WASP-13 | 100 | 9000 |
| 06.23 -29.90 | HIP29550 | 100 | 9000 |
| 04.36 +57.82 | TYC3727-1064-1 | 100 | 9000 |
| 18.71 +47.81 | KOI475.01 | 100 | 9000 |
| 10.98 +01.73 | HD95089 | 100 | 9000 |
| 18.71 +47.75 | KOI533.01 | 100 | 9000 |
| 08.06 -01.16 | HIP39417 | 100 | 9000 |
| 19.73 +44.58 | KOI138.01 | 100 | 9000 |
| 10.39 -00.90 | 24Sex | 100 | 9000 |
| 15.28 +71.82 | HIP74793 | 100 | 9000 |
| 09.25 +23.38 | HD79498 | 100 | 9000 |
| 18.69 +43.99 | KOI2411.01 | 100 | 9000 |
| 12.09 +76.91 | HIP58952 | 100 | 9000 |
| 19.62 +38.95 | KOI4.01 | 100 | 9000 |
| 19.12 +49.26 | KOI349.01 | 100 | 9000 |
| 19.25 +48.04 | KOI209.01 | 100 | 9000 |
| 19.05 +50.24 | KOI72.01 | 100 | 9000 |
| 18.75 +42.45 | KOI10.01 | 100 | 9000 |
| 19.70 +42.54 | KOI205.01 | 100 | 9000 |
| 11.58 +20.44 | HD100655 | 100 | 9000 |
| 19.43 +42.73 | KOI75.01 | 100 | 9000 |
| 19.06 +39.10 | KOI221.01 | 100 | 9000 |
| 19.80 +48.21 | KOI117.01 | 100 | 9000 |
| 06.41 -28.78 | HIP30503 | 100 | 9000 |
| 19.49 +47.88 | KOI749.01 | 100 | 9000 |
| 05.62 +20.73 | HIP26381 | 100 | 9000 |
| 19.17 +42.17 | KOI111.01 | 100 | 9000 |
| 10.17 +18.19 | HIP49813 | 100 | 9000 |
| 19.41 +51.33 | KOI602.01 | 100 | 9000 |
| 19.19 +47.63 | KOI474.01 | 100 | 8980 |
| 18.71 +43.92 | KOI2486.01 | 99 | 8919 |
| 19.46 +47.86 | KOI597.01 | 98 | 8849 |
| 10.31 +12.62 | HIP50473 | 97 | 8707 |
| 08.88 +28.33 | HIP43587 | 96 | 8631 |
| 08.31 +61.46 | HIP40687 | 95 | 8570 |
| 19.62 +50.08 | KOI777.01 | 95 | 8508 |
| 11.13 -30.17 | HIP54400 | 93 | 8398 |
| 19.83 +41.01 | KOI12.01 | 93 | 8338 |
| 19.61 +49.48 | KOI168.01 | 91 | 8201 |
| 19.28 +47.64 | KOI743.01 | 91 | 8150 |
| 19.72 +51.12 | KOI787.01 | 89 | 8054 |
| 19.21 +47.97 | KOI598.01 | 87 | 7825 |
| 11.70 +26.71 | HIP57087 | 86 | 7747 |
| 19.74 +42.14 | KOI563.01 | 86 | 7725 |
| 19.02 +46.67 | KOI135.01 | 83 | 7481 |
| 19.27 +48.02 | KOI1302.01 | 82 | 7419 |
| 04.70 +23.03 | 2MJ044144 | 82 | 7405 |
| 19.53 +48.60 | KOI124.01 | 82 | 7341 |
| 19.82 +41.89 | KOI94.01 | 81 | 7315 |
| 19.64 +43.85 | KOI520.01 | 81 | 7276 |
| 19.77 +50.68 | KOI1458.01 | 81 | 7260 |
| 06.72 -01.30 | CoRoT-12 | 80 | 7240 |
| 19.84 +43.25 | KOI1495.01 | 80 | 7232 |
| 19.74 +39.98 | KOI500.01 | 80 | 7202 |
| 19.20 +50.94 | KOI153.01 | 79 | 7128 |
| 08.26 +05.84 | HAT-P-30/WASP-51 | 79 | 7120 |
| 19.20 +42.36 | KOI163.01 | 79 | 7117 |
| 11.30 -19.05 | WASP-31 | 79 | 7103 |
| 19.67 +49.36 | KOI2356.01 | 79 | 7085 |
| 19.65 +43.06 | KOI886.01 | 79 | 7080 |
| 11.99 -20.35 | HIP58451 | 79 | 7080 |
| 11.03 -23.86 | WASP-34 | 79 | 7075 |
| 08.22 +04.79 | HAT-P-35 | 79 | 7072 |
| 07.76 +28.03 | HIP37826 | 79 | 7071 |
| 07.25 +14.26 | HAT-P-24 | 78 | 6993 |
| 18.99 +49.27 | KOI189.01 | 76 | 6840 |
| 18.88 +45.20 | KOI2074.01 | 76 | 6822 |
| 19.36 +40.28 | KOI123.01 | 76 | 6820 |
| 19.00 +46.57 | KOI250.01 | 76 | 6814 |
| 19.79 +48.24 | KOI17.01 | 76 | 6812 |
| 19.16 +43.38 | KOI269.01 | 75 | 6792 |
| 19.50 +46.20 | KOI165.01 | 75 | 6782 |
| 09.38 +50.60 | HIP45982 | 75 | 6780 |
| 05.38 +79.23 | HIP25110 | 75 | 6759 |
| 19.43 +37.91 | KOI794.01 | 74 | 6690 |
| 19.11 +49.41 | KOI2296.01 | 74 | 6670 |
| 19.22 +43.70 | KOI192.01 | 74 | 6650 |
| 19.21 +38.66 | KOI2430.01 | 73 | 6614 |
| 18.77 +42.45 | KOI2087.01 | 73 | 6536 |
| 19.49 +37.63 | KOI151.01 | 73 | 6532 |
| 18.96 +49.11 | KOI2632.01 | 72 | 6523 |
| 19.25 +41.15 | KOI85.01 | 72 | 6509 |
| 10.33 +19.84 | HD89484 | 72 | 6439 |
| 08.67 +64.33 | HIP42527 | 71 | 6427 |
| 19.87 +44.75 | KOI137.01 | 71 | 6409 |
| 05.67 +06.06 | HIP26664 | 71 | 6362 |
| 07.46 +24.34 | HAT-P-20 | 71 | 6346 |
| 18.74 +42.45 | KOI2114.01 | 70 | 6317 |
| 19.55 +44.87 | KOI172.01 | 70 | 6259 |
| 19.10 +49.42 | KOI483.01 | 69 | 6235 |
| 19.58 +41.90 | KOI270.01 | 69 | 6192 |
| 18.86 +48.34 | KOI194.01 | 69 | 6187 |
| 22.87 +35.45 | TYC2757-1152-1 | 68 | 6151 |
| 23.66 +77.63 | HIP116727 | 68 | 6137 |
| 16.17 +43.82 | HIP79248 | 68 | 6136 |
| 07.36 +58.27 | TYC3793-1994-1 | 68 | 6127 |
| 19.69 +41.22 | KOI191.01 | 68 | 6092 |
| 05.13 -13.99 | HD33142 | 68 | 6087 |
| 19.66 +48.98 | KOI107.01 | 67 | 6062 |
| 19.70 +51.01 | KOI279.01 | 67 | 6059 |
| 06.81 -00.67 | GSC04800-02187 | 67 | 6018 |
| 19.86 +48.24 | KOI341.01 | 67 | 6014 |
| 19.82 +43.72 | KOI459.01 | 66 | 5974 |
| 09.95 -15.90 | HIP48780 | 66 | 5933 |
| 20.03 +44.38 | KOI152.01 | 65 | 5889 |
| 18.97 +48.96 | KOI258.01 | 65 | 5884 |
| 18.88 +48.58 | KOI1432.01 | 65 | 5869 |
| 18.88 +48.35 | KOI46.01 | 65 | 5821 |
| 18.97 +46.06 | KOI110.01 | 65 | 5812 |
| 03.52 -23.82 | WASP-22 | 64 | 5792 |
| 19.94 +40.95 | KOI148.01 | 64 | 5748 |
| 11.74 +30.96 | HIP57274 | 63 | 5682 |
| 19.61 +48.35 | KOI156.01 | 63 | 5678 |
| 03.16 +30.67 | WASP-11/HAT-P-10 | 63 | 5677 |
| 19.48 +46.16 | KOI49.01 | 62 | 5603 |
| 19.16 +48.67 | KOI292.01 | 62 | 5583 |
| 18.77 +47.21 | KOI82.01 | 62 | 5579 |
| 18.75 +47.50 | KOI104.01 | 62 | 5558 |
| 08.65 +12.96 | HIP42446 | 62 | 5555 |
| 19.80 +50.41 | KOI265.01 | 61 | 5478 |
| 07.57 -22.30 | HIP36795 | 60 | 5423 |
| 19.89 +47.28 | KOI730.01 | 60 | 5410 |
| 19.63 +44.17 | KOI1070 | 60 | 5385 |
| 23.17 +57.03 | HD240210 | 60 | 5381 |
| 19.18 +47.33 | KOI98.01 | 60 | 5376 |
| 19.63 +45.98 | KOI196.01 | 60 | 5367 |
| 19.17 +51.38 | KOI2424.01 | 59 | 5328 |
| 05.29 +07.35 | HD34445 | 59 | 5300 |
| 00.63 +34.71 | HAT-P-19 | 59 | 5290 |
| 19.80 +43.21 | KOI150.01 | 59 | 5272 |
| 06.91 +24.25 | HIP33212 | 59 | 5272 |
| 19.48 +48.51 | KOI2248 | 58 | 5221 |
| 01.75 +20.08 | HIP8159 | 58 | 5206 |
| 19.52 +46.39 | KOI183.01 | 57 | 5168 |
| 19.79 +47.53 | KOI-428 | 57 | 5129 |
| 01.37 +76.71 | HIP6379 | 57 | 5114 |
| 19.99 +43.81 | KOI89.01 | 57 | 5097 |
| 19.91 +48.61 | KOI346.01 | 57 | 5093 |
| 19.45 +37.75 | KOI103.01 | 57 | 5086 |
| 18.43 +65.56 | HIP90344 | 56 | 5008 |
| 19.65 +43.54 | KOI171.01 | 56 | 5006 |
| 20.06 +44.34 | KOI116.01 | 55 | 4979 |
| 19.14 +46.90 | KOI528.01 | 55 | 4914 |
| 19.22 +42.41 | KOI513.01 | 54 | 4901 |
| 18.75 +42.40 | KOI2578.01 | 54 | 4901 |
| 18.96 +49.30 | KOI351.01 | 54 | 4891 |
| 19.76 +51.36 | KOI439.01 | 54 | 4845 |
| 11.35 -23.22 | HIP55409 | 54 | 4841 |
| 06.43 -31.48 | HIP30579 | 54 | 4828 |
| 19.48 +41.73 | KOI662.01 | 53 | 4811 |
| 03.19 +21.10 | HIP14810 | 53 | 4805 |
| 19.24 +49.74 | KOI1532.01 | 53 | 4801 |
| 19.92 +44.12 | KOI2573.01 | 53 | 4787 |
| 19.34 +39.82 | KOI1534.01 | 53 | 4787 |
| 13.00 +12.68 | PSRB1257+12 | 53 | 4779 |
| 20.10 +44.61 | KOI2123.01 | 53 | 4778 |
| 19.84 +47.40 | KOI332.01 | 53 | 4766 |
| 19.60 +39.05 | KOI393.01 | 53 | 4766 |
| 08.66 +47.35 | TYC3416-543-1 | 53 | 4742 |
| 19.22 +40.52 | KOI408.01 | 52 | 4721 |
| 13.74 +48.03 | TYC3466-819-1 | 52 | 4714 |
| 06.60 -27.62 | HIP31540 | 52 | 4711 |
| 08.90 +33.06 | HIP43674 | 52 | 4710 |
| 05.78 +01.17 | HIP27253 | 52 | 4699 |
| 19.04 +39.38 | KOI1337.01 | 52 | 4697 |
| 19.13 +46.87 | KOI13.01 | 52 | 4691 |
| 19.58 +47.91 | KOI2179 | 52 | 4689 |
| 19.32 +44.65 | KOI5.01 | 52 | 4689 |
| 08.38 +01.86 | HD70573 | 52 | 4687 |
| 19.43 +41.99 | KOI41.01 | 52 | 4678 |
| 19.48 +48.73 | KOI760.01 | 52 | 4659 |
| 04.84 -24.37 | HD30856 | 52 | 4658 |
| 19.96 +44.03 | KOI18.01 | 52 | 4653 |
| 19.44 +41.88 | KOI176.01 | 52 | 4650 |
| 19.85 +46.94 | KOI1591.01 | 51 | 4629 |
| 19.20 +50.65 | KOI141.01 | 51 | 4629 |
| 03.00 -20.80 | HD18742 | 51 | 4610 |
| 19.86 +46.88 | KOI1875.01 | 51 | 4603 |
| 19.34 +42.17 | KOI665.01 | 51 | 4600 |
| 11.30 -23.98 | HD98219 | 51 | 4597 |
| 09.40 +20.36 | HIP46076 | 51 | 4597 |
| 19.61 +38.71 | KOI386.01 | 51 | 4551 |
| 07.62 -13.91 | TYC5409-2156-1 | 50 | 4534 |
| 19.21 +50.03 | KOI262.01 | 50 | 4511 |
| 11.24 +25.71 | HD97658 | 50 | 4511 |
| 18.88 +41.34 | KOI284.01 | 50 | 4509 |
| 08.68 -23.46 | LHS2037 | 50 | 4494 |
| 19.56 +39.25 | KOI550.01 | 50 | 4483 |
| 06.51 +29.67 | 2MASSJ06303279+2940202 | 50 | 4467 |
| 20.11 +44.41 | KOI2529.01 | 49 | 4443 |
| 18.73 +44.09 | KOI1985.01 | 49 | 4439 |
| 19.19 +40.64 | KOI834.01 | 49 | 4436 |
| 10.40 -29.65 | HIP50921 | 49 | 4416 |
| 19.91 +48.58 | KOI214.01 | 49 | 4405 |
| 19.83 +47.16 | KOI1083.01 | 49 | 4381 |
| 18.83 +48.76 | KOI2333 | 49 | 4378 |
| 10.99 +40.43 | HIP53721 | 49 | 4370 |
| 04.52 +04.58 | HD28678 | 49 | 4366 |
| 19.54 +48.88 | KOI481.01 | 48 | 4360 |
| 19.69 +38.88 | KOI208.01 | 48 | 4352 |
| 19.32 +46.86 | KOI398.01 | 48 | 4347 |
| 07.40 +20.42 | TYC1355-214-1 | 48 | 4344 |
| 19.49 +40.59 | KOI248.01 | 48 | 4342 |
| 07.80 +50.23 | TYC3413-5-1 | 48 | 4338 |
| 19.73 +42.66 | KOI220.01 | 48 | 4330 |
| 19.82 +46.84 | KOI722.01 | 48 | 4325 |
| 18.74 +47.19 | KOI1085.01 | 48 | 4320 |
| 19.16 +43.74 | KOI1968.01 | 48 | 4320 |
| 19.19 +46.28 | KOI115.01 | 48 | 4298 |
| 19.27 +46.01 | KOI707.01 | 48 | 4288 |
| 12.22 +10.04 | HIP59610 | 48 | 4287 |
| 18.75 +47.77 | KOI263.01 | 48 | 4277 |
| 19.04 +49.96 | KOI253.01 | 47 | 4257 |
| 19.21 +44.07 | KOI318.01 | 47 | 4226 |
| 04.48 +19.18 | HIP20889 | 47 | 4221 |
| 19.47 +37.38 | KOI1001.01 | 47 | 4192 |
| 03.29 +31.13 | HIP15323 | 47 | 4190 |
| 19.71 +48.50 | KOI112.01 | 46 | 4174 |
| 19.28 +49.55 | KOI63.01 | 46 | 4171 |
| 18.89 +45.43 | KOI2257.01 | 46 | 4158 |
| 19.54 +47.73 | KOI1301.01 | 46 | 4155 |
| 19.91 +44.17 | KOI908.01 | 46 | 4139 |
| 19.75 +49.14 | KOI128.01 | 46 | 4124 |
| 22.05 +18.88 | HIP108859 | 46 | 4122 |
| 19.34 +48.34 | KOI2623.01 | 46 | 4114 |
| 18.84 +46.32 | KOI22.01 | 46 | 4112 |
| 12.54 +74.49 | HD109246 | 45 | 4093 |
| 19.19 +39.34 | KOI222.01 | 45 | 4088 |
| 19.91 +47.76 | KOI534.01 | 45 | 4080 |
| 19.08 +50.24 | KOI297.01 | 45 | 4070 |
| 19.83 +43.88 | KOI1560.01 | 45 | 4063 |
| 19.85 +47.86 | KOI429.01 | 45 | 4042 |
| 19.03 +38.93 | KOI443.01 | 45 | 4039 |
| 02.77 +49.65 | HD17092 | 45 | 4037 |
| 19.17 +51.06 | KOI2059.01 | 45 | 4031 |
| 13.47 +13.78 | HIP65721 | 45 | 4021 |
| 08.71 +04.58 | HIP42723 | 45 | 4007 |
| 19.50 +46.29 | KOI468.01 | 45 | 4005 |
| 18.77 +42.39 | KOI1362.01 | 44 | 4003 |
| 19.82 +48.32 | KOI291.01 | 44 | 3994 |
| 19.08 +39.68 | KOI223.01 | 44 | 3994 |
| 18.81 +43.04 | KOI313.01 | 44 | 3993 |
| 04.87 +06.48 | Gl179 | 44 | 3990 |
| 18.75 +43.41 | KOI684.01 | 44 | 3974 |
| 18.82 +43.67 | KOI898.01 | 44 | 3967 |
| 01.56 +29.27 | HIP7245 | 44 | 3958 |
| 19.58 +38.94 | KOI1199.01 | 44 | 3944 |
| 20.10 +44.16 | KOI1731.01 | 44 | 3932 |
| 18.75 +44.32 | KOI573.01 | 44 | 3919 |
| 05.16 +69.64 | HIP24003 | 43 | 3903 |
| 10.15 +34.24 | HIP49699 | 43 | 3901 |
| 03.21 -01.20 | HIP14954 | 43 | 3898 |
| 06.47 -00.02 | CoRoT-19 | 43 | 3878 |
| 15.59 +53.92 | HIP76311 | 43 | 3862 |
| 19.95 +41.38 | KOI306.01 | 43 | 3857 |
| 02.57 -12.38 | HIP11952 | 43 | 3837 |
| 19.89 +47.81 | KOI203.01 | 42 | 3822 |
| 11.76 +02.82 | HIP57370 | 42 | 3814 |
| 03.68 +31.83 | HIP17187 | 42 | 3814 |
| 09.60 +34.78 | HD82886 | 42 | 3811 |
| 19.27 +41.56 | KOI285.01 | 42 | 3808 |
| 13.96 +43.49 | HAT-P-12 | 42 | 3794 |
| 04.42 +39.46 | HAT-P-15 | 42 | 3789 |
| 11.10 +44.30 | HD96127 | 42 | 3780 |
| 19.68 +50.56 | KOI623.01 | 42 | 3765 |
| 18.90 +45.23 | KOI524.01 | 42 | 3764 |
| 06.73 -01.06 | TYC4799-1733-1 | 42 | 3760 |
| 02.83 +71.75 | HIP13192 | 42 | 3745 |
| 02.62 +24.65 | HIP12184 | 42 | 3742 |
| 19.84 +40.98 | KOI206.01 | 41 | 3722 |
| 18.89 +48.55 | KOI344.01 | 41 | 3711 |
| 19.24 +50.91 | KOI1947.01 | 41 | 3706 |
| 19.01 +41.28 | KOI2250.01 | 41 | 3698 |
| 18.83 +48.26 | KOI431.01 | 41 | 3688 |
| 04.72 +18.96 | HIP21932 | 41 | 3662 |
| 19.24 +41.09 | KOI97.01 | 41 | 3657 |
| 07.53 +17.09 | HIP36616 | 40 | 3635 |
| 19.18 +47.55 | KOI531.01 | 40 | 3609 |
| 19.16 +38.65 | KOI118.01 | 40 | 3601 |
| 19.11 +39.49 | KOI244.01 | 40 | 3599 |
| 01.88 -19.51 | HIP8770 | 40 | 3592 |
| 08.31 -12.63 | HIP40693 | 40 | 3572 |
| 19.23 +40.25 | KOI282.01 | 40 | 3564 |
| 09.48 +45.60 | HIP46471 | 40 | 3560 |
| 10.28 +19.89 | TYC1422-790-1 | 39 | 3551 |
| 19.67 +48.48 | KOI343.01 | 39 | 3545 |
| 19.77 +42.55 | KOI64.01 | 39 | 3544 |
| 19.50 +41.64 | KOI2221.01 | 39 | 3526 |
| 19.28 +41.71 | KOI2021.01 | 39 | 3518 |
| 00.26 -11.94 | WASP-44 | 39 | 3507 |
| 24.00 -22.43 | HIP118319 | 39 | 3505 |
| 02.71 +49.59 | BD+48738 | 39 | 3497 |
| 11.86 +57.64 | HD102956 | 39 | 3478 |
| 19.13 +48.38 | KOI757.01 | 39 | 3473 |
| 19.78 +41.20 | KOI2321.01 | 39 | 3467 |
| 13.57 +53.73 | HIP66192 | 38 | 3464 |
| 20.01 +45.09 | KOI44.01 | 38 | 3464 |
| 03.23 +25.20 | HAT-P-25 | 38 | 3463 |
| 19.46 +43.07 | KOI567.01 | 38 | 3454 |
| 19.82 +49.62 | KOI2550.01 | 38 | 3450 |
| 19.75 +48.22 | KOI756.01 | 38 | 3437 |
| 19.89 +47.49 | KOI738.01 | 38 | 3435 |
| 19.79 +43.71 | KOI418.01 | 38 | 3433 |
| 04.91 +12.35 | HD31253 | 38 | 3421 |
| 10.39 +00.90 | 24Sexb | 38 | 3409 |
| 19.81 +49.50 | KOI1856.01 | 38 | 3407 |
| 19.02 +41.26 | KOI2525.01 | 38 | 3407 |
| 19.36 +40.57 | KOI188.01 | 38 | 3407 |
| 19.19 +46.94 | KOI723.01 | 38 | 3400 |
| 18.96 +46.25 | KOI180.01 | 38 | 3385 |
| 19.47 +37.78 | KOI379.01 | 38 | 3378 |
| 19.35 +37.75 | KOI440.01 | 37 | 3361 |
| 11.42 +41.03 | HAT-P-21 | 37 | 3355 |
| 17.04 +47.08 | HIP83389 | 37 | 3328 |
| 19.80 +46.03 | KOI423.01 | 37 | 3328 |
| 19.73 +51.35 | KOI2543.01 | 37 | 3319 |
| 06.49 +10.93 | HIP30905 | 37 | 3314 |
| 19.25 +46.76 | KOI718.01 | 37 | 3314 |
| 19.77 +39.25 | KOI144.01 | 37 | 3313 |
| 20.08 +44.10 | KOI522.01 | 37 | 3301 |
| 19.53 +48.20 | KOI193.01 | 36 | 3277 |
| 06.75 +00.82 | CoRoT-Exo-5 | 36 | 3274 |
| 18.92 +45.96 | KOI936.01 | 36 | 3274 |
| 18.99 +45.97 | KOI249.01 | 36 | 3266 |
| 19.40 +47.36 | KOI733.01 | 36 | 3265 |
| 19.41 +39.95 | KOI232.01 | 36 | 3257 |
| 19.85 +47.95 | KOI751.01 | 36 | 3256 |
| 18.79 +42.78 | KOI369.01 | 36 | 3252 |
| 19.01 +46.03 | KOI271.01 | 36 | 3216 |
| 05.13 -26.80 | HIP23889 | 36 | 3215 |
| 19.36 +40.56 | KOI829.01 | 36 | 3210 |
| 19.38 +48.29 | KOI1306.01 | 36 | 3202 |
| 06.51 +58.16 | HIP31039 | 36 | 3200 |
| 18.97 +41.63 | KOI508.01 | 36 | 3196 |
| 19.02 +39.45 | KOI2652.01 | 35 | 3190 |
| 19.41 +38.88 | KOI442.01 | 35 | 3187 |
| 18.83 +43.89 | KOI316.01 | 35 | 3175 |
| 19.70 +44.53 | KOI99.01 | 35 | 3175 |
| 19.34 +48.25 | KOI2075.01 | 35 | 3168 |
| 19.95 +41.42 | KOI216.01 | 35 | 3148 |
| 19.82 +43.33 | KOI315.01 | 35 | 3145 |
| 19.57 +39.04 | KOI1203.01 | 35 | 3142 |
| 18.94 +45.51 | KOI323.01 | 35 | 3130 |
| 19.41 +50.58 | KOI484.01 | 35 | 3129 |
| 19.07 +49.61 | KOI1442.01 | 35 | 3124 |
| 19.10 +48.68 | KOI345.01 | 35 | 3107 |
| 14.95 +53.38 | HIP73146 | 34 | 3097 |
| 19.18 +42.87 | KOI456.01 | 34 | 3096 |
| 19.10 +49.65 | KOI1831 | 34 | 3088 |
| 19.80 +46.33 | KOI179.01 | 34 | 3063 |
| 18.98 +46.17 | KOI942.01 | 34 | 3052 |
| 11.70 +42.75 | HIP57050 | 34 | 3043 |
| 19.45 +47.68 | KOI335.01 | 34 | 3033 |
| 19.79 +49.16 | KOI330.01 | 34 | 3032 |
| 19.39 +51.22 | KOI1476.01 | 34 | 3029 |
| 19.84 +47.80 | KOI340.01 | 34 | 3028 |
| 18.96 +41.52 | KOI227.01 | 34 | 3023 |
| 20.02 +46.41 | KOI1849.01 | 34 | 3020 |
| 19.59 +45.19 | KOI698.01 | 33 | 3011 |
| 19.51 +38.35 | KOI490.01 | 33 | 2996 |
| 19.63 +38.79 | KOI2034.01 | 33 | 2989 |
| 19.68 +39.20 | KOI2172.01 | 33 | 2989 |
| 11.47 +43.97 | HD99706 | 33 | 2989 |
| 02.08 +25.41 | HIP9683 | 33 | 2979 |
| 19.60 +45.85 | KOI935.01 | 33 | 2966 |
| 00.66 +21.25 | HIP3093 | 33 | 2948 |
| 15.85 +35.66 | HIP77655 | 33 | 2943 |
| 18.89 +41.82 | KOI510.01 | 33 | 2933 |
| 19.36 +37.85 | KOI84.01 | 33 | 2925 |
| 19.33 +40.10 | KOI1973.01 | 32 | 2915 |
| 18.94 +39.78 | KOI446.01 | 32 | 2912 |
| 19.08 +49.87 | KOI1465.01 | 32 | 2907 |
| 19.06 +47.88 | KOI339.01 | 32 | 2905 |
| 19.29 +44.21 | KOI260.01 | 32 | 2904 |
| 15.33 +36.23 | TYC2569-1599-1 | 32 | 2900 |
| 19.48 +44.62 | KOI961.01 | 32 | 2897 |
| 19.65 +39.83 | KOI497.01 | 32 | 2890 |
| 19.01 +39.03 | KOI809.01 | 32 | 2884 |
| 14.55 +21.89 | TYC1482-882-1 | 32 | 2883 |
| 19.07 +45.05 | KOI523.01 | 32 | 2881 |
| 19.69 +39.18 | KOI1329.01 | 32 | 2868 |
| 19.83 +46.69 | KOI716.01 | 32 | 2867 |
| 13.31 -18.31 | HIP64924 | 32 | 2850 |
| 19.70 +40.24 | KOI638.01 | 32 | 2841 |
| 19.65 +39.07 | KOI494.01 | 32 | 2840 |
| 19.87 +48.40 | KOI478.01 | 31 | 2834 |
| 03.33 -28.85 | HIP15527 | 31 | 2834 |
| 02.62 +42.06 | HIP12191 | 31 | 2834 |
| 12.51 +21.95 | HD108863 | 31 | 2834 |
| 20.04 +45.79 | KOI2096.01 | 31 | 2831 |
| 19.52 +42.97 | KOI880.01 | 31 | 2829 |
| 18.98 +41.02 | KOI190.01 | 31 | 2818 |
| 01.42 +28.57 | HIP6643 | 31 | 2816 |
| 18.85 +46.24 | KOI586.01 | 31 | 2813 |
| 19.58 +46.13 | KOI708.01 | 31 | 2811 |
| 19.78 +41.76 | KOI509.01 | 31 | 2791 |
| 19.43 +38.67 | KOI69.01 | 31 | 2784 |
| 19.75 +42.34 | KOI417.01 | 31 | 2783 |
| 13.02 -27.52 | WASP-25 | 31 | 2772 |
| 12.51 +22.88 | HIP61028 | 31 | 2772 |
| 19.36 +48.82 | KOI252.01 | 31 | 2771 |
| 19.45 +42.52 | KOI670.01 | 31 | 2759 |
| 20.01 +45.76 | KOI204.01 | 31 | 2756 |
| 16.34 +41.05 | HIP80076 | 31 | 2754 |
| 23.52 +39.24 | HIP116076 | 31 | 2753 |
| 18.87 +44.28 | KOI1151.01 | 30 | 2734 |
| 18.75 +48.07 | KOI2108.01 | 30 | 2733 |
| 18.86 +41.32 | KOI657.01 | 30 | 2730 |
| 19.94 +43.50 | KOI131.01 | 30 | 2715 |
| 19.31 +48.71 | KOI276.01 | 30 | 2715 |
| 18.98 +44.96 | KOI693.01 | 30 | 2712 |
| 19.20 +39.09 | KOI392.01 | 30 | 2703 |
| 19.68 +42.56 | KOI1939.01 | 30 | 2700 |
| 02.16 +32.32 | HIP10085 | 30 | 2698 |
| 19.12 +49.98 | KOI775.01 | 30 | 2697 |
| 18.77 +47.47 | KOI1169.01 | 30 | 2689 |
| 19.38 +48.24 | KOI432.01 | 30 | 2677 |
| 19.61 +50.08 | KOI2467.01 | 30 | 2669 |
| 19.82 +46.02 | KOI941.01 | 30 | 2657 |
| 12.35 +17.79 | 11Comb | 30 | 2656 |
| 19.74 +50.10 | KOI543.01 | 29 | 2650 |
| 19.70 +43.49 | KOI685.01 | 29 | 2648 |
| 19.41 +42.64 | KOI672.01 | 29 | 2644 |
| 19.77 +46.58 | KOI590.01 | 29 | 2639 |
| 18.99 +44.59 | KOI691.01 | 29 | 2638 |
| 19.27 +40.06 | KOI108.01 | 29 | 2634 |
| 19.00 +50.15 | KOI247.01 | 29 | 2629 |
| 19.01 +49.57 | KOI256.01 | 29 | 2623 |
| 18.93 +49.11 | KOI1445.01 | 29 | 2619 |
| 19.48 +42.91 | KOI373.01 | 29 | 2617 |
| 19.54 +45.07 | KOI581.01 | 29 | 2615 |
| 19.48 +47.88 | KOI2227.01 | 29 | 2611 |
| 06.61 -19.26 | 7CMab | 29 | 2611 |
| 19.30 +49.47 | KOI195.01 | 29 | 2604 |
| 11.78 +03.47 | HD102329 | 29 | 2603 |
| 19.68 +43.96 | KOI162.01 | 29 | 2601 |
| 19.46 +48.14 | KOI752.01 | 29 | 2600 |
| 19.31 +44.35 | KOI127.01 | 29 | 2599 |
| 19.48 +42.04 | KOI1242.01 | 29 | 2597 |
| 18.99 +41.62 | KOI660.01 | 29 | 2595 |
| 19.98 +44.44 | KOI102.01 | 29 | 2592 |
| 19.96 +44.04 | Kepler-5 | 29 | 2588 |
| 19.11 +38.95 | KOI149.01 | 29 | 2587 |
| 19.61 +38.46 | KOI384.01 | 29 | 2585 |
| 19.11 +48.65 | KOI2533.01 | 29 | 2578 |
| 19.02 +46.60 | KOI1521.01 | 29 | 2577 |
| 03.80 +40.53 | HIP17747 | 29 | 2571 |
| 19.66 +44.08 | KOI1017.01 | 29 | 2569 |
| 19.84 +48.30 | KOI477.01 | 29 | 2568 |
| 19.83 +41.06 | KOI507.01 | 29 | 2568 |
| 19.89 +41.68 | KOI611.01 | 29 | 2567 |
| 19.41 +39.20 | KOI100.01 | 28 | 2558 |
| 19.76 +49.54 | KOI354.01 | 28 | 2557 |
| 19.15 +42.30 | KOI864.01 | 28 | 2545 |
| 19.00 +42.08 | KOI612.01 | 28 | 2544 |
| 20.97 +10.84 | HIP103527 | 28 | 2544 |
| 07.34 +37.14 | TYC2463-281-1 | 28 | 2542 |
| 17.16 +33.36 | HIP83949 | 28 | 2540 |
| 19.66 +46.29 | KOI217.01 | 28 | 2531 |
| 19.40 +37.77 | KOI166.01 | 28 | 2527 |
| 03.33 -28.78 | HD20781 | 28 | 2517 |
| 19.70 +50.52 | HIP96901 | 28 | 2517 |
| 14.79 -00.28 | HIP72339 | 28 | 2512 |
| 18.93 +49.23 | KOI1436.01 | 28 | 2511 |
| 19.76 +42.60 | KOI312.01 | 28 | 2511 |
| 19.83 +46.64 | KOI617.01 | 28 | 2499 |
| 19.44 +38.62 | KOI366.01 | 28 | 2485 |
| 18.91 +47.20 | KOI472.01 | 28 | 2485 |
| 20.06 +45.46 | KOI525.01 | 27 | 2472 |
| 22.45 -17.26 | HD212771 | 27 | 2471 |
| 19.68 +39.54 | KOI626.01 | 27 | 2467 |
| 19.94 +47.59 | KOI1161.01 | 27 | 2463 |
| 19.19 +42.44 | KOI1363.01 | 27 | 2462 |
| 19.49 +41.42 | KOI659.01 | 27 | 2461 |
| 15.42 +58.97 | HIP75458 | 27 | 2459 |
| 19.98 +40.86 | KOI371.01 | 27 | 2452 |
| 19.74 +41.60 | KOI212.01 | 27 | 2451 |
| 18.80 +48.54 | KOI972.01 | 27 | 2451 |
| 06.33 +41.09 | HIP30057 | 27 | 2450 |
| 19.86 +46.26 | KOI945.01 | 27 | 2447 |
| 19.79 +47.17 | KOI473.01 | 27 | 2443 |
| 13.21 +17.52 | HIP64426 | 27 | 2440 |
| 12.71 -30.64 | WASP-41b | 27 | 2428 |
| 19.81 +41.73 | KOI239.01 | 27 | 2424 |
| 19.57 +44.93 | KOI2552.01 | 27 | 2418 |
| 02.71 +38.62 | HIP12638 | 27 | 2404 |
| 18.96 +46.32 | KOI2037 | 27 | 2392 |
| 20.07 +44.71 | KOI917.01 | 27 | 2388 |
| 06.55 +05.46 | HIP31246 | 27 | 2386 |
| 06.48 +38.96 | HIP30860 | 26 | 2376 |
| 19.19 +42.54 | KOI255.01 | 26 | 2373 |
| 18.86 +47.57 | KOI289.01 | 26 | 2371 |
| 19.66 +43.01 | KOI1517.01 | 26 | 2354 |
| 19.81 +41.39 | KOI658.01 | 26 | 2354 |
| 19.66 +45.21 | KOI700.01 | 26 | 2353 |
| 19.21 +46.62 | KOI954.01 | 26 | 2351 |
| 19.85 +43.53 | KOI1499.01 | 26 | 2349 |
| 19.61 +39.13 | KOI1761.01 | 26 | 2349 |
| 19.57 +39.32 | KOI551.01 | 26 | 2345 |
| 19.93 +46.69 | KOI2483.01 | 26 | 2345 |
| 19.44 +38.03 | KOI1118.01 | 26 | 2340 |
| 19.21 +45.82 | KOI934.01 | 26 | 2333 |
| 19.24 +47.40 | KOI530.01 | 26 | 2328 |
| 00.74 -26.52 | HIP3479 | 26 | 2323 |
| 19.19 +45.59 | KOI584.01 | 26 | 2322 |
| 19.95 +40.82 | KOI840.01 | 26 | 2314 |
| 19.48 +47.46 | KOI736.01 | 26 | 2311 |
| 19.26 +49.76 | KOI541.01 | 26 | 2310 |
| 19.24 +46.60 | KOI589.01 | 26 | 2310 |
| 23.26 +58.04 | HD240237 | 25 | 2294 |
| 06.07 +44.26 | HIP28767 | 25 | 2294 |
| 19.18 +38.89 | KOI805.01 | 25 | 2291 |
| 18.98 +44.80 | KOI1278.01 | 25 | 2288 |
| 19.80 +42.78 | KOI238.01 | 25 | 2273 |
| 19.46 +44.86 | KOI921.01 | 25 | 2270 |
| 23.65 +42.47 | TYC3239-992-1 | 25 | 2268 |
| 19.28 +50.65 | KOI360.01 | 25 | 2262 |
| 19.88 +47.60 | KOI251.01 | 25 | 2259 |
| 19.07 +43.68 | KOI202.01 | 25 | 2249 |
| 19.33 +42.01 | KOI858.01 | 25 | 2249 |
| 16.97 +25.74 | HIP83043 | 25 | 2247 |
| 19.70 +49.77 | KOI542.01 | 25 | 2245 |
| 10.38 +50.13 | HAT-P-22 | 25 | 2240 |
| 18.98 +49.60 | KOI295.01 | 25 | 2238 |
| 18.87 +47.26 | KOI427.01 | 25 | 2232 |
| 19.37 +52.06 | KOI298.01 | 25 | 2231 |
| 18.79 +42.17 | KOI2005.01 | 25 | 2217 |
| 16.51 +38.35 | HIP80838 | 25 | 2217 |
| 19.67 +50.47 | KOI186.01 | 25 | 2214 |
| 19.41 +40.75 | KOI837.01 | 25 | 2209 |
| 19.02 +41.86 | KOI663.01 | 25 | 2209 |
| 22.96 +20.77 | HIP113357 | 24 | 2193 |
| 19.79 +46.63 | KOI470.01 | 24 | 2182 |
| 07.55 +33.84 | HAT-P-33 | 24 | 2178 |
| 14.60 +09.75 | HIP71395 | 24 | 2177 |
| 22.96 +38.67 | HAT-P-1 | 24 | 2177 |
| 22.20 +16.04 | HIP109577 | 24 | 2175 |
| 21.23 +14.69 | HIP104780 | 24 | 2175 |
| 19.87 +49.41 | KOI352.01 | 24 | 2175 |
| 14.21 +04.05 | HAT-P-26 | 24 | 2173 |
| 19.99 +45.44 | KOI421.01 | 24 | 2170 |
| 19.48 +41.09 | KOI841.01 | 24 | 2170 |
| 19.70 +43.23 | KOI2103.01 | 24 | 2165 |
| 19.88 +42.24 | KOI177.01 | 24 | 2160 |
| 18.18 +54.29 | HIP89047 | 24 | 2154 |
| 19.00 +50.08 | KOI296.01 | 24 | 2150 |
| 19.33 +51.28 | KOI361.01 | 24 | 2148 |
| 19.99 +44.09 | KOI572.01 | 24 | 2147 |
| 19.49 +43.88 | KOI155.01 | 24 | 2145 |
| 19.50 +43.08 | KOI676.01 | 24 | 2141 |
| 18.81 +42.35 | KOI512.01 | 24 | 2137 |
| 19.19 +49.26 | KOI350.01 | 24 | 2137 |
| 18.90 +48.55 | KOI479.01 | 24 | 2134 |
| 19.77 +44.11 | KOI907.01 | 24 | 2132 |
| 18.83 +46.32 | KOI588.01 | 24 | 2132 |
| 18.89 +43.79 | KOI92.01 | 24 | 2127 |
| 19.54 +41.36 | KOI200.01 | 24 | 2127 |
| 02.36 +32.25 | HAT-9-38 | 24 | 2123 |
| 19.99 +43.90 | KOI1014.01 | 24 | 2123 |
| 19.93 +44.86 | KOI105.01 | 24 | 2122 |
| 19.55 +45.05 | KOI1282.01 | 24 | 2121 |
| 19.80 +43.66 | KOI899.01 | 24 | 2118 |
| 19.04 +37.96 | KOI299.01 | 24 | 2118 |
| 02.78 -23.09 | HIP12961 | 24 | 2116 |
| 19.40 +48.44 | KOI1307.01 | 23 | 2110 |
| 19.58 +46.23 | KOI288.01 | 23 | 2106 |
| 19.53 +41.06 | KOI842.01 | 23 | 2105 |
| 01.10 -22.45 | HIP5158 | 23 | 2103 |
| 19.68 +40.59 | KOI645.01 | 23 | 2099 |
| 23.17 +18.40 | TYC1715-00565-1 | 23 | 2099 |
| 19.10 +41.81 | KOI853.01 | 23 | 2097 |
| 19.25 +42.55 | KOI187.01 | 23 | 2096 |
| 13.83 -13.23 | V*QSVir | 23 | 2095 |
| 19.74 +42.11 | KOI2014.01 | 23 | 2091 |
| 19.90 +48.79 | KOI762.01 | 23 | 2091 |
| 19.55 +41.62 | KOI307.01 | 23 | 2082 |
| 19.44 +38.49 | KOI800.01 | 23 | 2082 |
| 20.23 +65.16 | Qatar-1 | 23 | 2079 |
| 18.91 +48.39 | KOI1725.01 | 23 | 2077 |
| 19.67 +46.96 | KOI199.01 | 23 | 2073 |
| 19.24 +43.04 | KOI884.01 | 23 | 2067 |
| 19.94 +40.85 | KOI1540.01 | 23 | 2063 |
| 04.93 -23.24 | HD31527 | 23 | 2055 |
| 00.87 +34.73 | HAT-P-28 | 23 | 2054 |
| 19.75 +50.29 | KOI781.01 | 23 | 2046 |
| 19.36 +44.52 | KOI234.01 | 23 | 2032 |
| 18.88 +45.09 | KOI1274.01 | 23 | 2030 |
| 19.78 +41.26 | KOI558.01 | 23 | 2028 |
| 19.74 +40.02 | KOI1561.01 | 22 | 2024 |
| 19.88 +48.92 | KOI294.01 | 22 | 2024 |
| 18.97 +44.40 | KOI122.01 | 22 | 2022 |
| 22.07 +26.42 | V*V391Peg | 22 | 2014 |
| 19.21 +47.72 | KOI746.01 | 22 | 2010 |
| 19.37 +38.94 | KOI444.01 | 22 | 2009 |
| 19.80 +46.83 | KOI721.01 | 22 | 2006 |
| 18.90 +40.22 | KOI822.01 | 22 | 2006 |
| 06.51 +00.23 | CoRoT-20 | 22 | 2005 |
| 19.77 +49.70 | KOI355.01 | 22 | 2001 |
| 00.26 +01.20 | WASP-32 | 22 | 1998 |
| 19.60 +41.60 | KOI308.01 | 22 | 1995 |
| 19.58 +41.30 | KOI303.01 | 22 | 1995 |
| 20.09 +44.67 | KOI2181.01 | 22 | 1994 |
| 19.63 +44.15 | KOI1809 | 22 | 1993 |
| 19.26 +47.76 | KOI1162.01 | 22 | 1992 |
| 19.85 +49.64 | KOI356.01 | 22 | 1989 |
| 19.41 +40.67 | KOI142.01 | 22 | 1982 |
| 19.01 +42.03 | KOI857.01 | 22 | 1975 |
| 19.02 +47.85 | KOI275.01 | 22 | 1971 |
| 19.70 +40.65 | KOI504.01 | 22 | 1968 |
| 06.63 -32.34 | HIP31688 | 22 | 1964 |
| 14.35 -17.48 | HIP70123 | 22 | 1964 |
| 19.36 +48.79 | KOI480.01 | 22 | 1959 |
| 02.45 +37.55 | WASP-33 | 22 | 1951 |
| 19.62 +48.20 | KOI1303.01 | 22 | 1951 |
| 16.02 +33.30 | HIP78459 | 22 | 1949 |
| 17.88 -30.09 | OGLE-06-109L | 22 | 1946 |
| 19.95 +41.74 | KOI851.01 | 22 | 1940 |
| 19.21 +39.00 | KOI1216.01 | 21 | 1934 |
| 19.75 +47.36 | KOI331.01 | 21 | 1933 |
| 18.94 +44.52 | KOI245.01 | 21 | 1932 |
| 19.52 +41.06 | KOI254.01 | 21 | 1932 |
| 19.19 +43.36 | KOI2255.01 | 21 | 1922 |
| 19.43 +44.53 | KOI370.01 | 21 | 1921 |
| 18.80 +43.70 | KOI900.01 | 21 | 1920 |
| 19.47 +46.33 | KOI2049.01 | 21 | 1918 |
| 17.89 +37.21 | TrES-4b | 21 | 1912 |
| 19.16 +46.77 | KOI327.01 | 21 | 1910 |
| 02.28 +43.77 | HD13931 | 21 | 1910 |
| 18.78 +44.29 | KOI1150.01 | 21 | 1908 |
| 19.40 +49.04 | KOI246.01 | 21 | 1905 |
| 19.16 +38.71 | KOI975.01 | 21 | 1901 |
| 19.81 +43.71 | KOI687.01 | 21 | 1900 |
| 23.31 +18.65 | HIP115100 | 21 | 1900 |
| 19.66 +49.38 | KOI2399.01 | 21 | 1899 |
| 00.74 +20.45 | HIP3502 | 21 | 1896 |
| 19.86 +47.73 | KOI338.01 | 21 | 1887 |
| 19.06 +39.62 | KOI815.01 | 21 | 1887 |
| 19.62 +49.31 | KOI769.01 | 21 | 1886 |
| 19.14 +42.35 | KOI201.01 | 21 | 1878 |
| 18.88 +43.66 | KOI1515.01 | 21 | 1875 |
| 06.86 +40.87 | HIP32916 | 21 | 1875 |
| 17.87 +37.55 | TrES-3b | 21 | 1869 |
| 19.25 +39.56 | KOI496.01 | 21 | 1864 |
| 19.74 +45.98 | KOI938.01 | 21 | 1864 |
| 19.73 +40.30 | KOI823.01 | 21 | 1860 |
| 19.76 +48.23 | KOI535.01 | 21 | 1860 |
| 00.92 +00.79 | HIP4297 | 21 | 1856 |
| 19.01 +50.15 | KOI778.01 | 21 | 1855 |
| 18.76 +46.79 | KOI1537.01 | 21 | 1855 |
| 19.24 +40.94 | KOI283.01 | 20 | 1842 |
| 18.97 +40.93 | KOI610.01 | 20 | 1832 |
| 19.41 +42.51 | KOI871.01 | 20 | 1832 |
| 19.71 +49.13 | KOI2163.01 | 20 | 1828 |
| 18.98 +38.94 | KOI388.01 | 20 | 1827 |
| 18.95 +41.53 | KOI1727.01 | 20 | 1824 |
| 20.03 +45.61 | KOI1969.01 | 20 | 1821 |
| 19.07 +45.48 | KOI583.01 | 20 | 1816 |
| 07.78 +39.09 | GSC02959-00729 | 20 | 1815 |
| 19.17 +44.30 | KOI574.01 | 20 | 1812 |
| 19.71 +38.74 | KOI302.01 | 20 | 1809 |
| 19.80 +40.52 | KOI261.01 | 20 | 1809 |
| 19.46 +49.26 | KOI173.01 | 20 | 1807 |
| 19.85 +45.26 | KOI159.01 | 20 | 1804 |
| 00.31 -15.27 | WASP-26 | 20 | 1801 |
| 15.32 +41.73 | HD136418 | 20 | 1800 |
| 19.08 +45.33 | KOI466.01 | 20 | 1796 |
| 17.89 +56.39 | HD163607 | 20 | 1796 |
| 19.58 +48.83 | KOI348.01 | 20 | 1791 |
| 19.23 +41.10 | KOI506.01 | 20 | 1786 |
| 19.41 +43.86 | KOI240.01 | 20 | 1780 |
| 19.59 +41.66 | KOI850.01 | 20 | 1780 |
| 19.55 +48.81 | KOI1438.01 | 20 | 1780 |
| 14.80 +01.06 | WASP-37 | 20 | 1778 |
| 19.50 +44.38 | KOI575.01 | 20 | 1776 |
| 19.03 +47.83 | KOI1308.01 | 20 | 1774 |
| 19.00 +46.32 | KOI1498.01 | 20 | 1771 |
| 19.66 +48.65 | KOI538.01 | 20 | 1768 |
| 19.27 +47.04 | KOI725.01 | 20 | 1766 |
| 19.01 +44.03 | KOI904.01 | 20 | 1765 |
| 19.03 +49.28 | KOI765.01 | 20 | 1765 |
| 19.34 +46.27 | KOI467.01 | 20 | 1763 |
| 17.85 -29.79 | MOA-bin-1 | 20 | 1760 |
| 06.54 -00.61 | CoRoT-18 | 19 | 1754 |
| 19.81 +41.66 | KOI219.01 | 19 | 1751 |
| 19.83 +42.84 | KOI878.01 | 19 | 1745 |
| 19.78 +43.80 | KOI460.01 | 19 | 1743 |
| 19.06 +41.07 | KOI557.01 | 19 | 1743 |
| 19.27 +41.63 | KOI849.01 | 19 | 1739 |
| 18.80 +42.18 | KOI561.01 | 19 | 1738 |
| 19.14 +41.37 | KOI304.01 | 19 | 1738 |
| 19.15 +43.28 | KOI568.01 | 19 | 1737 |
| 19.44 +41.24 | KOI226.01 | 19 | 1728 |
| 19.03 +48.05 | KOI430.01 | 19 | 1728 |
| 19.77 +42.66 | KOI874.01 | 19 | 1726 |
| 21.47 -21.73 | HIP106006 | 19 | 1724 |
| 19.37 +40.69 | KOI409.01 | 19 | 1724 |
| 19.68 +38.65 | KOI492.01 | 19 | 1724 |
| 00.34 +31.99 | TYC2265-107-1 | 19 | 1723 |
| 18.90 +48.09 | KOI1311.01 | 19 | 1722 |
| 18.92 +49.48 | KOI772.01 | 19 | 1721 |
| 20.35 +59.45 | TrES-5 | 19 | 1720 |
| 19.05 +48.51 | KOI759.01 | 19 | 1715 |
| 19.62 +46.00 | KOI1835 | 19 | 1714 |
| 19.39 +48.42 | KOI601.01 | 19 | 1710 |
| 01.44 +34.58 | HD8673 | 19 | 1709 |
| 18.77 +46.81 | KOI1587.01 | 19 | 1709 |
| 19.46 +44.97 | KOI321.01 | 19 | 1708 |
| 19.64 +45.98 | KOI1922 | 19 | 1707 |
| 18.86 +43.19 | KOI458.01 | 19 | 1707 |
| 19.11 +39.21 | KOI280.01 | 19 | 1696 |
| 01.61 +41.41 | HIP7513 | 19 | 1694 |
| 19.34 +44.87 | KOI319.01 | 19 | 1691 |
| 19.71 +47.75 | KOI337.01 | 19 | 1691 |
| 18.98 +42.65 | KOI454.01 | 19 | 1691 |
| 19.17 +47.51 | KOI740.01 | 19 | 1691 |
| 00.48 -16.23 | HIP2247 | 19 | 1690 |
| 19.33 +40.14 | KOI2237.01 | 19 | 1689 |
| 19.44 +47.81 | KOI476.01 | 19 | 1689 |
| 18.98 +45.60 | KOI928.01 | 19 | 1687 |
| 19.99 +47.16 | KOI1078.01 | 19 | 1685 |
| 16.04 +28.17 | TYC2041-1657-1 | 19 | 1684 |
| 14.85 +05.95 | HAT-P-27/WASP-40 | 19 | 1683 |
| 15.15 +02.34 | WASP-24 | 19 | 1679 |
| 22.89 -14.26 | HIP113020 | 19 | 1679 |
| 19.82 +41.30 | KOI305.01 | 19 | 1673 |
| 18.88 +47.16 | KOI1086.01 | 19 | 1670 |
| 19.11 +39.08 | KOI1215.01 | 19 | 1669 |
| 18.29 +36.62 | TYC2634-1087-1 | 19 | 1669 |
| 18.76 +46.75 | KOI1815.01 | 19 | 1666 |
| 18.97 +46.25 | KOI2057.01 | 19 | 1666 |
| 19.67 +43.91 | KOI235.01 | 19 | 1665 |
| 19.54 +43.58 | KOI896.01 | 19 | 1665 |
| 15.88 +12.91 | NNSer(ab) | 18 | 1664 |
| 18.77 +47.40 | KOI2153 | 18 | 1664 |
| 19.58 +42.83 | KOI877.01 | 18 | 1664 |
| 19.32 +48.69 | KOI2442.01 | 18 | 1663 |
| 19.35 +52.15 | KOI487.01 | 18 | 1661 |
| 19.44 +42.44 | KOI869.01 | 18 | 1661 |
| 19.61 +47.41 | KOI737.01 | 18 | 1656 |
| 16.90 +11.97 | HD152581 | 18 | 1655 |
| 19.10 +50.30 | KOI1530.01 | 18 | 1654 |
| 19.34 +48.58 | KOI1420.01 | 18 | 1648 |
| 19.08 +48.62 | KOI1889 | 18 | 1646 |
| 19.39 +38.18 | KOI197.01 | 18 | 1646 |
| 19.63 +46.82 | KOI592.01 | 18 | 1646 |
| 19.27 +44.00 | KOI1015.01 | 18 | 1643 |
| 18.82 +42.46 | KOI1370.01 | 18 | 1637 |
| 19.22 +47.38 | KOI732.01 | 18 | 1636 |
| 14.31 -20.28 | TYC6147-229-1 | 18 | 1633 |
| 19.73 +43.85 | KOI237.01 | 18 | 1631 |
| 19.59 +50.23 | KOI780.01 | 18 | 1631 |
| 23.05 -00.43 | HD217786 | 18 | 1628 |
| 19.60 +47.00 | KOI426.01 | 18 | 1628 |
| 18.90 +45.73 | KOI1405.01 | 18 | 1624 |
| 19.14 +40.21 | KOI161.01 | 18 | 1623 |
| 19.71 +49.26 | KOI2171.01 | 18 | 1622 |
| 19.78 +46.84 | KOI720.01 | 18 | 1621 |
| 19.54 +45.55 | KOI526.01 | 18 | 1620 |
| 19.05 +48.68 | KOI537.01 | 18 | 1619 |
| 00.64 +42.46 | HAT-P-16 | 18 | 1615 |
| 18.85 +43.27 | KOI890.01 | 18 | 1611 |
| 19.24 +46.42 | KOI469.01 | 18 | 1609 |
| 19.66 +41.07 | KOI225.01 | 18 | 1609 |
| 18.95 +47.69 | KOI532.01 | 18 | 1606 |
| 19.73 +50.57 | KOI785.01 | 18 | 1606 |
| 19.92 +46.65 | KOI1468.01 | 18 | 1602 |
| 18.98 +40.72 | KOI257.01 | 18 | 1596 |
| 18.81 +46.01 | KOI1382.01 | 18 | 1587 |
| 02.02 +46.69 | HAT-P-32 | 18 | 1585 |
| 19.95 +41.15 | KOI2377.01 | 18 | 1584 |
| 19.03 +46.18 | KOI943.01 | 18 | 1583 |
| 01.01 +20.29 | HD5891b | 18 | 1582 |
| 18.89 +41.84 | KOI852.01 | 18 | 1580 |
| 18.87 +47.58 | KOI739.01 | 17 | 1574 |
| 18.92 +47.52 | KOI596.01 | 17 | 1571 |
| 19.88 +40.50 | KOI1589.01 | 17 | 1570 |
| 19.18 +39.24 | KOI281.01 | 17 | 1566 |
| 19.49 +37.58 | KOI1029.01 | 17 | 1566 |
| 19.70 +49.02 | KOI764.01 | 17 | 1565 |
| 18.88 +40.42 | KOI825.01 | 17 | 1564 |
| 19.48 +38.55 | KOI385.01 | 17 | 1561 |
| 19.44 +41.83 | KOI664.01 | 17 | 1561 |
| 19.38 +38.93 | KOI229.01 | 17 | 1560 |
| 18.92 +41.98 | KOI560.01 | 17 | 1556 |
| 19.02 +41.76 | KOI661.01 | 17 | 1556 |
| 18.92 +46.79 | KOI956.01 | 17 | 1556 |
| 19.48 +39.64 | KOI627.01 | 17 | 1554 |
| 19.79 +41.40 | KOI846.01 | 17 | 1553 |
| 17.43 +27.30 | HD158038 | 17 | 1553 |
| 19.37 +43.08 | KOI457.01 | 17 | 1546 |
| 19.32 +49.90 | KOI435.01 | 17 | 1545 |
| 18.85 +44.35 | KOI1141.01 | 17 | 1545 |
| 19.48 +41.02 | KOI413.01 | 17 | 1544 |
| 19.39 +42.09 | KOI368.01 | 17 | 1544 |
| 19.49 +42.63 | KOI566.01 | 17 | 1543 |
| 19.65 +49.14 | KOI1441.01 | 17 | 1540 |
| 18.98 +46.37 | KOI2535.01 | 17 | 1540 |
| 19.64 +41.88 | KOI511.01 | 17 | 1538 |
| 19.59 +44.64 | KOI578.01 | 17 | 1532 |
| 18.83 +43.98 | KOI274.01 | 17 | 1531 |
| 19.71 +44.55 | KOI912.01 | 17 | 1529 |
| 19.15 +38.86 | KOI387.01 | 17 | 1524 |
| 19.17 +38.23 | KOI273.01 | 17 | 1523 |
| 19.92 +46.69 | KOI1523.01 | 17 | 1523 |
| 19.30 +51.24 | KOI486.01 | 17 | 1522 |
| 18.80 +47.09 | KOI1608.01 | 17 | 1520 |
| 19.09 +42.68 | KOI1360.01 | 17 | 1517 |
| 19.57 +42.93 | KOI614.01 | 17 | 1516 |
| 21.11 +03.80 | HD200964 | 17 | 1509 |
| 18.96 +45.72 | KOI704.01 | 17 | 1508 |
| 12.55 +44.92 | HAT-P-36 | 17 | 1508 |
| 19.13 +39.38 | KOI403.01 | 17 | 1507 |
| 19.83 +40.23 | KOI552.01 | 17 | 1507 |
| 19.28 +50.24 | KOI779.01 | 17 | 1507 |
| 19.27 +51.74 | KOI791.01 | 17 | 1506 |
| 18.77 +43.18 | KOI2572.01 | 17 | 1498 |
| 19.36 +42.90 | KOI674.01 | 17 | 1489 |
| 19.13 +37.83 | KOI488.01 | 17 | 1488 |
| 19.93 +41.81 | KOI1230.01 | 16 | 1484 |
| 19.90 +40.08 | KOI501.01 | 16 | 1483 |
| 18.76 +44.30 | KOI1820 | 16 | 1476 |
| 19.34 +50.32 | KOI782.01 | 16 | 1475 |
| 19.21 +51.28 | KOI2331.01 | 16 | 1471 |
| 19.38 +38.71 | KOI242.01 | 16 | 1470 |
| 18.96 +39.91 | KOI367.01 | 16 | 1470 |
| 18.90 +49.45 | KOI1434.01 | 16 | 1470 |
| 18.88 +40.99 | KOI412.01 | 16 | 1467 |
| 19.32 +50.59 | KOI546.01 | 16 | 1464 |
| 18.99 +48.43 | KOI1312.01 | 16 | 1461 |
| 19.67 +46.99 | KOI471.01 | 16 | 1460 |
| 19.17 +44.24 | KOI1142.01 | 16 | 1459 |
| 19.41 +47.31 | KOI734.01 | 16 | 1457 |
| 18.99 +49.69 | KOI1437.01 | 16 | 1455 |
| 19.42 +49.23 | KOI277.01 | 16 | 1454 |
| 19.36 +41.04 | KOI650.01 | 16 | 1452 |
| 19.86 +44.67 | KOI2023.01 | 16 | 1452 |
| 19.07 +44.35 | KOI420.01 | 16 | 1451 |
| 19.48 +40.70 | KOI410.01 | 16 | 1449 |
| 19.40 +47.41 | KOI333.01 | 16 | 1447 |
| 19.37 +38.80 | KOI301.01 | 16 | 1446 |
| 19.78 +47.23 | KOI728.01 | 16 | 1443 |
| 19.20 +46.36 | KOI712.01 | 16 | 1442 |
| 19.54 +48.58 | KOI1440.01 | 16 | 1441 |
| 18.78 +47.57 | KOI1997.01 | 16 | 1439 |
| 18.90 +40.55 | KOI503.01 | 16 | 1439 |
| 11.45 +03.01 | HIP55848 | 16 | 1435 |
| 19.37 +39.73 | KOI1341.01 | 16 | 1435 |
| 14.04 -27.43 | HIP68581 | 16 | 1434 |
| 19.67 +39.95 | KOI499.01 | 16 | 1433 |
| 19.27 +49.73 | KOI774.01 | 16 | 1430 |
| 19.00 +45.07 | KOI1276.01 | 16 | 1430 |
| 19.54 +46.28 | KOI710.01 | 16 | 1429 |
| 19.48 +42.43 | KOI870.01 | 16 | 1429 |
| 19.97 +43.50 | KOI897.01 | 16 | 1428 |
| 19.10 +43.87 | KOI517.01 | 16 | 1424 |
| 19.78 +47.39 | KOI735.01 | 16 | 1424 |
| 06.81 +00.67 | CoRoT-4 | 16 | 1419 |
| 19.06 +44.18 | KOI419.01 | 16 | 1418 |
| 19.67 +46.84 | KOI593.01 | 16 | 1418 |
| 19.07 +40.92 | KOI505.01 | 16 | 1416 |
| 19.45 +41.94 | KOI855.01 | 16 | 1415 |
| 19.94 +43.72 | KOI2400.01 | 16 | 1415 |
| 19.95 +44.69 | KOI1275.01 | 16 | 1411 |
| 19.16 +46.06 | KOI1396.01 | 16 | 1408 |
| 17.09 +33.01 | HAT-P-18 | 16 | 1407 |
| 19.89 +46.28 | KOI587.01 | 16 | 1404 |
| 18.91 +43.38 | KOI891.01 | 16 | 1404 |
| 17.86 -29.88 | V*V5125 | 16 | 1402 |
| 19.40 +39.91 | KOI605.01 | 16 | 1399 |
| 19.50 +37.57 | KOI999.01 | 16 | 1398 |
| 19.01 +43.88 | KOI901.01 | 16 | 1398 |
| 19.30 +43.88 | KOI519.01 | 16 | 1395 |
| 18.96 +41.24 | KOI654.01 | 15 | 1393 |
| 18.77 +47.23 | KOI1870.01 | 15 | 1389 |
| 23.12 +21.13 | HIP114189 | 15 | 1387 |
| 19.27 +47.41 | KOI1880.01 | 15 | 1386 |
| 20.07 +44.38 | KOI1733.01 | 15 | 1382 |
| 19.45 +49.48 | KOI773.01 | 15 | 1380 |
| 19.72 +42.42 | KOI1364.01 | 15 | 1379 |
| 19.34 +46.58 | KOI949.01 | 15 | 1378 |
| 19.16 +46.20 | KOI709.01 | 15 | 1377 |
| 18.93 +45.02 | KOI694.01 | 15 | 1376 |
| 19.31 +44.14 | KOI906.01 | 15 | 1375 |
| 19.65 +45.97 | KOI2101.01 | 15 | 1373 |
| 10.37 +41.23 | HIP50786 | 15 | 1368 |
| 19.44 +42.59 | KOI2339 | 15 | 1367 |
| 19.89 +40.62 | KOI1599.01 | 15 | 1367 |
| 19.56 +39.94 | KOI633.01 | 15 | 1366 |
| 19.72 +42.94 | KOI882.01 | 15 | 1362 |
| 16.17 +26.74 | HD145457 | 15 | 1357 |
| 18.98 +46.25 | KOI2512.01 | 15 | 1357 |
| 19.65 +40.15 | KOI821.01 | 15 | 1357 |
| 19.29 +51.25 | KOI788.01 | 15 | 1357 |
| 19.57 +41.09 | KOI652.01 | 15 | 1354 |
| 19.80 +47.48 | KOI618.01 | 15 | 1351 |
| 19.41 +48.93 | KOI763.01 | 15 | 1348 |
| 19.04 +44.31 | KOI1146.01 | 15 | 1344 |
| 19.80 +48.13 | KOI753.01 | 15 | 1343 |
| 19.38 +44.09 | KOI521.01 | 15 | 1340 |
| 19.64 +39.31 | KOI813.01 | 15 | 1334 |
| 19.19 +42.83 | KOI876.01 | 15 | 1332 |
| 19.30 +46.62 | KOI955.01 | 15 | 1331 |
| 18.92 +45.34 | KOI2319.01 | 15 | 1331 |
| 19.41 +40.42 | KOI826.01 | 15 | 1328 |
| 19.49 +43.20 | KOI680.01 | 15 | 1327 |
| 18.82 +46.67 | KOI1505.01 | 15 | 1327 |
| 19.51 +48.85 | KOI977.01 | 15 | 1326 |
| 19.62 +40.19 | KOI635.01 | 15 | 1325 |
| 18.73 +36.56 | HIP91852 | 15 | 1325 |
| 19.74 +42.13 | KOI666.01 | 15 | 1325 |
| 18.88 +40.36 | KOI824.01 | 15 | 1324 |
| 19.14 +39.06 | KOI810.01 | 15 | 1321 |
| 19.09 +48.42 | KOI2177.01 | 15 | 1321 |
| 19.55 +41.97 | KOI856.01 | 15 | 1321 |
| 14.97 +44.04 | HD132563B | 15 | 1317 |
| 19.28 +47.65 | KOI1851.01 | 15 | 1317 |
| 19.99 +46.05 | KOI2029 | 15 | 1316 |
| 17.21 +63.35 | HD156279 | 15 | 1315 |
| 19.47 +39.77 | KOI629.01 | 15 | 1314 |
| 19.25 +39.71 | KOI628.01 | 15 | 1313 |
| 19.80 +47.55 | KOI741.01 | 15 | 1313 |
| 19.32 +40.62 | KOI607.01 | 15 | 1312 |
| 18.99 +44.66 | KOI913.01 | 15 | 1312 |
| 14.79 +00.28 | HD130322 | 15 | 1312 |
| 18.92 +41.35 | KOI844.01 | 15 | 1310 |
| 19.30 +51.69 | KOI548.01 | 15 | 1309 |
| 18.92 +44.71 | KOI981.01 | 15 | 1308 |
| 19.70 +42.81 | KOI987.01 | 15 | 1307 |
| 19.64 +50.67 | KOI547.01 | 15 | 1307 |
| 18.78 +46.95 | KOI2205.01 | 14 | 1304 |
| 19.77 +49.94 | KOI620.01 | 14 | 1303 |
| 19.25 +48.23 | KOI755.01 | 14 | 1303 |
| 19.38 +38.69 | KOI624.01 | 14 | 1303 |
| 18.81 +46.72 | KOI717.01 | 14 | 1302 |
| 19.36 +38.73 | KOI804.01 | 14 | 1299 |
| 19.69 +51.18 | KOI1474.01 | 14 | 1299 |
| 18.91 +45.41 | KOI1933.01 | 14 | 1299 |
| 18.90 +41.87 | KOI559.01 | 14 | 1298 |
| 19.51 +46.10 | KOI939.01 | 14 | 1298 |
| 19.18 +47.41 | KOI1160.01 | 14 | 1298 |
| 18.79 +42.30 | KOI2282.01 | 14 | 1294 |
| 19.01 +49.80 | KOI2376.01 | 14 | 1292 |
| 19.10 +39.53 | KOI625.01 | 14 | 1290 |
| 19.42 +45.75 | KOI585.01 | 14 | 1289 |
| 19.66 +39.23 | KOI2386.01 | 14 | 1285 |
| 19.48 +44.70 | KOI580.01 | 14 | 1285 |
| 21.64 +30.49 | HAT-P-17 | 14 | 1283 |
| 19.48 +40.76 | KOI838.01 | 14 | 1281 |
| 17.91 -30.38 | OGLE2005-BLG-390 | 14 | 1280 |
| 18.91 +47.86 | KOI747.01 | 14 | 1277 |
| 18.88 +43.89 | KOI1569.01 | 14 | 1274 |
| 19.32 +46.95 | KOI1595.01 | 14 | 1271 |
| 19.21 +40.88 | KOI609.01 | 14 | 1271 |
| 19.15 +34.60 | HIP94075 | 14 | 1270 |
| 19.63 +38.36 | KOI1089.01 | 14 | 1263 |
| 19.33 +40.53 | KOI644.01 | 14 | 1261 |
| 18.81 +42.23 | KOI667.01 | 14 | 1259 |
| 18.81 +44.42 | KOI1148.01 | 14 | 1259 |
| 19.70 +50.49 | KOI783.01 | 14 | 1256 |
| 19.70 +49.74 | KOI1535.01 | 14 | 1255 |
| 19.29 +39.95 | KOI632.01 | 14 | 1253 |
| 18.83 +43.73 | KOI1581.01 | 14 | 1248 |
| 19.63 +39.77 | KOI816.01 | 14 | 1247 |
| 19.67 +42.76 | KOI1377.01 | 14 | 1244 |
| 19.76 +51.32 | KOI790.01 | 14 | 1244 |
| 19.53 +45.91 | KOI937.01 | 14 | 1244 |
| 20.21 +18.11 | HAT-P-34 | 14 | 1244 |
| 13.79 +17.46 | HIP67275 | 14 | 1244 |
| 19.49 +46.23 | KOI1488.01 | 14 | 1239 |
| 00.32 +14.05 | HD1502b | 14 | 1235 |
| 19.08 +48.06 | KOI600.01 | 14 | 1229 |
| 19.40 +44.63 | KOI577.01 | 14 | 1224 |
| 17.34 +38.24 | HAT-P-14 | 14 | 1224 |
| 19.66 +45.57 | KOI703.01 | 14 | 1218 |
| 19.23 +41.30 | KOI845.01 | 13 | 1214 |
| 19.36 +38.34 | KOI1102.01 | 13 | 1214 |
| 19.58 +41.27 | KOI655.01 | 13 | 1212 |
| 19.59 +49.68 | KOI1990.01 | 13 | 1208 |
| 19.42 +47.70 | KOI1163.01 | 13 | 1208 |
| 19.00 +49.48 | KOI2318.01 | 13 | 1206 |
| 19.36 +40.44 | KOI1543.01 | 13 | 1204 |
| 19.03 +48.73 | KOI1419.01 | 13 | 1203 |
| 19.25 +40.03 | KOI818.01 | 13 | 1203 |
| 19.77 +47.18 | KOI991.01 | 13 | 1200 |
| 18.99 +39.24 | KOI1344.01 | 13 | 1200 |
| 19.34 +38.70 | KOI1221.01 | 13 | 1199 |
| 15.22 -25.31 | HIP74500 | 13 | 1199 |
| 19.41 +45.32 | KOI582.01 | 13 | 1194 |
| 19.58 +41.87 | KOI1241.01 | 13 | 1191 |
| 21.54 -20.96 | HD204941 | 13 | 1190 |
| 19.80 +48.48 | KOI758.01 | 13 | 1188 |
| 19.04 +45.08 | KOI695.01 | 13 | 1187 |
| 18.96 +44.48 | KOI910.01 | 13 | 1186 |
| 19.20 +46.47 | KOI714.01 | 13 | 1185 |
| 14.89 +18.24 | HD131496 | 13 | 1180 |
| 19.76 +45.57 | KOI931.01 | 13 | 1179 |
| 23.43 -20.62 | HD220689 | 13 | 1176 |
| 19.95 +40.23 | KOI641.01 | 13 | 1176 |
| 19.43 +42.18 | KOI1240.01 | 13 | 1171 |
| 19.41 +40.70 | KOI647.01 | 13 | 1169 |
| 18.98 +46.21 | KOI2637.01 | 13 | 1168 |
| 19.59 +42.53 | KOI671.01 | 13 | 1167 |
| 19.59 +42.62 | KOI673.01 | 13 | 1166 |
| 19.62 +38.31 | KOI799.01 | 13 | 1165 |
| 19.65 +46.21 | KOI944.01 | 13 | 1163 |
| 18.82 +43.24 | KOI2631.01 | 13 | 1160 |
| 19.43 +38.22 | KOI1115.01 | 13 | 1158 |
| 19.39 +50.05 | KOI776.01 | 13 | 1158 |
| 19.92 +46.50 | KOI2035.01 | 13 | 1158 |
| 19.78 +49.32 | KOI771.01 | 13 | 1157 |
| 19.58 +45.29 | KOI1391.01 | 13 | 1157 |
| 19.76 +51.27 | KOI1516.01 | 13 | 1156 |
| 18.89 +48.22 | KOI1946.01 | 13 | 1155 |
| 19.65 +39.52 | KOI814.01 | 13 | 1154 |
| 19.59 +38.42 | KOI801.01 | 13 | 1149 |
| 18.97 +46.45 | KOI2545.01 | 13 | 1148 |
| 19.79 +43.65 | KOI686.01 | 13 | 1145 |
| 19.25 +41.31 | KOI1054.01 | 13 | 1145 |
| 19.75 +48.97 | KOI1425.01 | 13 | 1145 |
| 16.00 -28.06 | TYC6787-1927-1 | 13 | 1143 |
| 19.36 +44.39 | KOI689.01 | 13 | 1142 |
| 18.90 +43.38 | KOI1519.01 | 13 | 1137 |
| 19.42 +38.24 | KOI797.01 | 13 | 1135 |
| 19.52 +47.49 | KOI1164.01 | 13 | 1135 |
| 19.01 +47.88 | KOI1314.01 | 13 | 1135 |
| 19.19 +39.08 | KOI1193.01 | 13 | 1132 |
| 19.60 +46.58 | KOI951.01 | 13 | 1128 |
| 19.13 +50.38 | KOI1512.01 | 13 | 1127 |
| 19.54 +41.14 | KOI1060.01 | 13 | 1125 |
| 19.66 +45.03 | KOI922.01 | 12 | 1124 |
| 19.24 +44.73 | KOI579.01 | 12 | 1120 |
| 19.04 +48.44 | KOI1310.01 | 12 | 1114 |
| 19.82 +49.80 | KOI1894.01 | 12 | 1110 |
| 19.81 +47.67 | KOI745.01 | 12 | 1108 |
| 19.69 +48.37 | KOI1315.01 | 12 | 1108 |
| 19.16 +42.19 | KOI1236.01 | 12 | 1108 |
| 19.57 +41.14 | KOI843.01 | 12 | 1107 |
| 19.36 +47.93 | KOI750.01 | 12 | 1102 |
| 19.70 +40.51 | KOI833.01 | 12 | 1100 |
| 19.37 +44.14 | KOI1072.01 | 12 | 1099 |
| 19.80 +49.23 | KOI767.01 | 12 | 1098 |
| 18.80 +43.67 | KOI1525.01 | 12 | 1094 |
| 19.03 +43.17 | KOI678.01 | 12 | 1090 |
| 19.46 +49.25 | KOI766.01 | 12 | 1090 |
| 19.15 +47.46 | KOI2271.01 | 12 | 1089 |
| 19.78 +47.33 | KOI1152.01 | 12 | 1088 |
| 11.77 +14.12 | HIP57428 | 12 | 1087 |
| 17.29 +29.23 | HD156668 | 12 | 1085 |
| 19.39 +44.65 | KOI692.01 | 12 | 1085 |
| 19.36 +40.69 | KOI554.01 | 12 | 1085 |
| 23.27 +31.46 | GSC02752-00114 | 12 | 1084 |
| 19.59 +38.56 | KOI802.01 | 12 | 1083 |
| 18.99 +43.95 | KOI2426.01 | 12 | 1079 |
| 21.25 -20.79 | HIP104903 | 12 | 1074 |
| 19.97 +46.90 | KOI2018.01 | 12 | 1074 |
| 19.29 +39.15 | KOI811.01 | 12 | 1073 |
| 19.61 +40.66 | KOI835.01 | 12 | 1073 |
| 19.44 +50.62 | KOI786.01 | 12 | 1072 |
| 00.76 +07.84 | HD4313 | 12 | 1072 |
| 19.17 +43.14 | KOI679.01 | 12 | 1071 |
| 19.61 +40.42 | KOI827.01 | 12 | 1071 |
| 19.39 +38.27 | KOI795.01 | 12 | 1070 |
| 19.10 +46.69 | KOI953.01 | 12 | 1069 |
| 19.89 +40.17 | KOI1597.01 | 12 | 1066 |
| 19.35 +44.04 | KOI688.01 | 12 | 1065 |
| 18.93 +44.81 | KOI918.01 | 12 | 1063 |
| 19.38 +38.57 | KOI1219.01 | 12 | 1063 |
| 19.04 +45.58 | KOI929.01 | 12 | 1063 |
| 19.37 +40.58 | KOI830.01 | 12 | 1059 |
| 19.02 +45.95 | KOI1409.01 | 12 | 1058 |
| 19.56 +44.79 | KOI916.01 | 12 | 1054 |
| 19.10 +38.37 | KOI1108.01 | 12 | 1054 |
| 18.97 +39.22 | KOI1339.01 | 12 | 1053 |
| 19.06 +48.31 | KOI2200.01 | 12 | 1051 |
| 19.79 +40.23 | KOI639.01 | 12 | 1048 |
| 18.71 +05.94 | CoRoT-11 | 12 | 1043 |
| 19.32 +40.80 | KOI649.01 | 12 | 1043 |
| 19.60 +50.53 | KOI784.01 | 12 | 1042 |
| 19.85 +46.49 | KOI1491.01 | 12 | 1040 |
| 19.07 +36.63 | TrES-1b | 12 | 1040 |
| 19.27 +45.15 | KOI697.01 | 12 | 1039 |
| 11.29 +17.96 | V*DPLeo | 11 | 1033 |
| 18.87 +48.13 | KOI1989.01 | 11 | 1033 |
| 19.02 +49.20 | KOI2639.01 | 11 | 1031 |
| 19.74 +47.92 | KOI599.01 | 11 | 1031 |
| 18.54 +06.95 | HD171028 | 11 | 1028 |
| 19.69 +43.04 | KOI2213.01 | 11 | 1027 |
| 20.00 +44.77 | KOI1754.01 | 11 | 1026 |
| 18.92 +44.61 | KOI1975.01 | 11 | 1026 |
| 18.90 +40.13 | KOI1923.01 | 11 | 1024 |
| 19.82 +40.29 | KOI640.01 | 11 | 1023 |
| 19.69 +48.60 | KOI1435.01 | 11 | 1023 |
| 19.50 +38.51 | KOI1198.01 | 11 | 1021 |
| 19.21 +46.25 | KOI2548.01 | 11 | 1021 |
| 18.95 +51.27 | HAT-P-37 | 11 | 1020 |
| 18.04 +26.31 | HIP88348 | 11 | 1019 |
| 19.50 +46.32 | KOI1970.01 | 11 | 1017 |
| 19.65 +45.39 | KOI1401.01 | 11 | 1017 |
| 18.36 -11.92 | HIP90004 | 11 | 1016 |
| 19.47 +38.05 | KOI1113.01 | 11 | 1016 |
| 19.04 +50.11 | KOI1833.01 | 11 | 1013 |
| 19.07 +45.01 | KOI1273.01 | 11 | 1013 |
| 19.45 +44.89 | KOI920.01 | 11 | 1007 |
| 19.29 +50.99 | KOI1506.01 | 11 | 1007 |
| 20.66 +11.25 | HIP101966 | 11 | 1006 |
| 19.76 +50.62 | KOI2407.01 | 11 | 1006 |
| 19.41 +55.47 | WASP-48 | 11 | 1004 |
| 19.43 +46.90 | KOI719.01 | 11 | 999 |
| 19.52 +46.74 | KOI2660.01 | 11 | 996 |
| 19.41 +45.12 | KOI923.01 | 11 | 995 |
| 19.44 +48.45 | KOI1300.01 | 11 | 994 |
| 19.55 +48.29 | KOI1299.01 | 11 | 994 |
| 18.96 +44.39 | KOI1149.01 | 11 | 993 |
| 19.94 +43.90 | KOI1549.01 | 11 | 993 |
| 19.91 +46.45 | KOI1751 | 11 | 993 |
| 19.71 +43.03 | KOI887.01 | 11 | 991 |
| 19.73 +49.79 | KOI2355.01 | 11 | 990 |
| 19.23 +41.03 | KOI1005.01 | 11 | 990 |
| 19.47 +42.93 | KOI1356.01 | 11 | 989 |
| 19.65 +39.40 | KOI1342.01 | 11 | 988 |
| 19.23 +45.80 | KOI2298.01 | 11 | 987 |
| 19.40 +47.48 | KOI1165.01 | 11 | 985 |
| 19.36 +50.90 | KOI2264.01 | 11 | 984 |
| 19.70 +50.08 | KOI1494.01 | 11 | 984 |
| 18.97 +47.08 | KOI2520.01 | 11 | 983 |
| 19.40 +39.22 | KOI1335.01 | 11 | 981 |
| 20.06 +44.03 | KOI1746.01 | 11 | 980 |
| 19.84 +49.40 | KOI2226.01 | 11 | 976 |
| 19.40 +38.54 | KOI976.01 | 11 | 975 |
| 18.93 +47.64 | KOI2484.01 | 11 | 974 |
| 19.56 +48.20 | KOI2228.01 | 11 | 974 |
| 19.69 +49.59 | KOI2361.01 | 11 | 971 |
| 19.95 +43.41 | KOI1842.01 | 11 | 967 |
| 19.85 +42.05 | KOI1227.01 | 11 | 961 |
| 18.94 +44.25 | KOI2193.01 | 11 | 961 |
| 19.69 +43.61 | KOI1511.01 | 11 | 958 |
| 19.72 +45.99 | KOI974.01 | 11 | 958 |
| 19.49 +46.63 | KOI2553.01 | 11 | 958 |
| 19.41 +45.24 | KOI924.01 | 11 | 958 |
| 19.89 +41.00 | KOI1793.01 | 11 | 956 |
| 19.67 +43.54 | KOI1456.01 | 11 | 954 |
| 19.36 +43.36 | KOI892.01 | 11 | 947 |
| 19.15 +41.64 | KOI1129.01 | 11 | 947 |
| 19.69 +50.48 | KOI1507.01 | 11 | 947 |
| 19.53 +43.35 | KOI893.01 | 11 | 947 |
| 19.03 +45.98 | KOI1410.01 | 11 | 945 |
| 19.04 +48.72 | KOI1863.01 | 10 | 941 |
| 19.18 +40.69 | KOI2471.01 | 10 | 940 |
| 19.76 +43.46 | KOI895.01 | 10 | 938 |
| 19.79 +41.10 | KOI1050.01 | 10 | 936 |
| 19.56 +48.44 | KOI1931 | 10 | 935 |
| 19.51 +38.36 | KOI1106.01 | 10 | 933 |
| 19.99 +43.73 | KOI2363.01 | 10 | 930 |
| 19.78 +47.00 | KOI2002.01 | 10 | 930 |
| 22.93 -26.66 | HIP113238 | 10 | 929 |
| 19.02 +45.37 | KOI1413.01 | 10 | 929 |
| 19.19 +48.77 | KOI1427.01 | 10 | 924 |
| 16.37 -20.72 | GSC06214-00210 | 10 | 923 |
| 19.55 +43.06 | KOI2357.01 | 10 | 922 |
| 19.08 +43.59 | KOI1533.01 | 10 | 921 |
| 19.74 +39.16 | KOI1323.01 | 10 | 921 |
| 15.88 +15.43 | HD142245 | 10 | 918 |
| 19.99 +43.67 | KOI2095.01 | 10 | 918 |
| 19.45 +37.53 | KOI2130.01 | 10 | 917 |
| 19.26 +46.99 | KOI1598.01 | 10 | 913 |
| 19.64 +49.87 | KOI2310.01 | 10 | 913 |
| 19.90 +40.64 | KOI1546.01 | 10 | 912 |
| 19.48 +47.16 | KOI1082.01 | 10 | 912 |
| 19.45 +38.57 | KOI1202.01 | 10 | 912 |
| 19.83 +42.88 | KOI1353.01 | 10 | 912 |
| 19.98 +46.20 | KOI1886.01 | 10 | 908 |
| 19.33 +42.12 | KOI861.01 | 10 | 907 |
| 19.44 +47.83 | KOI1285.01 | 10 | 906 |
| 19.25 +44.61 | KOI914.01 | 10 | 906 |
| 19.26 +48.21 | KOI1309.01 | 10 | 906 |
| 19.11 +46.87 | KOI1584.01 | 10 | 906 |
| 19.16 +49.45 | KOI2109.01 | 10 | 905 |
| 19.07 +44.66 | KOI1266.01 | 10 | 905 |
| 19.03 +49.99 | KOI2492.01 | 10 | 904 |
| 19.35 +45.56 | KOI1408.01 | 10 | 899 |
| 19.20 +51.35 | KOI2463.01 | 10 | 899 |
| 19.40 +36.84 | KOI984.01 | 10 | 899 |
| 19.32 +46.49 | KOI1448.01 | 10 | 898 |
| 20.00 +44.16 | KOI2144.01 | 10 | 895 |
| 19.01 +48.77 | KOI2485.01 | 10 | 895 |
| 19.52 +44.20 | KOI1144.01 | 10 | 893 |
| 19.90 +44.28 | KOI2293.01 | 10 | 891 |
| 19.73 +47.45 | KOI1175.01 | 10 | 891 |
| 02.21 +51.78 | HAT-P-29 | 10 | 890 |
| 18.91 +40.00 | KOI1959.01 | 10 | 889 |
| 19.41 +36.58 | KOI889.01 | 10 | 887 |
| 18.58 +35.66 | TYC2636-195-1 | 10 | 887 |
| 19.85 +39.92 | KOI1602.01 | 10 | 885 |
| 19.75 +50.67 | KOI2597 | 10 | 885 |
| 18.95 +45.26 | KOI1395.01 | 10 | 884 |
| 19.37 +41.20 | KOI1053.01 | 10 | 881 |
| 19.90 +47.01 | KOI1984.01 | 10 | 881 |
| 15.87 -18.44 | HIP77740 | 10 | 881 |
| 19.77 +48.08 | KOI1304.01 | 10 | 880 |
| 19.10 +44.80 | KOI1264.01 | 10 | 880 |
| 19.28 +42.60 | KOI872.01 | 10 | 880 |
| 19.09 +45.41 | KOI926.01 | 10 | 879 |
| 19.70 +44.01 | KOI960.01 | 10 | 875 |
| 19.99 +44.17 | KOI1066.01 | 10 | 874 |
| 19.90 +46.79 | KOI1779 | 10 | 872 |
| 19.37 +41.24 | KOI1052.01 | 10 | 872 |
| 19.56 +44.78 | KOI1279.01 | 10 | 872 |
| 19.90 +47.95 | KOI2317.01 | 10 | 868 |
| 18.72 +06.20 | CoRoT-9 | 10 | 867 |
| 19.43 +41.88 | KOI1246.01 | 10 | 866 |
| 18.92 +41.04 | KOI1928.01 | 10 | 865 |
| 19.34 +41.38 | KOI1061.01 | 10 | 860 |
| 19.46 +38.01 | KOI986.01 | 10 | 859 |
| 19.70 +43.88 | KOI903.01 | 10 | 858 |
| 19.65 +47.13 | KOI1081.01 | 9 | 853 |
| 13.93 -32.16 | RAVEJ135542.7-320934 | 9 | 853 |
| 18.93 +41.22 | KOI2031.01 | 9 | 852 |
| 20.66 +42.25 | HD197037 | 9 | 851 |
| 19.00 +39.36 | KOI1934.01 | 9 | 850 |
| 20.02 +44.14 | KOI2174 | 9 | 848 |
| 19.98 +45.97 | KOI1397.01 | 9 | 847 |
| 19.78 +34.42 | HIP97336 | 9 | 846 |
| 18.94 +40.52 | KOI2666.01 | 9 | 845 |
| 02.12 +23.46 | alfArib | 9 | 845 |
| 18.86 +42.67 | KOI1907.01 | 9 | 845 |
| 19.62 +40.72 | KOI1590.01 | 9 | 843 |
| 18.87 +40.56 | KOI2440.01 | 9 | 843 |
| 18.89 +41.52 | KOI2542.01 | 9 | 841 |
| 18.94 +41.18 | KOI1839.01 | 9 | 840 |
| 19.26 +46.06 | KOI2246.01 | 9 | 840 |
| 23.44 +08.64 | HD220773 | 9 | 839 |
| 19.54 +46.67 | KOI1998.01 | 9 | 838 |
| 19.29 +50.60 | KOI2072.01 | 9 | 837 |
| 19.17 +47.47 | KOI2058.01 | 9 | 835 |
| 19.44 +44.53 | KOI2204.01 | 9 | 834 |
| 19.85 +47.89 | KOI2149.01 | 9 | 831 |
| 19.69 +48.52 | KOI2580.01 | 9 | 831 |
| 19.52 +49.58 | KOI2365.01 | 9 | 829 |
| 19.78 +42.63 | KOI875.01 | 9 | 828 |
| 16.55 +02.08 | HIP81022 | 9 | 828 |
| 19.47 +00.12 | GSC00465-01645 | 9 | 823 |
| 16.16 -21.08 | 1RXS1609 | 9 | 822 |
| 19.28 +45.87 | KOI2242.01 | 9 | 822 |
| 18.86 +42.83 | KOI2561.01 | 9 | 820 |
| 19.95 +45.83 | KOI2137.01 | 9 | 819 |
| 18.90 +47.59 | KOI2583.01 | 9 | 819 |
| 19.24 +48.10 | KOI2236 | 9 | 818 |
| 19.83 +48.56 | KOI1444.01 | 9 | 818 |
| 19.72 +49.94 | KOI2582.01 | 9 | 815 |
| 19.25 +46.94 | KOI1572.01 | 9 | 812 |
| 19.63 +46.16 | KOI2364.01 | 9 | 809 |
| 19.72 +40.44 | KOI1593.01 | 9 | 804 |
| 19.78 +41.39 | KOI1059.01 | 9 | 803 |
| 19.27 +46.67 | KOI1526.01 | 9 | 800 |
| 19.00 +40.22 | KOI1843 | 9 | 800 |
| 19.23 +47.10 | KOI2313.01 | 9 | 798 |
| 18.82 +46.52 | KOI2457.01 | 9 | 798 |
| 19.89 +47.69 | KOI2266.01 | 9 | 797 |
| 19.55 +37.97 | KOI1094.01 | 9 | 796 |
| 19.02 +48.56 | KOI1800.01 | 9 | 795 |
| 18.96 +47.64 | KOI1877.01 | 9 | 795 |
| 18.87 +43.53 | KOI2247.01 | 9 | 795 |
| 19.25 +48.28 | KOI2104.01 | 9 | 795 |
| 19.40 +00.75 | CoRoT-10 | 9 | 793 |
| 18.81 +43.38 | KOI2307.01 | 9 | 792 |
| 18.99 +44.61 | KOI2466 | 9 | 791 |
| 19.90 +42.19 | KOI1357.01 | 9 | 786 |
| 18.87 +40.78 | KOI1930 | 9 | 786 |
| 19.53 +37.61 | KOI1020.01 | 9 | 782 |
| 19.13 +38.87 | KOI1207.01 | 9 | 780 |
| 19.67 +46.01 | KOI2079.01 | 9 | 780 |
| 19.46 +37.90 | KOI1024.01 | 9 | 777 |
| 18.93 +40.74 | KOI2668.01 | 9 | 777 |
| 19.76 +39.12 | KOI1336.01 | 9 | 776 |
| 19.41 +40.89 | KOI1588.01 | 9 | 775 |
| 19.87 +28.10 | HIP97769 | 9 | 774 |
| 19.39 +45.77 | KOI1385.01 | 9 | 774 |
| 19.37 +48.13 | KOI1316.01 | 9 | 774 |
| 19.14 +43.62 | KOI2527.01 | 9 | 774 |
| 18.87 +43.15 | KOI2316.01 | 9 | 773 |
| 18.80 +44.48 | KOI2291.01 | 9 | 769 |
| 19.42 +37.61 | KOI988.01 | 9 | 766 |
| 19.14 +47.12 | KOI1909 | 8 | 763 |
| 19.22 +51.48 | KOI2154.01 | 8 | 761 |
| 19.58 +44.66 | KOI1270.01 | 8 | 760 |
| 18.94 +47.81 | KOI2300.01 | 8 | 759 |
| 18.96 +40.57 | KOI2297.01 | 8 | 759 |
| 19.22 +47.93 | KOI1837.01 | 8 | 758 |
| 19.18 +49.16 | KOI1421.01 | 8 | 755 |
| 19.59 +42.21 | KOI863.01 | 8 | 753 |
| 18.97 +42.65 | KOI2083.01 | 8 | 751 |
| 18.95 +40.79 | KOI1962.01 | 8 | 750 |
| 19.96 +45.59 | KOI2522.01 | 8 | 750 |
| 19.71 +46.25 | KOI1481.01 | 8 | 750 |
| 18.90 +45.61 | KOI1393.01 | 8 | 745 |
| 17.26 +04.96 | LHS3275 | 8 | 744 |
| 19.77 +49.98 | KOI1906.01 | 8 | 742 |
| 19.45 +47.53 | KOI2147.01 | 8 | 740 |
| 19.50 +49.06 | KOI1802.01 | 8 | 740 |
| 19.73 +44.85 | KOI1283.01 | 8 | 739 |
| 19.58 +46.99 | KOI1720.01 | 8 | 738 |
| 20.06 +29.90 | HIP98767 | 8 | 737 |
| 19.43 +42.58 | KOI1747.01 | 8 | 735 |
| 19.47 +49.02 | KOI1857.01 | 8 | 735 |
| 19.77 +44.04 | KOI905.01 | 8 | 733 |
| 19.01 +45.38 | KOI2169 | 8 | 732 |
| 19.40 +39.00 | KOI1187.01 | 8 | 732 |
| 19.42 +50.91 | KOI2577.01 | 8 | 731 |
| 19.44 +48.01 | KOI2327.01 | 8 | 731 |
| 19.47 +42.38 | KOI867.01 | 8 | 729 |
| 19.14 +41.87 | KOI2408.01 | 8 | 729 |
| 19.35 +44.56 | KOI911.01 | 8 | 728 |
| 18.95 +45.07 | KOI2000.01 | 8 | 728 |
| 13.21 -31.87 | HIP64459 | 8 | 728 |
| 19.22 +46.34 | KOI2152.01 | 8 | 727 |
| 19.36 +37.73 | KOI1003.01 | 8 | 727 |
| 19.55 +37.85 | KOI1031.01 | 8 | 727 |
| 19.74 +47.44 | KOI1166.01 | 8 | 723 |
| 19.68 +38.77 | KOI1214.01 | 8 | 720 |
| 19.82 +43.14 | KOI1518.01 | 8 | 719 |
| 19.79 +40.14 | KOI1573.01 | 8 | 718 |
| 19.15 +39.14 | KOI1220.01 | 8 | 718 |
| 19.81 +41.78 | KOI1238.01 | 8 | 718 |
| 18.96 +40.85 | KOI2337.01 | 8 | 715 |
| 19.10 +39.27 | KOI2117.01 | 8 | 712 |
| 21.64 -31.74 | HIP106824 | 8 | 712 |
| 19.10 +42.66 | KOI2028 | 8 | 710 |
| 19.79 +41.76 | KOI2036 | 8 | 710 |
| 19.28 +44.15 | KOI1910.01 | 8 | 710 |
| 19.46 +42.70 | KOI873.01 | 8 | 709 |
| 19.59 +50.43 | KOI2106.01 | 8 | 709 |
| 19.40 +38.21 | KOI1117.01 | 8 | 705 |
| 19.82 +50.12 | KOI1828.01 | 8 | 700 |
| 19.61 +38.39 | KOI1101.01 | 8 | 700 |
| 19.48 +37.96 | KOI1022.01 | 8 | 700 |
| 19.43 +37.07 | KOI998.01 | 8 | 699 |
| 19.94 +45.87 | KOI2374 | 8 | 699 |
| 19.38 +37.35 | KOI1002.01 | 8 | 699 |
| 19.07 +39.63 | KOI2024.01 | 8 | 698 |
| 19.35 +50.03 | KOI2260.01 | 8 | 697 |
| 18.80 +43.63 | KOI1818.01 | 8 | 696 |
| 19.49 +47.64 | KOI2554 | 8 | 696 |
| 16.26 +10.03 | WASP-38 | 8 | 696 |
| 19.70 +42.17 | KOI1758.01 | 8 | 695 |
| 19.07 +45.38 | KOI1407.01 | 8 | 694 |
| 18.79 +44.16 | KOI2593.01 | 8 | 693 |
| 19.14 +49.05 | KOI1433.01 | 8 | 691 |
| 19.42 +47.33 | KOI1074.01 | 8 | 691 |
| 19.44 +38.35 | KOI1109.01 | 8 | 691 |
| 19.47 +38.63 | KOI1177.01 | 8 | 691 |
| 19.78 +42.97 | KOI883.01 | 8 | 691 |
| 18.82 +42.15 | KOI2306.01 | 8 | 691 |
| 19.01 +38.81 | KOI2536.01 | 8 | 691 |
| 19.86 +45.27 | KOI2286.01 | 8 | 688 |
| 18.88 +41.99 | KOI2451.01 | 8 | 687 |
| 19.83 +39.85 | KOI1475.01 | 8 | 686 |
| 19.32 +50.79 | KOI2588.01 | 8 | 686 |
| 19.97 +45.77 | KOI2067.01 | 8 | 686 |
| 19.43 +48.07 | KOI1305.01 | 8 | 686 |
| 19.25 +40.66 | KOI1601.01 | 8 | 682 |
| 18.65 +04.36 | CoRoT-23 | 8 | 681 |
| 18.88 +41.31 | KOI2148 | 8 | 681 |
| 18.97 +41.20 | KOI2448.01 | 8 | 681 |
| 19.81 +47.38 | KOI2390.01 | 8 | 680 |
| 18.81 +43.90 | KOI2328.01 | 8 | 680 |
| 19.41 +37.86 | KOI1030.01 | 8 | 679 |
| 19.82 +46.10 | KOI1470.01 | 8 | 679 |
| 19.11 +47.41 | KOI1905.01 | 8 | 679 |
| 19.98 +40.55 | KOI1935.01 | 8 | 679 |
| 19.33 +41.91 | KOI2445.01 | 8 | 678 |
| 16.10 -18.31 | UScoCTIO | 8 | 677 |
| 18.88 +41.71 | KOI2419.01 | 8 | 675 |
| 19.14 +46.64 | KOI1529.01 | 7 | 673 |
| 18.97 +39.13 | KOI1881.01 | 7 | 673 |
| 19.51 +38.40 | KOI2066.01 | 7 | 673 |
| 19.01 +40.90 | KOI2191.01 | 7 | 673 |
| 19.55 +48.28 | KOI2080 | 7 | 668 |
| 19.56 +41.35 | KOI1013.01 | 7 | 668 |
| 19.16 +38.09 | KOI1110.01 | 7 | 665 |
| 19.76 +42.39 | KOI1372.01 | 7 | 662 |
| 19.72 +50.91 | KOI1522.01 | 7 | 661 |
| 19.39 +38.06 | KOI1116.01 | 7 | 659 |
| 19.05 +41.61 | KOI2608.01 | 7 | 658 |
| 19.02 +41.99 | KOI1790.01 | 7 | 652 |
| 19.62 +28.50 | HIP96507 | 7 | 649 |
| 19.25 +40.39 | KOI1603.01 | 7 | 649 |
| 19.05 +45.69 | KOI2162.01 | 7 | 649 |
| 19.64 +43.37 | KOI1508.01 | 7 | 649 |
| 19.58 +46.38 | KOI2513.01 | 7 | 648 |
| 19.67 +40.14 | KOI2138.01 | 7 | 648 |
| 19.87 +40.95 | KOI2189.01 | 7 | 647 |
| 19.44 +47.04 | KOI1917.01 | 7 | 647 |
| 19.60 +46.31 | KOI2417.01 | 7 | 646 |
| 19.20 +46.12 | KOI1978 | 7 | 646 |
| 19.61 +46.71 | KOI1718.01 | 7 | 644 |
| 19.43 +41.70 | KOI1128.01 | 7 | 644 |
| 19.42 +44.93 | KOI1257.01 | 7 | 642 |
| 19.10 +38.41 | KOI1830.01 | 7 | 641 |
| 19.89 +45.07 | KOI2421.01 | 7 | 635 |
| 19.66 +46.49 | KOI2273.01 | 7 | 633 |
| 19.09 +38.74 | KOI1901.01 | 7 | 632 |
| 19.70 +42.48 | KOI2182.01 | 7 | 632 |
| 19.94 +45.44 | KOI2256.01 | 7 | 629 |
| 18.09 -28.89 | OGLE2003-BLG-235 | 7 | 627 |
| 16.47 -13.40 | HIP80687 | 7 | 627 |
| 19.58 +47.25 | KOI2342.01 | 7 | 626 |
| 19.01 +45.01 | KOI1853.01 | 7 | 624 |
| 23.21 -22.67 | USNO-B1.0 | 7 | 623 |
| 19.51 +46.73 | KOI1955 | 7 | 623 |
| 19.31 +40.71 | KOI1838.01 | 7 | 620 |
| 19.44 +41.31 | KOI2167.01 | 7 | 620 |
| 19.62 +47.28 | KOI1723.01 | 7 | 620 |
| 19.18 +48.98 | KOI2460.01 | 7 | 618 |
| 20.05 +28.31 | HIP98714 | 7 | 618 |
| 19.77 +44.26 | KOI1145.01 | 7 | 617 |
| 19.17 +38.40 | KOI1095.01 | 7 | 612 |
| 19.43 +47.21 | KOI2283.01 | 7 | 612 |
| 20.01 +22.71 | HIP98505 | 7 | 609 |
| 19.78 +49.21 | KOI1885.01 | 7 | 606 |
| 19.17 +38.89 | KOI2261 | 7 | 606 |
| 19.45 +44.75 | KOI1816.01 | 7 | 605 |
| 19.71 +44.53 | KOI2094.01 | 7 | 605 |
| 19.40 +42.28 | KOI1376.01 | 7 | 601 |
| 19.34 +46.71 | KOI1606.01 | 7 | 601 |
| 19.90 +41.89 | KOI1826.01 | 7 | 600 |
| 19.01 +44.11 | KOI1920.01 | 7 | 600 |
| 19.54 +45.66 | KOI2281.01 | 7 | 600 |
| 19.59 +47.03 | KOI1722.01 | 7 | 600 |
| 23.05 +00.43 | HD217786 | 7 | 596 |
| 19.78 +49.53 | KOI2225.01 | 7 | 592 |
| 19.29 +46.13 | KOI1866.01 | 7 | 592 |
| 19.19 +45.66 | KOI2443 | 7 | 592 |
| 19.83 +48.18 | KOI2378.01 | 7 | 591 |
| 19.85 +40.42 | KOI1576.01 | 7 | 591 |
| 17.07 -28.58 | HD154088 | 7 | 589 |
| 19.75 +39.37 | KOI1325.01 | 7 | 588 |
| 19.28 +49.94 | KOI1883.01 | 7 | 587 |
| 19.03 +39.71 | KOI2025 | 7 | 586 |
| 19.56 +42.82 | KOI1369.01 | 7 | 585 |
| 19.83 +46.82 | KOI1911.01 | 6 | 584 |
| 19.23 +43.76 | KOI1553.01 | 6 | 582 |
| 19.43 +42.25 | KOI1847.01 | 6 | 582 |
| 19.30 +42.50 | KOI2348.01 | 6 | 582 |
| 19.24 +42.61 | KOI2420.01 | 6 | 581 |
| 19.80 +46.39 | KOI1854.01 | 6 | 580 |
| 19.48 +41.58 | KOI2252.01 | 6 | 579 |
| 19.65 +46.98 | KOI1721.01 | 6 | 578 |
| 19.42 +47.69 | KOI2589.01 | 6 | 578 |
| 19.07 +45.14 | KOI2323.01 | 6 | 578 |
| 19.30 +50.94 | KOI2372.01 | 6 | 576 |
| 19.61 +46.17 | KOI2115.01 | 6 | 575 |
| 19.16 +47.77 | KOI2006 | 6 | 575 |
| 20.41 +16.76 | HAT-P-23 | 6 | 575 |
| 19.41 +45.78 | KOI2060.01 | 6 | 573 |
| 19.92 +40.76 | KOI2664.01 | 6 | 570 |
| 19.77 +44.01 | KOI1019.01 | 6 | 570 |
| 19.47 +38.82 | KOI1212.01 | 6 | 569 |
| 19.18 +42.24 | KOI2592.01 | 6 | 567 |
| 19.55 +39.91 | KOI1541.01 | 6 | 566 |
| 19.54 +45.25 | KOI2048.01 | 6 | 566 |
| 19.32 +38.26 | KOI1112.01 | 6 | 565 |
| 19.77 +49.10 | KOI2422.01 | 6 | 564 |
| 19.07 +44.76 | KOI2185.01 | 6 | 564 |
| 19.47 +49.24 | KOI2143.01 | 6 | 563 |
| 19.45 +37.69 | KOI2033.01 | 6 | 562 |
| 19.59 +38.92 | KOI1191.01 | 6 | 561 |
| 19.55 +42.37 | KOI1798.01 | 6 | 561 |
| 19.09 +42.41 | KOI1366.01 | 6 | 560 |
| 19.37 +47.73 | KOI2208.01 | 6 | 560 |
| 19.47 +44.88 | KOI1893.01 | 6 | 560 |
| 19.45 +01.38 | GSC00465-01282 | 6 | 560 |
| 19.29 +46.99 | KOI1783.01 | 6 | 560 |
| 19.41 +49.19 | KOI2090.01 | 6 | 559 |
| 19.02 +44.55 | KOI2402.01 | 6 | 557 |
| 19.44 +01.43 | CoRoT-8 | 6 | 556 |
| 19.46 +38.25 | KOI1950.01 | 6 | 556 |
| 19.27 +48.12 | KOI1288.01 | 6 | 555 |
| 19.63 +49.33 | KOI2110.01 | 6 | 554 |
| 20.47 +18.77 | HIP100970 | 6 | 553 |
| 19.51 +44.39 | KOI1879.01 | 6 | 553 |
| 19.19 +43.15 | KOI2524.01 | 6 | 551 |
| 17.93 -28.48 | MOA-2011-BLG-293L | 6 | 547 |
| 19.16 +45.70 | KOI1406.01 | 6 | 546 |
| 19.38 +37.25 | KOI993.01 | 6 | 543 |
| 19.49 +48.14 | KOI2373.01 | 6 | 543 |
| 19.52 +37.80 | KOI2009.01 | 6 | 542 |
| 19.61 +44.77 | KOI1258.01 | 6 | 541 |
| 19.18 +38.11 | KOI1788.01 | 6 | 541 |
| 19.24 +40.62 | KOI1567 | 6 | 541 |
| 19.48 +42.15 | KOI1244.01 | 6 | 541 |
| 19.09 +37.55 | KOI2201.01 | 6 | 541 |
| 17.38 -19.62 | HD157172 | 6 | 541 |
| 19.13 +44.99 | KOI1281.01 | 6 | 541 |
| 18.93 +04.27 | HIP92895 | 6 | 541 |
| 19.19 +40.55 | KOI2555.01 | 6 | 541 |
| 19.51 +39.12 | KOI1201.01 | 6 | 540 |
| 19.03 +45.05 | KOI2039.01 | 6 | 540 |
| 19.73 +45.81 | KOI1900.01 | 6 | 539 |
| 19.29 +46.46 | KOI1913.01 | 6 | 539 |
| 19.60 +41.59 | KOI2638.01 | 6 | 539 |
| 19.95 +45.23 | KOI1874 | 6 | 537 |
| 19.01 +43.83 | KOI2352 | 6 | 537 |
| 19.21 +40.37 | KOI2563.01 | 6 | 537 |
| 19.49 +44.65 | KOI2346.01 | 6 | 537 |
| 19.42 +46.97 | KOI2309.01 | 6 | 536 |
| 19.92 +45.47 | KOI2042.01 | 6 | 534 |
| 19.49 +38.28 | KOI1111.01 | 6 | 534 |
| 19.61 +42.37 | KOI2220 | 6 | 533 |
| 19.42 +40.16 | KOI1605.01 | 6 | 533 |
| 19.31 +41.59 | KOI2636.01 | 6 | 530 |
| 19.21 +45.68 | KOI2416.01 | 6 | 529 |
| 19.86 +40.28 | KOI1822.01 | 6 | 528 |
| 19.58 +45.82 | KOI1971.01 | 6 | 528 |
| 19.57 +46.74 | KOI1717.01 | 6 | 528 |
| 19.75 +47.68 | KOI1170.01 | 6 | 526 |
| 19.64 +46.39 | KOI2044.01 | 6 | 524 |
| 19.61 +46.31 | KOI2272.01 | 6 | 523 |
| 19.72 +40.01 | KOI1873.01 | 6 | 523 |
| 19.56 +39.21 | KOI1320.01 | 6 | 523 |
| 19.34 +43.09 | KOI1501.01 | 6 | 523 |
| 19.62 +42.25 | KOI1749.01 | 6 | 523 |
| 19.11 +47.43 | KOI2214.01 | 6 | 523 |
| 19.00 +43.91 | KOI1570.01 | 6 | 523 |
| 19.35 +40.55 | KOI2219.01 | 6 | 523 |
| 19.78 +46.37 | KOI2132.01 | 6 | 523 |
| 19.37 +51.70 | KOI1583.01 | 6 | 522 |
| 19.68 +40.90 | KOI2010.01 | 6 | 522 |
| 19.82 +40.55 | KOI2011.01 | 6 | 522 |
| 19.63 +46.69 | TYC3556-00408-1 | 6 | 522 |
| 19.09 +44.78 | KOI2111 | 6 | 522 |
| 19.50 +42.14 | KOI1245.01 | 6 | 522 |
| 19.01 +45.00 | KOI2159.01 | 6 | 521 |
| 19.48 +43.78 | KOI2274.01 | 6 | 521 |
| 19.71 +38.95 | KOI2289.01 | 6 | 521 |
| 19.80 +42.98 | KOI2392.01 | 6 | 521 |
| 19.35 +44.44 | KOI1914.01 | 6 | 521 |
| 19.40 +45.24 | KOI1412.01 | 6 | 520 |
| 19.43 +37.74 | KOI2603.01 | 6 | 520 |
| 19.54 +42.90 | KOI1862.01 | 6 | 520 |
| 19.68 +49.95 | KOI1980.01 | 6 | 519 |
| 19.35 +38.32 | KOI1096.01 | 6 | 519 |
| 19.47 +08.36 | HIP95740 | 6 | 518 |
| 19.43 +38.13 | KOI1824 | 6 | 514 |
| 19.85 +47.11 | KOI2097.01 | 6 | 514 |
| 19.51 +49.92 | KOI1531.01 | 6 | 513 |
| 19.79 +46.74 | KOI2135.01 | 6 | 512 |
| 19.06 +48.68 | KOI2206.01 | 6 | 512 |
| 19.32 +46.52 | KOI1520.01 | 6 | 511 |
| 19.14 +46.51 | KOI1459.01 | 6 | 511 |
| 19.55 +42.20 | KOI2092 | 6 | 510 |
| 19.50 +39.19 | KOI1222.01 | 6 | 510 |
| 19.55 +43.05 | KOI2063.01 | 6 | 510 |
| 19.43 +50.76 | KOI1536.01 | 6 | 510 |
| 19.61 +41.45 | KOI1051.01 | 6 | 509 |
| 19.43 +37.06 | KOI992.01 | 6 | 508 |
| 19.75 +43.82 | KOI2304.01 | 6 | 508 |
| 19.66 +46.44 | KOI2287.01 | 6 | 507 |
| 19.33 +47.02 | KOI2628.01 | 6 | 505 |
| 19.27 +51.76 | Kepler-16 | 6 | 504 |
| 19.42 +36.78 | KOI1010.01 | 6 | 504 |
| 19.34 +38.40 | KOI1114.01 | 6 | 504 |
| 19.33 +38.54 | KOI1204.01 | 6 | 504 |
| 19.73 +38.95 | KOI1205.01 | 6 | 504 |
| 19.49 +39.55 | KOI1338.01 | 6 | 504 |
| 19.37 +45.60 | KOI1403.01 | 6 | 504 |
| 19.34 +41.60 | KOI1812.01 | 6 | 504 |
| 19.69 +46.24 | KOI1819.01 | 6 | 504 |
| 19.15 +39.29 | KOI2210.01 | 6 | 504 |
| 19.02 +48.81 | KOI2497.01 | 6 | 504 |
| 19.20 +40.34 | KOI2532.01 | 6 | 504 |
| 19.59 +38.64 | KOI1194 | 6 | 503 |
| 19.34 +44.05 | KOI1858 | 6 | 503 |
| 19.48 +40.81 | KOI1814.01 | 6 | 503 |
| 19.22 +46.77 | KOI1872.01 | 6 | 503 |
| 19.48 +41.55 | KOI1967.01 | 6 | 503 |
| 19.04 +44.12 | KOI2133.01 | 6 | 503 |
| 19.08 +44.65 | KOI2658.01 | 6 | 503 |
| 19.76 +48.05 | KOI989.03 | 6 | 503 |
| 19.79 +47.35 | KOI1174.01 | 6 | 503 |
| 19.70 +42.93 | KOI1378.01 | 6 | 503 |
| 19.17 +42.81 | KOI1961.01 | 6 | 503 |
| 19.73 +44.72 | KOI2656.01 | 6 | 503 |
| 19.34 +49.92 | KOI1841.01 | 6 | 502 |
| 19.40 +39.14 | KOI2119.01 | 6 | 502 |
| 19.69 +38.74 | KOI2279 | 6 | 502 |
| 19.30 +39.27 | KOI1864.01 | 6 | 502 |
| 19.54 +43.07 | KOI1890.01 | 6 | 502 |
| 19.37 +44.93 | KOI2017.01 | 6 | 502 |
| 19.73 +39.92 | KOI1884 | 6 | 502 |
| 19.13 +44.71 | KOI2263.01 | 6 | 502 |
| 19.75 +47.00 | KOI1586.01 | 6 | 501 |
| 19.57 +39.03 | KOI1210.01 | 6 | 501 |
| 19.74 +45.63 | KOI2491.01 | 6 | 501 |
| 19.51 +44.09 | KOI1867 | 6 | 501 |
| 19.23 +46.59 | KOI1852.01 | 6 | 501 |
| 19.32 +38.31 | KOI2150.01 | 6 | 501 |
| 19.42 +44.01 | KOI2073 | 6 | 501 |
| 19.72 +42.54 | KOI1899.01 | 6 | 501 |
| 19.52 +48.90 | KOI2329.01 | 6 | 501 |
| 19.71 +42.55 | KOI2432.01 | 6 | 501 |
| 19.04 +45.64 | KOI2528.01 | 6 | 501 |
| 19.49 +37.68 | KOI2053 | 6 | 500 |
| 19.80 +41.38 | KOI2001.01 | 6 | 500 |
| 19.84 +47.17 | KOI1784.01 | 6 | 499 |
| 19.57 +42.38 | KOI1732.01 | 6 | 499 |
| 19.47 +38.84 | KOI1176.01 | 6 | 498 |
| 19.49 +40.91 | KOI1908 | 6 | 497 |
| 19.10 +46.75 | KOI2509.01 | 6 | 497 |
| 19.75 +41.32 | KOI1799.01 | 6 | 496 |
| 18.10 +26.43 | HAT-P-31 | 6 | 496 |
| 19.67 +50.11 | KOI1510.01 | 6 | 495 |
| 19.69 +40.56 | KOI1825.01 | 5 | 494 |
| 19.40 +38.55 | KOI1218.01 | 5 | 494 |
| 19.46 +44.56 | KOI2362.02 | 5 | 494 |
| 19.34 +47.26 | KOI2212.01 | 5 | 492 |
| 19.71 +42.56 | KOI2494.01 | 5 | 491 |
| 17.34 -19.33 | HIP84856 | 5 | 490 |
| 17.98 -29.19 | SWEEPS4 | 5 | 489 |
| 17.94 -29.54 | V*V5157 | 5 | 489 |
| 19.42 +42.10 | KOI1736.01 | 5 | 487 |
| 19.80 +44.33 | KOI2145.01 | 5 | 486 |
| 19.29 +40.00 | KOI2617.01 | 5 | 483 |
| 19.05 +40.86 | KOI2369 | 5 | 483 |
| 19.63 +46.95 | KOI2655.01 | 5 | 483 |
| 19.40 +42.46 | KOI1359 | 5 | 482 |
| 19.95 +40.34 | KOI1563.01 | 5 | 481 |
| 19.75 +50.28 | KOI2026.01 | 5 | 481 |
| 19.42 +50.57 | KOI1502.01 | 5 | 479 |
| 19.39 +47.44 | KOI2401.01 | 5 | 478 |
| 19.89 +45.32 | KOI1988.01 | 5 | 475 |
| 19.31 +46.31 | KOI1457.01 | 5 | 474 |
| 19.15 +46.07 | KOI1813.01 | 5 | 473 |
| 19.28 +49.31 | KOI1992.01 | 5 | 473 |
| 19.87 +40.28 | KOI2241.01 | 5 | 473 |
| 19.93 +40.14 | KOI2462.01 | 5 | 473 |
| 19.60 +46.69 | KOI1958.01 | 5 | 473 |
| 19.42 +49.26 | KOI2195.01 | 5 | 471 |
| 19.94 +43.44 | KOI2523.01 | 5 | 471 |
| 19.74 +48.96 | KOI1972.01 | 5 | 470 |
| 19.11 +47.10 | KOI1888.01 | 5 | 469 |
| 16.39 -26.53 | PSRB1620-26 | 5 | 469 |
| 19.93 +44.13 | KOI1986.01 | 5 | 469 |
| 19.76 +49.18 | KOI2069.01 | 5 | 469 |
| 19.40 +49.20 | KOI1428.01 | 5 | 468 |
| 19.52 +45.73 | KOI2657.01 | 5 | 468 |
| 19.85 +45.42 | KOI2470.01 | 5 | 464 |
| 19.58 +44.62 | KOI2022 | 5 | 464 |
| 19.49 +43.03 | KOI2598.01 | 5 | 463 |
| 19.38 +40.14 | KOI2406.01 | 5 | 462 |
| 19.24 +50.79 | KOI2004.01 | 5 | 458 |
| 19.31 +49.78 | KOI1965.01 | 5 | 456 |
| 19.49 +49.65 | KOI1424.01 | 5 | 455 |
| 19.28 +50.94 | KOI2387.01 | 5 | 454 |
| 19.77 +46.97 | KOI2409.01 | 5 | 451 |
| 19.79 +49.08 | KOI2168 | 5 | 451 |
| 18.05 -28.56 | HIP88414 | 5 | 451 |
| 19.26 +39.15 | KOI2519.01 | 5 | 451 |
| 18.95 +32.90 | HD176051 | 5 | 451 |
| 19.55 +49.32 | KOI2047.01 | 5 | 450 |
| 19.37 +51.06 | KOI2541.01 | 5 | 450 |
| 19.05 +44.86 | KOI2410 | 5 | 449 |
| 19.42 +44.49 | KOI2624.01 | 5 | 448 |
| 19.62 +47.22 | KOI1724.01 | 5 | 438 |
| 19.66 +44.15 | KOI1726.01 | 5 | 438 |
| 19.08 +41.04 | Kepler-12 | 5 | 435 |
| 19.37 +38.50 | KOI1845 | 5 | 432 |
| 19.93 +43.75 | KOI1552.01 | 5 | 429 |
| 19.57 +45.38 | KOI2564.01 | 5 | 428 |
| 19.64 +46.51 | KOI2600.01 | 5 | 425 |
| 19.09 +25.92 | HIP93746 | 5 | 424 |
| 22.96 -29.62 | HIP113368 | 5 | 422 |
| 19.73 +45.38 | KOI1402.01 | 5 | 417 |
| 19.27 +47.12 | KOI2056.01 | 5 | 413 |
| 19.15 +38.41 | KOI2157.01 | 5 | 412 |
| 19.52 +43.43 | KOI1921.01 | 5 | 411 |
| 19.46 +45.39 | KOI2175.01 | 5 | 411 |
| 19.75 +42.53 | KOI2215.01 | 5 | 411 |
| 19.64 +45.67 | KOI2107.01 | 5 | 411 |
| 19.14 +45.18 | KOI1404.01 | 5 | 410 |
| 18.46 -29.82 | HIP90485 | 5 | 410 |
| 20.25 -27.03 | Gl785 | 5 | 408 |
| 19.63 +40.56 | KOI1557.01 | 5 | 406 |
| 19.40 +47.99 | KOI2046.01 | 5 | 405 |
| 19.05 +42.81 | KOI2051 | 4 | 403 |
| 19.90 +08.46 | HIP97938 | 4 | 396 |
| 19.79 +50.04 | KOI2371.01 | 4 | 396 |
| 19.05 +45.71 | KOI2397.01 | 4 | 396 |
| 19.37 +38.14 | KOI2032.01 | 4 | 395 |
| 19.46 +40.95 | KOI2091.01 | 4 | 393 |
| 19.77 +44.13 | KOI2344.01 | 4 | 393 |
| 11.03 +23.86 | WASP-34 | 4 | 392 |
| 19.46 +41.53 | KOI1882.01 | 4 | 392 |
| 19.76 +39.60 | KOI2547.01 | 4 | 392 |
| 19.35 +46.70 | KOI2586.01 | 4 | 392 |
| 19.14 +42.42 | KOI2294.01 | 4 | 392 |
| 19.53 +38.30 | KOI2490.01 | 4 | 392 |
| 19.33 +45.21 | KOI2517.01 | 4 | 392 |
| 19.19 +45.34 | KOI2530.01 | 4 | 392 |
| 19.29 +51.41 | KOI1577.01 | 4 | 391 |
| 16.21 -18.88 | HIP79431 | 4 | 389 |
| 18.03 +00.10 | HD164509 | 4 | 389 |
| 19.37 +45.21 | KOI1387.01 | 4 | 388 |
| 20.52 +06.43 | GSC00522-01199 | 4 | 387 |
| 19.28 +47.06 | KOI2602.01 | 4 | 386 |
| 19.78 +48.27 | KOI2305.01 | 4 | 384 |
| 19.50 +45.97 | KOI2125.01 | 4 | 377 |
| 19.45 +38.07 | KOI2081.01 | 4 | 374 |
| 19.63 +39.40 | KOI1331.01 | 4 | 373 |
| 19.74 +45.99 | KOI2121.01 | 4 | 372 |
| 19.70 +45.81 | KOI2129.01 | 4 | 372 |
| 19.25 +39.77 | KOI1805 | 4 | 372 |
| 19.61 +42.81 | KOI1974.01 | 4 | 371 |
| 19.31 +43.46 | KOI2324.01 | 4 | 371 |
| 19.17 +49.15 | KOI2366.01 | 4 | 371 |
| 19.70 +42.86 | KOI2516.01 | 4 | 371 |
| 19.45 +38.15 | KOI2335.01 | 4 | 370 |
| 19.26 +42.27 | KOI2086 | 4 | 370 |
| 19.48 +37.89 | KOI2064.01 | 4 | 370 |
| 19.68 +42.73 | KOI2487.01 | 4 | 370 |
| 19.38 +43.86 | KOI2615.01 | 4 | 370 |
| 19.24 +39.82 | KOI1996.01 | 4 | 369 |
| 19.11 +37.89 | KOI2156.01 | 4 | 369 |
| 19.20 +39.23 | KOI2202.01 | 4 | 369 |
| 20.27 +04.58 | HIP99894 | 4 | 368 |
| 19.72 +47.60 | KOI2040.01 | 4 | 367 |
| 19.62 +50.34 | KOI2433 | 4 | 365 |
| 19.37 +44.56 | KOI2016.01 | 4 | 361 |
| 19.08 +38.26 | KOI1878.01 | 4 | 359 |
| 19.72 +45.46 | KOI1399.01 | 4 | 355 |
| 19.08 +37.41 | KOI2634.01 | 4 | 354 |
| 19.45 +45.44 | KOI2640.01 | 4 | 354 |
| 19.06 +38.21 | KOI2186 | 4 | 354 |
| 19.75 +43.94 | KOI2504.01 | 4 | 354 |
| 19.77 +44.33 | KOI2507.01 | 4 | 354 |
| 19.44 +42.04 | KOI2559.01 | 4 | 354 |
| 19.80 +39.93 | KOI2120.01 | 4 | 353 |
| 19.68 +45.76 | KOI2222.01 | 4 | 353 |
| 19.46 +44.90 | KOI2473.01 | 4 | 353 |
| 19.06 +45.74 | KOI1398.01 | 4 | 353 |
| 19.51 +44.08 | KOI2209.01 | 4 | 353 |
| 19.71 +39.76 | KOI2560.01 | 4 | 353 |
| 19.72 +47.94 | KOI1797.01 | 4 | 351 |
| 19.40 +45.29 | KOI2038 | 4 | 351 |
| 19.26 -24.18 | HIP94645 | 4 | 351 |
| 19.60 +42.88 | KOI2450.01 | 4 | 351 |
| 19.40 +45.19 | KOI2595.01 | 4 | 351 |
| 19.07 +38.01 | KOI1942.01 | 4 | 351 |
| 19.74 +42.77 | KOI2354.01 | 4 | 351 |
| 19.77 +46.86 | KOI2299.01 | 4 | 349 |
| 20.23 -00.87 | HIP99711 | 4 | 348 |
| 19.16 +45.92 | KOI2383.01 | 4 | 347 |
| 19.58 +46.85 | KOI1925.01 | 4 | 347 |
| 19.83 +46.66 | KOI2423.01 | 4 | 346 |
| 19.72 +43.43 | KOI1840.01 | 4 | 344 |
| 19.32 +49.83 | KOI1832 | 4 | 344 |
| 19.05 +41.34 | KOI2045 | 4 | 342 |
| 16.19 -27.08 | HIP79346 | 4 | 341 |
| 19.70 +40.78 | KOI2587.01 | 4 | 336 |
| 19.17 +46.14 | KOI2076.01 | 4 | 334 |
| 19.70 +45.24 | KOI2233.01 | 4 | 334 |
| 19.52 +48.22 | KOI2368.01 | 4 | 334 |
| 19.05 +44.87 | KOI2544.01 | 4 | 334 |
| 19.77 +44.14 | KOI2569.01 | 4 | 334 |
| 19.35 -23.62 | HD181342 | 4 | 333 |
| 18.74 +06.66 | CoRoT-6 | 4 | 333 |
| 19.65 +43.46 | KOI1808.01 | 4 | 333 |
| 19.70 +40.73 | KOI2160.01 | 4 | 333 |
| 19.39 +45.41 | KOI2367.01 | 4 | 333 |
| 19.44 +41.21 | KOI2498.01 | 4 | 333 |
| 18.10 -30.73 | OGLE2005-BLG-169 | 4 | 333 |
| 19.59 +42.78 | KOI2113 | 4 | 333 |
| 19.42 +38.36 | KOI2413 | 4 | 332 |
| 19.80 +50.00 | KOI2269.01 | 4 | 332 |
| 19.64 +38.46 | KOI2444.01 | 4 | 332 |
| 18.13 -27.15 | MOA-2007-BLG-192-L | 4 | 332 |
| 19.66 +39.09 | KOI2084.01 | 4 | 332 |
| 19.45 +38.62 | KOI2126.01 | 4 | 332 |
| 19.80 +47.75 | KOI2131.01 | 4 | 332 |
| 19.08 +42.38 | KOI2477.01 | 4 | 332 |
| 19.32 +39.52 | KOI2303.01 | 4 | 331 |
| 18.12 -26.82 | MOA-2009-BLG-319 | 4 | 331 |
| 19.44 +38.41 | KOI2278 | 4 | 331 |
| 18.49 +11.70 | HIP90593 | 4 | 331 |
| 19.69 +46.60 | KOI2012.01 | 4 | 331 |
| 19.64 +39.78 | KOI2393.01 | 4 | 331 |
| 19.17 +44.18 | KOI1069.02 | 4 | 331 |
| 19.82 +46.35 | KOI2654.01 | 4 | 330 |
| 19.15 +39.38 | KOI2216.01 | 4 | 327 |
| 19.12 +43.19 | KOI1480.01 | 4 | 325 |
| 19.25 +31.86 | HD180314 | 4 | 323 |
| 19.13 +38.32 | KOI2314.01 | 4 | 321 |
| 19.19 +45.98 | KOI2472.01 | 4 | 315 |
| 19.24 +46.22 | KOI2217.01 | 4 | 315 |
| 19.39 +45.35 | KOI2604.01 | 4 | 315 |
| 19.06 +38.52 | KOI2662.01 | 4 | 315 |
| 18.16 -29.22 | MOA2007-BLG-400-L | 4 | 315 |
| 19.26 +39.59 | KOI1332 | 4 | 315 |
| 19.11 +43.04 | KOI1848.01 | 4 | 315 |
| 19.09 +43.32 | KOI1898.01 | 4 | 315 |
| 19.40 +45.72 | KOI2223.01 | 4 | 315 |
| 19.32 -23.56 | HD180902 | 3 | 314 |
| 19.49 +38.66 | KOI2194.01 | 3 | 314 |
| 19.82 +46.90 | KOI2312.01 | 3 | 314 |
| 19.33 +41.50 | KOI2398.01 | 3 | 314 |
| 19.43 +45.59 | KOI2625.01 | 3 | 314 |
| 19.67 +45.12 | KOI2641.01 | 3 | 314 |
| 19.17 +49.52 | KOI1781 | 3 | 312 |
| 19.23 +44.80 | KOI1891 | 3 | 312 |
| 19.60 +39.57 | KOI2229.01 | 3 | 312 |
| 19.36 +37.77 | KOI2280.01 | 3 | 312 |
| 19.07 +44.78 | KOI2414 | 3 | 311 |
| 19.26 +44.84 | KOI2340.01 | 3 | 311 |
| 19.40 +42.27 | KOI2062.01 | 3 | 311 |
| 19.73 +43.16 | KOI2302.01 | 3 | 311 |
| 19.25 +39.52 | KOI2245.01 | 3 | 310 |
| 19.50 +42.65 | KOI2325.01 | 3 | 310 |
| 19.46 +41.77 | KOI2404.01 | 3 | 310 |
| 19.62 +44.79 | KOI1271.01 | 3 | 310 |
| 19.74 +45.43 | KOI1915.01 | 3 | 309 |
| 19.72 +47.68 | KOI2320.01 | 3 | 308 |
| 19.43 +40.13 | KOI1609.01 | 3 | 307 |
| 19.64 +46.62 | KOI2461.01 | 3 | 307 |
| 19.53 +43.18 | KOI2601.01 | 3 | 306 |
| 19.46 +44.09 | KOI2105.01 | 3 | 303 |
| 19.43 +39.13 | KOI2013.01 | 3 | 301 |
| 19.64 +46.08 | KOI2082.01 | 3 | 296 |
| 19.44 +39.23 | KOI1860 | 3 | 295 |
| 19.35 +41.33 | KOI1916 | 3 | 295 |
| 17.98 -29.20 | SWEEPS11 | 3 | 295 |
| 19.67 +42.73 | KOI1379.01 | 3 | 295 |
| 19.22 +46.61 | KOI1489.01 | 3 | 295 |
| 19.64 +43.34 | KOI1528.01 | 3 | 295 |
| 19.61 +40.46 | KOI1547.01 | 3 | 295 |
| 19.45 +39.45 | KOI1738.01 | 3 | 295 |
| 19.88 +40.53 | KOI1762.01 | 3 | 295 |
| 19.38 +42.24 | KOI1868.01 | 3 | 295 |
| 19.07 +37.97 | KOI1976.01 | 3 | 295 |
| 19.65 +38.36 | KOI2052.01 | 3 | 295 |
| 19.29 +39.34 | KOI2134.01 | 3 | 295 |
| 19.83 +39.88 | KOI2243.01 | 3 | 295 |
| 19.62 +45.20 | KOI2253.01 | 3 | 295 |
| 19.66 +38.59 | KOI2276.01 | 3 | 295 |
| 19.82 +41.31 | KOI2315.01 | 3 | 295 |
| 19.48 +40.79 | KOI2332.01 | 3 | 295 |
| 19.84 +40.54 | KOI2370.01 | 3 | 295 |
| 19.67 +46.47 | KOI2379.01 | 3 | 295 |
| 19.49 +40.84 | KOI2415.01 | 3 | 295 |
| 19.71 +44.69 | KOI2436.01 | 3 | 295 |
| 19.86 +40.16 | KOI2437.01 | 3 | 295 |
| 19.79 +44.37 | KOI2468.01 | 3 | 295 |
| 19.60 +46.72 | KOI2579.01 | 3 | 295 |
| 19.48 +41.41 | KOI2627.01 | 3 | 295 |
| 19.09 +37.41 | KOI2534 | 3 | 294 |
| 19.48 +37.24 | KOI1007.01 | 3 | 294 |
| 19.60 +38.90 | KOI1206.01 | 3 | 294 |
| 19.83 +40.68 | KOI1585.01 | 3 | 294 |
| 19.55 +41.52 | KOI1750.01 | 3 | 294 |
| 19.36 +37.72 | KOI1944.01 | 3 | 294 |
| 19.83 +41.56 | KOI2019.01 | 3 | 294 |
| 19.28 +46.81 | KOI2285.01 | 3 | 294 |
| 19.35 +41.52 | KOI2350.01 | 3 | 294 |
| 19.50 +37.32 | KOI2351.01 | 3 | 294 |
| 19.44 +50.83 | KOI2449.01 | 3 | 294 |
| 19.43 +44.68 | KOI2488.01 | 3 | 294 |
| 19.25 +42.21 | KOI2506.01 | 3 | 294 |
| 19.76 +44.64 | Kepler-34(AB) | 3 | 294 |
| 19.27 +49.77 | KOI1945 | 3 | 294 |
| 19.48 +38.74 | KOI1861.01 | 3 | 294 |
| 19.27 +42.88 | KOI1979.01 | 3 | 294 |
| 19.30 +41.32 | KOI2071.01 | 3 | 294 |
| 19.44 +39.40 | KOI2122.01 | 3 | 294 |
| 19.36 +44.11 | KOI2238.01 | 3 | 294 |
| 19.36 +41.43 | KOI2343.01 | 3 | 294 |
| 19.24 +51.50 | KOI2353.01 | 3 | 294 |
| 19.63 +44.76 | KOI2503.01 | 3 | 294 |
| 19.35 +41.78 | KOI1127 | 3 | 293 |
| 19.66 +39.03 | KOI1760 | 3 | 293 |
| 19.82 +49.98 | KOI2173 | 3 | 293 |
| 19.48 +40.92 | KOI2183 | 3 | 293 |
| 19.55 +38.03 | KOI1103.01 | 3 | 293 |
| 19.13 +42.47 | KOI1367.01 | 3 | 293 |
| 19.20 +41.13 | KOI1787.01 | 3 | 293 |
| 19.64 +45.06 | KOI1850.01 | 3 | 293 |
| 19.13 +42.58 | KOI2128.01 | 3 | 293 |
| 19.39 +42.29 | KOI2146.01 | 3 | 293 |
| 19.84 +40.35 | KOI2158.01 | 3 | 293 |
| 19.48 +38.89 | KOI2308.01 | 3 | 293 |
| 19.68 +40.98 | KOI2358.01 | 3 | 293 |
| 19.61 +38.11 | KOI2380.01 | 3 | 293 |
| 19.54 +42.32 | KOI2439.01 | 3 | 293 |
| 19.51 +37.93 | KOI2489.01 | 3 | 293 |
| 19.54 +42.35 | KOI2571.01 | 3 | 293 |
| 19.23 +40.27 | KOI1801.01 | 3 | 293 |
| 19.49 +43.91 | KOI2338.01 | 3 | 293 |
| 19.06 +42.73 | KOI2521 | 3 | 292 |
| 19.61 +38.23 | KOI2055.01 | 3 | 292 |
| 19.53 +41.93 | KOI2093.01 | 3 | 292 |
| 19.47 +49.26 | KOI2482.01 | 3 | 292 |
| 19.74 +47.14 | KOI1773.01 | 3 | 292 |
| 19.46 +44.12 | KOI2140.01 | 3 | 292 |
| 19.54 +39.40 | KOI2166.01 | 3 | 292 |
| 19.55 +42.87 | KOI2180.01 | 3 | 292 |
| 19.68 +45.97 | KOI1977 | 3 | 292 |
| 19.65 +45.31 | KOI1892.01 | 3 | 292 |
| 19.77 +44.94 | KOI1904.01 | 3 | 292 |
| 19.62 +40.21 | KOI1924.01 | 3 | 292 |
| 19.38 +43.61 | KOI1964.01 | 3 | 292 |
| 19.55 +42.72 | KOI2164.01 | 3 | 292 |
| 19.45 +38.09 | KOI2232.01 | 3 | 292 |
| 19.30 +39.16 | KOI2295.01 | 3 | 292 |
| 19.38 +43.26 | KOI2458.01 | 3 | 292 |
| 19.65 +39.60 | KOI2481.01 | 3 | 292 |
| 19.67 +40.57 | KOI2556.01 | 3 | 292 |
| 19.38 +40.26 | KOI2607.01 | 3 | 292 |
| 19.29 +46.97 | KOI1940 | 3 | 291 |
| 19.61 +45.15 | KOI2224 | 3 | 291 |
| 19.46 +42.43 | KOI1960.01 | 3 | 291 |
| 19.32 +43.47 | KOI1952 | 3 | 291 |
| 19.49 +38.27 | KOI2099.01 | 3 | 291 |
| 19.53 +41.28 | KOI2479.01 | 3 | 291 |
| 19.48 +42.01 | KOI1239 | 3 | 290 |
| 19.61 +38.22 | KOI1786.01 | 3 | 290 |
| 19.24 +43.37 | KOI2585.01 | 3 | 290 |
| 19.15 +39.33 | KOI2311 | 3 | 290 |
| 19.40 +50.31 | KOI2647.01 | 3 | 290 |
| 19.24 +40.99 | KOI1897.01 | 3 | 289 |
| 19.63 +50.67 | KOI2155.01 | 3 | 289 |
| 19.52 +44.18 | KOI2347.01 | 3 | 289 |
| 19.68 +45.84 | KOI2078.01 | 3 | 289 |
| 19.46 +39.31 | KOI1895 | 3 | 288 |
| 19.41 +44.06 | KOI2116.01 | 3 | 287 |
| 19.77 +43.36 | KOI1829.01 | 3 | 285 |
| 19.09 +37.39 | KOI2538.01 | 3 | 283 |
| 19.63 +44.75 | KOI2453.01 | 3 | 280 |
| 19.51 +46.42 | KOI2613.01 | 3 | 279 |
| 20.23 +00.87 | HD192263 | 3 | 276 |
| 19.64 +40.83 | KOI2345.01 | 3 | 272 |
| 19.45 +42.34 | KOI1982.01 | 3 | 270 |
| 19.53 +16.47 | HIP96078 | 2 | 157 |
| 18.58 -28.07 | HIP91085 | 2 | 156 |

HabCat

| | | | |
|---|---|---|---|
| 09.34 +33.76 | HIP45822 | 100 | 9000 |
| 12.08 +76.76 | HIP58924 | 97 | 8722 |
| 11.02 -23.87 | HIP53873 | 69 | 6208 |
| 05.36 +79.05 | HIP25022 | 67 | 5997 |
| 23.60 +00.45 | HIP116454 | 56 | 5015 |
| 15.25 +71.75 | HIP74644 | 55 | 4918 |
| 07.45 +24.16 | HIP36180 | 52 | 4661 |
| 07.55 -22.30 | HIP36731 | 50 | 4500 |
| 18.43 +65.33 | HIP90334 | 43 | 3911 |
| 05.13 +69.53 | HIP23869 | 42 | 3745 |
| 04.70 +23.06 | HIP21868 | 35 | 3140 |
| 01.76 +20.31 | HIP8221 | 33 | 2983 |
| 04.45 +19.12 | HIP20769 | 32 | 2882 |
| 06.74 -00.93 | HIP32270 | 28 | 2521 |
| 04.43 +57.50 | HIP20690 | 27 | 2472 |
| 17.07 +47.11 | HIP83503 | 27 | 2430 |
| 09.39 +33.91 | HIP46022 | 27 | 2399 |
| 03.19 +31.26 | HIP14840 | 25 | 2289 |
| 04.68 +23.81 | HIP21788 | 25 | 2277 |
| 11.49 +43.80 | HIP56064 | 25 | 2222 |
| 03.22 +20.73 | HIP14974 | 24 | 2168 |
| 02.79 +49.81 | HIP13034 | 24 | 2145 |
| 10.18 +18.57 | HIP49869 | 24 | 2141 |
| 07.53 +17.32 | HIP36637 | 23 | 2100 |
| 19.63 +43.63 | HIP96561 | 23 | 2079 |
| 16.35 +40.96 | HIP80093 | 23 | 2078 |
| 10.29 +12.29 | HIP50396 | 23 | 2045 |
| 17.15 +33.22 | HIP83899 | 23 | 2032 |
| 11.60 +20.11 | HIP56565 | 21 | 1912 |
| 19.25 +51.15 | HIP94565 | 21 | 1870 |
| 21.23 +14.95 | HIP104815 | 21 | 1856 |
| 06.80 +00.30 | HIP32608 | 20 | 1838 |
| 05.62 +20.36 | HIP26387 | 20 | 1790 |
| 02.78 +50.12 | HIP12984 | 20 | 1788 |
| 14.53 +22.08 | HIP71072 | 20 | 1774 |
| 23.15 +18.41 | HIP114320 | 20 | 1756 |
| 08.82 +28.98 | HIP43274 | 19 | 1723 |
| 02.81 +72.34 | HIP13115 | 19 | 1722 |
| 22.83 +35.50 | HIP112712 | 18 | 1633 |
| 03.06 -21.36 | HIP14249 | 18 | 1629 |
| 04.42 +39.05 | HIP20633 | 18 | 1599 |
| 03.39 -29.37 | HIP15764 | 17 | 1571 |
| 09.27 +33.91 | HIP45484 | 17 | 1552 |
| 06.28 -29.87 | HIP29836 | 17 | 1541 |
| 04.54 +04.14 | HIP21186 | 17 | 1494 |
| 03.28 +25.59 | HIP15273 | 17 | 1493 |
| 08.94 +28.67 | HIP43882 | 16 | 1479 |
| 07.36 +27.07 | HIP35659 | 16 | 1413 |
| 01.52 +29.41 | HIP7090 | 15 | 1384 |
| 04.92 +11.94 | HIP22851 | 15 | 1382 |
| 04.50 +19.84 | HIP21008 | 15 | 1354 |
| 22.87 -14.43 | HIP112920 | 15 | 1340 |
| 08.70 +65.29 | HIP42694 | 15 | 1333 |
| 04.55 +04.97 | HIP21226 | 15 | 1320 |
| 06.30 -29.64 | HIP29923 | 14 | 1224 |
| 08.20 +51.32 | HIP40131 | 14 | 1224 |
| 03.28 -28.49 | HIP15242 | 13 | 1213 |
| 13.03 +13.17 | HIP63577 | 13 | 1197 |
| 00.92 +00.93 | HIP4312 | 13 | 1167 |
| 11.11 -29.97 | HIP54285 | 13 | 1161 |
| 04.46 +05.62 | HIP20822 | 13 | 1160 |
| 13.00 -27.84 | HIP63450 | 13 | 1153 |
| 13.52 +53.31 | HIP65946 | 13 | 1148 |
| 05.63 +06.11 | HIP26444 | 13 | 1143 |
| 07.18 +29.85 | HIP34677 | 13 | 1129 |
| 07.01 +48.48 | HIP33755 | 13 | 1127 |
| 08.73 +04.64 | HIP42862 | 13 | 1126 |
| 09.24 +33.82 | HIP45325 | 12 | 1121 |
| 03.09 +30.19 | HIP14353 | 12 | 1117 |
| 05.59 +21.29 | HIP26250 | 12 | 1104 |
| 05.41 +63.39 | HIP25300 | 12 | 1095 |
| 03.04 -19.91 | HIP14138 | 12 | 1094 |
| 08.69 -23.89 | HIP42618 | 12 | 1085 |
| 22.04 +18.30 | HIP108815 | 12 | 1072 |
| 03.08 -21.41 | HIP14314 | 12 | 1065 |
| 06.48 +68.19 | HIP30833 | 12 | 1039 |
| 00.80 +34.59 | HIP3723 | 12 | 1038 |
| 10.11 +34.35 | HIP49529 | 12 | 1036 |
| 00.32 +13.86 | HIP1563 | 11 | 1033 |
| 07.15 +30.39 | HIP34521 | 11 | 1026 |
| 07.03 +48.38 | HIP33852 | 11 | 1017 |
| 04.54 +05.63 | HIP21155 | 11 | 1016 |
| 08.91 +63.76 | HIP43745 | 11 | 1015 |
| 12.05 +77.69 | HIP58737 | 11 | 1009 |
| 08.82 +72.68 | HIP43273 | 11 | 1008 |
| 04.71 +18.14 | HIP21926 | 11 | 995 |
| 08.68 +72.25 | HIP42609 | 11 | 994 |
| 02.61 -12.26 | HIP12139 | 11 | 993 |
| 04.66 +19.12 | HIP21687 | 11 | 990 |
| 05.15 -12.92 | HIP23951 | 11 | 990 |
| 06.35 +67.72 | HIP30189 | 11 | 987 |
| 06.77 +33.03 | HIP32415 | 11 | 974 |
| 05.30 +05.99 | HIP24728 | 11 | 968 |
| 05.74 -20.13 | HIP27075 | 11 | 966 |
| 08.42 +31.05 | HIP41292 | 11 | 965 |
| 06.52 +05.88 | HIP31069 | 11 | 963 |
| 05.62 +31.33 | HIP26419 | 11 | 956 |
| 12.32 +17.73 | HIP60091 | 11 | 947 |
| 05.66 +31.22 | HIP26650 | 10 | 937 |
| 07.84 +51.37 | HIP38277 | 10 | 933 |
| 08.71 +31.72 | HIP42763 | 10 | 921 |
| 22.18 +15.71 | HIP109490 | 10 | 919 |
| 14.23 +04.30 | HIP69498 | 10 | 918 |
| 02.69 +24.72 | HIP12532 | 10 | 916 |
| 08.37 -26.44 | HIP41005 | 10 | 915 |
| 09.35 +25.16 | HIP45869 | 10 | 914 |
| 23.87 +77.60 | HIP117681 | 10 | 913 |
| 07.45 +53.21 | HIP36172 | 10 | 908 |
| 08.98 +09.46 | HIP44071 | 10 | 908 |
| 07.94 +59.68 | HIP38791 | 10 | 904 |
| 08.69 +47.48 | HIP42627 | 10 | 904 |
| 07.11 +64.42 | HIP34294 | 10 | 899 |
| 05.11 +35.56 | HIP23765 | 10 | 899 |
| 08.56 +51.01 | HIP41993 | 10 | 898 |
| 09.46 +60.37 | HIP46394 | 10 | 887 |
| 03.30 +25.25 | HIP15332 | 10 | 881 |
| 04.72 +24.09 | HIP21961 | 10 | 881 |
| 01.49 +29.54 | HIP6950 | 10 | 877 |
| 05.21 -26.55 | HIP24250 | 10 | 873 |
| 11.70 +42.45 | HIP57060 | 10 | 864 |
| 01.71 +20.45 | HIP7966 | 10 | 859 |
| 06.85 +40.50 | HIP32892 | 10 | 858 |
| 08.93 +61.17 | HIP43841 | 9 | 854 |
| 03.62 +31.12 | HIP16900 | 9 | 853 |
| 07.46 +52.37 | HIP36248 | 9 | 853 |
| 13.51 +13.84 | HIP65914 | 9 | 843 |
| 06.74 +34.17 | HIP32262 | 9 | 840 |
| 23.98 -23.02 | HIP118236 | 9 | 839 |
| 05.68 +05.01 | HIP26705 | 9 | 837 |
| 07.39 +57.58 | HIP35867 | 9 | 835 |
| 08.80 +15.73 | HIP43217 | 9 | 834 |
| 10.03 -15.42 | HIP49127 | 9 | 832 |
| 03.54 +10.74 | HIP16504 | 9 | 831 |
| 05.60 +35.51 | HIP26290 | 9 | 829 |
| 08.71 +63.40 | HIP42731 | 9 | 825 |
| 06.83 +35.14 | HIP32723 | 9 | 822 |
| 07.31 +27.25 | HIP35377 | 9 | 811 |
| 19.65 +42.02 | HIP96642 | 9 | 810 |
| 03.12 -20.65 | HIP14498 | 9 | 808 |
| 05.72 +54.26 | HIP26933 | 9 | 807 |
| 03.97 +20.68 | HIP18544 | 9 | 795 |
| 08.60 +14.12 | HIP42165 | 9 | 788 |
| 09.23 -27.52 | HIP45301 | 9 | 787 |
| 06.24 +84.99 | HIP29600 | 9 | 785 |
| 20.19 +18.48 | HIP99494 | 9 | 777 |
| 06.35 +84.18 | HIP30166 | 9 | 773 |
| 03.78 +59.80 | HIP17636 | 9 | 773 |
| 06.60 +58.11 | HIP31552 | 9 | 770 |
| 09.13 +07.48 | HIP44780 | 9 | 769 |
| 09.14 +07.62 | HIP44878 | 9 | 769 |
| 06.70 -15.22 | HIP32076 | 9 | 768 |
| 07.51 +29.38 | HIP36480 | 9 | 768 |
| 22.96 -26.80 | HIP113358 | 9 | 768 |
| 20.93 +10.88 | HIP103316 | 9 | 768 |
| 05.05 -26.64 | HIP23468 | 8 | 764 |
| 10.37 +50.49 | HIP50815 | 8 | 763 |
| 04.14 +22.02 | HIP19299 | 8 | 761 |
| 08.05 +33.60 | HIP39392 | 8 | 761 |
| 06.10 +67.64 | HIP28905 | 8 | 757 |
| 17.24 +29.57 | HIP84309 | 8 | 757 |
| 08.03 -17.17 | HIP39293 | 8 | 756 |
| 09.30 +13.66 | HIP45609 | 8 | 754 |
| 06.57 -16.01 | HIP31341 | 8 | 754 |
| 06.12 +68.80 | HIP28996 | 8 | 753 |
| 02.68 +39.08 | HIP12509 | 8 | 753 |
| 03.07 +11.14 | HIP14298 | 8 | 753 |
| 07.88 +50.17 | HIP38443 | 8 | 750 |
| 11.01 -31.84 | HIP53818 | 8 | 750 |
| 05.36 +86.87 | HIP25029 | 8 | 750 |
| 03.25 +20.34 | HIP15100 | 8 | 749 |
| 08.39 +02.18 | HIP41142 | 8 | 749 |
| 09.49 -27.80 | HIP46541 | 8 | 749 |
| 07.97 +33.96 | HIP38949 | 8 | 747 |
| 06.58 -15.33 | HIP31401 | 8 | 745 |
| 09.58 +30.56 | HIP46993 | 8 | 745 |
| 07.13 +29.83 | HIP34414 | 8 | 744 |
| 05.70 +34.82 | HIP26849 | 8 | 744 |
| 05.84 +75.71 | HIP27576 | 8 | 743 |
| 07.75 +02.14 | HIP37798 | 8 | 741 |
| 04.37 +39.93 | HIP20395 | 8 | 740 |
| 22.84 +36.70 | HIP112768 | 8 | 737 |
| 06.21 +50.60 | HIP29484 | 8 | 735 |
| 04.30 -24.06 | HIP20060 | 8 | 734 |
| 07.04 +79.16 | HIP33932 | 8 | 733 |
| 06.05 -16.04 | HIP28678 | 8 | 732 |
| 07.92 +32.28 | HIP38658 | 8 | 731 |
| 11.32 -23.26 | HIP55295 | 8 | 731 |
| 05.69 +34.81 | HIP26799 | 8 | 730 |
| 22.84 +34.86 | HIP112800 | 8 | 729 |
| 09.74 +16.41 | HIP47784 | 8 | 728 |
| 16.90 +11.91 | HIP82694 | 8 | 727 |
| 09.80 -27.98 | HIP48073 | 8 | 727 |
| 06.42 -00.95 | HIP30545 | 8 | 727 |
| 13.45 +14.37 | HIP65589 | 8 | 725 |
| 05.37 +59.28 | HIP25094 | 8 | 725 |
| 09.86 -27.42 | HIP48353 | 8 | 725 |
| 08.33 +27.22 | HIP40843 | 8 | 725 |
| 12.46 +75.40 | HIP60764 | 8 | 724 |
| 11.56 +19.56 | HIP56399 | 8 | 724 |
| 08.77 +04.50 | HIP43070 | 8 | 721 |
| 10.14 -19.76 | HIP49668 | 8 | 720 |
| 07.82 +52.52 | HIP38149 | 8 | 719 |
| 07.76 +74.35 | HIP37830 | 8 | 717 |
| 09.65 +50.25 | HIP47331 | 8 | 717 |
| 04.93 +07.40 | HIP22916 | 8 | 716 |
| 04.79 +31.59 | HIP22241 | 8 | 713 |
| 09.54 +05.88 | HIP46834 | 8 | 712 |
| 21.65 -31.91 | HIP106880 | 8 | 712 |
| 08.15 +47.13 | HIP39917 | 8 | 712 |
| 09.50 +05.66 | HIP46580 | 8 | 711 |
| 19.49 +47.00 | HIP95802 | 8 | 709 |
| 09.19 +46.62 | HIP45116 | 8 | 708 |
| 06.31 +25.90 | HIP29967 | 8 | 705 |
| 11.12 +34.84 | HIP54369 | 8 | 704 |
| 06.96 +22.89 | HIP33497 | 8 | 696 |
| 07.91 +33.15 | HIP38637 | 8 | 695 |
| 09.04 +61.52 | HIP44399 | 8 | 695 |
| 10.02 +61.61 | HIP49075 | 8 | 691 |
| 16.09 +43.61 | HIP78796 | 8 | 691 |
| 01.84 -20.04 | HIP8573 | 8 | 690 |
| 07.48 +58.13 | HIP36344 | 8 | 686 |
| 07.87 +78.23 | HIP38399 | 8 | 686 |
| 09.64 +18.64 | HIP47278 | 8 | 686 |
| 05.63 +78.36 | HIP26486 | 8 | 684 |
| 03.40 -29.13 | HIP15815 | 8 | 681 |
| 08.57 +16.88 | HIP42057 | 8 | 680 |
| 06.42 +11.23 | HIP30552 | 8 | 680 |
| 18.75 +44.36 | HIP91990 | 8 | 678 |
| 09.21 +48.31 | HIP45174 | 8 | 678 |
| 06.98 +05.54 | HIP33608 | 8 | 675 |
| 08.68 +46.43 | HIP42599 | 8 | 675 |
| 03.60 -23.52 | HIP16783 | 7 | 674 |
| 07.55 +78.69 | HIP36715 | 7 | 674 |
| 19.75 +51.60 | HIP97168 | 7 | 673 |
| 05.45 -16.52 | HIP25497 | 7 | 671 |
| 06.92 +05.91 | HIP33275 | 7 | 671 |
| 07.39 -20.02 | HIP35851 | 7 | 668 |
| 04.92 +05.64 | HIP22889 | 7 | 668 |
| 05.91 +76.46 | HIP27909 | 7 | 667 |
| 05.65 +20.03 | HIP26568 | 7 | 666 |
| 00.23 -10.95 | HIP1119 | 7 | 664 |
| 09.79 +01.58 | HIP48016 | 7 | 664 |
| 03.29 +10.14 | HIP15288 | 7 | 663 |
| 03.13 +25.55 | HIP14548 | 7 | 661 |
| 04.71 -29.71 | HIP21891 | 7 | 659 |
| 18.49 +07.36 | HIP90621 | 7 | 659 |
| 20.65 +10.98 | HIP101920 | 7 | 654 |
| 10.34 +51.13 | HIP50616 | 7 | 653 |
| 04.34 +12.40 | HIP20253 | 7 | 652 |
| 04.76 +21.46 | HIP22118 | 7 | 651 |
| 23.25 +56.30 | HIP114788 | 7 | 649 |
| 23.46 -19.89 | HIP115812 | 7 | 643 |
| 16.19 +26.89 | HIP79305 | 7 | 642 |
| 13.60 +52.73 | HIP66340 | 7 | 640 |
| 04.33 +21.07 | HIP20181 | 7 | 640 |
| 08.01 -00.71 | HIP39175 | 7 | 639 |
| 07.89 +51.30 | HIP38522 | 7 | 637 |
| 04.17 +25.20 | HIP19468 | 7 | 636 |
| 11.67 +25.78 | HIP56928 | 7 | 636 |
| 04.32 +21.30 | HIP20136 | 7 | 635 |
| 04.35 +83.90 | HIP20311 | 7 | 631 |
| 05.95 +10.48 | HIP28135 | 7 | 631 |
| 03.09 +29.54 | HIP14364 | 7 | 631 |
| 09.44 +51.65 | HIP46275 | 7 | 628 |
| 06.76 +35.53 | HIP32364 | 7 | 626 |
| 06.20 +69.84 | HIP29402 | 7 | 622 |
| 07.44 +60.55 | HIP36129 | 7 | 620 |
| 04.38 +09.77 | HIP20447 | 7 | 617 |
| 11.04 -24.72 | HIP53941 | 7 | 616 |
| 04.42 +25.75 | HIP20626 | 7 | 615 |
| 08.19 +55.49 | HIP40094 | 7 | 614 |
| 06.11 +63.84 | HIP28935 | 7 | 614 |
| 07.37 +29.55 | HIP35732 | 7 | 614 |
| 06.50 -28.66 | HIP30976 | 7 | 609 |
| 23.48 +39.62 | HIP115880 | 7 | 606 |
| 05.39 +17.32 | HIP25220 | 7 | 605 |
| 06.62 +34.77 | HIP31619 | 7 | 605 |
| 08.97 -14.37 | HIP44059 | 7 | 605 |
| 02.97 +19.23 | HIP13855 | 7 | 603 |
| 08.39 +21.85 | HIP41130 | 7 | 602 |
| 16.18 -21.29 | HIP79300 | 7 | 601 |
| 05.60 -15.38 | HIP26302 | 7 | 600 |
| 11.75 +27.41 | HIP57300 | 7 | 599 |
| 04.81 -23.28 | HIP22360 | 7 | 596 |
| 03.46 +34.81 | HIP16107 | 7 | 593 |
| 16.20 +43.17 | HIP79359 | 7 | 591 |
| 09.09 +19.44 | HIP44616 | 7 | 591 |
| 05.81 +58.42 | HIP27414 | 7 | 590 |
| 13.95 +44.28 | HIP68134 | 7 | 590 |
| 08.35 +01.39 | HIP40903 | 7 | 589 |
| 07.60 -20.05 | HIP36951 | 7 | 588 |
| 09.46 +15.74 | HIP46397 | 7 | 588 |
| 11.25 +26.26 | HIP54945 | 7 | 588 |
| 14.18 +04.40 | HIP69280 | 7 | 588 |
| 09.20 +20.43 | HIP45131 | 7 | 587 |
| 05.06 +25.77 | HIP23523 | 7 | 587 |
| 08.47 +22.10 | HIP41513 | 7 | 587 |
| 06.28 +17.44 | HIP29828 | 7 | 587 |
| 19.78 +51.02 | HIP97316 | 7 | 587 |
| 06.79 +33.78 | HIP32551 | 7 | 586 |
| 04.60 +21.69 | HIP21442 | 7 | 586 |
| 03.03 +16.52 | HIP14098 | 6 | 584 |
| 05.04 +30.00 | HIP23438 | 6 | 582 |
| 07.93 +49.27 | HIP38731 | 6 | 582 |
| 00.39 +31.85 | HIP1853 | 6 | 582 |
| 09.21 +33.60 | HIP45196 | 6 | 582 |
| 15.33 +40.98 | HIP75039 | 6 | 581 |
| 03.48 -24.90 | HIP16225 | 6 | 580 |
| 05.51 -25.13 | HIP25791 | 6 | 580 |
| 18.10 +25.68 | HIP88651 | 6 | 579 |
| 06.83 +60.34 | HIP32769 | 6 | 578 |
| 04.88 +75.71 | HIP22703 | 6 | 578 |
| 04.48 +55.00 | HIP20910 | 6 | 576 |
| 09.03 +06.50 | HIP44341 | 6 | 575 |
| 06.87 +49.94 | HIP33000 | 6 | 573 |
| 03.16 +48.62 | HIP14673 | 6 | 573 |
| 06.07 +46.90 | HIP28759 | 6 | 573 |
| 13.84 +47.82 | HIP67526 | 6 | 572 |
| 07.63 +61.53 | HIP37150 | 6 | 572 |
| 07.97 +28.18 | HIP38941 | 6 | 571 |
| 02.57 +42.79 | HIP11949 | 6 | 570 |
| 12.02 -21.53 | HIP58629 | 6 | 569 |
| 09.89 -20.80 | HIP48480 | 6 | 569 |
| 13.67 +47.47 | HIP66713 | 6 | 568 |
| 08.68 +55.67 | HIP42575 | 6 | 568 |
| 08.80 +31.77 | HIP43180 | 6 | 568 |
| 03.49 +09.69 | HIP16255 | 6 | 567 |
| 07.76 +00.82 | HIP37885 | 6 | 567 |
| 09.61 +19.44 | HIP47140 | 6 | 567 |
| 07.24 +28.44 | HIP35022 | 6 | 567 |
| 07.79 +73.99 | HIP38033 | 6 | 565 |
| 08.16 +48.37 | HIP39972 | 6 | 565 |
| 08.24 +13.02 | HIP40375 | 6 | 564 |
| 19.03 +41.49 | HIP93427 | 6 | 564 |
| 04.68 -29.38 | HIP21798 | 6 | 564 |
| 06.23 +14.02 | HIP29544 | 6 | 564 |
| 08.85 +32.53 | HIP43439 | 6 | 564 |
| 05.18 +25.96 | HIP24136 | 6 | 564 |
| 09.12 +22.98 | HIP44768 | 6 | 564 |
| 09.11 +30.28 | HIP44687 | 6 | 564 |
| 07.26 +70.51 | HIP35134 | 6 | 562 |
| 07.38 +06.37 | HIP35788 | 6 | 561 |
| 10.64 +34.87 | HIP52052 | 6 | 561 |
| 10.91 +40.12 | HIP53350 | 6 | 560 |
| 06.95 +30.76 | HIP33428 | 6 | 560 |
| 09.47 +15.17 | HIP46449 | 6 | 560 |
| 04.68 +30.29 | HIP21783 | 6 | 560 |
| 08.10 +34.08 | HIP39612 | 6 | 560 |
| 05.88 -30.78 | HIP27779 | 6 | 559 |
| 07.37 +21.14 | HIP35703 | 6 | 559 |
| 06.22 +13.14 | HIP29493 | 6 | 559 |
| 05.58 +17.14 | HIP26157 | 6 | 557 |
| 10.18 +75.14 | HIP49868 | 6 | 556 |
| 09.76 +13.27 | HIP47865 | 6 | 555 |
| 23.64 +41.62 | HIP116669 | 6 | 555 |
| 05.82 +21.04 | HIP27495 | 6 | 555 |
| 09.04 +25.89 | HIP44387 | 6 | 554 |
| 08.35 +54.99 | HIP40930 | 6 | 554 |
| 23.60 +41.97 | HIP116443 | 6 | 553 |
| 04.57 +00.41 | HIP21307 | 6 | 552 |
| 10.88 -18.64 | HIP53156 | 6 | 552 |
| 08.89 -17.36 | HIP43655 | 6 | 550 |
| 09.94 +62.79 | HIP48714 | 6 | 549 |
| 10.11 +68.66 | HIP49497 | 6 | 548 |
| 03.15 +49.61 | HIP14632 | 6 | 547 |
| 03.86 -28.84 | HIP18075 | 6 | 547 |
| 10.33 -28.96 | HIP50568 | 6 | 547 |
| 09.40 +19.48 | HIP46104 | 6 | 546 |
| 06.80 +31.70 | HIP32568 | 6 | 546 |
| 09.01 +50.93 | HIP44252 | 6 | 546 |
| 09.35 +60.40 | HIP45863 | 6 | 544 |
| 04.20 +21.54 | HIP19615 | 6 | 544 |
| 09.60 -21.66 | HIP47103 | 6 | 543 |
| 03.68 +18.10 | HIP17174 | 6 | 543 |
| 08.32 +01.34 | HIP40774 | 6 | 542 |
| 07.32 +51.97 | HIP35491 | 6 | 541 |
| 08.54 +22.41 | HIP41891 | 6 | 541 |
| 09.51 +46.40 | HIP46629 | 6 | 541 |
| 18.10 +53.96 | HIP88626 | 6 | 541 |
| 19.82 +27.73 | HIP97509 | 6 | 541 |
| 09.06 +76.40 | HIP44463 | 6 | 541 |
| 05.93 +28.03 | HIP28020 | 6 | 540 |
| 20.04 +28.81 | HIP98651 | 6 | 540 |
| 09.11 +50.63 | HIP44715 | 6 | 540 |
| 03.56 -15.13 | HIP16607 | 6 | 539 |
| 09.52 -26.80 | HIP46703 | 6 | 539 |
| 06.92 +25.38 | HIP33277 | 6 | 538 |
| 10.32 +51.41 | HIP50513 | 6 | 538 |
| 03.94 +46.57 | HIP18423 | 6 | 537 |
| 05.11 +34.54 | HIP23797 | 6 | 537 |
| 09.84 +24.56 | HIP48248 | 6 | 537 |
| 06.43 +33.25 | HIP30573 | 6 | 536 |
| 07.37 +60.33 | HIP35697 | 6 | 536 |
| 04.10 +10.97 | HIP19112 | 6 | 536 |
| 08.32 -11.88 | HIP40786 | 6 | 536 |
| 07.39 +52.28 | HIP35824 | 6 | 535 |
| 19.04 +38.57 | HIP93469 | 6 | 534 |
| 06.90 +28.39 | HIP33129 | 6 | 533 |
| 03.87 +39.80 | HIP18090 | 6 | 531 |
| 09.92 +24.64 | HIP48642 | 6 | 531 |
| 10.87 +33.04 | HIP53120 | 6 | 530 |
| 06.14 +50.75 | HIP29113 | 6 | 528 |
| 05.70 +31.87 | HIP26863 | 6 | 528 |
| 10.24 +49.77 | HIP50165 | 6 | 528 |
| 06.74 +66.30 | HIP32264 | 6 | 527 |
| 10.35 -28.74 | HIP50681 | 6 | 527 |
| 15.53 +53.32 | HIP76019 | 6 | 527 |
| 10.19 +49.45 | HIP49908 | 6 | 527 |
| 10.53 +00.48 | HIP51559 | 6 | 527 |
| 10.28 +25.30 | HIP50321 | 6 | 526 |
| 17.10 -28.29 | HIP83657 | 6 | 526 |
| 06.71 +66.20 | HIP32117 | 6 | 525 |
| 10.26 +47.37 | HIP50245 | 6 | 525 |
| 06.34 +66.63 | HIP30152 | 6 | 525 |
| 10.81 +33.77 | HIP52862 | 6 | 524 |
| 05.61 +32.00 | HIP26326 | 6 | 523 |
| 06.55 -27.40 | HIP31201 | 6 | 523 |
| 07.74 +77.78 | HIP37768 | 6 | 523 |
| 08.61 +46.16 | HIP42246 | 6 | 522 |
| 18.47 +06.84 | HIP90532 | 6 | 521 |
| 07.53 +60.50 | HIP36590 | 6 | 520 |
| 10.48 +00.52 | HIP51285 | 6 | 519 |
| 09.92 -27.26 | HIP48659 | 6 | 519 |
| 09.37 +05.58 | HIP45942 | 6 | 519 |
| 04.99 +48.78 | HIP23198 | 6 | 518 |
| 06.97 +60.79 | HIP33536 | 6 | 517 |
| 16.06 +34.01 | HIP78680 | 6 | 514 |
| 07.07 +76.74 | HIP34075 | 6 | 514 |
| 23.56 +76.61 | HIP116255 | 6 | 513 |
| 04.85 +09.92 | HIP22548 | 6 | 512 |
| 07.20 +52.27 | HIP34796 | 6 | 511 |
| 17.29 +37.61 | HIP84562 | 6 | 510 |
| 05.68 +32.90 | HIP26744 | 6 | 510 |
| 09.76 +03.54 | HIP47897 | 6 | 510 |
| 05.76 +35.62 | HIP27155 | 6 | 509 |
| 07.27 +27.14 | HIP35191 | 6 | 508 |
| 09.53 +34.41 | HIP46757 | 6 | 508 |
| 08.77 +35.77 | HIP43024 | 6 | 508 |
| 06.90 +60.87 | HIP33142 | 6 | 506 |
| 10.69 +12.93 | HIP52306 | 6 | 503 |
| 00.03 +66.31 | HIP142 | 6 | 503 |
| 06.30 +35.33 | HIP29910 | 6 | 501 |
| 10.91 +26.67 | HIP53367 | 6 | 501 |
| 15.51 +53.97 | HIP75930 | 6 | 501 |
| 16.45 +38.84 | HIP80568 | 6 | 501 |
| 03.15 +15.33 | HIP14614 | 6 | 500 |
| 16.51 +37.53 | HIP80861 | 6 | 500 |
| 06.39 +32.33 | HIP30411 | 6 | 499 |
| 05.95 +51.12 | HIP28176 | 6 | 498 |
| 09.12 +46.67 | HIP44746 | 6 | 496 |
| 00.89 +01.91 | HIP4203 | 6 | 496 |
| 11.10 +35.20 | HIP54239 | 6 | 496 |
| 10.41 +74.77 | HIP50972 | 6 | 495 |
| 08.53 +50.62 | HIP41844 | 5 | 494 |
| 10.05 +58.10 | HIP49246 | 5 | 494 |
| 10.51 +59.75 | HIP51468 | 5 | 493 |
| 19.46 +08.97 | HIP95696 | 5 | 493 |
| 19.52 +08.40 | HIP95983 | 5 | 493 |
| 05.12 -12.49 | HIP23831 | 5 | 493 |
| 10.16 +51.01 | HIP49798 | 5 | 492 |
| 07.91 +60.13 | HIP38619 | 5 | 492 |
| 03.85 +01.01 | HIP17996 | 5 | 492 |
| 10.08 +67.42 | HIP49387 | 5 | 491 |
| 09.81 +25.17 | HIP48107 | 5 | 489 |
| 06.77 +73.68 | HIP32450 | 5 | 489 |
| 08.66 +17.48 | HIP42471 | 5 | 489 |
| 05.02 +56.27 | HIP23332 | 5 | 487 |
| 09.24 +00.26 | HIP45317 | 5 | 486 |
| 05.41 +22.31 | HIP25274 | 5 | 486 |
| 08.95 -22.53 | HIP43921 | 5 | 485 |
| 09.96 -22.91 | HIP48826 | 5 | 483 |
| 20.10 +28.92 | HIP99017 | 5 | 483 |
| 18.91 +51.31 | HIP92835 | 5 | 482 |
| 08.21 +56.01 | HIP40185 | 5 | 481 |
| 08.45 +28.93 | HIP41443 | 5 | 480 |
| 11.01 -25.14 | HIP53805 | 5 | 480 |
| 10.78 +04.65 | HIP52705 | 5 | 479 |
| 09.02 +09.05 | HIP44272 | 5 | 479 |
| 04.56 +05.39 | HIP21272 | 5 | 478 |
| 04.09 +19.96 | HIP19093 | 5 | 478 |
| 08.12 -29.40 | HIP39710 | 5 | 477 |
| 09.45 +14.93 | HIP46339 | 5 | 476 |
| 08.25 -28.69 | HIP40424 | 5 | 475 |
| 04.37 +12.51 | HIP20397 | 5 | 474 |
| 02.95 +48.70 | HIP13741 | 5 | 473 |
| 19.40 +44.93 | HIP95362 | 5 | 473 |
| 06.57 +14.75 | HIP31345 | 5 | 471 |
| 08.96 +62.68 | HIP43999 | 5 | 470 |
| 16.41 -26.81 | HIP80400 | 5 | 470 |
| 06.72 +34.06 | HIP32168 | 5 | 469 |
| 11.17 -29.41 | HIP54597 | 5 | 469 |
| 09.04 +54.52 | HIP44392 | 5 | 468 |
| 03.04 -28.73 | HIP14127 | 5 | 467 |
| 07.85 +49.67 | HIP38334 | 5 | 465 |
| 02.83 +75.02 | HIP13179 | 5 | 464 |
| 10.34 +41.13 | HIP50621 | 5 | 463 |
| 00.96 +34.17 | HIP4495 | 5 | 461 |
| 05.38 +46.99 | HIP25166 | 5 | 460 |
| 05.52 +06.46 | HIP25846 | 5 | 460 |
| 03.74 +34.97 | HIP17458 | 5 | 460 |
| 10.93 +65.73 | HIP53435 | 5 | 457 |
| 04.29 +59.11 | HIP20023 | 5 | 455 |
| 04.44 +16.85 | HIP20719 | 5 | 455 |
| 00.17 -11.79 | HIP847HD602 | 5 | 455 |
| 08.76 +15.05 | HIP42950 | 5 | 454 |
| 02.78 +37.74 | HIP12967 | 5 | 452 |
| 03.51 -15.85 | HIP16331 | 5 | 450 |
| 11.74 +14.96 | HIP57249 | 5 | 450 |
| 08.14 +19.22 | HIP39816 | 5 | 449 |
| 07.99 +58.70 | HIP39055 | 5 | 447 |
| 09.61 +32.12 | HIP47170 | 5 | 447 |
| 04.17 +24.09 | HIP19459 | 5 | 446 |
| 09.47 +31.39 | HIP46446 | 5 | 446 |
| 00.23 +00.87 | HIP1121 | 5 | 446 |
| 11.02 +65.18 | HIP53885 | 5 | 445 |
| 19.09 +33.87 | HIP93754 | 5 | 445 |
| 02.23 +32.98 | HIP10373 | 5 | 444 |
| 06.84 -29.95 | HIP32804 | 5 | 444 |
| 09.52 +31.74 | HIP46707 | 5 | 442 |
| 04.80 +22.17 | HIP22291 | 5 | 442 |
| 05.33 +17.78 | HIP24861 | 5 | 441 |
| 09.74 -26.89 | HIP47794 | 5 | 440 |
| 04.06 +81.58 | HIP18936 | 5 | 437 |
| 01.80 +21.21 | HIP8382 | 5 | 437 |
| 06.83 -29.56 | HIP32749 | 5 | 436 |
| 07.56 +47.70 | HIP36761 | 5 | 434 |
| 16.08 +33.45 | HIP78788 | 5 | 434 |
| 19.78 +39.31 | HIP97343 | 5 | 433 |
| 06.86 +23.36 | HIP32906 | 5 | 433 |
| 13.76 +18.34 | HIP67157 | 5 | 433 |
| 17.46 +26.79 | HIP85436 | 5 | 433 |
| 14.94 +43.32 | HIP73078 | 5 | 432 |
| 13.86 -12.39 | HIP67644 | 5 | 431 |
| 14.83 +06.82 | HIP72535 | 5 | 431 |
| 16.16 +26.00 | HIP79193 | 5 | 431 |
| 15.87 +15.24 | HIP77718 | 5 | 430 |
| 20.10 +30.33 | HIP98981 | 5 | 430 |
| 15.88 +15.94 | HIP77775 | 5 | 429 |
| 17.28 +62.58 | HIP84541 | 5 | 429 |
| 17.42 +28.15 | HIP85262 | 5 | 429 |
| 09.62 +31.94 | HIP47191 | 5 | 428 |
| 09.50 +09.83 | HIP46568 | 5 | 428 |
| 08.12 +70.58 | HIP39718 | 5 | 428 |
| 08.86 +64.42 | HIP43477 | 5 | 428 |
| 14.41 -17.45 | HIP70472 | 5 | 428 |
| 19.65 +43.87 | HIP96634 | 5 | 427 |
| 11.16 +34.91 | HIP54548 | 5 | 427 |
| 05.86 -30.47 | HIP27688 | 5 | 425 |
| 06.14 +09.48 | HIP29132 | 5 | 424 |
| 15.37 +42.10 | HIP75198 | 5 | 424 |
| 06.45 -25.86 | HIP30711 | 5 | 424 |
| 09.36 +19.44 | HIP45886 | 5 | 424 |
| 07.89 +32.20 | HIP38513 | 5 | 423 |
| 10.21 +48.41 | HIP50012 | 5 | 423 |
| 07.80 +49.37 | HIP38084 | 5 | 423 |
| 18.33 +35.74 | HIP89794 | 5 | 423 |
| 06.83 +60.16 | HIP32725 | 5 | 422 |
| 07.33 +46.97 | HIP35520 | 5 | 419 |
| 03.80 +10.50 | HIP17722 | 5 | 419 |
| 04.46 -24.46 | HIP20807 | 5 | 419 |
| 04.60 +21.54 | HIP21436 | 5 | 419 |
| 05.65 +09.92 | HIP26596 | 5 | 419 |
| 06.16 +17.83 | HIP29189 | 5 | 419 |
| 06.81 -23.20 | HIP32662 | 5 | 419 |
| 06.93 +53.44 | HIP33340 | 5 | 419 |
| 07.08 +35.58 | HIP34136 | 5 | 419 |
| 07.32 +66.74 | HIP35449 | 5 | 419 |
| 08.15 +34.89 | HIP39881 | 5 | 418 |
| 01.49 +34.09 | HIP6938 | 5 | 418 |
| 01.59 +42.04 | HIP7420 | 5 | 418 |
| 01.68 +41.93 | HIP7828 | 5 | 418 |
| 06.10 +64.08 | HIP28914 | 5 | 418 |
| 00.58 +42.69 | HIP2742 | 5 | 417 |
| 01.46 +35.40 | HIP6817 | 5 | 417 |
| 06.89 +51.65 | HIP33090 | 5 | 416 |
| 06.36 -27.51 | HIP30244 | 5 | 415 |
| 00.88 -00.38 | HIP4131 | 5 | 415 |
| 08.48 +06.96 | HIP41569 | 5 | 412 |
| 00.74 +07.77 | HIP3466 | 5 | 412 |
| 07.42 +85.74 | HIP36022 | 5 | 411 |
| 11.15 +34.30 | HIP54491 | 5 | 411 |
| 17.24 +04.09 | HIP84316 | 5 | 411 |
| 00.28 +32.45 | HIP1344 | 5 | 410 |
| 00.72 +21.79 | HIP3377 | 5 | 410 |
| 03.90 +08.56 | HIP18253 | 5 | 410 |
| 04.88 +11.27 | HIP22712 | 5 | 410 |
| 07.31 +48.24 | HIP35399 | 5 | 410 |
| 07.58 +28.73 | HIP36846 | 5 | 410 |
| 02.98 +15.15 | HIP13871 | 5 | 410 |
| 00.60 +21.33 | HIP2829 | 5 | 409 |
| 08.78 +66.51 | HIP43119 | 5 | 409 |
| 03.18 -13.57 | HIP14783 | 5 | 408 |
| 03.01 +84.65 | HIP14015 | 5 | 406 |
| 04.00 +30.43 | HIP18676 | 5 | 406 |
| 04.79 +11.79 | HIP22264 | 5 | 406 |
| 04.89 +50.61 | HIP22755 | 5 | 406 |
| 07.61 +05.86 | HIP37031 | 5 | 406 |
| 08.04 +04.15 | HIP39325 | 5 | 406 |
| 10.70 +12.11 | HIP52339 | 5 | 406 |
| 07.30 -31.69 | HIP35332 | 5 | 406 |
| 19.38 +47.71 | HIP95274 | 5 | 406 |
| 04.99 +35.56 | HIP23212 | 5 | 406 |
| 05.01 +05.10 | HIP23277 | 5 | 406 |
| 06.17 +82.11 | HIP29277 | 5 | 406 |
| 19.04 +43.12 | HIP93511 | 5 | 406 |
| 04.48 +67.85 | HIP20891 | 5 | 406 |
| 04.33 +12.62 | HIP20187 | 5 | 405 |
| 04.13 +21.26 | HIP19262 | 5 | 405 |
| 09.78 -16.55 | HIP47985 | 4 | 404 |
| 17.80 +38.23 | HIP87135 | 4 | 404 |
| 19.49 +00.53 | HIP95829 | 4 | 404 |
| 02.79 +19.84 | HIP13021 | 4 | 403 |
| 06.29 +32.26 | HIP29878 | 4 | 402 |
| 10.51 -21.24 | HIP51443 | 4 | 402 |
| 10.61 +35.12 | HIP51917 | 4 | 399 |
| 03.84 -23.83 | HIP17956 | 4 | 399 |
| 07.02 -21.96 | HIP33799 | 4 | 399 |
| 23.98 -23.67 | HIP118239 | 4 | 398 |
| 04.54 +64.25 | HIP21164 | 4 | 398 |
| 03.24 +09.13 | HIP15074 | 4 | 398 |
| 04.42 +63.63 | HIP20637 | 4 | 398 |
| 05.31 -17.57 | HIP24781 | 4 | 397 |
| 03.06 -14.56 | HIP14250 | 4 | 396 |
| 03.05 -26.98 | HIP14193 | 4 | 396 |
| 03.06 -26.73 | HIP14253 | 4 | 396 |
| 03.09 +00.85 | HIP14358 | 4 | 396 |
| 06.41 -27.31 | HIP30478 | 4 | 395 |
| 03.04 +35.60 | HIP14123 | 4 | 395 |
| 04.94 +64.40 | HIP22961 | 4 | 395 |
| 08.71 +85.68 | HIP42716 | 4 | 395 |
| 10.34 +04.65 | HIP50656 | 4 | 394 |
| 02.11 +24.33 | HIP9829 | 4 | 393 |
| 03.06 +01.51 | HIP14231 | 4 | 393 |
| 06.09 +71.03 | HIP28868 | 4 | 393 |
| 03.76 +58.86 | HIP17568 | 4 | 393 |
| 05.53 +24.99 | HIP25929 | 4 | 393 |
| 03.08 +04.00 | HIP14352 | 4 | 393 |
| 06.74 +55.87 | HIP32280 | 4 | 393 |
| 13.14 +16.97 | HIP64113 | 4 | 392 |
| 16.45 +40.91 | HIP80549 | 4 | 392 |
| 10.37 +51.42 | HIP50783 | 4 | 392 |
| 05.80 +07.96 | HIP27377 | 4 | 392 |
| 06.68 +17.64 | HIP31950 | 4 | 392 |
| 07.86 +09.39 | HIP38374 | 4 | 392 |
| 08.89 +50.38 | HIP43632 | 4 | 392 |
| 06.18 +10.32 | HIP29316 | 4 | 392 |
| 06.84 +60.93 | HIP32806 | 4 | 392 |
| 03.05 -14.40 | HIP14177 | 4 | 392 |
| 05.31 -14.02 | HIP24777 | 4 | 392 |
| 06.13 -25.74 | HIP29052 | 4 | 392 |
| 08.25 +52.94 | HIP40402 | 4 | 392 |
| 10.93 -26.35 | HIP53416 | 4 | 392 |
| 05.64 +24.99 | HIP26525 | 4 | 392 |
| 11.12 +09.62 | HIP54346 | 4 | 392 |
| 10.12 +69.86 | HIP49596 | 4 | 391 |
| 05.94 +33.22 | HIP28102 | 4 | 391 |
| 08.00 +53.37 | HIP39092 | 4 | 391 |
| 03.03 +00.50 | HIP14113 | 4 | 391 |
| 06.13 +67.98 | HIP29067 | 4 | 391 |
| 05.06 -17.37 | HIP23512 | 4 | 391 |
| 06.24 -17.48 | HIP29601 | 4 | 391 |
| 10.04 -13.30 | HIP49202 | 4 | 391 |
| 10.30 -27.19 | HIP50458 | 4 | 391 |
| 04.20 +49.75 | HIP19581 | 4 | 390 |
| 07.91 +50.74 | HIP38616 | 4 | 390 |
| 04.09 +19.44 | HIP19082 | 4 | 390 |
| 05.43 +61.78 | HIP25393 | 4 | 390 |
| 06.20 +06.78 | HIP29432 | 4 | 390 |
| 09.40 +80.13 | HIP46085 | 4 | 390 |
| 09.29 +16.55 | HIP45579 | 4 | 390 |
| 05.01 +48.81 | HIP23307 | 4 | 390 |
| 08.83 -19.22 | HIP43356 | 4 | 389 |
| 09.90 +14.74 | HIP48524 | 4 | 389 |
| 06.46 +48.78 | HIP30745 | 4 | 389 |
| 08.26 +46.11 | HIP40434 | 4 | 389 |
| 07.57 -16.19 | HIP36809 | 4 | 389 |
| 08.41 +17.18 | HIP41226 | 4 | 389 |
| 08.27 +63.73 | HIP40517 | 4 | 389 |
| 10.60 +57.94 | HIP51881 | 4 | 389 |
| 09.43 +20.38 | HIP46249 | 4 | 388 |
| 03.74 +11.14 | HIP17469 | 4 | 388 |
| 06.22 +52.91 | HIP29532 | 4 | 388 |
| 11.18 +11.22 | HIP54622 | 4 | 388 |
| 07.51 -15.99 | HIP36512 | 4 | 387 |
| 05.58 +21.62 | HIP26159 | 4 | 387 |
| 09.89 +50.75 | HIP48509 | 4 | 387 |
| 10.55 +62.86 | HIP51648 | 4 | 387 |
| 09.88 +29.00 | HIP48450 | 4 | 387 |
| 07.70 +49.22 | HIP37494 | 4 | 387 |
| 04.75 +14.63 | HIP22063 | 4 | 386 |
| 06.20 +71.13 | HIP29421 | 4 | 386 |
| 06.04 +32.59 | HIP28626 | 4 | 386 |
| 09.97 +46.95 | HIP48862 | 4 | 386 |
| 10.34 +53.78 | HIP50606 | 4 | 386 |
| 08.94 +70.97 | HIP43891 | 4 | 386 |
| 09.31 +19.42 | HIP45678 | 4 | 386 |
| 03.26 +14.30 | HIP15153 | 4 | 386 |
| 08.90 +62.58 | HIP43696 | 4 | 385 |
| 06.17 +17.93 | HIP29248 | 4 | 385 |
| 05.69 +15.21 | HIP26807 | 4 | 385 |
| 08.64 +17.06 | HIP42382 | 4 | 385 |
| 11.25 +34.61 | HIP54936 | 4 | 385 |
| 05.11 +14.08 | HIP23789 | 4 | 385 |
| 07.59 +54.85 | HIP36915 | 4 | 385 |
| 04.35 +08.41 | HIP20273 | 4 | 384 |
| 04.79 -22.56 | HIP22252 | 4 | 384 |
| 03.52 +13.58 | HIP16399 | 4 | 384 |
| 07.83 +08.18 | HIP38213 | 4 | 384 |
| 20.10 +29.94 | HIP99015 | 4 | 384 |
| 04.52 +14.69 | HIP21090 | 4 | 384 |
| 06.49 +39.65 | HIP30911 | 4 | 384 |
| 06.83 +40.18 | HIP32719 | 4 | 384 |
| 10.18 +16.04 | HIP49846 | 4 | 384 |
| 10.82 +20.89 | HIP52891 | 4 | 384 |
| 08.01 +49.40 | HIP39167 | 4 | 384 |
| 07.49 -17.34 | HIP36392 | 4 | 383 |
| 09.99 +47.94 | HIP48986 | 4 | 383 |
| 19.62 +40.59 | HIP96528 | 4 | 383 |
| 03.83 +58.77 | HIP17887 | 4 | 383 |
| 06.00 +00.06 | HIP28395 | 4 | 383 |
| 06.38 +74.85 | HIP30355 | 4 | 383 |
| 03.05 +22.37 | HIP14211 | 4 | 383 |
| 07.72 +39.30 | HIP37620 | 4 | 383 |
| 15.51 +53.81 | HIP75932 | 4 | 383 |
| 03.96 +21.55 | HIP18502 | 4 | 383 |
| 10.49 +32.01 | HIP51352 | 4 | 383 |
| 10.51 +34.14 | HIP51478 | 4 | 383 |
| 05.26 +12.88 | HIP24527 | 4 | 382 |
| 05.27 +16.94 | HIP24558 | 4 | 382 |
| 05.44 +16.70 | HIP25419 | 4 | 382 |
| 06.92 +30.16 | HIP33287 | 4 | 382 |
| 08.31 +18.96 | HIP40700 | 4 | 382 |
| 04.78 +32.63 | HIP22217 | 4 | 382 |
| 02.98 +35.56 | HIP13875 | 4 | 381 |
| 08.77 +07.81 | HIP43036 | 4 | 381 |
| 02.90 +49.36 | HIP13500 | 4 | 381 |
| 08.37 +60.05 | HIP41031 | 4 | 381 |
| 09.21 +20.67 | HIP45204 | 4 | 381 |
| 04.41 -29.21 | HIP20583 | 4 | 381 |
| 07.60 +32.89 | HIP36952 | 4 | 380 |
| 09.62 +01.88 | HIP47220 | 4 | 380 |
| 07.09 +71.46 | HIP34171 | 4 | 380 |
| 21.10 +04.43 | HIP104137 | 4 | 380 |
| 04.47 +13.87 | HIP20850 | 4 | 380 |
| 03.47 -26.61 | HIP16180 | 4 | 379 |
| 06.89 -27.87 | HIP33073 | 4 | 379 |
| 06.91 -27.56 | HIP33197 | 4 | 379 |
| 07.77 -23.14 | HIP37906 | 4 | 379 |
| 09.37 +01.87 | HIP45967 | 4 | 379 |
| 09.96 +09.98 | HIP48836 | 4 | 379 |
| 07.71 +22.22 | HIP37579 | 4 | 379 |
| 04.13 +05.69 | HIP19271 | 4 | 379 |
| 04.65 +20.78 | HIP21639 | 4 | 379 |
| 05.17 +30.57 | HIP24065 | 4 | 379 |
| 04.83 -14.79 | HIP22424 | 4 | 378 |
| 07.17 +01.64 | HIP34633 | 4 | 378 |
| 09.37 -22.59 | HIP45953 | 4 | 378 |
| 09.96 +54.48 | HIP48821 | 4 | 378 |
| 07.07 +68.90 | HIP34114 | 4 | 378 |
| 10.27 +15.40 | HIP50282 | 4 | 378 |
| 07.58 +23.93 | HIP36887 | 4 | 378 |
| 03.98 -00.94 | HIP18617 | 4 | 378 |
| 04.46 +07.06 | HIP20794 | 4 | 378 |
| 04.64 -20.84 | HIP21596 | 4 | 378 |
| 07.73 +77.88 | HIP37681 | 4 | 378 |
| 08.07 -28.42 | HIP39491 | 4 | 378 |
| 08.98 +25.40 | HIP44089 | 4 | 378 |
| 06.05 -30.41 | HIP28641 | 4 | 378 |
| 06.33 +52.39 | HIP30061 | 4 | 377 |
| 04.00 +14.31 | HIP18692 | 4 | 377 |
| 10.91 +30.53 | HIP53324 | 4 | 377 |
| 08.16 +21.90 | HIP39950 | 4 | 377 |
| 07.72 +18.18 | HIP37622 | 4 | 377 |
| 04.03 -30.47 | HIP18821 | 4 | 377 |
| 09.51 +18.25 | HIP46662 | 4 | 377 |
| 07.51 +19.81 | HIP36481 | 4 | 377 |
| 10.23 +22.46 | HIP50130 | 4 | 376 |
| 03.90 +13.51 | HIP18220 | 4 | 376 |
| 10.24 +15.07 | HIP50155 | 4 | 376 |
| 10.95 +48.70 | HIP53528 | 4 | 375 |
| 22.95 -26.16 | HIP113332 | 4 | 375 |
| 09.64 +01.84 | HIP47292 | 4 | 375 |
| 03.38 +17.73 | HIP15724 | 4 | 375 |
| 06.74 -14.73 | HIP32303 | 4 | 374 |
| 07.05 -13.22 | HIP33973 | 4 | 374 |
| 05.23 +03.58 | HIP24368 | 4 | 374 |
| 06.37 +34.60 | HIP30270 | 4 | 374 |
| 03.65 +52.82 | HIP17033 | 4 | 374 |
| 04.82 +30.31 | HIP22410 | 4 | 374 |
| 13.97 -27.57 | HIP68242 | 4 | 374 |
| 14.00 -26.64 | HIP68374 | 4 | 374 |
| 03.92 -01.26 | HIP18317 | 4 | 374 |
| 08.01 +32.12 | HIP39143 | 4 | 374 |
| 04.97 +09.52 | HIP23077 | 4 | 373 |
| 05.25 +20.88 | HIP24456 | 4 | 373 |
| 09.02 +10.73 | HIP44303 | 4 | 373 |
| 08.13 +21.58 | HIP39780 | 4 | 373 |
| 10.17 +34.13 | HIP49830 | 4 | 373 |
| 08.54 +23.75 | HIP41883 | 4 | 372 |
| 03.50 +18.74 | HIP16271 | 4 | 371 |
| 05.51 +08.40 | HIP25814 | 4 | 371 |
| 03.87 -22.69 | HIP18087 | 4 | 371 |
| 07.20 -24.89 | HIP34785 | 4 | 371 |
| 07.34 -19.82 | HIP35555 | 4 | 371 |
| 07.54 -24.57 | HIP36667 | 4 | 371 |
| 03.09 +06.20 | HIP14384 | 4 | 371 |
| 03.89 +52.56 | HIP18207 | 4 | 370 |
| 04.19 -26.76 | HIP19568 | 4 | 370 |
| 06.85 -29.45 | HIP32842 | 4 | 370 |
| 07.46 +05.23 | HIP36208 | 4 | 370 |
| 07.63 -24.60 | HIP37109 | 4 | 370 |
| 11.20 -26.14 | HIP54704 | 4 | 370 |
| 03.58 +14.18 | HIP16671 | 4 | 370 |
| 03.90 +01.00 | HIP18239 | 4 | 369 |
| 04.20 -27.55 | HIP19600 | 4 | 369 |
| 07.22 -24.23 | HIP34879 | 4 | 369 |
| 11.16 -25.99 | HIP54530 | 4 | 369 |
| 06.79 -18.26 | HIP32530 | 4 | 369 |
| 08.10 +15.35 | HIP39650 | 4 | 369 |
| 07.02 +59.83 | HIP33805 | 4 | 369 |
| 08.02 -26.46 | HIP39196 | 4 | 369 |
| 08.11 +34.77 | HIP39667 | 4 | 369 |
| 11.61 +21.06 | HIP56638 | 4 | 369 |
| 04.03 +33.15 | HIP18786 | 4 | 369 |
| 04.39 +60.82 | HIP20475 | 4 | 368 |
| 06.85 +61.53 | HIP32879 | 4 | 368 |
| 07.47 +04.15 | HIP36310 | 4 | 368 |
| 16.09 -18.72 | HIP78836 | 4 | 368 |
| 06.02 +32.23 | HIP28502 | 4 | 367 |
| 03.10 -13.43 | HIP14413 | 4 | 367 |
| 04.48 -13.66 | HIP20915 | 4 | 367 |
| 06.40 +64.91 | HIP30437 | 4 | 367 |
| 10.68 +55.65 | HIP52256 | 4 | 367 |
| 11.09 +65.81 | HIP54199 | 4 | 367 |
| 06.33 +24.57 | HIP30102 | 4 | 367 |
| 09.42 +24.73 | HIP46202 | 4 | 366 |
| 03.13 -12.04 | HIP14563 | 4 | 366 |
| 04.18 -26.61 | HIP19494 | 4 | 366 |
| 04.36 -24.29 | HIP20351 | 4 | 366 |
| 05.56 -24.34 | HIP26060 | 4 | 366 |
| 06.26 -00.51 | HIP29716 | 4 | 366 |
| 07.97 -15.02 | HIP38969 | 4 | 366 |
| 07.68 +46.99 | HIP37425 | 4 | 366 |
| 03.46 -27.41 | HIP16127 | 4 | 365 |
| 03.08 -12.17 | HIP14330 | 4 | 365 |
| 03.12 -13.76 | HIP14501 | 4 | 365 |
| 08.99 +07.41 | HIP44137 | 4 | 365 |
| 10.47 -20.68 | HIP51282 | 4 | 365 |
| 06.93 +55.64 | HIP33322 | 4 | 365 |
| 07.78 -01.16 | HIP37960 | 4 | 365 |
| 07.95 -14.59 | HIP38862 | 4 | 365 |
| 08.10 +71.21 | HIP39645 | 4 | 365 |
| 08.50 +50.67 | HIP41672 | 4 | 365 |
| 08.50 -31.02 | HIP41706 | 4 | 365 |
| 09.81 +35.74 | HIP48136 | 4 | 365 |
| 09.98 +46.47 | HIP48914 | 4 | 365 |
| 10.02 +46.15 | HIP49102 | 4 | 365 |
| 10.15 -15.49 | HIP49728 | 4 | 365 |
| 10.68 +54.72 | HIP52251 | 4 | 365 |
| 10.19 +18.32 | HIP49928 | 4 | 365 |
| 05.91 +26.64 | HIP27907 | 4 | 364 |
| 07.90 -25.30 | HIP38594 | 4 | 364 |
| 04.22 +16.77 | HIP19696 | 4 | 364 |
| 08.40 -11.44 | HIP41178 | 4 | 364 |
| 19.61 +28.23 | HIP96437 | 4 | 364 |
| 06.41 +38.57 | HIP30486 | 4 | 364 |
| 05.21 +16.31 | HIP24292 | 4 | 363 |
| 09.75 -27.36 | HIP47823 | 4 | 363 |
| 08.60 +31.82 | HIP42179 | 4 | 362 |
| 05.09 +52.81 | HIP23694 | 4 | 362 |
| 06.48 +26.97 | HIP30880 | 4 | 362 |
| 05.18 +15.28 | HIP24104 | 4 | 361 |
| 06.28 +25.50 | HIP29833 | 4 | 361 |
| 07.64 +73.60 | HIP37205 | 4 | 361 |
| 08.15 +49.72 | HIP39875 | 4 | 361 |
| 17.33 +28.47 | HIP84797 | 4 | 361 |
| 09.16 +10.49 | HIP44991 | 4 | 360 |
| 10.78 +51.78 | HIP52724 | 4 | 360 |
| 04.92 +10.21 | HIP22893 | 4 | 360 |
| 07.97 +52.89 | HIP38952 | 4 | 360 |
| 10.01 +80.40 | HIP49059 | 4 | 360 |
| 09.27 +26.47 | HIP45490 | 4 | 360 |
| 08.98 +25.93 | HIP44100 | 4 | 360 |
| 05.51 +63.95 | HIP25800 | 4 | 360 |
| 08.20 +51.91 | HIP40170 | 4 | 360 |
| 03.92 +53.56 | HIP18351 | 4 | 360 |
| 10.47 -21.70 | HIP51237 | 4 | 360 |
| 08.48 +53.86 | HIP41598 | 4 | 360 |
| 04.36 -20.92 | HIP20342 | 4 | 359 |
| 04.54 +77.62 | HIP21165 | 4 | 359 |
| 03.69 +75.40 | HIP17213 | 4 | 359 |
| 07.66 +33.46 | HIP37312 | 4 | 359 |
| 07.68 +71.67 | HIP37393 | 4 | 359 |
| 08.43 -29.93 | HIP41317 | 4 | 359 |
| 05.72 +05.50 | HIP26936 | 4 | 359 |
| 08.66 +24.62 | HIP42479 | 4 | 359 |
| 03.64 +52.60 | HIP16973 | 4 | 359 |
| 06.28 +25.21 | HIP29810 | 4 | 358 |
| 10.72 +52.32 | HIP52439 | 4 | 358 |
| 06.97 +22.48 | HIP33537 | 4 | 358 |
| 09.16 +54.40 | HIP44957 | 4 | 358 |
| 09.30 +26.01 | HIP45612 | 4 | 358 |
| 09.72 +12.60 | HIP47683 | 4 | 358 |
| 10.94 +27.74 | HIP53471 | 4 | 358 |
| 04.10 +47.22 | HIP19107 | 4 | 358 |
| 04.59 +60.74 | HIP21384 | 4 | 357 |
| 04.60 +01.45 | HIP21411 | 4 | 357 |
| 06.14 +11.46 | HIP29111 | 4 | 357 |
| 08.63 +60.59 | HIP42338 | 4 | 357 |
| 09.27 +47.69 | HIP45476 | 4 | 357 |
| 10.13 +57.25 | HIP49614 | 4 | 357 |
| 10.37 +59.55 | HIP50760 | 4 | 357 |
| 04.44 +06.05 | HIP20705 | 4 | 356 |
| 08.43 +08.01 | HIP41340 | 4 | 356 |
| 08.46 -30.64 | HIP41477 | 4 | 356 |
| 04.21 +26.05 | HIP19637 | 4 | 356 |
| 09.36 +13.11 | HIP45879 | 4 | 356 |
| 09.68 +66.15 | HIP47499 | 4 | 356 |
| 04.25 +26.47 | HIP19803 | 4 | 356 |
| 03.20 +50.38 | HIP14853 | 4 | 356 |
| 05.58 +18.12 | HIP26183 | 4 | 356 |
| 08.29 +69.69 | HIP40615 | 4 | 356 |
| 04.81 +05.54 | HIP22362 | 4 | 356 |
| 10.82 +06.74 | HIP52917 | 4 | 356 |
| 04.77 +78.34 | HIP22160 | 4 | 355 |
| 10.07 +70.87 | HIP49344 | 4 | 355 |
| 05.88 +03.05 | HIP27799 | 4 | 354 |
| 06.36 -00.54 | HIP30243 | 4 | 354 |
| 08.58 +32.92 | HIP42104 | 4 | 354 |
| 10.25 +09.21 | HIP50203 | 4 | 354 |
| 10.45 +46.94 | HIP51163 | 4 | 354 |
| 06.59 +34.95 | HIP31431 | 4 | 354 |
| 08.23 +34.11 | HIP40294 | 4 | 354 |
| 09.03 +17.33 | HIP44358 | 4 | 354 |
| 04.89 +23.02 | HIP22751 | 4 | 353 |
| 07.71 +65.00 | HIP37587 | 4 | 353 |
| 16.24 +40.96 | HIP79591 | 4 | 353 |
| 04.89 +24.18 | HIP22729 | 4 | 352 |
| 10.59 +05.81 | HIP51829 | 4 | 352 |
| 11.06 +54.53 | HIP54040 | 4 | 352 |
| 05.13 +26.33 | HIP23884 | 4 | 352 |
| 08.40 +50.17 | HIP41175 | 4 | 352 |
| 14.73 +00.05 | HIP72014 | 4 | 352 |
| 06.33 +16.01 | HIP30067 | 4 | 352 |
| 06.34 +02.26 | HIP30112 | 4 | 352 |
| 07.80 +70.34 | HIP38047 | 4 | 352 |
| 09.42 +59.99 | HIP46170 | 4 | 352 |
| 14.56 +09.33 | HIP71190 | 4 | 352 |
| 08.62 +17.25 | HIP42277 | 4 | 351 |
| 10.44 +46.17 | HIP51096 | 4 | 351 |
| 05.83 +27.12 | HIP27527 | 4 | 351 |
| 06.27 +56.93 | HIP29777 | 4 | 351 |
| 09.14 +30.32 | HIP44856 | 4 | 351 |
| 09.75 +68.70 | HIP47838 | 4 | 351 |
| 09.99 +50.98 | HIP48954 | 4 | 351 |
| 03.12 +16.08 | HIP14500 | 4 | 351 |
| 04.68 +29.73 | HIP21804 | 4 | 351 |
| 05.11 +26.50 | HIP23772 | 4 | 351 |
| 10.54 +46.53 | HIP51599 | 4 | 351 |
| 08.55 +71.56 | HIP41934 | 4 | 351 |
| 04.85 +32.89 | HIP22559 | 4 | 350 |
| 05.87 -31.50 | HIP27708 | 4 | 350 |
| 07.18 +53.18 | HIP34685 | 4 | 349 |
| 07.68 +72.37 | HIP37427 | 4 | 349 |
| 09.81 +69.77 | HIP48139 | 4 | 349 |
| 10.44 +48.08 | HIP51128 | 4 | 348 |
| 10.95 +56.63 | HIP53519 | 4 | 348 |
| 04.20 +25.98 | HIP19583 | 4 | 348 |
| 08.05 +27.93 | HIP39393 | 4 | 348 |
| 06.60 +13.99 | HIP31495 | 4 | 348 |
| 09.51 +52.15 | HIP46613 | 4 | 347 |
| 07.76 +64.72 | HIP37866 | 4 | 347 |
| 08.84 +17.69 | HIP43418 | 4 | 347 |
| 10.46 -21.44 | HIP51196 | 4 | 347 |
| 10.38 +07.92 | HIP50853 | 4 | 347 |
| 11.13 +32.35 | HIP54403 | 4 | 347 |
| 08.77 +55.06 | HIP43032 | 4 | 347 |
| 10.02 +51.93 | HIP49082 | 4 | 347 |
| 10.42 +30.37 | HIP50997 | 4 | 347 |
| 07.88 +22.56 | HIP38492 | 4 | 346 |
| 05.72 +52.49 | HIP26974 | 4 | 346 |
| 07.92 +51.86 | HIP38673 | 4 | 346 |
| 09.00 +24.93 | HIP44208 | 4 | 346 |
| 15.27 +37.07 | HIP74730 | 4 | 346 |
| 04.38 +51.94 | HIP20434 | 4 | 346 |
| 09.36 +05.50 | HIP45882 | 4 | 346 |
| 10.55 -26.81 | HIP51657 | 4 | 345 |
| 08.11 +55.89 | HIP39685 | 4 | 345 |
| 04.51 +28.13 | HIP21010 | 4 | 345 |
| 08.21 +50.01 | HIP40190 | 4 | 345 |
| 09.17 +18.95 | HIP45005 | 4 | 345 |
| 09.27 +14.13 | HIP45488 | 4 | 345 |
| 05.61 +25.59 | HIP26330 | 4 | 345 |
| 06.54 -29.58 | HIP31187 | 4 | 344 |
| 03.92 +19.38 | HIP18347 | 4 | 344 |
| 05.03 +31.63 | HIP23409 | 4 | 344 |
| 10.94 +40.25 | HIP53483 | 4 | 344 |
| 03.23 +51.78 | HIP15037 | 4 | 343 |
| 07.85 +32.98 | HIP38340 | 4 | 343 |
| 06.60 -29.40 | HIP31543 | 4 | 343 |
| 09.08 +22.83 | HIP44586 | 4 | 342 |
| 11.07 +53.38 | HIP54094 | 4 | 342 |
| 03.15 -11.73 | HIP14645 | 4 | 342 |
| 06.61 +14.00 | HIP31581 | 4 | 342 |
| 19.92 +47.15 | HIP97992 | 4 | 342 |
| 09.63 +24.07 | HIP47258 | 4 | 342 |
| 05.72 +59.70 | HIP26968 | 4 | 341 |
| 04.15 +04.99 | HIP19373 | 4 | 339 |
| 04.86 +29.83 | HIP22571 | 4 | 339 |
| 08.33 +55.20 | HIP40847 | 4 | 338 |
| 10.23 +08.41 | HIP50089 | 4 | 338 |
| 10.61 +14.05 | HIP51949 | 4 | 338 |
| 09.98 -27.07 | HIP48930 | 4 | 338 |
| 06.65 +15.21 | HIP31802 | 4 | 337 |
| 08.99 +66.24 | HIP44163 | 4 | 337 |
| 08.23 +10.85 | HIP40280 | 4 | 336 |
| 08.38 +50.95 | HIP41087 | 4 | 336 |
| 03.37 +01.77 | HIP15708 | 4 | 335 |
| 00.32 +13.58 | HIP1541 | 4 | 334 |
| 17.05 +25.64 | HIP83412 | 4 | 334 |
| 12.37 +18.37 | HIP60326 | 4 | 334 |
| 03.29 -27.70 | HIP15312 | 4 | 333 |
| 09.39 +32.92 | HIP46023 | 4 | 333 |
| 09.78 +35.22 | HIP47997 | 4 | 333 |
| 11.01 +23.71 | HIP53813 | 4 | 333 |
| 03.81 +57.77 | HIP17790 | 4 | 333 |
| 05.97 +51.13 | HIP28250 | 4 | 333 |
| 10.54 +31.42 | HIP51606 | 4 | 333 |
| 19.78 +48.93 | HIP97341 | 4 | 333 |
| 21.07 +03.98 | HIP103983 | 4 | 333 |
| 04.50 +16.67 | HIP20978 | 4 | 333 |
| 10.21 -19.15 | HIP50013 | 4 | 332 |
| 06.64 +72.85 | HIP31723 | 4 | 332 |
| 06.69 +14.85 | HIP32009 | 4 | 332 |
| 03.41 +53.17 | HIP15892 | 4 | 331 |
| 06.67 +46.61 | HIP31926 | 4 | 331 |
| 10.89 +65.43 | HIP53263 | 4 | 331 |
| 03.73 +14.62 | HIP17426 | 4 | 329 |
| 09.11 +66.71 | HIP44695 | 4 | 328 |
| 09.99 +65.34 | HIP48969 | 4 | 328 |
| 09.35 +21.67 | HIP45844 | 4 | 328 |
| 06.17 -30.91 | HIP29255 | 4 | 328 |
| 20.52 +05.98 | HIP101247 | 4 | 328 |
| 03.31 +10.31 | HIP15401 | 4 | 328 |
| 11.07 -00.03 | HIP54085 | 4 | 328 |
| 05.48 +09.44 | HIP25666 | 4 | 327 |
| 00.72 -25.92 | HIP3411 | 4 | 326 |
| 12.83 +74.22 | HIP62598 | 4 | 325 |
| 04.50 +27.28 | HIP20964 | 4 | 325 |
| 08.06 +07.12 | HIP39405 | 4 | 325 |
| 09.78 +16.53 | HIP47972 | 4 | 325 |
| 05.36 +23.28 | HIP25051 | 4 | 324 |
| 10.26 +06.93 | HIP50255 | 4 | 324 |
| 04.98 -27.06 | HIP23127 | 4 | 324 |
| 06.68 +51.98 | HIP31965 | 4 | 324 |
| 11.78 +13.69 | HIP57442 | 4 | 323 |
| 10.88 -17.03 | HIP53172 | 4 | 322 |
| 09.78 +01.72 | HIP47966 | 4 | 322 |
| 09.79 +02.32 | HIP48025 | 4 | 322 |
| 05.77 +25.48 | HIP27236 | 4 | 322 |
| 01.69 +20.73 | HIP7910 | 4 | 321 |
| 08.86 -29.46 | HIP43517 | 4 | 321 |
| 10.76 +32.16 | HIP52609 | 4 | 321 |
| 10.74 +13.00 | HIP52519 | 4 | 320 |
| 03.99 +18.66 | HIP18643 | 4 | 319 |
| 09.61 -22.43 | HIP47156 | 4 | 319 |
| 10.31 +10.13 | HIP50496 | 4 | 319 |
| 10.48 +00.84 | HIP51317 | 4 | 319 |
| 06.33 -30.45 | HIP30095 | 4 | 319 |
| 09.52 +13.49 | HIP46680 | 4 | 319 |
| 10.80 +65.46 | HIP52805 | 4 | 319 |
| 09.26 +54.30 | HIP45452 | 4 | 318 |
| 10.21 +71.58 | HIP50014 | 4 | 318 |
| 00.55 +42.00 | HIP2604 | 4 | 317 |
| 08.55 -31.50 | HIP41926 | 4 | 316 |
| 08.96 +67.30 | HIP44015 | 4 | 315 |
| 12.97 +11.58 | HIP63290 | 4 | 315 |
| 09.72 +54.99 | HIP47682 | 4 | 315 |
| 09.95 +57.92 | HIP48788 | 4 | 315 |
| 09.68 +71.76 | HIP47469 | 4 | 315 |
| 03.31 +32.67 | HIP15406 | 3 | 314 |
| 02.71 +50.64 | HIP12643 | 3 | 314 |
| 07.92 +78.74 | HIP38697 | 3 | 314 |
| 19.35 +44.16 | HIP95098 | 3 | 314 |
| 20.08 +29.46 | HIP98886 | 3 | 314 |
| 14.23 -20.21 | HIP69528 | 3 | 314 |
| 11.79 +76.23 | HIP57545 | 3 | 313 |
| 10.27 +48.74 | HIP50296 | 3 | 313 |
| 12.47 +23.14 | HIP60830 | 3 | 313 |
| 10.22 +51.48 | HIP50035 | 3 | 312 |
| 10.28 +25.86 | HIP50355 | 3 | 312 |
| 10.97 +61.95 | HIP53624 | 3 | 312 |
| 06.39 +32.97 | HIP30367 | 3 | 311 |
| 13.15 +16.57 | HIP64140 | 3 | 311 |
| 10.77 +35.64 | HIP52675 | 3 | 310 |
| 10.78 +34.91 | HIP52735 | 3 | 310 |
| 16.03 -28.37 | HIP78523 | 3 | 310 |
| 10.93 +25.69 | HIP53452 | 3 | 309 |
| 15.25 +36.83 | HIP74643 | 3 | 309 |
| 09.71 +55.67 | HIP47629 | 3 | 308 |
| 10.56 +08.59 | HIP51696 | 3 | 308 |
| 07.61 +48.71 | HIP37051 | 3 | 306 |
| 03.14 +12.38 | HIP14606 | 3 | 306 |
| 11.01 +61.04 | HIP53828 | 3 | 306 |
| 19.15 +25.32 | HIP94085 | 3 | 306 |
| 10.96 +61.43 | HIP53558 | 3 | 306 |
| 07.52 +48.37 | HIP36550 | 3 | 306 |
| 08.73 +51.40 | HIP42832 | 3 | 305 |
| 02.51 -12.22 | HIP11694 | 3 | 304 |
| 00.01 -23.45 | HIP24 | 3 | 304 |
| 08.20 +33.62 | HIP40159 | 3 | 304 |
| 09.65 -22.96 | HIP47353 | 3 | 303 |
| 03.05 -28.03 | HIP14171 | 3 | 302 |
| 09.83 +34.11 | HIP48202 | 3 | 302 |
| 09.58 +24.90 | HIP47026 | 3 | 301 |
| 06.12 +76.17 | HIP29005 | 3 | 301 |
| 05.01 +30.90 | HIP23302 | 3 | 301 |
| 07.12 -21.46 | HIP34361 | 3 | 301 |
| 05.27 -29.87 | HIP24585 | 3 | 300 |
| 03.76 -27.86 | HIP17544 | 3 | 298 |
| 04.40 +18.00 | HIP20563 | 3 | 297 |
| 06.54 -27.03 | HIP31148 | 3 | 297 |
| 09.50 +32.04 | HIP46590 | 3 | 297 |
| 03.25 +12.43 | HIP15117 | 3 | 297 |
| 04.23 +82.92 | HIP19739 | 3 | 296 |
| 00.64 +43.00 | HIP3022 | 3 | 296 |
| 02.50 -13.25 | HIP11614 | 3 | 295 |
| 11.46 +03.98 | HIP55915 | 3 | 295 |
| 16.43 -14.02 | HIP80495 | 3 | 295 |
| 19.02 +32.68 | HIP93400 | 3 | 295 |
| 20.63 +42.85 | HIP101818 | 3 | 295 |
| 20.67 +41.25 | HIP101989 | 3 | 295 |
| 23.33 +17.68 | HIP115185 | 3 | 295 |
| 10.32 +09.21 | HIP50539 | 3 | 295 |
| 14.06 -27.01 | HIP68688 | 3 | 295 |
| 00.35 -15.67 | HIP1654 | 3 | 294 |
| 18.40 -30.32 | HIP90158 | 3 | 294 |
| 18.44 -30.39 | HIP90397 | 3 | 294 |
| 20.46 +19.62 | HIP100906 | 3 | 294 |
| 11.40 +03.62 | HIP55638 | 3 | 294 |
| 12.56 +44.10 | HIP61310 | 3 | 294 |
| 17.33 -18.59 | HIP84813 | 3 | 294 |
| 22.19 +15.10 | HIP109557 | 3 | 294 |
| 22.44 -18.01 | HIP110758 | 3 | 294 |
| 16.47 -13.10 | HIP80658 | 3 | 294 |
| 16.26 -27.21 | HIP79658 | 3 | 293 |
| 18.04 +00.10 | HIP88324 | 3 | 293 |
| 17.19 +04.69 | HIP84082 | 3 | 292 |
| 23.27 +30.67 | HIP114886 | 3 | 292 |
| 07.18 +63.32 | HIP34655 | 3 | 292 |
| 07.14 -30.00 | HIP34469 | 3 | 292 |
| 16.35 -19.95 | HIP80109 | 3 | 292 |
| 04.43 -13.28 | HIP20697 | 3 | 291 |
| 08.88 +09.19 | HIP43618 | 3 | 291 |
| 07.43 +48.51 | HIP36064 | 3 | 290 |
| 10.28 +41.28 | HIP50325 | 3 | 287 |
| 09.28 +46.18 | HIP45515 | 3 | 286 |
| 04.82 -28.72 | HIP22373 | 3 | 285 |
| 06.83 +32.11 | HIP32752 | 3 | 285 |
| 05.32 +58.11 | HIP24801 | 3 | 284 |
| 03.70 -28.79 | HIP17258 | 3 | 284 |
| 10.07 +50.16 | HIP49345 | 3 | 284 |
| 09.33 +24.42 | HIP45756 | 3 | 284 |
| 10.93 +29.32 | HIP53442 | 3 | 284 |
| 09.36 +18.75 | HIP45917 | 3 | 283 |
| 04.16 +17.50 | HIP19420 | 3 | 282 |
| 08.79 +14.16 | HIP43164 | 3 | 282 |
| 10.37 -24.21 | HIP50779 | 3 | 282 |
| 04.87 -29.43 | HIP22643 | 3 | 281 |
| 04.85 -29.78 | HIP22515 | 3 | 280 |
| 08.19 +47.58 | HIP40089 | 3 | 280 |
| 04.36 +16.80 | HIP20352 | 3 | 279 |
| 05.55 +20.04 | HIP26018 | 3 | 279 |
| 04.70 -29.63 | HIP21860 | 3 | 278 |
| 04.73 +11.17 | HIP21983 | 3 | 277 |
| 10.29 +18.80 | HIP50375 | 3 | 277 |
| 05.02 -23.71 | HIP23347 | 3 | 275 |
| 21.62 +29.47 | HIP106736 | 3 | 275 |
| 10.79 +64.38 | HIP52790 | 3 | 275 |
| 13.38 -18.03 | HIP65275 | 3 | 275 |
| 05.43 -30.67 | HIP25408 | 3 | 274 |
| 10.37 -28.85 | HIP50807 | 3 | 274 |
| 11.03 +44.36 | HIP53914 | 3 | 274 |
| 14.05 +43.23 | HIP68658 | 3 | 274 |
| 21.61 +29.54 | HIP106674 | 3 | 274 |
| 07.62 +57.55 | HIP37061 | 3 | 274 |
| 05.45 -31.57 | HIP25490 | 3 | 274 |
| 16.23 -26.29 | HIP79546 | 3 | 274 |
| 19.99 +42.17 | HIP98381 | 3 | 274 |
| 01.83 -18.94 | HIP8507 | 3 | 274 |
| 14.59 +10.43 | HIP71330 | 3 | 274 |
| 11.40 +49.38 | HIP55652 | 3 | 274 |
| 11.42 +49.38 | HIP55744 | 3 | 274 |
| 05.48 -29.35 | HIP25632 | 3 | 273 |
| 07.95 +21.71 | HIP38871 | 3 | 273 |
| 09.56 +24.26 | HIP46923 | 3 | 273 |
| 11.36 +18.19 | HIP55486 | 3 | 273 |
| 15.14 +03.18 | HIP74088 | 3 | 273 |
| 02.74 +07.39 | HIP12771 | 3 | 271 |
| 02.79 -17.56 | HIP13025 | 3 | 271 |
| 02.81 +08.92 | HIP13110 | 3 | 271 |
| 05.58 +28.10 | HIP26196 | 3 | 271 |
| 15.11 +02.66 | HIP73925 | 3 | 271 |
| 00.51 -15.92 | HIP2386 | 3 | 270 |
| 15.38 +35.40 | HIP75246 | 3 | 270 |
| 11.03 +20.83 | HIP53898 | 3 | 270 |
| 03.50 -19.62 | HIP16286 | 3 | 270 |
| 11.62 +25.42 | HIP56690 | 3 | 269 |
| 11.02 +19.96 | HIP53866 | 3 | 268 |
| 02.78 -18.04 | HIP12951 | 3 | 267 |
| 04.91 +17.61 | HIP22805 | 3 | 266 |
| 08.45 -26.92 | HIP41415 | 3 | 265 |
| 01.80 -13.20 | HIP8360 | 3 | 265 |
| 01.80 -12.77 | HIP8361 | 3 | 265 |
| 02.09 +19.19 | HIP9725 | 3 | 265 |
| 02.29 +27.14 | HIP10682 | 3 | 265 |
| 02.32 +35.22 | HIP10789 | 3 | 265 |
| 02.77 +11.78 | HIP12929 | 3 | 265 |
| 02.84 +13.71 | HIP13223 | 3 | 265 |
| 02.85 +01.68 | HIP13302 | 3 | 265 |
| 02.87 +01.93 | HIP13386 | 3 | 265 |
| 02.88 -26.72 | HIP13405 | 3 | 265 |
| 03.23 +10.45 | HIP15054 | 3 | 265 |
| 03.75 -21.58 | HIP17508 | 3 | 265 |
| 04.16 +09.31 | HIP19441 | 3 | 265 |
| 04.20 +24.43 | HIP19586 | 3 | 265 |
| 04.37 +45.61 | HIP20413 | 3 | 265 |
| 04.80 +59.24 | HIP22303 | 3 | 265 |
| 05.64 +35.98 | HIP26506 | 3 | 265 |
| 06.32 +31.36 | HIP30020 | 3 | 265 |
| 18.87 +49.56 | HIP92582 | 3 | 265 |
| 19.27 +47.09 | HIP94701 | 3 | 265 |
| 19.44 +50.99 | HIP95568 | 3 | 265 |
| 19.92 +41.87 | HIP98001 | 3 | 265 |
| 02.29 +26.28 | HIP10668 | 3 | 264 |
| 02.36 +07.18 | HIP11005 | 3 | 264 |
| 02.39 +06.60 | HIP11151 | 3 | 264 |
| 02.70 +45.49 | HIP12580 | 3 | 264 |
| 02.74 +10.96 | HIP12787 | 3 | 264 |
| 02.75 +45.60 | HIP12818 | 3 | 264 |
| 02.84 -26.04 | HIP13220 | 3 | 264 |
| 02.84 +02.44 | HIP13234 | 3 | 264 |
| 02.87 +26.97 | HIP13398 | 3 | 264 |
| 02.93 +26.87 | HIP13642 | 3 | 264 |
| 04.18 +08.54 | HIP19532 | 3 | 264 |
| 04.39 +45.08 | HIP20494 | 3 | 264 |
| 04.90 +17.03 | HIP22782 | 3 | 264 |
| 05.47 +46.51 | HIP25571 | 3 | 264 |
| 06.76 +44.23 | HIP32382 | 3 | 264 |
| 07.20 +10.51 | HIP34767 | 3 | 264 |
| 19.12 +45.92 | HIP93921 | 3 | 263 |
| 04.49 +17.86 | HIP20948 | 3 | 263 |
| 04.66 +45.62 | HIP21701 | 3 | 262 |
| 09.93 +62.62 | HIP48698 | 3 | 262 |
| 09.92 +68.94 | HIP48635 | 3 | 261 |
| 07.24 +32.68 | HIP34977 | 3 | 260 |
| 04.67 +67.59 | HIP21750 | 3 | 259 |
| 06.81 +43.83 | HIP32665 | 3 | 259 |
| 04.56 +19.01 | HIP21261 | 3 | 258 |
| 07.52 +02.17 | HIP36557 | 3 | 257 |
| 05.48 -17.43 | HIP25670 | 3 | 256 |
| 05.95 -13.87 | HIP28160 | 3 | 256 |
| 10.52 +15.91 | HIP51479 | 3 | 255 |
| 09.17 +75.61 | HIP45012 | 3 | 255 |
| 16.23 -21.40 | HIP79524 | 3 | 255 |
| 22.90 +19.89 | HIP113086 | 3 | 255 |
| 09.25 +32.75 | HIP45374 | 3 | 255 |
| 11.83 +02.23 | HIP57681 | 3 | 254 |
| 07.59 +17.35 | HIP36903 | 3 | 254 |
| 16.52 +01.47 | HIP80899 | 3 | 254 |
| 19.47 +42.78 | HIP95703 | 3 | 254 |
| 03.69 -27.92 | HIP17221 | 3 | 252 |
| 09.41 +75.23 | HIP46158 | 3 | 251 |
| 10.95 +65.28 | HIP53535 | 3 | 249 |
| 05.29 +22.83 | HIP24668 | 3 | 249 |
| 03.04 +79.76 | HIP14159 | 3 | 248 |
| 00.22 +84.04 | HIP1052 | 3 | 247 |
| 00.64 +84.67 | HIP3006 | 3 | 247 |
| 02.45 +74.01 | HIP11399 | 3 | 247 |
| 07.84 +38.53 | HIP38289 | 3 | 247 |
| 13.52 -18.07 | HIP65956 | 3 | 247 |
| 13.86 +37.72 | HIP67631 | 3 | 247 |
| 13.90 +37.02 | HIP67886 | 3 | 247 |
| 13.94 +05.29 | HIP68081 | 3 | 247 |
| 13.95 +04.13 | HIP68119 | 3 | 247 |
| 13.97 +04.65 | HIP68227 | 3 | 247 |
| 14.30 +36.05 | HIP69885 | 3 | 247 |
| 14.76 +37.74 | HIP72175 | 3 | 247 |
| 14.87 +15.31 | HIP72724 | 3 | 247 |
| 14.92 +15.32 | HIP73003 | 3 | 247 |
| 15.02 +45.43 | HIP73470 | 3 | 247 |
| 15.06 +45.65 | HIP73665 | 3 | 247 |
| 15.06 +46.16 | HIP73668 | 3 | 247 |
| 15.19 +19.45 | HIP74325 | 3 | 247 |
| 15.20 +11.48 | HIP74366 | 3 | 247 |
| 15.22 +20.31 | HIP74457 | 3 | 247 |
| 15.24 +10.81 | HIP74589 | 3 | 247 |
| 15.24 +11.52 | HIP74608 | 3 | 247 |
| 15.31 +25.77 | HIP74933 | 3 | 247 |
| 15.34 +16.19 | HIP75048 | 3 | 247 |
| 15.41 +21.39 | HIP75446 | 3 | 247 |
| 15.48 +21.18 | HIP75766 | 3 | 247 |
| 15.48 +15.89 | HIP75776 | 3 | 247 |
| 15.51 -24.30 | HIP75934 | 3 | 247 |
| 15.60 +28.12 | HIP76374 | 3 | 247 |
| 15.60 +28.41 | HIP76380 | 3 | 247 |
| 15.62 +04.48 | HIP76465 | 3 | 247 |
| 23.60 +83.76 | HIP116481 | 3 | 247 |
| 00.10 +78.74 | HIP487 | 3 | 247 |
| 05.10 +30.67 | HIP23703 | 3 | 247 |
| 09.80 +34.43 | HIP48081 | 3 | 246 |
| 15.34 +15.38 | HIP75090 | 3 | 246 |
| 00.17 +84.15 | HIP822 | 3 | 246 |
| 10.50 +14.14 | HIP51392 | 3 | 246 |
| 07.99 +20.84 | HIP39064 | 3 | 245 |
| 13.82 +37.29 | HIP67460 | 3 | 245 |
| 02.26 +74.14 | HIP10550 | 3 | 245 |
| 15.28 +25.75 | HIP74773 | 3 | 245 |
| 15.34 +16.56 | HIP75066 | 3 | 245 |
| 04.52 -19.47 | HIP21093 | 3 | 245 |
| 11.30 -29.40 | HIP55176 | 3 | 244 |
| 02.49 +37.97 | HIP11591 | 3 | 241 |
| 07.89 -27.46 | HIP38550 | 3 | 241 |
| 07.32 +20.96 | HIP35459 | 3 | 240 |
| 10.09 -30.24 | HIP49401 | 3 | 240 |
| 10.11 -30.12 | HIP49502 | 3 | 240 |
| 10.95 -17.45 | HIP53509 | 3 | 240 |
| 09.64 +03.88 | HIP47282 | 3 | 239 |
| 05.30 -28.17 | HIP24696 | 3 | 238 |
| 03.22 -28.97 | HIP15005 | 3 | 238 |
| 09.94 -25.34 | HIP48757 | 3 | 238 |
| 07.80 -27.52 | HIP38072 | 3 | 238 |
| 18.94 +47.66 | HIP92983 | 3 | 236 |
| 14.84 +17.18 | HIP72580 | 3 | 236 |
| 15.15 -25.05 | HIP74136 | 3 | 236 |
| 16.04 +34.44 | HIP78550 | 3 | 236 |
| 16.97 +48.00 | HIP83069 | 3 | 236 |
| 19.83 +08.03 | HIP97571 | 3 | 236 |
| 21.56 -21.97 | HIP106454 | 3 | 236 |
| 22.88 -15.31 | HIP112959 | 3 | 236 |
| 19.73 +48.93 | HIP97071 | 3 | 235 |
| 19.76 +39.50 | HIP97236 | 3 | 235 |
| 01.48 +35.77 | HIP6903 | 3 | 234 |
| 05.20 -28.22 | HIP24213 | 3 | 234 |
| 10.79 +58.71 | HIP52749 | 3 | 233 |
| 10.14 +33.51 | HIP49680 | 3 | 233 |
| 18.95 +51.72 | HIP93029 | 3 | 233 |
| 07.22 +65.25 | HIP34907 | 3 | 232 |
| 02.80 -17.55 | HIP13091 | 3 | 232 |
| 03.17 +13.00 | HIP14711 | 3 | 230 |
| 04.16 +69.54 | HIP19422 | 3 | 230 |
| 05.96 +33.26 | HIP28212 | 3 | 229 |
| 07.14 +52.69 | HIP34434 | 3 | 229 |
| 08.45 +21.96 | HIP41431 | 3 | 229 |
| 08.79 +29.11 | HIP43145 | 3 | 228 |
| 06.32 +66.65 | HIP30033 | 3 | 228 |
| 06.90 +78.01 | HIP33156 | 3 | 227 |
| 14.81 +16.14 | HIP72414 | 3 | 227 |
| 10.90 +30.29 | HIP53275 | 3 | 226 |
| 11.26 -25.70 | HIP55014 | 3 | 225 |
| 19.31 +43.24 | HIP94898 | 3 | 225 |
| 09.49 +05.05 | HIP46542 | 2 | 224 |
| 08.87 +50.41 | HIP43542 | 2 | 222 |
| 10.63 +05.09 | HIP52014 | 2 | 222 |
| 03.38 -30.19 | HIP15725 | 2 | 222 |
| 08.45 +32.20 | HIP41423 | 2 | 220 |
| 08.69 +35.97 | HIP42623 | 2 | 219 |
| 06.85 +62.23 | HIP32889 | 2 | 217 |
| 09.01 +53.33 | HIP44209 | 2 | 217 |
| 02.33 +31.34 | HIP10868 | 2 | 216 |
| 08.80 +50.28 | HIP43194 | 2 | 216 |
| 10.28 +13.44 | HIP50324 | 2 | 216 |
| 17.13 +32.11 | HIP83827 | 2 | 216 |
| 20.62 +12.05 | HIP101710 | 2 | 216 |
| 04.86 -31.57 | HIP22589 | 2 | 216 |
| 10.33 +53.53 | HIP50553 | 2 | 216 |
| 09.15 +50.78 | HIP44907 | 2 | 215 |
| 22.93 -30.38 | HIP113221 | 2 | 215 |
| 23.21 -21.94 | HIP114584 | 2 | 215 |
| 10.46 +12.98 | HIP51208 | 2 | 215 |
| 10.82 -31.06 | HIP52888 | 2 | 215 |
| 11.15 +53.68 | HIP54508 | 2 | 214 |
| 03.43 -22.06 | HIP15989 | 2 | 213 |
| 09.08 +11.57 | HIP44579 | 2 | 213 |
| 06.00 +28.75 | HIP28411 | 2 | 212 |
| 10.49 +27.50 | HIP51340 | 2 | 211 |
| 11.07 +08.17 | HIP54100 | 2 | 211 |
| 03.01 +15.03 | HIP14048 | 2 | 211 |
| 08.53 +48.38 | HIP41854 | 2 | 211 |
| 08.54 +48.24 | HIP41867 | 2 | 211 |
| 06.05 +47.81 | HIP28634 | 2 | 210 |
| 06.71 +47.74 | HIP32135 | 2 | 210 |
| 19.68 +44.53 | HIP96816 | 2 | 210 |
| 02.95 +07.65 | HIP13757 | 2 | 210 |
| 02.96 +07.78 | HIP13787 | 2 | 210 |
| 02.99 +12.00 | HIP13921 | 2 | 210 |
| 02.99 +06.34 | HIP13932 | 2 | 210 |
| 03.00 +11.98 | HIP13966 | 2 | 210 |
| 03.00 +07.10 | HIP13974 | 2 | 210 |
| 03.03 +13.24 | HIP14111 | 2 | 210 |
| 03.25 -26.45 | HIP15095 | 2 | 210 |
| 03.27 +58.17 | HIP15220 | 2 | 210 |
| 03.34 +08.45 | HIP15563 | 2 | 210 |
| 03.42 -12.01 | HIP15916 | 2 | 210 |
| 03.71 +21.47 | HIP17316 | 2 | 210 |
| 03.80 +10.17 | HIP17762 | 2 | 210 |
| 03.84 -24.04 | HIP17958 | 2 | 210 |
| 04.02 +04.54 | HIP18761 | 2 | 210 |
| 04.11 +17.66 | HIP19187 | 2 | 210 |
| 04.37 +30.42 | HIP20415 | 2 | 210 |
| 04.44 +52.50 | HIP20722 | 2 | 210 |
| 04.54 +34.14 | HIP21197 | 2 | 210 |
| 04.60 +80.29 | HIP21407 | 2 | 210 |
| 04.61 +11.91 | HIP21475 | 2 | 210 |
| 04.82 +27.01 | HIP22382 | 2 | 210 |
| 05.03 +16.19 | HIP23423 | 2 | 210 |
| 05.07 +12.17 | HIP23597 | 2 | 210 |
| 05.08 +58.32 | HIP23658 | 2 | 210 |
| 05.29 +82.25 | HIP24680 | 2 | 210 |
| 05.40 +30.24 | HIP25237 | 2 | 210 |
| 05.41 +23.59 | HIP25309 | 2 | 210 |
| 05.46 +10.78 | HIP25554 | 2 | 210 |
| 05.82 +23.62 | HIP27496 | 2 | 210 |
| 05.83 +32.97 | HIP27528 | 2 | 210 |
| 06.02 +27.28 | HIP28528 | 2 | 210 |
| 06.10 +35.39 | HIP28908 | 2 | 210 |
| 06.59 +18.07 | HIP31435 | 2 | 210 |
| 06.61 +72.01 | HIP31575 | 2 | 210 |
| 06.64 +64.74 | HIP31729 | 2 | 210 |
| 06.87 +10.80 | HIP32962 | 2 | 210 |
| 07.47 +09.50 | HIP36290 | 2 | 210 |
| 08.13 +60.69 | HIP39764 | 2 | 210 |
| 08.24 +79.73 | HIP40337 | 2 | 210 |
| 08.28 +75.74 | HIP40549 | 2 | 210 |
| 09.07 +34.34 | HIP44525 | 2 | 210 |
| 09.85 -21.07 | HIP48323 | 2 | 210 |
| 10.07 +32.65 | HIP49315 | 2 | 210 |
| 10.12 +46.11 | HIP49576 | 2 | 210 |
| 10.13 +46.03 | HIP49634 | 2 | 210 |
| 10.32 +32.39 | HIP50506 | 2 | 210 |
| 10.35 +32.40 | HIP50660 | 2 | 210 |
| 10.60 +29.10 | HIP51859 | 2 | 210 |
| 10.65 +21.60 | HIP52153 | 2 | 210 |
| 10.79 -22.29 | HIP52776 | 2 | 210 |
| 11.02 -19.82 | HIP53882 | 2 | 210 |
| 11.03 -20.06 | HIP53902 | 2 | 210 |
| 11.20 +58.90 | HIP54687 | 2 | 210 |
| 11.20 -01.19 | HIP54728 | 2 | 210 |
| 11.23 +59.14 | HIP54828 | 2 | 210 |
| 11.26 +34.61 | HIP55015 | 2 | 210 |
| 11.32 +14.27 | HIP55254 | 2 | 210 |
| 11.35 +07.06 | HIP55427 | 2 | 210 |
| 11.51 +25.45 | HIP56164 | 2 | 210 |
| 11.89 +16.85 | HIP57952 | 2 | 210 |
| 13.23 -32.25 | HIP64570 | 2 | 210 |
| 13.57 -17.35 | HIP66231 | 2 | 209 |
| 03.00 +00.62 | HIP13957 | 2 | 209 |
| 02.72 -16.94 | HIP12697 | 2 | 209 |
| 03.05 +52.25 | HIP14182 | 2 | 209 |
| 03.22 -23.58 | HIP14989 | 2 | 209 |
| 03.25 +08.98 | HIP15099 | 2 | 209 |
| 03.40 +00.24 | HIP15856 | 2 | 209 |
| 03.50 +51.51 | HIP16294 | 2 | 209 |
| 03.83 +03.21 | HIP17927 | 2 | 209 |
| 03.88 +26.68 | HIP18163 | 2 | 209 |
| 04.10 +04.69 | HIP19125 | 2 | 209 |
| 04.12 +69.47 | HIP19203 | 2 | 209 |
| 04.14 +12.19 | HIP19316 | 2 | 209 |
| 04.26 +66.45 | HIP19865 | 2 | 209 |
| 04.43 +63.67 | HIP20669 | 2 | 209 |
| 05.01 +01.01 | HIP23266 | 2 | 209 |
| 05.12 -20.12 | HIP23806 | 2 | 209 |
| 05.76 +62.24 | HIP27188 | 2 | 209 |
| 05.98 +18.98 | HIP28312 | 2 | 209 |
| 06.18 -21.86 | HIP29295 | 2 | 209 |
| 06.24 +30.78 | HIP29617 | 2 | 209 |
| 07.47 +15.53 | HIP36278 | 2 | 209 |
| 07.73 +03.49 | HIP37645 | 2 | 209 |
| 07.84 +69.74 | HIP38285 | 2 | 209 |
| 07.96 +18.80 | HIP38899 | 2 | 209 |
| 08.38 +09.46 | HIP41064 | 2 | 209 |
| 08.41 -00.14 | HIP41230 | 2 | 209 |
| 09.39 +66.14 | HIP46066 | 2 | 209 |
| 09.60 +76.55 | HIP47098 | 2 | 209 |
| 10.28 -18.31 | HIP50337 | 2 | 209 |
| 10.64 -23.55 | HIP52045 | 2 | 209 |
| 11.21 +56.16 | HIP54756 | 2 | 209 |
| 11.50 +88.75 | HIP56124 | 2 | 209 |
| 03.02 -17.23 | HIP14073 | 2 | 208 |
| 03.62 +03.27 | HIP16895 | 2 | 208 |
| 04.06 -22.82 | HIP18918 | 2 | 208 |
| 09.77 -21.37 | HIP47931 | 2 | 208 |
| 06.97 +80.93 | HIP33548 | 2 | 207 |
| 04.01 +20.30 | HIP18721 | 2 | 206 |
| 05.06 +02.73 | HIP23548 | 2 | 206 |
| 07.21 -31.08 | HIP34841 | 2 | 206 |
| 05.93 -29.92 | HIP28021 | 2 | 205 |
| 03.35 +01.16 | HIP15599 | 2 | 205 |
| 02.65 -13.00 | HIP12341 | 2 | 204 |
| 09.86 +18.41 | HIP48366 | 2 | 202 |
| 03.19 -13.52 | HIP14832 | 2 | 202 |
| 05.73 -19.65 | HIP27014 | 2 | 202 |
| 09.08 +34.40 | HIP44559 | 2 | 202 |
| 06.52 +02.91 | HIP31083 | 2 | 201 |
| 11.10 +07.14 | HIP54274 | 2 | 201 |
| 19.45 +46.45 | HIP95631 | 2 | 200 |
| 08.53 +47.14 | HIP41815 | 2 | 199 |
| 09.93 +84.38 | HIP48675 | 2 | 199 |
| 11.01 +00.38 | HIP53812 | 2 | 199 |
| 10.86 +07.94 | HIP53071 | 2 | 198 |
| 11.20 -29.45 | HIP54690 | 2 | 198 |
| 08.22 -14.95 | HIP40229 | 2 | 197 |
| 02.57 +12.18 | HIP11941 | 2 | 197 |
| 02.84 -00.69 | HIP13260 | 2 | 197 |
| 02.98 +26.77 | HIP13891 | 2 | 197 |
| 03.04 -22.81 | HIP14148 | 2 | 197 |
| 03.07 +50.96 | HIP14268 | 2 | 197 |
| 03.31 +18.17 | HIP15381 | 2 | 197 |
| 03.42 -21.07 | HIP15932 | 2 | 197 |
| 03.42 +17.74 | HIP15954 | 2 | 197 |
| 03.43 +45.46 | HIP15998 | 2 | 197 |
| 03.55 +15.55 | HIP16532 | 2 | 197 |
| 03.65 +04.77 | HIP17021 | 2 | 197 |
| 03.72 +00.45 | HIP17389 | 2 | 197 |
| 03.86 +44.95 | HIP18071 | 2 | 197 |
| 03.91 -26.60 | HIP18302 | 2 | 197 |
| 03.93 +63.14 | HIP18395 | 2 | 197 |
| 03.96 +12.96 | HIP18504 | 2 | 197 |
| 04.16 -11.99 | HIP19409 | 2 | 197 |
| 04.53 -29.74 | HIP21125 | 2 | 197 |
| 04.59 +32.58 | HIP21397 | 2 | 197 |
| 04.86 +26.86 | HIP22600 | 2 | 197 |
| 04.88 -25.99 | HIP22668 | 2 | 197 |
| 05.07 +19.09 | HIP23579 | 2 | 197 |
| 05.22 +24.40 | HIP24315 | 2 | 197 |
| 05.22 +03.69 | HIP24336 | 2 | 197 |
| 05.39 +54.80 | HIP25189 | 2 | 197 |
| 05.46 +18.85 | HIP25523 | 2 | 197 |
| 05.48 +10.25 | HIP25677 | 2 | 197 |
| 05.70 -15.63 | HIP26834 | 2 | 197 |
| 05.89 -27.46 | HIP27814 | 2 | 197 |
| 05.89 +11.81 | HIP27820 | 2 | 197 |
| 05.89 +22.07 | HIP27828 | 2 | 197 |
| 06.38 +74.75 | HIP30315 | 2 | 197 |
| 06.59 +76.87 | HIP31473 | 2 | 197 |
| 06.60 +26.71 | HIP31535 | 2 | 197 |
| 06.78 +57.77 | HIP32470 | 2 | 197 |
| 07.05 +09.02 | HIP33995 | 2 | 197 |
| 07.15 +15.42 | HIP34511 | 2 | 197 |
| 07.18 +19.80 | HIP34650 | 2 | 197 |
| 07.43 -15.67 | HIP36081 | 2 | 197 |
| 07.60 +32.84 | HIP36993 | 2 | 197 |
| 07.95 +01.51 | HIP38853 | 2 | 197 |
| 08.21 +72.59 | HIP40219 | 2 | 197 |
| 08.30 -25.36 | HIP40652 | 2 | 197 |
| 08.43 +00.95 | HIP41297 | 2 | 197 |
| 08.49 +52.05 | HIP41651 | 2 | 197 |
| 08.65 +06.96 | HIP42418 | 2 | 197 |
| 08.75 +44.18 | HIP42934 | 2 | 197 |
| 08.92 -18.81 | HIP43777 | 2 | 197 |
| 09.31 -24.46 | HIP45685 | 2 | 197 |
| 09.60 +09.87 | HIP47091 | 2 | 197 |
| 09.67 -13.00 | HIP47440 | 2 | 197 |
| 09.73 -13.09 | HIP47738 | 2 | 197 |
| 09.73 +33.39 | HIP47739 | 2 | 197 |
| 09.76 +45.55 | HIP47891 | 2 | 197 |
| 09.80 +37.60 | HIP48070 | 2 | 197 |
| 11.38 +24.73 | HIP55547 | 2 | 197 |
| 09.24 +52.69 | HIP120005 | 2 | 197 |
| 00.41 -15.74 | HIP1950 | 2 | 197 |
| 15.92 -27.90 | HIP77943 | 2 | 197 |
| 22.83 -26.50 | HIP112739 | 2 | 197 |
| 22.88 -25.51 | HIP112977 | 2 | 197 |
| 07.27 +35.02 | HIP35198 | 2 | 196 |
| 01.01 -22.72 | HIP4700 | 2 | 196 |
| 09.51 +15.26 | HIP46635 | 2 | 196 |
| 10.75 +64.90 | HIP52587 | 2 | 196 |
| 14.53 +10.24 | HIP71022 | 2 | 196 |
| 22.88 -28.80 | HIP112978 | 2 | 196 |
| 06.86 +69.04 | HIP32945 | 2 | 196 |
| 08.32 +14.20 | HIP40761 | 2 | 196 |
| 08.35 +14.07 | HIP40910 | 2 | 196 |
| 08.85 +47.98 | HIP43429 | 2 | 196 |
| 09.34 -20.70 | HIP45801 | 2 | 196 |
| 10.28 +47.31 | HIP50366 | 2 | 196 |
| 19.01 +42.25 | HIP93350 | 2 | 196 |
| 06.97 +24.92 | HIP33566 | 2 | 195 |
| 16.12 -17.88 | HIP78975 | 2 | 195 |
| 16.20 -18.23 | HIP79377 | 2 | 195 |
| 07.21 +20.97 | HIP34845 | 2 | 194 |
| 03.01 -27.64 | HIP14035 | 2 | 193 |
| 08.53 +53.56 | HIP41832 | 2 | 192 |
| 10.10 +50.92 | HIP49452 | 2 | 192 |
| 03.32 -31.56 | HIP15469 | 2 | 192 |
| 05.94 -22.27 | HIP28083 | 2 | 192 |
| 09.30 +00.02 | HIP45635 | 2 | 191 |
| 10.06 +47.13 | HIP49266 | 2 | 191 |
| 05.22 -18.56 | HIP24339 | 2 | 190 |
| 04.12 +57.34 | HIP19247 | 2 | 190 |
| 03.05 +12.35 | HIP14175 | 2 | 189 |
| 09.58 +33.68 | HIP47016 | 2 | 189 |
| 10.28 +46.79 | HIP50362 | 2 | 189 |
| 19.45 -00.91 | HIP95622 | 2 | 189 |
| 03.82 -00.84 | HIP17863 | 2 | 189 |
| 04.35 +64.89 | HIP20279 | 2 | 189 |
| 04.55 +64.99 | HIP21229 | 2 | 189 |
| 02.92 +16.31 | HIP13601 | 2 | 189 |
| 10.99 +18.00 | HIP53731 | 2 | 188 |
| 05.14 -18.17 | HIP23932 | 2 | 188 |
| 05.54 +14.88 | HIP25966 | 2 | 188 |
| 05.55 +15.94 | HIP26027 | 2 | 188 |
| 06.71 -30.13 | HIP32155 | 2 | 188 |
| 03.10 -23.99 | HIP14426 | 2 | 188 |
| 03.48 -25.70 | HIP16187 | 2 | 188 |
| 03.61 -13.32 | HIP16834 | 2 | 188 |
| 03.61 -14.10 | HIP16847 | 2 | 188 |
| 03.91 +54.90 | HIP18256 | 2 | 188 |
| 04.27 +08.72 | HIP19912 | 2 | 188 |
| 04.63 -28.18 | HIP21548 | 2 | 188 |
| 04.75 -20.01 | HIP22064 | 2 | 188 |
| 04.81 -20.43 | HIP22333 | 2 | 188 |
| 04.83 -22.26 | HIP22436 | 2 | 188 |
| 05.21 -12.42 | HIP24259 | 2 | 188 |
| 05.87 -16.07 | HIP27716 | 2 | 188 |
| 05.89 -16.27 | HIP27849 | 2 | 188 |
| 05.93 -00.89 | HIP28044 | 2 | 188 |
| 06.29 -24.44 | HIP29888 | 2 | 188 |
| 07.19 -12.04 | HIP34710 | 2 | 188 |
| 07.81 -30.25 | HIP38120 | 2 | 188 |
| 08.12 +33.25 | HIP39735 | 2 | 188 |
| 09.44 +02.51 | HIP46294 | 2 | 188 |
| 10.46 -23.20 | HIP51199 | 2 | 188 |
| 10.48 -23.32 | HIP51324 | 2 | 188 |
| 10.83 -17.98 | HIP52954 | 2 | 188 |
| 18.84 +49.77 | HIP92467 | 2 | 188 |
| 06.74 -30.17 | HIP32281 | 2 | 188 |
| 03.11 +01.97 | HIP14445 | 2 | 187 |
| 03.15 -24.89 | HIP14623 | 2 | 187 |
| 03.23 -24.88 | HIP15016 | 2 | 187 |
| 03.89 -21.01 | HIP18191 | 2 | 187 |
| 04.68 -28.20 | HIP21778 | 2 | 187 |
| 05.25 -12.20 | HIP24465 | 2 | 187 |
| 05.26 +13.38 | HIP24511 | 2 | 187 |
| 05.79 -27.13 | HIP27330 | 2 | 187 |
| 09.02 -25.53 | HIP44291 | 2 | 187 |
| 09.14 -25.84 | HIP44874 | 2 | 187 |
| 10.80 +52.78 | HIP52826 | 2 | 187 |
| 11.10 +02.40 | HIP54281 | 2 | 187 |
| 03.21 -25.89 | HIP14928 | 2 | 186 |
| 03.21 -26.95 | HIP14931 | 2 | 186 |
| 03.62 +73.08 | HIP16887 | 2 | 186 |
| 03.63 +18.48 | HIP16944 | 2 | 186 |
| 05.79 +29.66 | HIP27336 | 2 | 186 |
| 05.81 +52.54 | HIP27429 | 2 | 186 |
| 06.51 +64.45 | HIP31018 | 2 | 186 |
| 07.88 +30.17 | HIP38480 | 2 | 186 |
| 08.47 -27.00 | HIP41522 | 2 | 186 |
| 08.52 -26.68 | HIP41759 | 2 | 186 |
| 08.52 +48.01 | HIP41762 | 2 | 186 |
| 08.56 -23.36 | HIP42011 | 2 | 186 |
| 08.76 +84.69 | HIP42982 | 2 | 186 |
| 09.56 -30.67 | HIP46901 | 2 | 186 |
| 09.71 -21.13 | HIP47607 | 2 | 186 |
| 09.75 -20.67 | HIP47856 | 2 | 186 |
| 10.47 -24.27 | HIP51263 | 2 | 186 |
| 10.64 +13.95 | HIP52070 | 2 | 186 |
| 10.28 +21.91 | HIP50364 | 2 | 185 |
| 04.66 +09.87 | HIP21710 | 2 | 185 |
| 07.20 +29.34 | HIP34800 | 2 | 185 |
| 04.19 +29.83 | HIP19572 | 2 | 184 |
| 05.73 +59.94 | HIP26996 | 2 | 184 |
| 06.83 +55.88 | HIP32754 | 2 | 184 |
| 06.97 +58.60 | HIP33542 | 2 | 184 |
| 08.42 +48.55 | HIP41251 | 2 | 184 |
| 09.45 +48.68 | HIP46361 | 2 | 184 |
| 10.58 +13.61 | HIP51792 | 2 | 184 |
| 10.61 +08.17 | HIP51922 | 2 | 184 |
| 17.85 +37.70 | HIP87349 | 2 | 184 |
| 04.24 +03.02 | HIP19788 | 2 | 184 |
| 04.28 +02.65 | HIP19927 | 2 | 184 |
| 04.63 +60.68 | HIP21571 | 2 | 184 |
| 04.90 -17.77 | HIP22762 | 2 | 184 |
| 05.12 -00.79 | HIP23809 | 2 | 184 |
| 05.24 -15.83 | HIP24419 | 2 | 184 |
| 05.34 +05.66 | HIP24933 | 2 | 184 |
| 05.43 -19.69 | HIP25396 | 2 | 184 |
| 05.55 -26.72 | HIP26013 | 2 | 184 |
| 06.16 -25.28 | HIP29199 | 2 | 184 |
| 06.28 -22.72 | HIP29843 | 2 | 184 |
| 06.66 -23.81 | HIP31836 | 2 | 184 |
| 06.83 -19.38 | HIP32742 | 2 | 184 |
| 07.41 -01.24 | HIP35959 | 2 | 184 |
| 07.92 -15.50 | HIP38702 | 2 | 184 |
| 08.51 +23.49 | HIP41743 | 2 | 184 |
| 08.56 +07.70 | HIP42000 | 2 | 184 |
| 08.70 +25.04 | HIP42703 | 2 | 184 |
| 09.05 +02.06 | HIP44424 | 2 | 184 |
| 09.06 +20.06 | HIP44477 | 2 | 184 |
| 09.50 -26.01 | HIP46603 | 2 | 184 |
| 09.73 +10.52 | HIP47704 | 2 | 184 |
| 09.97 +09.69 | HIP48866 | 2 | 184 |
| 10.03 +10.22 | HIP49143 | 2 | 184 |
| 10.37 +11.31 | HIP50782 | 2 | 184 |
| 10.37 +12.15 | HIP50802 | 2 | 184 |
| 10.39 -23.47 | HIP50897 | 2 | 184 |
| 10.55 +26.29 | HIP51617 | 2 | 184 |
| 10.60 -26.68 | HIP51885 | 2 | 184 |
| 10.87 -00.16 | HIP53123 | 2 | 184 |
| 10.92 -00.81 | HIP53383 | 2 | 184 |
| 11.14 -26.15 | HIP54446 | 2 | 184 |
| 11.15 +01.60 | HIP54502 | 2 | 184 |
| 11.16 +02.46 | HIP54541 | 2 | 184 |
| 09.95 +28.10 | HIP48777 | 2 | 184 |
| 03.07 +78.70 | HIP14276 | 2 | 184 |
| 03.17 +15.37 | HIP14705 | 2 | 184 |
| 03.87 +19.60 | HIP18091 | 2 | 184 |
| 04.05 +18.67 | HIP18909 | 2 | 184 |
| 05.27 -16.35 | HIP24551 | 2 | 184 |
| 05.54 -27.66 | HIP25946 | 2 | 184 |
| 05.76 +52.16 | HIP27165 | 2 | 184 |
| 06.34 +65.50 | HIP30130 | 2 | 184 |
| 07.37 +68.27 | HIP35706 | 2 | 184 |
| 08.72 +08.54 | HIP42807 | 2 | 184 |
| 09.07 +03.03 | HIP44518 | 2 | 184 |
| 09.13 +64.98 | HIP44788 | 2 | 184 |
| 09.34 -27.74 | HIP45832 | 2 | 184 |
| 09.37 +06.69 | HIP45943 | 2 | 184 |
| 09.55 +08.17 | HIP46862 | 2 | 184 |
| 11.13 +16.88 | HIP54398 | 2 | 184 |
| 11.14 +09.02 | HIP54417 | 2 | 184 |
| 03.98 -01.21 | HIP18582 | 2 | 184 |
| 04.21 +70.20 | HIP19655 | 2 | 184 |
| 08.72 +24.14 | HIP42783 | 2 | 184 |
| 10.58 +26.16 | HIP51770 | 2 | 184 |
| 04.06 +19.46 | HIP18946 | 2 | 184 |
| 11.87 +18.76 | HIP57866 | 2 | 184 |
| 02.96 +35.74 | HIP13796 | 2 | 183 |
| 03.09 +32.10 | HIP14370 | 2 | 183 |
| 03.17 +18.35 | HIP14721 | 2 | 183 |
| 03.21 +15.04 | HIP14909 | 2 | 183 |
| 03.39 -14.88 | HIP15784 | 2 | 183 |
| 03.39 -14.22 | HIP15786 | 2 | 183 |
| 03.40 -26.22 | HIP15834 | 2 | 183 |
| 03.43 +02.03 | HIP15977 | 2 | 183 |
| 03.59 -00.74 | HIP16727 | 2 | 183 |
| 03.64 +13.61 | HIP16991 | 2 | 183 |
| 03.69 +03.61 | HIP17207 | 2 | 183 |
| 03.69 +23.49 | HIP17225 | 2 | 183 |
| 03.70 -17.15 | HIP17288 | 2 | 183 |
| 03.71 -16.08 | HIP17355 | 2 | 183 |
| 03.72 +01.52 | HIP17375 | 2 | 183 |
| 04.02 -15.86 | HIP18768 | 2 | 183 |
| 04.20 +14.93 | HIP19617 | 2 | 183 |
| 04.22 +02.87 | HIP19712 | 2 | 183 |
| 04.24 +14.63 | HIP19781 | 2 | 183 |
| 04.33 -28.79 | HIP20184 | 2 | 183 |
| 04.39 -00.97 | HIP20477 | 2 | 183 |
| 04.45 -18.66 | HIP20761 | 2 | 183 |
| 04.53 +25.89 | HIP21134 | 2 | 183 |
| 04.53 -12.63 | HIP21143 | 2 | 183 |
| 04.64 +02.07 | HIP21623 | 2 | 183 |
| 04.73 -12.29 | HIP22002 | 2 | 183 |
| 04.85 +24.29 | HIP22557 | 2 | 183 |
| 04.90 +23.98 | HIP22801 | 2 | 183 |
| 04.94 +23.30 | HIP22972 | 2 | 183 |
| 05.00 -27.83 | HIP23243 | 2 | 183 |
| 05.04 -21.26 | HIP23452 | 2 | 183 |
| 05.07 -00.65 | HIP23575 | 2 | 183 |
| 05.22 +73.49 | HIP24329 | 2 | 183 |
| 05.47 -30.03 | HIP25598 | 2 | 183 |
| 05.47 +31.05 | HIP25614 | 2 | 183 |
| 05.49 +31.43 | HIP25711 | 2 | 183 |
| 05.78 -21.02 | HIP27256 | 2 | 183 |
| 05.95 +73.19 | HIP28143 | 2 | 183 |
| 06.00 -30.86 | HIP28396 | 2 | 183 |
| 06.27 -14.43 | HIP29754 | 2 | 183 |
| 06.43 +01.15 | HIP30585 | 2 | 183 |
| 06.68 +34.96 | HIP31972 | 2 | 183 |
| 06.84 +30.33 | HIP32834 | 2 | 183 |
| 06.85 -17.85 | HIP32868 | 2 | 183 |
| 07.10 -01.02 | HIP34239 | 2 | 183 |
| 07.12 +03.45 | HIP34341 | 2 | 183 |
| 07.44 +82.30 | HIP36103 | 2 | 183 |
| 07.52 -27.41 | HIP36571 | 2 | 183 |
| 07.70 -11.85 | HIP37520 | 2 | 183 |
| 08.31 -15.20 | HIP40724 | 2 | 183 |
| 08.98 +01.86 | HIP44109 | 2 | 183 |
| 08.98 +67.07 | HIP44112 | 2 | 183 |
| 09.00 -12.46 | HIP44190 | 2 | 183 |
| 09.22 +17.48 | HIP45217 | 2 | 183 |
| 09.38 -27.17 | HIP46007 | 2 | 183 |
| 09.40 +00.14 | HIP46087 | 2 | 183 |
| 09.63 +49.33 | HIP47264 | 2 | 183 |
| 09.67 +16.97 | HIP47455 | 2 | 183 |
| 09.93 +01.19 | HIP48681 | 2 | 183 |
| 10.52 -26.87 | HIP51494 | 2 | 183 |
| 10.66 +07.43 | HIP52164 | 2 | 183 |
| 10.70 +07.51 | HIP52360 | 2 | 183 |
| 10.88 -11.42 | HIP53186 | 2 | 183 |
| 05.48 -19.14 | HIP25679 | 2 | 183 |
| 11.08 +32.85 | HIP54183 | 2 | 183 |
| 04.39 -27.66 | HIP20489 | 2 | 183 |
| 04.10 +47.29 | HIP19134 | 2 | 183 |
| 03.18 +78.38 | HIP14794 | 2 | 182 |
| 03.23 -12.26 | HIP15050 | 2 | 182 |
| 03.25 -12.71 | HIP15121 | 2 | 182 |
| 03.55 -18.41 | HIP16546 | 2 | 182 |
| 03.68 +02.06 | HIP17192 | 2 | 182 |
| 03.71 +04.14 | HIP17330 | 2 | 182 |
| 03.83 -25.50 | HIP17919 | 2 | 182 |
| 03.87 -24.96 | HIP18093 | 2 | 182 |
| 03.87 -22.88 | HIP18115 | 2 | 182 |
| 03.91 +25.53 | HIP18273 | 2 | 182 |
| 03.98 -15.10 | HIP18569 | 2 | 182 |
| 04.11 -30.92 | HIP19190 | 2 | 182 |
| 04.31 +16.09 | HIP20082 | 2 | 182 |
| 04.60 -21.34 | HIP21414 | 2 | 182 |
| 04.76 -13.17 | HIP22134 | 2 | 182 |
| 04.82 -00.87 | HIP22408 | 2 | 182 |
| 04.83 -15.81 | HIP22432 | 2 | 182 |
| 04.87 +66.50 | HIP22615 | 2 | 182 |
| 05.08 -11.70 | HIP23627 | 2 | 182 |
| 05.12 -17.30 | HIP23816 | 2 | 182 |
| 05.21 -24.92 | HIP24257 | 2 | 182 |
| 05.27 -30.95 | HIP24566 | 2 | 182 |
| 05.32 -14.77 | HIP24834 | 2 | 182 |
| 05.58 -23.47 | HIP26175 | 2 | 182 |
| 05.66 -13.87 | HIP26654 | 2 | 182 |
| 05.77 -21.83 | HIP27235 | 2 | 182 |
| 06.17 -25.63 | HIP29241 | 2 | 182 |
| 06.55 -14.10 | HIP31244 | 2 | 182 |
| 07.40 +67.35 | HIP35900 | 2 | 182 |
| 07.48 -27.90 | HIP36332 | 2 | 182 |
| 07.87 -30.83 | HIP38424 | 2 | 182 |
| 08.18 +52.14 | HIP40041 | 2 | 182 |
| 08.20 +02.29 | HIP40169 | 2 | 182 |
| 08.52 +09.20 | HIP41776 | 2 | 182 |
| 08.54 -19.47 | HIP41877 | 2 | 182 |
| 08.90 +54.31 | HIP43718 | 2 | 182 |
| 09.40 +02.74 | HIP46073 | 2 | 182 |
| 09.73 -25.96 | HIP47732 | 2 | 182 |
| 10.11 +03.47 | HIP49508 | 2 | 182 |
| 10.24 -27.04 | HIP50176 | 2 | 182 |
| 10.42 -19.80 | HIP51010 | 2 | 182 |
| 10.57 +65.01 | HIP51721 | 2 | 182 |
| 10.66 +10.33 | HIP52198 | 2 | 182 |
| 10.68 +05.47 | HIP52259 | 2 | 182 |
| 10.72 -18.86 | HIP52463 | 2 | 182 |
| 10.76 -27.52 | HIP52630 | 2 | 182 |
| 10.86 -25.62 | HIP53095 | 2 | 182 |
| 11.03 +59.53 | HIP53927 | 2 | 182 |
| 11.16 +10.16 | HIP54556 | 2 | 182 |
| 03.17 +12.05 | HIP14729 | 2 | 182 |
| 03.51 +14.30 | HIP16359 | 2 | 182 |
| 04.29 +69.50 | HIP19998 | 2 | 182 |
| 05.26 +24.52 | HIP24494 | 2 | 182 |
| 06.13 +08.22 | HIP29063 | 2 | 182 |
| 06.86 +05.75 | HIP32933 | 2 | 182 |
| 07.40 +68.63 | HIP35883 | 2 | 182 |
| 07.59 +23.05 | HIP36898 | 2 | 182 |
| 08.00 +13.80 | HIP39127 | 2 | 182 |
| 09.91 +00.94 | HIP48591 | 2 | 182 |
| 03.24 -31.96 | HIP15070 | 2 | 181 |
| 03.27 +35.05 | HIP15205 | 2 | 181 |
| 03.57 +50.58 | HIP16642 | 2 | 181 |
| 04.45 +20.99 | HIP20785 | 2 | 181 |
| 05.33 -22.54 | HIP24862 | 2 | 181 |
| 05.89 +33.74 | HIP27860 | 2 | 181 |
| 06.04 -26.66 | HIP28594 | 2 | 181 |
| 07.58 +01.59 | HIP36886 | 2 | 181 |
| 07.64 +34.45 | HIP37177 | 2 | 181 |
| 09.55 +04.17 | HIP46871 | 2 | 181 |
| 09.84 +50.72 | HIP48270 | 2 | 181 |
| 09.91 +28.69 | HIP48596 | 2 | 181 |
| 10.24 +35.06 | HIP50141 | 2 | 181 |
| 11.15 +27.99 | HIP54492 | 2 | 181 |
| 17.99 +55.69 | HIP88086 | 2 | 181 |
| 03.27 +11.63 | HIP15206 | 2 | 181 |
| 10.65 +55.51 | HIP52137 | 2 | 181 |
| 08.86 +04.94 | HIP43516 | 2 | 180 |
| 03.07 +05.24 | HIP14261 | 2 | 180 |
| 03.38 -22.96 | HIP15734 | 2 | 180 |
| 03.51 +08.57 | HIP16342 | 2 | 180 |
| 03.69 +50.82 | HIP17231 | 2 | 180 |
| 03.71 +51.17 | HIP17341 | 2 | 180 |
| 04.45 +50.70 | HIP20768 | 2 | 180 |
| 04.46 +14.42 | HIP20827 | 2 | 180 |
| 05.19 +10.12 | HIP24177 | 2 | 180 |
| 05.44 +61.24 | HIP25414 | 2 | 180 |
| 05.44 +27.00 | HIP25450 | 2 | 180 |
| 05.71 +53.52 | HIP26871 | 2 | 180 |
| 05.72 -25.62 | HIP26977 | 2 | 180 |
| 05.78 -25.71 | HIP27292 | 2 | 180 |
| 05.85 +25.23 | HIP27640 | 2 | 180 |
| 06.37 +26.75 | HIP30305 | 2 | 180 |
| 06.78 +46.11 | HIP32510 | 2 | 180 |
| 06.91 -28.37 | HIP33222 | 2 | 180 |
| 07.10 +83.61 | HIP34235 | 2 | 180 |
| 07.19 +56.31 | HIP34754 | 2 | 180 |
| 07.98 +26.02 | HIP38985 | 2 | 180 |
| 09.47 +80.58 | HIP46417 | 2 | 180 |
| 09.47 +82.69 | HIP46432 | 2 | 180 |
| 09.82 +00.78 | HIP48153 | 2 | 180 |
| 10.52 +64.29 | HIP51511 | 2 | 180 |
| 10.60 +18.05 | HIP51874 | 2 | 180 |
| 04.10 +11.97 | HIP19117 | 2 | 180 |
| 04.94 +15.81 | HIP22967 | 2 | 180 |
| 05.25 +24.96 | HIP24489 | 2 | 180 |
| 06.77 +51.83 | HIP32441 | 2 | 180 |
| 09.27 +14.88 | HIP45474 | 2 | 180 |
| 09.63 +64.41 | HIP47229 | 2 | 180 |
| 10.90 +46.16 | HIP53283 | 2 | 180 |
| 03.87 +25.16 | HIP18106 | 2 | 179 |
| 03.89 +30.77 | HIP18204 | 2 | 179 |
| 04.28 +04.54 | HIP19971 | 2 | 179 |
| 04.65 +11.27 | HIP21684 | 2 | 179 |
| 04.72 +75.81 | HIP21970 | 2 | 179 |
| 04.92 +15.20 | HIP22886 | 2 | 179 |
| 04.94 +46.45 | HIP22965 | 2 | 179 |
| 06.53 +50.65 | HIP31141 | 2 | 179 |
| 06.56 +75.55 | HIP31272 | 2 | 179 |
| 06.63 +31.78 | HIP31702 | 2 | 179 |
| 06.90 +27.30 | HIP33149 | 2 | 179 |
| 07.60 +49.00 | HIP36975 | 2 | 179 |
| 09.15 +03.96 | HIP44906 | 2 | 179 |
| 09.64 +23.55 | HIP47288 | 2 | 179 |
| 09.84 +00.10 | HIP48255 | 2 | 179 |
| 10.01 +28.17 | HIP49060 | 2 | 179 |
| 10.11 +65.33 | HIP49517 | 2 | 179 |
| 10.61 +11.02 | HIP51945 | 2 | 179 |
| 10.83 +15.07 | HIP52932 | 2 | 179 |
| 10.97 +59.28 | HIP53634 | 2 | 179 |
| 10.97 +17.18 | HIP53644 | 2 | 179 |
| 10.99 +50.98 | HIP53695 | 2 | 179 |
| 03.71 +35.53 | HIP17326 | 2 | 179 |
| 08.01 +51.73 | HIP39179 | 2 | 179 |
| 08.21 +46.01 | HIP40195 | 2 | 179 |
| 09.93 +70.04 | HIP48691 | 2 | 179 |
| 03.65 +09.65 | HIP17032 | 2 | 179 |
| 05.17 +09.50 | HIP24069 | 2 | 179 |
| 03.14 +57.01 | HIP14592 | 2 | 179 |
| 03.21 +34.99 | HIP14907 | 2 | 179 |
| 03.68 -31.24 | HIP17175 | 2 | 179 |
| 03.81 +51.82 | HIP17800 | 2 | 179 |
| 03.82 +50.84 | HIP17827 | 2 | 179 |
| 04.28 +03.70 | HIP19946 | 2 | 179 |
| 04.48 +21.92 | HIP20917 | 2 | 179 |
| 04.63 +30.53 | HIP21573 | 2 | 179 |
| 04.67 +09.50 | HIP21761 | 2 | 179 |
| 05.25 +20.06 | HIP24480 | 2 | 179 |
| 06.16 +19.09 | HIP29221 | 2 | 179 |
| 06.29 +17.99 | HIP29875 | 2 | 179 |
| 06.77 +67.23 | HIP32419 | 2 | 179 |
| 07.30 +28.49 | HIP35336 | 2 | 179 |
| 08.26 +09.98 | HIP40449 | 2 | 179 |
| 08.94 +53.06 | HIP43870 | 2 | 179 |
| 09.33 +82.61 | HIP45771 | 2 | 179 |
| 10.63 +18.35 | HIP52015 | 2 | 179 |
| 10.83 -29.99 | HIP52939 | 2 | 179 |
| 10.85 +56.07 | HIP53042 | 2 | 179 |
| 10.85 +46.80 | HIP53051 | 2 | 179 |
| 10.91 +76.07 | HIP53327 | 2 | 179 |
| 11.03 +17.36 | HIP53901 | 2 | 179 |
| 03.23 +18.30 | HIP15062 | 2 | 179 |
| 05.72 +20.07 | HIP26951 | 2 | 179 |
| 07.64 +55.21 | HIP37178 | 2 | 179 |
| 09.53 +07.30 | HIP46761 | 2 | 179 |
| 10.55 -19.20 | HIP51620 | 2 | 179 |
| 10.73 +27.40 | HIP52491 | 2 | 179 |
| 09.87 +14.48 | HIP48410 | 2 | 178 |
| 03.66 +23.29 | HIP17091 | 2 | 178 |
| 04.44 +35.74 | HIP20707 | 2 | 178 |
| 06.27 +70.78 | HIP29761 | 2 | 178 |
| 07.79 +31.94 | HIP37991 | 2 | 178 |
| 07.86 +20.84 | HIP38352 | 2 | 178 |
| 08.28 +54.07 | HIP40536 | 2 | 178 |
| 08.62 +20.49 | HIP42298 | 2 | 178 |
| 08.75 +29.67 | HIP42938 | 2 | 178 |
| 09.48 +20.71 | HIP46478 | 2 | 178 |
| 10.83 +23.16 | HIP52956 | 2 | 178 |
| 11.26 +10.94 | HIP54968 | 2 | 178 |
| 03.25 +56.30 | HIP15111 | 2 | 178 |
| 03.66 +49.94 | HIP17104 | 2 | 178 |
| 07.73 +52.65 | HIP37694 | 2 | 178 |
| 09.91 +69.45 | HIP48573 | 2 | 178 |
| 10.65 +32.30 | HIP52141 | 2 | 178 |
| 11.06 +32.89 | HIP54056 | 2 | 178 |
| 00.46 -17.33 | HIP2196 | 2 | 177 |
| 02.90 +13.33 | HIP13530 | 2 | 177 |
| 03.47 +78.72 | HIP16173 | 2 | 177 |
| 04.72 +20.63 | HIP21953 | 2 | 177 |
| 06.36 +56.18 | HIP30230 | 2 | 177 |
| 08.06 +55.31 | HIP39444 | 2 | 177 |
| 08.07 +30.57 | HIP39459 | 2 | 177 |
| 08.43 +17.83 | HIP41326 | 2 | 177 |
| 09.36 +08.21 | HIP45912 | 2 | 177 |
| 09.60 +51.13 | HIP47123 | 2 | 177 |
| 09.87 +49.62 | HIP48391 | 2 | 177 |
| 10.70 +16.28 | HIP52363 | 2 | 177 |
| 10.80 +14.21 | HIP52824 | 2 | 177 |
| 10.91 +22.95 | HIP53337 | 2 | 177 |
| 11.09 +45.01 | HIP54212 | 2 | 177 |
| 11.09 +28.02 | HIP54217 | 2 | 177 |
| 11.89 -20.35 | HIP57986 | 2 | 177 |
| 20.20 -26.36 | HIP99536 | 2 | 177 |
| 20.59 +06.13 | HIP101579 | 2 | 177 |
| 21.29 -21.74 | HIP105083 | 2 | 177 |
| 23.38 -21.54 | HIP115436 | 2 | 177 |
| 03.59 +14.17 | HIP16748 | 2 | 177 |
| 03.81 +30.66 | HIP17796 | 2 | 177 |
| 04.50 +29.04 | HIP20989 | 2 | 177 |
| 08.31 +05.18 | HIP40704 | 2 | 177 |
| 10.22 +31.38 | HIP50059 | 2 | 177 |
| 10.29 +68.81 | HIP50399 | 2 | 177 |
| 22.87 +20.06 | HIP112927 | 2 | 177 |
| 03.66 +09.07 | HIP17059 | 2 | 176 |
| 09.75 +58.44 | HIP47840 | 2 | 176 |
| 02.19 +23.22 | HIP10201 | 2 | 176 |
| 05.49 +21.32 | HIP25694 | 2 | 176 |
| 05.79 +35.56 | HIP27361 | 2 | 176 |
| 07.64 +54.43 | HIP37216 | 2 | 176 |
| 09.37 +22.54 | HIP45954 | 2 | 176 |
| 09.39 +08.36 | HIP46036 | 2 | 176 |
| 10.65 +47.73 | HIP52110 | 2 | 176 |
| 10.74 +47.21 | HIP52505 | 2 | 176 |
| 18.43 +12.47 | HIP90305 | 2 | 176 |
| 20.19 +19.37 | HIP99462 | 2 | 176 |
| 22.43 -18.36 | HIP110735 | 2 | 176 |
| 23.19 +32.09 | HIP114511 | 2 | 176 |
| 04.16 +11.47 | HIP19427 | 2 | 176 |
| 04.99 +31.59 | HIP23205 | 2 | 176 |
| 10.27 +11.67 | HIP50297 | 2 | 176 |
| 23.46 +76.71 | HIP115777 | 2 | 176 |
| 08.03 +51.70 | HIP39295 | 2 | 175 |
| 03.52 +19.78 | HIP16401 | 2 | 175 |
| 03.59 +05.73 | HIP16739 | 2 | 175 |
| 04.37 -25.91 | HIP20418 | 2 | 175 |
| 05.18 +17.40 | HIP24142 | 2 | 175 |
| 06.61 +32.15 | HIP31576 | 2 | 175 |
| 07.01 +27.27 | HIP33765 | 2 | 175 |
| 07.15 +65.96 | HIP34503 | 2 | 175 |
| 07.41 +22.93 | HIP35930 | 2 | 175 |
| 09.66 +58.87 | HIP47410 | 2 | 175 |
| 09.69 +13.21 | HIP47513 | 2 | 175 |
| 10.24 +30.58 | HIP50185 | 2 | 175 |
| 10.48 +34.86 | HIP51320 | 2 | 175 |
| 10.57 +50.75 | HIP51708 | 2 | 175 |
| 10.62 +17.26 | HIP51965 | 2 | 175 |
| 10.63 +32.69 | HIP52027 | 2 | 175 |
| 10.89 +51.01 | HIP53222 | 2 | 175 |
| 04.04 +15.13 | HIP18843 | 2 | 175 |
| 08.31 +18.32 | HIP40733 | 2 | 175 |
| 08.43 +17.05 | HIP41319 | 2 | 175 |
| 09.28 +32.18 | HIP45517 | 2 | 175 |
| 09.62 +22.69 | HIP47201 | 2 | 175 |
| 09.93 +65.77 | HIP48678 | 2 | 175 |
| 10.59 +05.10 | HIP51855 | 2 | 175 |
| 03.63 -31.01 | HIP16953 | 2 | 174 |
| 03.67 +34.12 | HIP17129 | 2 | 174 |
| 03.70 +22.86 | HIP17289 | 2 | 174 |
| 01.94 +45.97 | HIP9040 | 2 | 174 |
| 03.47 +78.48 | HIP16148 | 2 | 174 |
| 03.63 +05.64 | HIP16930 | 2 | 174 |
| 03.69 +13.52 | HIP17238 | 2 | 174 |
| 03.71 +11.96 | HIP17354 | 2 | 174 |
| 04.39 -25.39 | HIP20478 | 2 | 174 |
| 05.05 +13.73 | HIP23498 | 2 | 174 |
| 05.63 +74.69 | HIP26437 | 2 | 174 |
| 05.77 +74.61 | HIP27246 | 2 | 174 |
| 05.87 +87.22 | HIP27724 | 2 | 174 |
| 07.50 +78.21 | HIP36432 | 2 | 174 |
| 08.58 +69.69 | HIP42096 | 2 | 174 |
| 09.27 +68.23 | HIP45501 | 2 | 174 |
| 09.32 +00.90 | HIP45737 | 2 | 174 |
| 10.09 +17.57 | HIP49406 | 2 | 174 |
| 10.40 +49.10 | HIP50918 | 2 | 174 |
| 10.65 +15.98 | HIP52126 | 2 | 174 |
| 10.68 +30.20 | HIP52275 | 2 | 174 |
| 10.83 +20.01 | HIP52934 | 2 | 174 |
| 06.29 +07.44 | HIP29868 | 2 | 174 |
| 07.04 +37.72 | HIP33903 | 2 | 174 |
| 04.08 -27.74 | HIP19042 | 2 | 174 |
| 04.30 +52.37 | HIP20028 | 2 | 174 |
| 04.74 +46.97 | HIP22032 | 2 | 174 |
| 04.86 +32.30 | HIP22603 | 2 | 174 |
| 04.96 +07.31 | HIP23057 | 2 | 174 |
| 05.52 +27.77 | HIP25871 | 2 | 174 |
| 05.57 +76.40 | HIP26097 | 2 | 174 |
| 06.34 +80.36 | HIP30129 | 2 | 174 |
| 07.53 +68.62 | HIP36635 | 2 | 174 |
| 08.12 +51.99 | HIP39727 | 2 | 174 |
| 09.48 +64.35 | HIP46468 | 2 | 174 |
| 10.16 +31.59 | HIP49770 | 2 | 174 |
| 10.64 +76.17 | HIP52063 | 2 | 174 |
| 10.71 +63.04 | HIP52383 | 2 | 174 |
| 10.79 -27.76 | HIP52756 | 2 | 174 |
| 10.97 +15.48 | HIP53616 | 2 | 174 |
| 11.14 +06.57 | HIP54423 | 2 | 174 |
| 03.62 +18.60 | HIP16896 | 2 | 174 |
| 05.76 -12.43 | HIP27158 | 2 | 174 |
| 10.08 +35.10 | HIP49358 | 2 | 174 |
| 10.87 +09.06 | HIP53117 | 2 | 174 |
| 03.57 +24.34 | HIP16639 | 2 | 173 |
| 03.97 +58.62 | HIP18567 | 2 | 173 |
| 05.47 +16.44 | HIP25589 | 2 | 173 |
| 08.03 -12.80 | HIP39298 | 2 | 173 |
| 08.67 +21.06 | HIP42518 | 2 | 173 |
| 08.77 -31.35 | HIP43080 | 2 | 173 |
| 09.59 +64.46 | HIP47032 | 2 | 173 |
| 10.56 +06.37 | HIP51667 | 2 | 173 |
| 10.59 +31.78 | HIP51840 | 2 | 173 |
| 10.94 +31.04 | HIP53473 | 2 | 173 |
| 11.01 +15.45 | HIP53822 | 2 | 173 |
| 10.87 +58.37 | HIP53127 | 2 | 173 |
| 04.53 -31.69 | HIP21113 | 2 | 172 |
| 08.82 -30.22 | HIP43287 | 2 | 172 |
| 04.47 -31.46 | HIP20870 | 2 | 172 |
| 04.71 +12.21 | HIP21918 | 2 | 172 |
| 07.43 +69.45 | HIP36051 | 2 | 172 |
| 08.08 +15.36 | HIP39515 | 2 | 172 |
| 08.08 +23.61 | HIP39546 | 2 | 172 |
| 08.46 +59.40 | HIP41461 | 2 | 172 |
| 10.54 +51.85 | HIP51613 | 2 | 172 |
| 11.08 +56.99 | HIP54162 | 2 | 172 |
| 10.93 +16.15 | HIP53453 | 2 | 172 |
| 03.40 -32.07 | HIP15817 | 2 | 171 |
| 04.16 +54.83 | HIP19400 | 2 | 171 |
| 04.55 -19.36 | HIP21220 | 2 | 171 |
| 04.72 +09.62 | HIP21954 | 2 | 171 |
| 05.74 -29.92 | HIP27090 | 2 | 171 |
| 06.08 +66.94 | HIP28808 | 2 | 171 |
| 07.46 +22.04 | HIP36215 | 2 | 171 |
| 08.18 -13.80 | HIP40035 | 2 | 171 |
| 08.74 +49.23 | HIP42904 | 2 | 171 |
| 08.82 +50.67 | HIP43311 | 2 | 171 |
| 09.39 +22.31 | HIP46018 | 2 | 171 |
| 10.45 +26.64 | HIP51157 | 2 | 171 |
| 11.12 +25.54 | HIP54347 | 2 | 171 |
| 03.33 +12.00 | HIP15489 | 2 | 171 |
| 06.31 +06.73 | HIP29991 | 2 | 171 |
| 08.23 +63.10 | HIP40298 | 2 | 171 |
| 10.22 +16.53 | HIP50040 | 2 | 171 |
| 07.21 +27.22 | HIP34860 | 2 | 170 |
| 09.02 +15.27 | HIP44295 | 2 | 170 |
| 10.42 +24.61 | HIP50998 | 2 | 170 |
| 04.01 +52.95 | HIP18712 | 2 | 170 |
| 04.08 +54.98 | HIP19043 | 2 | 170 |
| 04.27 +07.56 | HIP19896 | 2 | 170 |
| 04.76 +47.70 | HIP22124 | 2 | 170 |
| 05.15 +33.25 | HIP23971 | 2 | 170 |
| 05.36 +34.74 | HIP25030 | 2 | 170 |
| 05.38 +28.75 | HIP25141 | 2 | 170 |
| 05.47 +09.64 | HIP25578 | 2 | 170 |
| 05.73 -21.28 | HIP27021 | 2 | 170 |
| 07.07 +68.29 | HIP34115 | 2 | 170 |
| 08.67 +49.83 | HIP42545 | 2 | 170 |
| 10.19 +54.58 | HIP49917 | 2 | 170 |
| 10.60 +15.87 | HIP51897 | 2 | 170 |
| 10.68 +59.34 | HIP52278 | 2 | 170 |
| 09.59 +03.91 | HIP47058 | 2 | 170 |
| 07.25 +57.28 | HIP35069 | 2 | 170 |
| 08.12 +07.38 | HIP39721 | 2 | 170 |
| 03.53 +23.65 | HIP16440 | 2 | 170 |
| 03.99 +22.99 | HIP18625 | 2 | 170 |
| 05.76 +25.73 | HIP27173 | 2 | 170 |
| 06.40 +06.01 | HIP30421 | 2 | 170 |
| 07.10 +27.47 | HIP34222 | 2 | 170 |
| 07.58 +22.34 | HIP36853 | 2 | 170 |
| 07.99 +30.72 | HIP39031 | 2 | 170 |
| 08.80 -31.24 | HIP43193 | 2 | 170 |
| 08.84 +22.89 | HIP43381 | 2 | 170 |
| 09.55 +51.68 | HIP46853 | 2 | 170 |
| 09.56 +46.23 | HIP46903 | 2 | 170 |
| 10.77 +25.76 | HIP52667 | 2 | 170 |
| 15.25 +18.74 | HIP74621 | 2 | 170 |
| 15.27 +19.16 | HIP74735 | 2 | 170 |
| 15.28 +19.65 | HIP74781 | 2 | 170 |
| 20.54 +34.92 | HIP101359 | 2 | 170 |
| 20.55 +34.02 | HIP101424 | 2 | 170 |
| 05.79 +15.46 | HIP27310 | 2 | 169 |
| 06.50 +32.31 | HIP30969 | 2 | 169 |
| 03.37 +27.16 | HIP15720 | 2 | 169 |
| 03.40 +16.72 | HIP15823 | 2 | 169 |
| 05.14 +52.37 | HIP23913 | 2 | 169 |
| 06.46 +32.62 | HIP30751 | 2 | 169 |
| 07.41 +32.59 | HIP35962 | 2 | 169 |
| 07.77 +17.01 | HIP37922 | 2 | 169 |
| 09.32 +17.71 | HIP45699 | 2 | 169 |
| 09.45 +18.23 | HIP46357 | 2 | 169 |
| 15.30 +18.85 | HIP74888 | 2 | 169 |
| 04.40 -28.77 | HIP20509 | 2 | 168 |
| 06.08 +49.85 | HIP28795 | 2 | 168 |
| 06.31 +33.48 | HIP29987 | 2 | 168 |
| 07.15 +26.63 | HIP34487 | 2 | 168 |
| 10.23 +55.15 | HIP50114 | 2 | 168 |
| 10.36 +48.04 | HIP50742 | 2 | 168 |
| 10.72 +25.77 | HIP52456 | 2 | 168 |
| 19.99 +28.24 | HIP98386 | 2 | 168 |
| 10.45 +18.06 | HIP51171 | 2 | 167 |
| 11.08 +33.00 | HIP54134 | 2 | 167 |
| 00.86 +00.91 | HIP4007 | 2 | 167 |
| 03.37 +52.62 | HIP15675 | 2 | 167 |
| 03.92 +12.49 | HIP18322 | 2 | 167 |
| 04.06 +22.94 | HIP18955 | 2 | 167 |
| 04.19 +46.92 | HIP19553 | 2 | 167 |
| 04.45 -29.19 | HIP20790 | 2 | 167 |
| 04.83 +28.94 | HIP22460 | 2 | 167 |
| 05.01 +14.38 | HIP23286 | 2 | 167 |
| 06.13 +50.58 | HIP29080 | 2 | 167 |
| 06.54 +54.85 | HIP31162 | 2 | 167 |
| 07.10 +23.28 | HIP34228 | 2 | 167 |
| 07.53 +19.97 | HIP36607 | 2 | 167 |
| 08.50 +28.08 | HIP41700 | 2 | 167 |
| 08.72 +49.99 | HIP42786 | 2 | 167 |
| 09.11 +15.22 | HIP44723 | 2 | 167 |
| 10.43 +52.62 | HIP51062 | 2 | 167 |
| 10.50 +52.26 | HIP51417 | 2 | 167 |
| 11.04 +30.42 | HIP53969 | 2 | 167 |
| 11.04 +21.97 | HIP53985 | 2 | 167 |
| 11.12 +22.05 | HIP54375 | 2 | 167 |
| 10.40 +18.54 | HIP50936 | 2 | 166 |
| 04.08 +20.86 | HIP19054 | 2 | 166 |
| 07.39 +33.18 | HIP35825 | 2 | 166 |
| 08.11 +24.15 | HIP39686 | 2 | 166 |
| 09.84 +34.62 | HIP48280 | 2 | 166 |
| 10.96 +47.76 | HIP53581 | 2 | 166 |
| 11.06 +31.22 | HIP54068 | 2 | 166 |
| 05.21 +30.22 | HIP24258 | 2 | 165 |
| 05.04 +13.91 | HIP23443 | 2 | 165 |
| 10.16 +24.54 | HIP49749 | 2 | 165 |
| 08.24 +33.18 | HIP40384 | 2 | 165 |
| 09.22 +10.67 | HIP45221 | 2 | 165 |
| 10.97 +48.29 | HIP53637 | 2 | 165 |
| 03.98 +00.75 | HIP18612 | 2 | 164 |
| 10.30 +11.28 | HIP50424 | 2 | 164 |
| 03.94 -25.18 | HIP18432 | 2 | 164 |
| 08.74 -28.19 | HIP42872 | 2 | 164 |
| 07.07 +05.21 | HIP34095 | 2 | 163 |
| 03.89 +07.77 | HIP18174 | 2 | 163 |
| 08.35 -27.64 | HIP40897 | 2 | 163 |
| 10.01 +27.27 | HIP49046 | 2 | 163 |
| 13.93 -31.61 | HIP68039 | 2 | 163 |
| 04.65 +10.92 | HIP21676 | 2 | 162 |
| 10.76 +06.96 | HIP52598 | 2 | 162 |
| 11.88 +18.93 | HIP57949 | 2 | 162 |
| 09.16 -00.40 | HIP44953 | 2 | 161 |
| 03.77 +26.22 | HIP17609 | 2 | 161 |
| 03.99 -00.41 | HIP18642 | 2 | 161 |
| 05.99 +02.48 | HIP28356 | 2 | 160 |
| 10.94 +18.83 | HIP53459 | 2 | 160 |
| 09.97 +51.57 | HIP48887 | 2 | 159 |
| 09.95 +52.01 | HIP48773 | 2 | 158 |
| 10.12 +52.16 | HIP49555 | 2 | 158 |
| 00.29 +00.59 | HIP1385 | 2 | 157 |
| 00.36 -16.21 | HIP1718 | 2 | 157 |
| 00.68 -27.28 | HIP3181 | 2 | 157 |
| 00.71 +08.71 | HIP3364 | 2 | 157 |
| 00.95 +19.88 | HIP4459 | 2 | 157 |
| 01.10 -23.59 | HIP5137 | 2 | 157 |
| 01.41 +29.84 | HIP6573 | 2 | 157 |
| 02.39 -29.04 | HIP11125 | 2 | 157 |
| 02.41 -28.65 | HIP11244 | 2 | 157 |
| 02.84 +36.95 | HIP13257 | 2 | 157 |
| 04.44 +70.09 | HIP20724 | 2 | 157 |
| 08.26 +11.43 | HIP40452 | 2 | 157 |
| 08.64 +29.63 | HIP42371 | 2 | 157 |
| 11.47 +02.78 | HIP55982 | 2 | 157 |
| 13.31 -17.13 | HIP64944 | 2 | 157 |
| 14.47 -16.11 | HIP70756 | 2 | 157 |
| 14.89 +07.03 | HIP72872 | 2 | 157 |
| 14.97 +18.66 | HIP73228 | 2 | 157 |
| 15.23 +02.25 | HIP74513 | 2 | 157 |
| 15.31 -25.55 | HIP74912 | 2 | 157 |
| 15.41 +57.75 | HIP75431 | 2 | 157 |
| 15.96 +32.41 | HIP78161 | 2 | 157 |
| 16.10 -19.38 | HIP78885 | 2 | 157 |
| 16.18 +10.04 | HIP79270 | 2 | 157 |
| 16.32 -25.76 | HIP79969 | 2 | 157 |
| 16.41 -13.64 | HIP80366 | 2 | 157 |
| 16.93 +47.59 | HIP82836 | 2 | 157 |
| 17.08 -27.38 | HIP83578 | 2 | 157 |
| 17.21 +29.91 | HIP84202 | 2 | 157 |
| 17.22 +64.62 | HIP84244 | 2 | 157 |
| 18.49 -27.97 | HIP90611 | 2 | 157 |
| 18.51 +35.73 | HIP90708 | 2 | 157 |
| 18.63 +34.69 | HIP91326 | 2 | 157 |
| 18.83 +36.68 | HIP92395 | 2 | 157 |
| 19.51 +16.43 | HIP95963 | 2 | 157 |
| 20.20 +59.69 | HIP99530 | 2 | 157 |
| 20.32 -26.17 | HIP100182 | 2 | 157 |
| 21.29 -19.79 | HIP105089 | 2 | 157 |
| 22.12 +25.35 | HIP109176 | 2 | 157 |
| 22.99 -01.18 | HIP113496 | 2 | 157 |
| 23.15 +20.01 | HIP114282 | 2 | 157 |
| 23.23 +19.26 | HIP114683 | 2 | 157 |
| 23.24 +19.26 | HIP114749 | 2 | 157 |
| 00.74 -25.52 | HIP3484 | 2 | 157 |
| 02.68 -22.96 | HIP12499 | 2 | 157 |
| 02.83 +66.94 | HIP13204 | 2 | 157 |
| 03.79 +39.16 | HIP17719 | 2 | 157 |
| 04.08 +37.56 | HIP19026 | 2 | 157 |
| 04.40 +40.96 | HIP20550 | 2 | 157 |
| 10.20 -18.62 | HIP49973 | 2 | 157 |
| 11.83 +04.05 | HIP57708 | 2 | 157 |
| 13.04 -26.79 | HIP63618 | 2 | 157 |
| 14.00 +44.81 | HIP68403 | 2 | 157 |
| 14.84 +52.41 | HIP72563 | 2 | 157 |
| 16.14 -18.24 | HIP79048 | 2 | 157 |
| 11.73 -29.75 | HIP57217 | 2 | 157 |
| 11.74 -29.88 | HIP57271 | 2 | 157 |
| 00.66 -26.47 | HIP3125 | 2 | 156 |
| 02.44 -29.14 | HIP11363 | 2 | 156 |
| 03.20 -02.36 | HIP14878 | 2 | 156 |
| 15.16 +03.17 | HIP74190 | 2 | 156 |
| 15.66 +54.87 | HIP76697 | 2 | 156 |
| 15.97 -26.97 | HIP78191 | 2 | 156 |
| 18.06 -26.84 | HIP88465 | 2 | 156 |
| 18.10 -29.92 | HIP88650 | 2 | 156 |
| 18.86 +05.04 | HIP92566 | 2 | 156 |
| 20.23 -01.21 | HIP99665 | 2 | 156 |
| 21.97 +19.08 | HIP108443 | 2 | 156 |
| 23.09 +20.24 | HIP114034 | 2 | 156 |
| 09.50 +46.84 | HIP46599 | 2 | 156 |
| 11.21 +24.65 | HIP54741 | 2 | 156 |
| 01.82 -19.01 | HIP8493 | 2 | 156 |
| 08.79 +28.16 | HIP43129 | 2 | 156 |
| 19.05 +51.59 | HIP93530 | 2 | 156 |
| 03.65 +24.47 | HIP17044 | 2 | 156 |
| 03.75 -00.76 | HIP17492 | 2 | 156 |
| 03.82 -17.97 | HIP17852 | 2 | 156 |
| 04.14 +33.64 | HIP19337 | 2 | 156 |
| 04.16 +33.77 | HIP19416 | 2 | 156 |
| 04.65 -15.22 | HIP21668 | 2 | 156 |
| 04.77 +64.61 | HIP22191 | 2 | 156 |
| 05.07 +28.88 | HIP23563 | 2 | 156 |
| 05.21 -31.31 | HIP24282 | 2 | 156 |
| 05.77 +32.87 | HIP27234 | 2 | 156 |
| 07.48 +19.65 | HIP36319 | 2 | 156 |
| 08.08 +24.33 | HIP39528 | 2 | 156 |
| 08.15 -20.50 | HIP39908 | 2 | 156 |
| 08.48 -15.87 | HIP41609 | 2 | 156 |
| 08.62 +23.25 | HIP42253 | 2 | 156 |
| 08.65 -13.26 | HIP42401 | 2 | 156 |
| 08.95 -31.74 | HIP43942 | 2 | 156 |
| 09.17 +59.25 | HIP45040 | 2 | 156 |
| 09.82 +07.82 | HIP48165 | 2 | 156 |
| 10.31 +17.40 | HIP50469 | 2 | 156 |
| 10.41 +77.47 | HIP50960 | 2 | 156 |
| 11.05 +73.63 | HIP54008 | 2 | 156 |
| 11.73 +18.52 | HIP57202 | 2 | 156 |
| 12.02 +08.80 | HIP58607 | 2 | 156 |
| 13.27 +46.22 | HIP64771 | 2 | 156 |
| 13.92 -29.09 | HIP67960 | 2 | 156 |
| 08.11 -00.11 | HIP39669 | 2 | 155 |
| 09.95 +04.24 | HIP48768 | 2 | 155 |
| 15.77 +36.45 | HIP77245 | 2 | 155 |
| 16.60 +01.38 | HIP81279 | 2 | 155 |
| 08.88 -26.95 | HIP43597 | 2 | 154 |
| 04.98 +17.99 | HIP23139 | 2 | 154 |
| 03.19 -12.87 | HIP14808 | 2 | 154 |
| 08.12 -17.81 | HIP39750 | 2 | 154 |
| 08.91 +22.21 | HIP43727 | 2 | 154 |
| 10.33 +40.48 | HIP50588 | 2 | 154 |
| 01.78 +18.68 | HIP8303 | 2 | 154 |
| 06.62 +19.75 | HIP31626 | 2 | 154 |
| 04.95 +83.30 | HIP23010 | 2 | 153 |
| 06.12 -31.90 | HIP28999 | 2 | 153 |
| 05.90 +12.42 | HIP27878 | 2 | 152 |
| 06.08 +25.18 | HIP28791 | 2 | 152 |
| 08.49 +34.13 | HIP41642 | 2 | 152 |
| 19.19 +32.73 | HIP94273 | 2 | 152 |
| 03.45 +51.71 | HIP16052 | 2 | 151 |
| 04.12 -27.43 | HIP19232 | 2 | 151 |
| 15.34 +40.49 | HIP75083 | 2 | 151 |
| 09.57 +24.87 | HIP46935 | 2 | 150 |
| 04.05 +01.85 | HIP18877 | 2 | 149 |
| 03.57 +24.88 | HIP16635 | 2 | 149 |
| 05.27 +27.85 | HIP24579 | 2 | 149 |
| 08.72 +48.19 | HIP42773 | 2 | 148 |
| 03.32 +49.31 | HIP15460 | 2 | 147 |
| 10.27 +28.68 | HIP50316 | 2 | 147 |
| 10.99 +07.22 | HIP53725 | 2 | 145 |
| 04.86 -18.85 | HIP22608 | 2 | 143 |
| 06.28 -23.91 | HIP29831 | 2 | 143 |
| 09.61 +07.12 | HIP47158 | 2 | 143 |
| 05.83 +60.57 | HIP27541 | 2 | 142 |
| 05.53 +64.32 | HIP25883 | 2 | 141 |
| 08.90 +08.42 | HIP43692 | 2 | 141 |
| 18.28 -11.41 | HIP89583 | 2 | 141 |
| 10.90 +20.53 | HIP53287 | 2 | 140 |
| 08.84 +53.84 | HIP43382 | 2 | 140 |
| 05.21 +18.15 | HIP24298 | 2 | 139 |
| 05.55 +56.25 | HIP26006 | 2 | 139 |
| 10.44 -21.76 | HIP51114 | 2 | 139 |
| 01.76 -12.26 | HIP8197 | 2 | 138 |
| 02.01 +19.19 | HIP9379 | 2 | 138 |
| 04.97 +13.03 | HIP23085 | 2 | 138 |
| 08.17 +35.45 | HIP40001 | 2 | 138 |
| 10.61 +15.19 | HIP51959 | 2 | 138 |
| 14.47 +21.38 | HIP70728 | 2 | 138 |
| 19.02 +04.11 | HIP93364 | 2 | 138 |
| 22.04 +18.27 | HIP108816 | 2 | 138 |
| 06.28 +32.24 | HIP29815 | 2 | 138 |
| 07.64 -27.15 | HIP37171 | 2 | 138 |
| 08.27 +01.30 | HIP40501 | 2 | 138 |
| 08.58 +50.47 | HIP42109 | 2 | 138 |
| 10.24 +08.62 | HIP50172 | 2 | 138 |
| 10.26 +25.83 | HIP50271 | 2 | 138 |
| 10.27 +56.76 | HIP50314 | 2 | 138 |
| 10.33 +26.57 | HIP50582 | 2 | 138 |
| 10.33 +28.44 | HIP50585 | 2 | 138 |
| 10.35 +26.70 | HIP50699 | 2 | 138 |
| 10.35 +26.89 | HIP50700 | 2 | 138 |
| 10.67 -00.90 | HIP52228 | 2 | 138 |
| 10.68 -00.67 | HIP52277 | 2 | 138 |
| 10.89 +27.26 | HIP53243 | 2 | 138 |
| 10.89 +28.17 | HIP53247 | 2 | 138 |
| 10.95 +25.27 | HIP53538 | 2 | 138 |
| 11.00 +77.69 | HIP53774 | 2 | 138 |
| 11.02 +24.07 | HIP53851 | 2 | 138 |
| 11.11 -14.74 | HIP54304 | 2 | 138 |
| 11.31 -26.60 | HIP55221 | 2 | 138 |
| 11.34 -26.16 | HIP55385 | 2 | 138 |
| 11.40 +13.90 | HIP55661 | 2 | 138 |
| 11.41 +31.01 | HIP55695 | 2 | 138 |
| 11.44 +10.42 | HIP55820 | 2 | 138 |
| 11.48 +10.04 | HIP56028 | 2 | 138 |
| 11.52 +09.62 | HIP56185 | 2 | 138 |
| 11.79 +57.93 | HIP57518 | 2 | 138 |
| 11.83 +03.86 | HIP57694 | 2 | 138 |
| 16.04 -17.86 | HIP78566 | 2 | 138 |
| 16.18 -20.01 | HIP79286 | 2 | 138 |
| 18.12 -26.29 | HIP88739 | 2 | 138 |
| 20.19 -27.33 | HIP99499 | 2 | 138 |
| 21.58 -20.25 | HIP106568 | 2 | 138 |
| 22.92 -30.83 | HIP113179 | 2 | 138 |
| 23.06 +21.60 | HIP113884 | 2 | 138 |
| 23.91 -21.77 | HIP117886 | 2 | 138 |
| 08.80 +34.60 | HIP43219 | 2 | 137 |
| 08.85 +15.35 | HIP43454 | 2 | 137 |
| 07.85 +38.59 | HIP38295 | 2 | 137 |
| 09.25 +46.65 | HIP45382 | 2 | 137 |
| 09.90 +57.68 | HIP48537 | 2 | 137 |
| 10.86 -16.40 | HIP53084 | 2 | 137 |
| 11.00 +25.29 | HIP53747 | 2 | 137 |
| 11.69 +13.79 | HIP57007 | 2 | 137 |
| 17.02 -28.40 | HIP83246 | 2 | 137 |
| 23.08 +20.86 | HIP113990 | 2 | 137 |
| 06.92 +73.32 | HIP33235 | 2 | 136 |
| 06.98 +51.87 | HIP33580 | 2 | 136 |
| 07.66 -26.47 | HIP37309 | 2 | 136 |
| 10.13 +72.20 | HIP49639 | 2 | 136 |
| 10.32 +72.38 | HIP50517 | 2 | 136 |
| 10.36 +57.42 | HIP50725 | 2 | 136 |
| 10.92 +41.46 | HIP53405 | 2 | 136 |
| 11.36 +31.09 | HIP55481 | 2 | 136 |
| 19.53 +17.78 | HIP96037 | 2 | 136 |
| 11.77 +30.23 | HIP57395 | 2 | 136 |
| 15.49 -23.70 | HIP75801 | 2 | 136 |
| 16.07 -19.10 | HIP78704 | 2 | 136 |
| 05.47 +21.44 | HIP25613 | 2 | 135 |
| 19.42 +02.46 | HIP95470 | 2 | 135 |
| 06.29 +32.71 | HIP29874 | 2 | 135 |
| 16.19 -18.00 | HIP79341 | 2 | 135 |
| 11.09 +01.24 | HIP54205 | 1 | 134 |
| 09.16 +14.46 | HIP44997 | 1 | 133 |
| 03.89 +15.58 | HIP18176 | 1 | 132 |
| 08.68 +66.05 | HIP42593 | 1 | 132 |
| 10.16 +15.35 | HIP49771 | 1 | 132 |
| 07.85 +25.11 | HIP38312 | 1 | 132 |
| 14.85 +37.38 | HIP72620 | 1 | 130 |
| 15.28 +25.22 | HIP74757 | 1 | 130 |
| 15.66 +27.63 | HIP76674 | 1 | 130 |
| 19.77 +14.74 | HIP97274 | 1 | 130 |
| 09.58 +51.42 | HIP47009 | 1 | 129 |
| 10.46 +16.35 | HIP51221 | 1 | 128 |
| 10.16 -21.67 | HIP49785 | 1 | 127 |
| 03.92 -24.88 | HIP18331 | 1 | 125 |
| 17.40 +37.28 | HIP85137 | 1 | 124 |
| 15.55 +18.60 | HIP76131 | 1 | 124 |
| 03.24 +12.66 | HIP15083 | 1 | 123 |
| 04.95 +73.84 | HIP23016 | 1 | 123 |
| 09.69 +11.56 | HIP47515 | 1 | 123 |
| 06.83 +45.71 | HIP32763 | 1 | 122 |
| 04.91 +45.68 | HIP22817 | 1 | 121 |
| 10.90 +34.13 | HIP53305 | 1 | 120 |
| 10.50 +20.80 | HIP51419 | 1 | 119 |
| 09.86 +26.31 | HIP48378 | 1 | 118 |
| 21.65 +31.69 | HIP106915 | 1 | 118 |
| 02.37 +26.27 | HIP11053 | 1 | 118 |
| 02.77 +36.27 | HIP12919 | 1 | 118 |
| 02.92 +47.67 | HIP13580 | 1 | 118 |
| 04.83 +07.64 | HIP22426 | 1 | 118 |
| 05.27 -29.44 | HIP24573 | 1 | 118 |
| 05.58 +26.97 | HIP26192 | 1 | 118 |
| 09.47 +12.67 | HIP46453 | 1 | 118 |
| 09.61 +35.60 | HIP47182 | 1 | 118 |
| 11.42 +40.00 | HIP55718 | 1 | 118 |
| 11.55 +18.00 | HIP56321 | 1 | 118 |
| 11.55 +18.82 | HIP56360 | 1 | 118 |
| 11.58 +18.68 | HIP56467 | 1 | 118 |
| 11.65 +23.81 | HIP56825 | 1 | 118 |
| 11.71 +23.91 | HIP57112 | 1 | 118 |
| 13.40 +12.95 | HIP65358 | 1 | 118 |
| 13.41 +14.86 | HIP65427 | 1 | 118 |
| 17.29 +04.08 | HIP84594 | 1 | 118 |
| 19.28 +47.00 | HIP94755 | 1 | 118 |
| 20.28 -25.74 | HIP99937 | 1 | 118 |
| 22.50 -16.47 | HIP111063 | 1 | 118 |
| 04.27 -31.99 | HIP19885 | 1 | 118 |
| 02.60 -10.92 | HIP12118 | 1 | 117 |
| 05.80 +46.98 | HIP27372 | 1 | 117 |
| 10.87 -32.02 | HIP53114 | 1 | 117 |
| 11.49 +49.85 | HIP56053 | 1 | 117 |
| 03.80 +32.20 | HIP17732 | 1 | 117 |
| 13.42 +53.30 | HIP65462 | 1 | 117 |
| 00.31 -10.95 | HIP1509 | 1 | 116 |
| 04.31 +20.36 | HIP20084 | 1 | 116 |
| 10.89 +02.34 | HIP53245 | 1 | 116 |
| 14.67 +09.88 | HIP71714 | 1 | 116 |
| 15.35 +25.56 | HIP75124 | 1 | 116 |
| 10.82 -30.64 | HIP52905 | 1 | 116 |
| 04.24 +09.41 | HIP19757 | 1 | 116 |
| 08.76 +16.95 | HIP42968 | 1 | 116 |
| 03.29 -31.24 | HIP15320 | 1 | 115 |
| 03.61 +00.40 | HIP16852 | 1 | 115 |
| 04.25 -25.01 | HIP19807 | 1 | 113 |
| 15.57 -25.11 | HIP76201 | 1 | 111 |
| 19.77 +32.02 | HIP97292 | 1 | 111 |
| 19.79 +31.80 | HIP97387 | 1 | 111 |
| 19.50 +39.60 | HIP95859 | 1 | 110 |
| 19.82 +41.58 | HIP97527 | 1 | 107 |
| 19.70 +50.53 | HIP96895 | 1 | 107 |
| 12.23 +05.00 | HIP59655 | 1 | 106 |
| 12.25 +04.91 | HIP59762 | 1 | 106 |
| 11.16 -24.60 | HIP54532 | 1 | 104 |
| 07.66 +01.29 | HIP37323 | 1 | 102 |
| 11.15 +29.02 | HIP54480 | 1 | 102 |
| 05.67 +16.39 | HIP26689 | 1 | 101 |
| 15.06 +15.42 | HIP73673 | 1 | 101 |
| 10.48 -27.37 | HIP51297 | 1 | 101 |
| 18.71 +44.27 | HIP91739 | 1 | 101 |
| 04.88 +14.62 | HIP22688 | 1 | 100 |
| 10.92 +13.18 | HIP53388 | 1 | 100 |

| Tycho Backup Stars | | | |
|---|---|---|---|
| 19.73 +44.30 | TYC3148-02216-1 | 100 | 9000 |
| 19.73 +44.32 | TYC3148-01352-1 | 100 | 9000 |
| 08.59 -30.03 | TYC7136-02291-1 | 100 | 9000 |
| 19.73 +44.63 | TYC3148-01239-1 | 100 | 8966 |
| 19.73 +44.51 | TYC3148-01711-1 | 99 | 8954 |
| 12.53 -27.79 | TYC6690-00337-1 | 96 | 8620 |
| 09.90 +14.03 | TYC0835-01166-1 | 95 | 8520 |
| 12.54 -28.11 | TYC6690-00183-1 | 93 | 8406 |
| 09.57 -12.23 | TYC5468-00893-1 | 92 | 8250 |
| 15.27 +71.99 | TYC4414-00018-1 | 88 | 7941 |
| 10.97 -31.15 | TYC7200-00307-1 | 75 | 6758 |
| 10.16 +18.26 | TYC1422-00422-1 | 67 | 6066 |
| 10.30 +12.73 | TYC0843-00519-1 | 66 | 5948 |
| 09.23 +23.37 | TYC1951-01057-1 | 66 | 5930 |
| 19.57 +41.61 | TYC3143-01690-1 | 64 | 5783 |
| 19.59 +41.58 | TYC3143-02054-1 | 64 | 5782 |
| 09.93 -23.93 | TYC6604-00972-1 | 64 | 5772 |
| 01.57 +00.57 | TYC0028-00282-1 | 61 | 5470 |
| 10.98 +01.69 | TYC0255-00844-1 | 60 | 5367 |
| 04.47 -25.10 | TYC6467-02337-1 | 59 | 5329 |
| 11.58 +20.38 | TYC1440-01386-1 | 58 | 5253 |
| 10.38 -01.02 | TYC4905-00050-1 | 54 | 4900 |
| 01.57 +00.53 | TYC0028-00261-1 | 51 | 4600 |
| 06.21 -29.92 | TYC6517-02129-1 | 50 | 4502 |
| 18.86 +48.93 | TYC3548-01765-1 | 50 | 4455 |
| 08.20 +04.84 | TYC0203-01259-1 | 49 | 4366 |
| 10.39 -01.06 | TYC4905-01054-1 | 48 | 4340 |
| 11.02 -24.03 | TYC6635-00676-1 | 48 | 4289 |
| 11.13 -29.92 | TYC6648-01076-1 | 47 | 4242 |
| 19.19 +43.99 | TYC3133-00494-1 | 47 | 4207 |
| 19.28 +47.97 | TYC3546-00173-1 | 47 | 4206 |
| 09.94 -16.02 | TYC6045-01617-1 | 47 | 4196 |
| 15.30 +72.05 | TYC4414-00083-1 | 47 | 4186 |
| 11.70 +26.77 | TYC1984-01884-1 | 46 | 4157 |
| 08.63 +64.33 | TYC4130-00014-1 | 45 | 4080 |
| 07.18 +30.49 | TYC2438-01050-1 | 45 | 4065 |
| 04.48 -25.32 | TYC6467-02353-1 | 43 | 3830 |
| 09.58 -12.09 | TYC5469-01215-1 | 42 | 3808 |
| 09.93 -24.11 | TYC6604-01164-1 | 42 | 3790 |
| 11.29 -24.04 | TYC6649-00641-1 | 42 | 3777 |
| 11.99 -20.43 | TYC6100-01416-1 | 41 | 3685 |
| 11.14 -30.24 | TYC7201-00103-1 | 41 | 3657 |
| 13.73 +48.14 | TYC3466-00860-1 | 41 | 3654 |
| 07.17 +30.21 | TYC2438-00858-1 | 40 | 3609 |
| 07.17 +30.16 | TYC2438-00607-1 | 39 | 3551 |
| 08.05 -01.32 | TYC4846-00420-1 | 39 | 3486 |
| 09.24 +23.49 | TYC1951-01130-1 | 38 | 3459 |
| 08.04 -01.11 | TYC4846-00010-1 | 38 | 3459 |
| 10.97 -31.20 | TYC7200-00346-1 | 38 | 3437 |
| 07.35 +58.27 | TYC3793-01990-1 | 38 | 3418 |
| 11.74 +31.13 | TYC2524-01781-1 | 38 | 3402 |
| 08.66 +12.98 | TYC0805-00340-1 | 37 | 3373 |
| 09.39 +50.78 | TYC3431-01373-1 | 37 | 3306 |
| 06.42 -31.38 | TYC7073-01688-1 | 37 | 3293 |
| 06.71 -01.27 | TYC4799-00429-1 | 36 | 3278 |
| 08.87 +28.25 | TYC1949-00942-1 | 36 | 3262 |
| 08.27 +05.71 | TYC0208-00705-1 | 36 | 3262 |
| 23.23 +56.99 | TYC4006-00868-1 | 36 | 3231 |
| 08.65 +47.47 | TYC3416-00053-1 | 35 | 3169 |
| 10.31 +12.76 | TYC0843-00558-1 | 35 | 3151 |
| 19.82 +40.85 | TYC3140-00087-1 | 35 | 3133 |
| 10.98 +01.83 | TYC0255-00996-1 | 35 | 3123 |
| 10.33 +19.87 | TYC1423-00174-1 | 35 | 3106 |
| 08.07 -01.06 | TYC4846-01324-1 | 34 | 3091 |
| 10.32 +19.77 | TYC1423-00118-1 | 34 | 3079 |
| 06.39 -28.76 | TYC6518-00881-1 | 34 | 3041 |
| 07.25 +14.19 | TYC0774-00931-1 | 34 | 3034 |
| 07.74 +27.95 | TYC1920-00194-1 | 33 | 3000 |
| 08.26 +05.74 | TYC0208-00539-1 | 33 | 2994 |
| 08.28 +61.39 | TYC4126-01422-1 | 33 | 2977 |
| 05.60 +20.82 | TYC1310-02663-1 | 33 | 2963 |
| 09.92 -24.08 | TYC6604-01257-1 | 33 | 2935 |
| 18.67 +43.76 | TYC3130-01388-1 | 32 | 2922 |
| 18.77 +41.90 | TYC3126-01761-1 | 32 | 2899 |
| 10.39 +00.84 | TYC0246-00230-1 | 32 | 2893 |
| 06.21 -29.73 | TYC6517-02064-1 | 32 | 2888 |
| 10.96 +01.63 | TYC0255-00983-1 | 32 | 2884 |
| 11.57 +20.47 | TYC1440-01031-1 | 32 | 2853 |
| 04.90 -16.17 | TYC5899-00123-1 | 30 | 2739 |
| 11.29 -18.82 | TYC6087-00677-1 | 30 | 2730 |
| 11.13 -30.26 | TYC7201-00585-1 | 30 | 2726 |
| 19.32 +44.78 | TYC3133-00013-1 | 30 | 2722 |
| 09.91 -23.99 | TYC6604-01219-1 | 30 | 2714 |
| 07.26 +14.24 | TYC0774-00723-1 | 30 | 2685 |
| 06.22 -29.82 | TYC6517-01817-1 | 30 | 2657 |
| 19.50 +41.70 | TYC3142-00281-1 | 29 | 2646 |
| 23.59 +77.67 | TYC4606-01972-1 | 29 | 2621 |
| 07.33 +58.31 | TYC3793-01804-1 | 29 | 2615 |
| 18.97 +49.30 | TYC3549-01385-1 | 29 | 2585 |
| 09.94 -15.91 | TYC6045-01343-1 | 29 | 2585 |
| 09.37 +50.57 | TYC3431-00689-1 | 28 | 2563 |
| 04.89 -16.14 | TYC5899-00751-1 | 28 | 2555 |
| 06.73 -01.17 | TYC4799-02061-1 | 28 | 2551 |
| 10.16 +18.09 | TYC1422-01432-1 | 28 | 2517 |
| 11.01 +40.46 | TYC3010-01704-1 | 28 | 2503 |
| 11.69 +26.61 | TYC1984-01840-1 | 28 | 2496 |
| 11.58 +20.21 | TYC1440-00023-1 | 27 | 2469 |
| 11.30 -19.12 | TYC6087-01174-1 | 27 | 2466 |
| 09.39 +20.44 | TYC1408-00169-1 | 27 | 2437 |
| 19.27 +47.85 | TYC3546-01892-1 | 27 | 2437 |
| 06.46 +00.07 | TYC0133-00413-1 | 27 | 2435 |
| 11.49 +07.60 | TYC0856-01256-1 | 27 | 2434 |
| 05.66 +06.22 | TYC0127-00376-1 | 27 | 2421 |
| 19.81 +43.95 | TYC3148-00041-1 | 27 | 2416 |
| 06.91 +24.31 | TYC1894-02667-1 | 27 | 2410 |
| 07.79 +50.27 | TYC3413-00011-1 | 27 | 2409 |
| 10.13 +34.41 | TYC2506-01090-1 | 27 | 2406 |
| 08.21 +04.66 | TYC0203-00498-1 | 27 | 2402 |
| 03.51 -23.72 | TYC6446-00192-1 | 27 | 2400 |
| 09.88 +13.79 | TYC0835-00976-1 | 26 | 2364 |
| 05.11 -14.14 | TYC5342-00921-1 | 26 | 2352 |
| 08.89 +28.39 | TYC1949-01779-1 | 26 | 2346 |
| 06.41 -28.99 | TYC6518-02371-1 | 26 | 2321 |
| 13.74 +47.92 | TYC3466-01158-1 | 26 | 2306 |
| 19.92 +48.29 | TYC3562-01082-1 | 26 | 2301 |
| 06.46 -00.12 | TYC4785-01173-1 | 26 | 2299 |
| 04.35 +57.97 | TYC3727-00501-1 | 25 | 2291 |
| 05.60 +20.77 | TYC1309-01844-1 | 25 | 2269 |
| 11.34 -23.35 | TYC6650-01088-1 | 25 | 2253 |
| 16.19 +44.27 | TYC3067-00519-1 | 25 | 2239 |
| 01.73 -15.74 | TYC5855-01494-1 | 25 | 2223 |
| 19.81 +40.87 | TYC3140-00139-1 | 24 | 2179 |
| 05.11 -13.94 | TYC5342-01261-1 | 24 | 2151 |
| 08.60 -29.91 | TYC6582-01986-1 | 24 | 2148 |
| 10.38 -00.76 | TYC4905-01303-1 | 24 | 2145 |
| 06.61 -27.45 | TYC6516-01196-1 | 24 | 2136 |
| 07.77 +27.87 | TYC1920-02118-1 | 24 | 2135 |
| 01.76 -16.07 | TYC5855-00850-1 | 24 | 2119 |
| 11.75 +30.78 | TYC2524-02177-1 | 24 | 2116 |
| 04.46 -25.11 | TYC6460-01767-1 | 23 | 2114 |
| 08.72 +04.43 | TYC0219-01563-1 | 23 | 2109 |
| 09.59 +34.97 | TYC2497-00518-1 | 23 | 2091 |
| 19.31 +41.74 | TYC3129-00260-1 | 23 | 2087 |
| 04.39 +57.87 | TYC3727-00989-1 | 23 | 2087 |
| 08.24 +05.95 | TYC0207-00953-1 | 23 | 2081 |
| 04.35 +57.79 | TYC3727-00442-1 | 23 | 2060 |
| 11.50 +07.44 | TYC0277-00060-1 | 23 | 2060 |
| 19.20 +43.89 | TYC3133-01162-1 | 23 | 2058 |
| 19.41 +51.10 | TYC3555-00176-1 | 23 | 2051 |
| 04.91 -16.03 | TYC5899-00099-1 | 23 | 2050 |
| 12.35 +17.70 | TYC1445-01265-1 | 23 | 2040 |
| 11.50 +07.21 | TYC0270-00681-1 | 23 | 2038 |
| 11.70 +26.53 | TYC1984-01952-1 | 23 | 2031 |
| 19.86 +43.12 | TYC3145-00177-1 | 23 | 2031 |
| 11.75 +02.83 | TYC0275-00047-1 | 22 | 2019 |
| 08.88 +28.56 | TYC1949-01285-1 | 22 | 2014 |
| 08.64 +12.93 | TYC0805-00484-1 | 22 | 1993 |
| 05.61 +20.71 | TYC1310-02615-1 | 22 | 1983 |
| 19.07 +39.23 | TYC3120-00459-1 | 22 | 1981 |
| 19.16 +43.55 | TYC3132-00837-1 | 22 | 1979 |
| 01.32 +76.63 | TYC4493-01400-1 | 22 | 1966 |
| 08.88 +33.00 | TYC2488-01261-1 | 22 | 1950 |
| 19.18 +48.01 | TYC3546-01144-1 | 22 | 1949 |
| 12.52 +74.36 | TYC4400-00005-1 | 22 | 1946 |
| 03.50 -23.75 | TYC6446-00227-1 | 22 | 1944 |
| 10.17 +18.48 | TYC1422-01164-1 | 22 | 1940 |
| 08.60 -30.26 | TYC7136-02269-1 | 21 | 1930 |
| 09.24 +23.26 | TYC1951-01599-1 | 21 | 1925 |
| 06.52 +29.80 | TYC1891-00134-1 | 21 | 1900 |
| 01.72 -15.73 | TYC5855-00806-1 | 21 | 1893 |
| 11.24 +25.87 | TYC1981-01893-1 | 21 | 1886 |
| 16.14 +43.92 | TYC3067-00520-1 | 21 | 1883 |
| 05.60 +20.81 | TYC1310-02546-1 | 21 | 1875 |
| 09.93 -24.05 | TYC6604-00717-1 | 21 | 1864 |
| 04.91 +12.45 | TYC0692-01201-1 | 21 | 1862 |
| 11.08 +44.11 | TYC3012-00091-1 | 21 | 1853 |
| 15.59 +53.99 | TYC3869-00405-1 | 20 | 1838 |
| 06.51 +29.59 | TYC1891-01367-1 | 20 | 1825 |
| 10.98 -31.04 | TYC7200-00382-1 | 20 | 1813 |
| 11.08 +44.24 | TYC3012-00416-1 | 20 | 1794 |
| 05.33 +79.09 | TYC4519-01525-1 | 20 | 1791 |
| 23.57 +77.83 | TYC4606-00116-1 | 20 | 1783 |
| 04.41 +39.30 | TYC2879-00164-1 | 20 | 1774 |
| 06.91 +24.45 | TYC1898-00612-1 | 20 | 1768 |
| 18.94 +48.84 | TYC3549-00353-1 | 20 | 1768 |
| 05.67 +06.19 | TYC0127-00988-1 | 20 | 1756 |
| 23.21 +57.08 | TYC4006-00592-1 | 19 | 1747 |
| 06.42 -31.70 | TYC7073-02243-1 | 19 | 1745 |
| 08.29 +61.65 | TYC4126-00728-1 | 19 | 1742 |
| 19.21 +38.70 | TYC3121-00261-1 | 19 | 1741 |
| 18.68 +47.76 | TYC3531-01961-1 | 19 | 1741 |
| 10.96 -31.00 | TYC7200-00362-1 | 19 | 1736 |
| 07.78 +50.32 | TYC3413-00228-1 | 19 | 1716 |
| 13.58 +53.68 | TYC3850-00376-1 | 19 | 1705 |
| 19.31 +41.68 | TYC3129-00114-1 | 19 | 1698 |
| 09.59 +34.61 | TYC2497-00328-1 | 19 | 1683 |
| 11.29 -23.81 | TYC6649-00513-1 | 19 | 1679 |
| 06.47 +10.95 | TYC0736-01297-1 | 19 | 1675 |
| 10.38 -29.54 | TYC6631-00901-1 | 19 | 1670 |
| 11.30 -19.21 | TYC6087-01292-1 | 19 | 1666 |
| 19.59 +47.82 | TYC3560-00797-1 | 18 | 1663 |
| 13.31 -18.06 | TYC6116-00429-1 | 18 | 1654 |
| 19.07 +39.16 | TYC3120-00487-1 | 18 | 1640 |
| 07.18 +30.33 | TYC2438-00481-1 | 18 | 1633 |
| 07.75 +27.81 | TYC1920-01510-1 | 18 | 1631 |
| 10.38 +00.87 | TYC0246-01167-1 | 18 | 1624 |
| 11.42 +41.28 | TYC3015-02271-1 | 18 | 1618 |
| 05.30 +07.43 | TYC0112-01685-1 | 18 | 1614 |
| 10.27 +20.03 | TYC1425-01524-1 | 18 | 1597 |
| 13.31 -18.23 | TYC6116-00386-1 | 18 | 1597 |
| 06.48 +11.08 | TYC0736-00227-1 | 18 | 1594 |
| 11.34 -22.99 | TYC6650-00741-1 | 18 | 1593 |
| 19.21 +42.12 | TYC3129-00584-1 | 18 | 1587 |
| 09.40 +20.55 | TYC1408-01026-1 | 18 | 1579 |
| 18.97 +49.29 | TYC3549-01408-1 | 18 | 1575 |
| 11.76 +30.92 | TYC2524-01993-1 | 17 | 1571 |
| 19.68 +43.08 | TYC3144-00810-1 | 17 | 1564 |
| 04.83 -24.61 | TYC6465-00846-1 | 17 | 1558 |
| 04.52 +04.70 | TYC0090-00964-1 | 17 | 1555 |
| 05.79 +01.14 | TYC0116-00505-1 | 17 | 1552 |
| 09.46 +45.70 | TYC3425-00724-1 | 17 | 1547 |
| 19.32 +44.56 | TYC3133-00041-1 | 17 | 1543 |
| 18.77 +41.78 | TYC3126-01443-1 | 17 | 1542 |
| 12.98 +12.50 | TYC0886-00903-1 | 17 | 1542 |
| 10.98 -31.18 | TYC7200-00324-1 | 17 | 1538 |
| 13.97 +43.38 | TYC3033-00696-1 | 17 | 1535 |
| 11.83 +57.68 | TYC3838-00639-1 | 17 | 1529 |
| 07.34 +37.18 | TYC2463-00947-1 | 17 | 1529 |
| 10.38 -29.81 | TYC6631-01235-1 | 17 | 1529 |
| 08.21 +04.57 | TYC0203-00379-1 | 17 | 1510 |
| 19.79 +50.53 | TYC3565-00011-1 | 17 | 1506 |
| 05.67 +06.02 | TYC0127-00094-1 | 17 | 1496 |
| 19.11 +49.55 | TYC3549-02818-1 | 17 | 1494 |
| 04.53 +04.55 | TYC0090-00701-1 | 17 | 1491 |
| 15.60 +53.87 | TYC3869-00559-1 | 17 | 1488 |
| 11.50 +07.47 | TYC0277-00096-1 | 16 | 1475 |
| 09.32 +33.99 | TYC2496-00821-1 | 16 | 1475 |
| 11.34 -23.34 | TYC6650-00332-1 | 16 | 1472 |
| 11.68 +42.92 | TYC3016-01671-1 | 16 | 1468 |
| 18.95 +48.67 | TYC3545-00029-1 | 16 | 1457 |
| 08.37 +01.77 | TYC0196-02506-1 | 16 | 1454 |
| 08.29 -12.69 | TYC5434-01527-1 | 16 | 1451 |
| 19.92 +48.41 | TYC3562-00324-1 | 16 | 1448 |
| 19.50 +48.60 | TYC3547-00273-1 | 16 | 1444 |
| 19.68 +49.31 | TYC3564-02934-1 | 16 | 1443 |
| 19.23 +50.07 | TYC3550-01345-1 | 16 | 1440 |
| 19.95 +44.18 | TYC3149-00123-1 | 16 | 1438 |
| 04.91 -16.07 | TYC5899-00135-1 | 16 | 1437 |
| 19.22 +50.94 | TYC3554-00022-1 | 16 | 1434 |
| 12.70 -30.49 | TYC7247-00459-1 | 16 | 1432 |
| 06.50 +58.22 | TYC3777-00784-1 | 16 | 1430 |
| 11.98 -20.26 | TYC6100-00607-1 | 16 | 1428 |
| 08.90 +32.88 | TYC2488-00807-1 | 16 | 1426 |
| 03.15 +30.49 | TYC2340-01534-1 | 16 | 1425 |
| 09.32 +33.95 | TYC2496-01616-1 | 16 | 1424 |
| 07.21 +30.30 | TYC2451-01763-1 | 16 | 1419 |
| 11.00 +40.27 | TYC3009-00603-1 | 16 | 1419 |
| 04.36 +57.75 | TYC3727-00775-1 | 16 | 1401 |
| 00.66 +21.26 | TYC1193-00917-1 | 16 | 1396 |
| 08.37 +01.74 | TYC0196-01894-1 | 15 | 1393 |
| 18.87 +48.73 | TYC3544-00236-1 | 15 | 1384 |
| 11.69 +42.82 | TYC3016-01428-1 | 15 | 1362 |
| 11.77 +03.67 | TYC0275-00807-1 | 15 | 1362 |
| 09.90 -24.15 | TYC6603-00433-1 | 15 | 1361 |
| 06.63 -32.29 | TYC7091-00470-1 | 15 | 1353 |
| 04.46 -25.23 | TYC6460-01673-1 | 15 | 1352 |
| 10.33 +20.00 | TYC1423-00095-1 | 15 | 1334 |
| 03.15 +30.50 | TYC2339-01503-1 | 15 | 1328 |
| 18.77 +41.95 | TYC3126-01808-1 | 15 | 1323 |
| 08.31 +61.41 | TYC4126-00798-1 | 15 | 1322 |
| 00.24 -12.12 | TYC5264-00008-1 | 15 | 1319 |
| 05.29 +07.15 | TYC0112-00620-1 | 14 | 1303 |
| 04.42 +39.51 | TYC2883-01121-1 | 14 | 1303 |
| 06.42 -28.70 | TYC6519-00616-1 | 14 | 1293 |
| 16.18 +43.92 | TYC3067-00802-1 | 14 | 1289 |
| 19.15 +51.23 | TYC3554-00077-1 | 14 | 1285 |
| 03.17 +21.12 | TYC1231-00048-1 | 14 | 1284 |
| 11.22 +25.38 | TYC1981-00590-1 | 14 | 1280 |
| 16.04 +28.28 | TYC2041-00152-1 | 14 | 1278 |
| 22.84 +35.48 | TYC2757-01244-1 | 14 | 1277 |
| 01.74 -16.06 | TYC5855-01748-1 | 14 | 1277 |
| 08.29 +61.35 | TYC4126-01960-1 | 14 | 1264 |
| 13.02 -27.46 | TYC6706-00675-1 | 14 | 1259 |
| 01.72 -15.78 | TYC5855-01232-1 | 14 | 1258 |
| 13.01 +12.82 | TYC0889-00884-1 | 14 | 1257 |
| 10.96 +01.78 | TYC0255-00937-1 | 14 | 1248 |
| 10.38 -00.69 | TYC4905-00976-1 | 14 | 1243 |
| 15.84 +35.61 | TYC2578-01247-1 | 14 | 1236 |
| 02.61 +41.86 | TYC2840-00232-1 | 14 | 1235 |
| 09.92 -24.15 | TYC6604-01481-1 | 14 | 1230 |
| 03.68 +31.60 | TYC2355-00453-1 | 14 | 1221 |
| 23.51 +39.17 | TYC3231-00543-1 | 14 | 1215 |
| 04.49 +19.29 | TYC1273-00994-1 | 13 | 1207 |
| 14.56 +21.72 | TYC1482-00083-1 | 13 | 1205 |
| 09.92 -24.03 | TYC6604-01301-1 | 13 | 1202 |
| 13.19 +17.40 | TYC1451-00117-1 | 13 | 1202 |
| 18.94 +48.92 | TYC3549-02580-1 | 13 | 1199 |
| 12.00 -20.56 | TYC6100-00666-1 | 13 | 1198 |
| 19.86 +48.07 | TYC3561-00236-1 | 13 | 1195 |
| 15.33 +36.10 | TYC2569-00778-1 | 13 | 1190 |
| 07.53 +17.16 | TYC1364-00062-1 | 13 | 1185 |
| 14.59 +09.68 | TYC0910-00047-1 | 13 | 1176 |
| 18.70 +47.83 | TYC3531-01409-1 | 13 | 1171 |
| 18.17 +54.34 | TYC3903-01434-1 | 13 | 1171 |
| 06.22 -29.89 | TYC6517-02269-1 | 13 | 1169 |
| 07.24 +14.24 | TYC0774-00551-1 | 13 | 1169 |
| 22.85 +35.29 | TYC2757-01774-1 | 13 | 1167 |
| 04.69 +23.11 | TYC1830-00746-1 | 13 | 1167 |
| 19.43 +51.09 | TYC3555-00262-1 | 13 | 1165 |
| 08.90 +32.83 | TYC2488-01445-1 | 13 | 1165 |
| 03.20 -00.85 | TYC4708-00037-1 | 13 | 1161 |
| 14.20 +04.19 | TYC0320-01153-1 | 13 | 1159 |
| 06.40 -28.83 | TYC6518-01150-1 | 13 | 1154 |
| 01.75 +19.86 | TYC1208-00081-1 | 13 | 1153 |
| 01.71 -15.96 | TYC5855-00684-1 | 13 | 1150 |
| 11.31 -19.24 | TYC6087-01208-1 | 13 | 1150 |
| 07.54 +33.88 | TYC2461-01112-1 | 13 | 1147 |
| 10.95 -31.33 | TYC7200-00393-1 | 13 | 1145 |
| 06.81 -00.53 | TYC4800-01885-1 | 13 | 1145 |
| 05.12 -26.85 | TYC6481-00518-1 | 13 | 1140 |
| 19.80 +41.23 | TYC3140-00135-1 | 13 | 1135 |
| 23.17 +56.90 | TYC4006-00837-1 | 13 | 1135 |
| 03.29 +30.97 | TYC2340-01113-1 | 13 | 1135 |
| 06.07 +44.28 | TYC2937-01755-1 | 13 | 1132 |
| 23.20 +57.22 | TYC4006-00664-1 | 12 | 1124 |
| 20.24 +65.13 | TYC4240-00034-1 | 12 | 1122 |
| 11.40 +41.17 | TYC3013-00072-1 | 12 | 1118 |
| 10.98 -30.96 | TYC7200-00586-1 | 12 | 1112 |
| 19.22 +50.11 | TYC3550-00019-1 | 12 | 1106 |
| 02.70 +49.43 | TYC3304-00147-1 | 12 | 1102 |
| 01.55 +29.38 | TYC1755-01222-1 | 12 | 1100 |
| 11.96 -20.73 | TYC6100-00515-1 | 12 | 1099 |
| 10.96 -30.97 | TYC7200-00344-1 | 12 | 1099 |
| 04.71 +18.74 | TYC1271-00323-1 | 12 | 1095 |
| 01.27 +76.80 | TYC4493-01316-1 | 12 | 1094 |
| 06.49 +38.98 | TYC2927-00392-1 | 12 | 1091 |
| 08.67 -23.37 | TYC6571-02617-1 | 12 | 1090 |
| 06.90 +24.36 | TYC1894-02736-1 | 12 | 1083 |
| 23.98 -22.36 | TYC6411-00069-1 | 12 | 1083 |
| 08.67 +64.57 | TYC4130-01639-1 | 12 | 1083 |
| 07.61 -13.87 | TYC5409-03255-1 | 12 | 1079 |
| 19.16 +43.64 | TYC3132-00485-1 | 12 | 1079 |
| 13.97 +43.24 | TYC3033-00836-1 | 12 | 1072 |
| 04.91 +12.51 | TYC0692-01309-1 | 12 | 1070 |
| 08.65 +47.57 | TYC3419-01146-1 | 12 | 1070 |
| 01.32 +76.80 | TYC4493-01420-1 | 12 | 1067 |
| 16.50 +38.48 | TYC3063-01794-1 | 12 | 1066 |
| 13.85 -13.31 | TYC5559-00543-1 | 12 | 1061 |
| 22.88 +35.53 | TYC2757-01212-1 | 12 | 1056 |
| 03.22 -01.05 | TYC4708-00337-1 | 12 | 1049 |
| 08.26 +05.72 | TYC0208-00859-1 | 12 | 1049 |
| 08.59 -29.80 | TYC6582-01510-1 | 12 | 1047 |
| 13.46 +13.93 | TYC0898-00042-1 | 12 | 1042 |
| 08.19 +51.40 | TYC3414-00371-1 | 12 | 1040 |
| 04.86 -24.32 | TYC6466-01076-1 | 12 | 1040 |
| 01.73 +20.19 | TYC1211-00806-1 | 12 | 1039 |
| 11.28 -18.80 | TYC6087-00735-1 | 12 | 1037 |
| 07.33 +37.17 | TYC2463-01473-1 | 12 | 1036 |
| 15.33 +41.93 | TYC3048-00031-1 | 11 | 1034 |
| 23.65 +42.28 | TYC3239-00346-1 | 11 | 1033 |
| 15.40 +58.85 | TYC3874-01184-1 | 11 | 1028 |
| 19.30 +42.03 | TYC3129-00082-1 | 11 | 1027 |
| 04.69 +22.98 | TYC1830-00612-1 | 11 | 1022 |
| 03.02 -20.77 | TYC5870-00842-1 | 11 | 1022 |
| 02.07 +25.25 | TYC1761-00712-1 | 11 | 1021 |
| 18.69 +47.82 | TYC3531-01946-1 | 11 | 1019 |
| 06.75 +00.94 | TYC0148-01733-1 | 11 | 1019 |
| 04.94 -23.06 | TYC6466-00367-1 | 11 | 1011 |
| 19.68 +43.29 | TYC3148-01170-1 | 11 | 1008 |
| 04.71 +18.90 | TYC1275-01382-1 | 11 | 1001 |
| 04.88 -16.80 | TYC5899-00046-1 | 11 | 999 |
| 12.51 +21.82 | TYC1447-02471-1 | 11 | 999 |
| 11.74 +02.87 | TYC0275-00132-1 | 11 | 999 |
| 19.15 +43.71 | TYC3132-00453-1 | 11 | 997 |
| 11.25 +25.11 | TYC1981-01211-1 | 11 | 992 |
| 09.93 -16.10 | TYC6045-01214-1 | 11 | 991 |
| 05.28 +07.47 | TYC0112-00821-1 | 11 | 990 |
| 19.13 +48.70 | TYC3546-00178-1 | 11 | 989 |
| 15.14 +02.30 | TYC0339-00520-1 | 11 | 989 |
| 11.42 +41.20 | TYC3013-00700-1 | 11 | 988 |
| 06.50 +00.22 | TYC0133-00057-1 | 11 | 987 |
| 06.54 +05.53 | TYC0154-01987-1 | 11 | 984 |
| 09.57 -12.56 | TYC5468-00125-1 | 11 | 983 |
| 19.62 +38.73 | TYC3135-00124-1 | 11 | 981 |
| 05.77 +00.95 | TYC0116-00156-1 | 11 | 981 |
| 19.67 +43.18 | TYC3147-01729-1 | 11 | 978 |
| 04.83 -24.22 | TYC6465-01333-1 | 11 | 978 |
| 14.79 -00.49 | TYC4986-00365-1 | 11 | 976 |
| 01.74 +19.87 | TYC1208-00019-1 | 11 | 975 |
| 19.92 +41.36 | TYC3145-01366-1 | 11 | 969 |
| 18.37 +65.62 | TYC4222-00125-1 | 11 | 966 |
| 19.12 +48.72 | TYC3545-02636-1 | 11 | 963 |
| 19.62 +38.78 | TYC3135-00173-1 | 11 | 962 |
| 19.47 +37.44 | TYC2666-00005-1 | 11 | 961 |
| 16.17 +44.06 | TYC3067-00380-1 | 11 | 957 |
| 01.87 -19.59 | TYC5858-01898-1 | 11 | 952 |
| 04.34 +57.98 | TYC3727-01113-1 | 11 | 952 |
| 02.70 +49.27 | TYC3304-00047-1 | 11 | 948 |
| 00.24 -11.90 | TYC5264-00093-1 | 11 | 945 |
| 05.65 +06.04 | TYC0127-00158-1 | 10 | 943 |
| 02.60 +24.54 | TYC1771-00760-1 | 10 | 942 |
| 18.69 +47.78 | TYC3531-01323-1 | 10 | 939 |
| 09.38 +50.03 | TYC3431-01525-1 | 10 | 939 |
| 00.75 +20.39 | TYC1194-00122-1 | 10 | 938 |
| 18.67 +43.78 | TYC3130-01742-1 | 10 | 934 |
| 08.31 -12.40 | TYC5435-00918-1 | 10 | 934 |
| 00.25 +01.17 | TYC0002-00376-1 | 10 | 930 |
| 19.69 +43.08 | TYC3144-01510-1 | 10 | 925 |
| 13.48 +13.88 | TYC0898-00024-1 | 10 | 925 |
| 10.39 -29.58 | TYC6631-00271-1 | 10 | 924 |
| 12.05 +76.94 | TYC4550-00077-1 | 10 | 921 |
| 05.78 +01.29 | TYC0116-00907-1 | 10 | 920 |
| 19.62 +43.77 | TYC3147-00456-1 | 10 | 919 |
| 03.17 +30.75 | TYC2340-01087-1 | 10 | 918 |
| 16.18 +44.02 | TYC3067-00398-1 | 10 | 913 |
| 01.00 +20.49 | TYC1195-00727-1 | 10 | 909 |
| 07.60 -13.89 | TYC5409-01206-1 | 10 | 905 |
| 19.15 +51.41 | TYC3554-00223-1 | 10 | 897 |
| 02.58 -12.53 | TYC5291-00664-1 | 10 | 893 |
| 14.78 -00.37 | TYC4986-00501-1 | 10 | 891 |
| 19.92 +41.35 | TYC3145-02010-1 | 10 | 890 |
| 22.03 +18.92 | TYC1688-01684-1 | 10 | 884 |
| 04.49 -25.24 | TYC6467-02595-1 | 10 | 878 |
| 11.45 +07.28 | TYC0270-00097-1 | 10 | 877 |
| 06.59 -27.53 | TYC6516-01355-1 | 10 | 876 |
| 09.94 -15.96 | TYC6045-01454-1 | 10 | 872 |
| 15.30 +71.96 | TYC4414-01354-1 | 10 | 871 |
| 14.80 +01.17 | TYC0326-01289-1 | 10 | 871 |
| 02.45 +37.47 | TYC2335-00839-1 | 10 | 870 |
| 02.07 +25.40 | TYC1761-00724-1 | 10 | 865 |
| 03.67 +31.93 | TYC2359-01113-1 | 10 | 863 |
| 22.04 +18.90 | TYC1688-01766-1 | 10 | 859 |
| 10.37 -01.15 | TYC4905-00527-1 | 10 | 859 |
| 12.09 +77.11 | TYC4550-00728-1 | 10 | 859 |
| 11.09 +44.46 | TYC3012-02312-1 | 10 | 855 |
| 09.46 +45.75 | TYC3425-00502-1 | 10 | 855 |
| 19.81 +40.90 | TYC3140-00291-1 | 9 | 854 |
| 09.52 -12.64 | TYC5468-00209-1 | 9 | 853 |
| 14.78 +00.27 | TYC0326-00361-1 | 9 | 852 |
| 11.98 -20.33 | TYC6100-00294-1 | 9 | 845 |
| 19.19 +43.77 | TYC3133-01476-1 | 9 | 845 |
| 07.17 +30.36 | TYC2438-00557-1 | 9 | 844 |
| 15.88 +12.81 | TYC0951-00035-1 | 9 | 843 |
| 19.62 +44.20 | TYC3147-00024-1 | 9 | 840 |
| 15.87 +35.70 | TYC2578-00661-1 | 9 | 840 |
| 07.45 +24.37 | TYC1910-00089-1 | 9 | 839 |
| 00.72 -26.69 | TYC6423-01750-1 | 9 | 839 |
| 13.71 +47.87 | TYC3466-00773-1 | 9 | 838 |
| 10.96 +01.61 | TYC0255-00818-1 | 9 | 836 |
| 22.95 +38.73 | TYC3215-01743-1 | 9 | 831 |
| 22.94 +20.80 | TYC1717-00015-1 | 9 | 831 |
| 18.95 +48.69 | TYC3545-00049-1 | 9 | 829 |
| 11.77 +03.35 | TYC0275-00325-1 | 9 | 827 |
| 01.41 +28.56 | TYC1754-00185-1 | 9 | 826 |
| 07.17 +30.01 | TYC2438-00472-1 | 9 | 824 |
| 08.40 +01.81 | TYC0197-00217-1 | 9 | 824 |
| 02.84 +71.68 | TYC4321-02076-1 | 9 | 819 |
| 03.19 +21.12 | TYC1231-00397-1 | 9 | 818 |
| 11.53 +07.59 | TYC0856-00011-1 | 9 | 806 |
| 02.27 +43.64 | TYC2842-00260-1 | 9 | 804 |
| 19.73 +41.12 | TYC3140-00631-1 | 9 | 804 |
| 19.18 +43.60 | TYC3133-01572-1 | 9 | 804 |
| 20.04 +46.47 | TYC3558-00381-1 | 9 | 801 |
| 19.96 +43.97 | TYC3149-00183-1 | 9 | 799 |
| 06.33 +40.88 | TYC2930-01675-1 | 9 | 798 |
| 10.99 -30.97 | TYC7200-01038-1 | 9 | 796 |
| 04.85 +06.43 | TYC0096-00398-1 | 9 | 792 |
| 07.52 +17.07 | TYC1364-01550-1 | 9 | 790 |
| 16.96 +25.76 | TYC2063-00387-1 | 9 | 786 |
| 12.51 +23.02 | TYC1989-02505-1 | 9 | 786 |
| 01.44 +76.60 | TYC4494-01396-1 | 9 | 786 |
| 11.28 -19.31 | TYC6087-01041-1 | 9 | 784 |
| 19.49 +41.79 | TYC3142-00367-1 | 9 | 783 |
| 13.01 +12.68 | TYC0889-00782-1 | 9 | 783 |
| 00.65 +21.28 | TYC1193-01006-1 | 9 | 781 |
| 02.72 +38.73 | TYC2845-00035-1 | 9 | 779 |
| 11.76 +02.65 | TYC0275-00051-1 | 9 | 778 |
| 08.65 +64.46 | TYC4130-01092-1 | 9 | 778 |
| 03.22 -01.15 | TYC4708-00159-1 | 9 | 778 |
| 23.25 +56.85 | TYC4006-00413-1 | 9 | 777 |
| 05.12 -26.97 | TYC6481-00980-1 | 9 | 777 |
| 01.55 +29.11 | TYC1755-01231-1 | 9 | 776 |
| 03.18 +21.24 | TYC1231-00216-1 | 9 | 773 |
| 07.18 +30.30 | TYC2438-00812-1 | 9 | 770 |
| 08.05 -01.14 | TYC4846-01348-1 | 9 | 770 |
| 19.21 +44.04 | TYC3133-00568-1 | 9 | 769 |
| 10.94 -30.96 | TYC7199-01554-1 | 9 | 768 |
| 11.52 +07.77 | TYC0856-00001-1 | 9 | 767 |
| 04.83 -24.50 | TYC6465-01864-1 | 9 | 766 |
| 06.39 -28.39 | TYC6518-01789-1 | 8 | 763 |
| 02.17 +32.10 | TYC2313-00037-1 | 8 | 760 |
| 11.85 +57.64 | TYC3839-00656-1 | 8 | 754 |
| 23.31 +18.54 | TYC1716-00776-1 | 8 | 754 |
| 19.90 +48.25 | TYC3562-00117-1 | 8 | 748 |
| 01.43 +28.48 | TYC1754-00355-1 | 8 | 747 |
| 18.30 +36.56 | TYC2634-01090-1 | 8 | 744 |
| 22.85 +35.47 | TYC2757-01036-1 | 8 | 743 |
| 02.00 +46.60 | TYC3280-01274-1 | 8 | 742 |
| 19.06 +50.33 | TYC3549-00024-1 | 8 | 741 |
| 09.33 +34.05 | TYC2496-01043-1 | 8 | 741 |
| 10.26 +19.85 | TYC1422-00107-1 | 8 | 738 |
| 08.29 -12.62 | TYC5434-03048-1 | 8 | 737 |
| 04.91 -23.31 | TYC6466-01525-1 | 8 | 736 |
| 04.88 -16.11 | TYC5899-00525-1 | 8 | 735 |
| 14.03 -27.31 | TYC6737-00012-1 | 8 | 735 |
| 11.60 +20.17 | TYC1440-00266-1 | 8 | 734 |
| 06.40 -28.90 | TYC6518-01444-1 | 8 | 732 |
| 00.63 +34.90 | TYC2283-01195-1 | 8 | 731 |
| 14.95 +53.31 | TYC3861-00208-1 | 8 | 726 |
| 07.55 +33.79 | TYC2461-01478-1 | 8 | 721 |
| 00.65 +21.12 | TYC1193-01072-1 | 8 | 720 |
| 08.32 +61.46 | TYC4126-01721-1 | 8 | 718 |
| 11.49 +07.69 | TYC0856-00026-1 | 8 | 714 |
| 04.86 +06.50 | TYC0096-00388-1 | 8 | 713 |
| 17.86 +37.42 | TYC2620-00812-1 | 8 | 713 |
| 09.57 -12.88 | TYC5468-00018-1 | 8 | 712 |
| 14.88 +18.44 | TYC1481-00160-1 | 8 | 712 |
| 14.91 +53.03 | TYC3861-00004-1 | 8 | 711 |
| 02.78 -23.24 | TYC6434-00461-1 | 8 | 709 |
| 20.04 +46.34 | TYC3558-01477-1 | 8 | 709 |
| 02.31 +32.29 | TYC2314-00419-1 | 8 | 709 |
| 01.87 -19.40 | TYC5858-02018-1 | 8 | 709 |
| 07.24 +14.27 | TYC0774-00493-1 | 8 | 708 |
| 23.22 +57.36 | TYC4006-00432-1 | 8 | 708 |
| 19.49 +47.80 | TYC3547-00925-1 | 8 | 703 |
| 23.29 +18.83 | TYC1715-01582-1 | 8 | 701 |
| 11.46 +44.11 | TYC3015-00785-1 | 8 | 698 |
| 18.46 +65.61 | TYC4222-00523-1 | 8 | 696 |
| 19.14 +49.32 | TYC3550-00408-1 | 8 | 687 |
| 08.53 +64.06 | TYC4130-00085-1 | 8 | 685 |
| 00.47 -16.37 | TYC5840-00520-1 | 8 | 682 |
| 11.72 +26.86 | TYC1984-01846-1 | 8 | 682 |
| 08.04 -01.60 | TYC4846-00610-1 | 8 | 682 |
| 03.29 +31.37 | TYC2340-00824-1 | 8 | 680 |
| 15.21 -25.35 | TYC6765-01339-1 | 8 | 677 |
| 13.98 +43.43 | TYC3033-00750-1 | 8 | 677 |
| 17.89 +56.53 | TYC3910-00087-1 | 8 | 676 |
| 01.54 +29.30 | TYC1755-01710-1 | 7 | 673 |
| 23.23 +57.14 | TYC4006-00446-1 | 7 | 673 |
| 01.85 -19.64 | TYC5857-01795-1 | 7 | 672 |
| 02.97 -20.89 | TYC5870-00211-1 | 7 | 668 |
| 07.75 +27.84 | TYC1920-00034-1 | 7 | 667 |
| 00.24 +01.09 | TYC0002-00038-1 | 7 | 667 |
| 23.51 +39.34 | TYC3231-00827-1 | 7 | 667 |
| 23.64 +42.77 | TYC3239-00080-1 | 7 | 666 |
| 02.57 -12.34 | TYC5288-00249-1 | 7 | 665 |
| 06.60 -27.51 | TYC6516-00517-1 | 7 | 665 |
| 02.44 +37.60 | TYC2831-00408-1 | 7 | 664 |
| 15.31 +41.82 | TYC3048-00121-1 | 7 | 664 |
| 02.82 +71.84 | TYC4321-01462-1 | 7 | 664 |
| 06.20 -30.12 | TYC7072-00068-1 | 7 | 663 |
| 02.56 -12.46 | TYC5288-00182-1 | 7 | 662 |
| 16.00 +33.35 | TYC2575-01436-1 | 7 | 662 |
| 09.38 +20.56 | TYC1408-00576-1 | 7 | 662 |
| 22.84 +35.28 | TYC2757-00224-1 | 7 | 660 |
| 06.63 -19.41 | TYC5956-01832-1 | 7 | 660 |
| 07.38 +20.56 | TYC1355-00359-1 | 7 | 660 |
| 00.62 +34.74 | TYC2283-01005-1 | 7 | 656 |
| 23.05 -00.57 | TYC5242-00637-1 | 7 | 655 |
| 08.21 +51.48 | TYC3414-00624-1 | 7 | 655 |
| 08.82 +72.45 | TYC4378-00444-1 | 7 | 654 |
| 22.86 +35.55 | TYC2757-00314-1 | 7 | 653 |
| 23.20 +57.11 | TYC4006-00956-1 | 7 | 653 |
| 04.33 +57.97 | TYC3727-01082-1 | 7 | 649 |
| 19.50 +41.59 | TYC3142-00259-1 | 7 | 646 |
| 19.70 +42.16 | TYC3144-01710-1 | 7 | 646 |
| 10.42 -29.49 | TYC6631-00401-1 | 7 | 644 |
| 11.47 +07.54 | TYC0856-00097-1 | 7 | 643 |
| 13.49 +13.82 | TYC0898-00001-1 | 7 | 642 |
| 03.13 +30.44 | TYC2339-01638-1 | 7 | 642 |
| 11.29 -19.24 | TYC6087-01720-1 | 7 | 642 |
| 09.56 -12.90 | TYC5468-00220-1 | 7 | 641 |
| 07.55 -22.29 | TYC5991-00100-1 | 7 | 641 |
| 11.50 +07.32 | TYC0270-00103-1 | 7 | 640 |
| 02.30 +43.78 | TYC2842-00024-1 | 7 | 639 |
| 03.48 -23.91 | TYC6446-00366-1 | 7 | 638 |
| 06.70 -01.26 | TYC4799-00670-1 | 7 | 636 |
| 14.36 -17.57 | TYC6143-00953-1 | 7 | 635 |
| 15.88 +13.07 | TYC0951-00131-1 | 7 | 634 |
| 08.65 +47.37 | TYC3416-00491-1 | 7 | 634 |
| 15.32 +35.95 | TYC2569-00678-1 | 7 | 634 |
| 12.56 +44.86 | TYC3020-02075-1 | 7 | 631 |
| 04.90 -16.70 | TYC5899-00089-1 | 7 | 629 |
| 02.61 +24.43 | TYC1771-01308-1 | 7 | 628 |
| 06.59 -28.04 | TYC6516-01310-1 | 7 | 627 |
| 14.31 -20.45 | TYC6147-00138-1 | 7 | 626 |
| 11.23 +25.50 | TYC1981-00482-1 | 7 | 626 |
| 19.22 +49.91 | TYC3550-00513-1 | 7 | 625 |
| 07.39 +20.52 | TYC1355-00075-1 | 7 | 623 |
| 23.23 +57.61 | TYC4006-00026-1 | 7 | 623 |
| 17.09 +32.82 | TYC2594-00297-1 | 7 | 623 |
| 06.80 +00.62 | TYC0148-01932-1 | 7 | 623 |
| 12.51 +21.94 | TYC1447-02439-1 | 7 | 623 |
| 02.18 +32.36 | TYC2313-00861-1 | 7 | 623 |
| 06.51 +58.05 | TYC3773-01476-1 | 7 | 623 |
| 08.67 -23.51 | TYC6571-01649-1 | 7 | 623 |
| 02.99 -20.73 | TYC5870-00559-1 | 7 | 621 |
| 03.79 +40.52 | TYC2867-01284-1 | 7 | 621 |
| 07.78 +39.10 | TYC2959-01875-1 | 7 | 621 |
| 08.37 +01.89 | TYC0200-01407-1 | 7 | 619 |
| 15.31 +36.32 | TYC2569-00652-1 | 7 | 618 |
| 19.02 +39.14 | TYC3120-00043-1 | 7 | 614 |
| 18.85 +47.66 | TYC3544-00655-1 | 7 | 614 |
| 19.15 +34.42 | TYC2648-00027-1 | 7 | 614 |
| 08.06 -01.22 | TYC4846-01208-1 | 7 | 610 |
| 03.64 +31.94 | TYC2359-01119-1 | 7 | 610 |
| 07.61 -14.00 | TYC5409-01422-1 | 7 | 609 |
| 19.28 +47.58 | TYC3546-00132-1 | 7 | 608 |
| 00.59 +34.85 | TYC2270-00862-1 | 7 | 606 |
| 19.30 +41.66 | TYC3129-00264-1 | 7 | 606 |
| 03.23 +25.36 | TYC1788-00773-1 | 7 | 605 |
| 07.38 +20.22 | TYC1355-01000-1 | 7 | 605 |
| 18.78 +42.76 | TYC3126-00596-1 | 7 | 603 |
| 07.35 +58.29 | TYC3793-01976-1 | 7 | 603 |
| 08.21 +04.93 | TYC0203-00559-1 | 7 | 603 |
| 06.41 -28.74 | TYC6518-02294-1 | 7 | 603 |
| 19.12 +46.71 | TYC3541-03164-1 | 7 | 600 |
| 09.76 +03.68 | TYC0239-02065-1 | 7 | 600 |
| 01.56 +29.08 | TYC1755-01177-1 | 7 | 600 |
| 06.42 -31.55 | TYC7073-02260-1 | 7 | 598 |
| 14.21 +03.87 | TYC0320-00932-1 | 7 | 598 |
| 19.69 +50.57 | TYC3564-02971-1 | 7 | 591 |
| 01.44 +76.54 | TYC4494-00872-1 | 7 | 590 |
| 06.30 +41.16 | TYC2930-01003-1 | 7 | 587 |
| 10.17 +18.14 | TYC1422-01205-1 | 6 | 584 |
| 22.44 -17.52 | TYC6388-00176-1 | 6 | 583 |
| 19.80 +50.28 | TYC3565-00005-1 | 6 | 582 |
| 07.60 -14.32 | TYC5409-01521-1 | 6 | 581 |
| 16.40 +41.04 | TYC3065-00009-1 | 6 | 581 |
| 04.65 +23.15 | TYC1830-01020-1 | 6 | 578 |
| 17.88 +37.39 | TYC2620-00175-1 | 6 | 578 |
| 19.59 +47.91 | TYC3560-00009-1 | 6 | 575 |
| 06.70 -01.41 | TYC4799-01064-1 | 6 | 570 |
| 23.18 +57.16 | TYC4006-00812-1 | 6 | 569 |
| 14.56 +21.83 | TYC1482-00602-1 | 6 | 569 |
| 02.81 -23.09 | TYC6437-00199-1 | 6 | 569 |
| 06.04 +44.47 | TYC2937-01643-1 | 6 | 568 |
| 22.45 -17.31 | TYC6385-00640-1 | 6 | 565 |
| 23.75 +77.90 | TYC4606-00002-1 | 6 | 565 |
| 19.47 +48.17 | TYC3547-00737-1 | 6 | 565 |
| 19.93 +48.40 | TYC3562-00846-1 | 6 | 565 |
| 11.85 +57.65 | TYC3838-00184-1 | 6 | 564 |
| 00.60 +34.90 | TYC2283-00922-1 | 6 | 564 |
| 02.99 -20.79 | TYC5870-00658-1 | 6 | 563 |
| 23.12 +21.01 | TYC1718-00021-1 | 6 | 561 |
| 08.31 +61.22 | TYC4126-01641-1 | 6 | 561 |
| 01.11 -22.36 | TYC5853-00754-1 | 6 | 558 |
| 00.01 -22.67 | TYC6412-00467-1 | 6 | 555 |
| 10.40 -00.90 | TYC4905-00923-1 | 6 | 554 |
| 07.75 +01.48 | TYC0179-02079-1 | 6 | 553 |
| 02.14 +32.12 | TYC2313-00073-1 | 6 | 553 |
| 17.89 +56.28 | TYC3910-00089-1 | 6 | 553 |
| 19.12 +41.93 | TYC3128-00078-1 | 6 | 552 |
| 09.48 +45.77 | TYC3425-00791-1 | 6 | 552 |
| 00.74 -26.59 | TYC6423-01914-1 | 6 | 551 |
| 04.33 +57.63 | TYC3727-00194-1 | 6 | 551 |
| 23.19 +57.26 | TYC4006-00996-1 | 6 | 551 |
| 02.70 +38.67 | TYC2845-00038-1 | 6 | 550 |
| 06.64 -32.36 | TYC7091-00061-1 | 6 | 549 |
| 18.42 +65.59 | TYC4222-00097-1 | 6 | 548 |
| 05.59 +20.91 | TYC1309-00782-1 | 6 | 546 |
| 06.48 +57.97 | TYC3773-00113-1 | 6 | 546 |
| 00.34 +31.76 | TYC2261-00814-1 | 6 | 546 |
| 13.44 +13.61 | TYC0897-00537-1 | 6 | 546 |
| 08.07 -01.16 | TYC4846-00072-1 | 6 | 546 |
| 04.88 -16.15 | TYC5899-00775-1 | 6 | 546 |
| 06.55 -00.87 | TYC4798-00434-1 | 6 | 546 |
| 06.71 -01.31 | TYC4799-00357-1 | 6 | 544 |
| 14.36 -17.43 | TYC6143-00975-1 | 6 | 544 |
| 21.48 -21.65 | TYC6373-00756-1 | 6 | 544 |
| 02.72 +49.63 | TYC3304-01234-1 | 6 | 544 |
| 23.28 +58.04 | TYC4006-00013-1 | 6 | 543 |
| 06.86 -01.13 | TYC4800-01195-1 | 6 | 540 |
| 01.53 +29.50 | TYC1755-01788-1 | 6 | 539 |
| 02.35 +32.04 | TYC2314-00807-1 | 6 | 537 |
| 10.37 -29.56 | TYC6631-00715-1 | 6 | 536 |
| 06.79 -01.14 | TYC4800-01609-1 | 6 | 535 |
| 19.36 +50.92 | TYC3555-00627-1 | 6 | 533 |
| 08.29 +62.35 | TYC4126-00046-1 | 6 | 532 |
| 09.55 -12.64 | TYC5468-00017-1 | 6 | 532 |
| 07.26 +14.36 | TYC0774-00211-1 | 6 | 532 |
| 22.83 +35.44 | TYC2757-01198-1 | 6 | 532 |
| 13.97 +44.05 | TYC3033-00573-1 | 6 | 530 |
| 07.60 -13.96 | TYC5409-02008-1 | 6 | 530 |
| 04.67 +22.94 | TYC1830-01253-1 | 6 | 529 |
| 01.43 +34.77 | TYC2300-01392-1 | 6 | 528 |
| 12.22 +10.26 | TYC0869-01162-1 | 6 | 528 |
| 02.62 +42.09 | TYC2840-01556-1 | 6 | 526 |
| 16.98 +25.83 | TYC2063-00699-1 | 6 | 525 |
| 22.85 +35.27 | TYC2757-00232-1 | 6 | 524 |
| 04.87 -16.35 | TYC5899-00587-1 | 6 | 524 |
| 10.15 +34.48 | TYC2506-01242-1 | 6 | 523 |
| 22.89 +35.42 | TYC2757-01418-1 | 6 | 523 |
| 13.58 +54.19 | TYC3850-01262-1 | 6 | 523 |
| 19.09 +48.74 | TYC3545-00141-1 | 6 | 522 |
| 00.77 +07.80 | TYC0604-01209-1 | 6 | 522 |
| 15.98 -27.97 | TYC6787-01843-1 | 6 | 522 |
| 19.06 +36.71 | TYC2652-00062-1 | 6 | 520 |
| 03.82 +40.56 | TYC2867-01488-1 | 6 | 520 |
| 17.33 +38.08 | TYC3073-01348-1 | 6 | 520 |
| 02.20 +51.61 | TYC3293-00844-1 | 6 | 520 |
| 06.60 -19.22 | TYC5956-01728-1 | 6 | 520 |
| 06.59 -27.75 | TYC6516-01579-1 | 6 | 520 |
| 01.89 -19.36 | TYC5858-01789-1 | 6 | 519 |
| 14.95 +44.07 | TYC3050-00938-1 | 6 | 519 |
| 06.50 +38.89 | TYC2927-00588-1 | 6 | 518 |
| 19.24 +49.77 | TYC3550-01642-1 | 6 | 515 |
| 22.93 +38.51 | TYC3215-00194-1 | 6 | 514 |
| 23.51 +77.69 | TYC4606-00350-1 | 6 | 514 |
| 13.56 +53.74 | TYC3850-00209-1 | 6 | 514 |
| 19.72 +44.19 | TYC3148-01648-1 | 6 | 513 |
| 19.80 +41.14 | TYC3140-00183-1 | 6 | 512 |
| 02.62 +24.49 | TYC1771-01291-1 | 6 | 511 |
| 01.46 +76.82 | TYC4494-00678-1 | 6 | 511 |
| 18.48 +65.53 | TYC4222-00511-1 | 6 | 511 |
| 12.23 +09.74 | TYC0866-00416-1 | 6 | 510 |
| 08.03 -01.09 | TYC4846-00326-1 | 6 | 509 |
| 16.17 +26.91 | TYC2038-01374-1 | 6 | 508 |
| 02.77 +49.54 | TYC3304-01056-1 | 6 | 508 |
| 05.15 +69.54 | TYC4346-01038-1 | 6 | 508 |
| 23.62 +77.62 | TYC4606-02124-1 | 6 | 508 |
| 11.46 +44.16 | TYC3015-00681-1 | 6 | 507 |
| 00.60 +34.60 | TYC2270-00115-1 | 6 | 506 |
| 07.38 +20.61 | TYC1355-00306-1 | 6 | 506 |
| 08.63 +64.90 | TYC4130-00145-1 | 6 | 505 |
| 02.54 -12.56 | TYC5284-01027-1 | 6 | 505 |
| 12.67 -30.56 | TYC7247-00715-1 | 6 | 505 |
| 04.85 +06.37 | TYC0096-00540-1 | 6 | 505 |
| 14.92 +53.46 | TYC3861-00027-1 | 6 | 504 |
| 19.17 +43.65 | TYC3132-00993-1 | 6 | 503 |
| 06.44 -31.44 | TYC7073-02040-1 | 6 | 503 |
| 10.39 -00.69 | TYC4905-00060-1 | 6 | 503 |
| 09.21 +46.64 | TYC3424-00052-1 | 6 | 502 |
| 19.08 +50.18 | TYC3549-00916-1 | 6 | 501 |
| 17.86 +37.52 | TYC3089-00889-1 | 6 | 501 |
| 09.90 -23.68 | TYC6603-01132-1 | 6 | 500 |
| 23.49 +38.57 | TYC3230-00002-1 | 6 | 497 |
| 06.44 +57.49 | TYC3772-00097-1 | 6 | 497 |
| 13.21 -31.78 | TYC7263-02039-1 | 6 | 497 |
| 22.03 +18.97 | TYC1688-01808-1 | 6 | 496 |
| 03.19 -01.38 | TYC4708-00506-1 | 6 | 496 |
| 10.37 -00.75 | TYC4905-01305-1 | 6 | 496 |
| 11.68 +27.21 | TYC1984-02259-1 | 6 | 496 |
| 09.57 -11.98 | TYC5468-00860-1 | 6 | 496 |
| 19.37 +50.81 | TYC3555-00362-1 | 5 | 492 |
| 08.65 +12.97 | TYC0805-00082-1 | 5 | 490 |
| 02.99 -20.87 | TYC5870-00583-1 | 5 | 490 |
| 01.45 +34.52 | TYC2300-01346-1 | 5 | 490 |
| 07.40 +20.26 | TYC1355-01255-1 | 5 | 489 |
| 23.59 +00.32 | TYC0585-00977-1 | 5 | 489 |
| 23.98 -22.38 | TYC6411-00208-1 | 5 | 489 |
| 21.18 +14.69 | TYC1116-00165-1 | 5 | 489 |
| 10.98 +40.73 | TYC3009-00092-1 | 5 | 486 |
| 02.64 +24.46 | TYC1771-01287-1 | 5 | 485 |
| 08.69 +04.63 | TYC0219-01841-1 | 5 | 485 |
| 08.37 +01.98 | TYC0200-01480-1 | 5 | 483 |
| 21.44 -21.60 | TYC6372-00064-1 | 5 | 483 |
| 19.96 +44.38 | TYC3149-00126-1 | 5 | 483 |
| 01.55 +29.22 | TYC1755-01508-1 | 5 | 481 |
| 16.15 +26.85 | TYC2038-01319-1 | 5 | 481 |
| 23.16 +18.37 | TYC1715-00943-1 | 5 | 480 |
| 19.55 +47.89 | TYC3560-01165-1 | 5 | 480 |
| 06.71 -01.38 | TYC4799-00267-1 | 5 | 480 |
| 06.81 -00.60 | TYC4800-00150-1 | 5 | 480 |
| 06.81 -00.54 | TYC4800-01719-1 | 5 | 480 |
| 08.29 -12.63 | TYC5434-02277-1 | 5 | 480 |
| 19.73 +44.41 | TYC3148-01653-1 | 5 | 479 |
| 22.89 -14.51 | TYC5819-00261-1 | 5 | 479 |
| 07.39 +20.25 | TYC1355-01195-1 | 5 | 478 |
| 08.98 +10.37 | TYC0814-01894-1 | 5 | 478 |
| 19.43 +48.42 | TYC3547-00153-1 | 5 | 478 |
| 16.02 +33.24 | TYC2576-02039-1 | 5 | 478 |
| 04.35 +57.64 | TYC3727-00350-1 | 5 | 474 |
| 04.86 -16.24 | TYC5899-00848-1 | 5 | 474 |
| 01.76 -16.48 | TYC5855-00787-1 | 5 | 473 |
| 19.17 +51.40 | TYC3554-00028-1 | 5 | 473 |
| 01.76 -16.44 | TYC5855-01119-1 | 5 | 473 |
| 13.00 +12.82 | TYC0889-00886-1 | 5 | 472 |
| 22.44 -17.34 | TYC6385-00770-1 | 5 | 472 |
| 03.26 +31.02 | TYC2340-00783-1 | 5 | 471 |
| 11.82 +57.74 | TYC3838-00272-1 | 5 | 471 |
| 08.58 -30.39 | TYC7136-02133-1 | 5 | 470 |
| 06.74 +00.88 | TYC0147-01749-1 | 5 | 468 |
| 07.48 +24.46 | TYC1914-00689-1 | 5 | 467 |
| 16.04 +27.97 | TYC2041-01807-1 | 5 | 467 |
| 01.61 +41.65 | TYC2822-00286-1 | 5 | 467 |
| 01.60 +41.57 | TYC2822-01880-1 | 5 | 467 |
| 06.09 +44.15 | TYC2937-00230-1 | 5 | 467 |
| 06.86 +40.67 | TYC2946-00088-1 | 5 | 467 |
| 06.85 +41.08 | TYC2946-00477-1 | 5 | 467 |
| 17.04 +46.96 | TYC3501-00011-1 | 5 | 467 |
| 14.93 +53.41 | TYC3861-00266-1 | 5 | 467 |
| 15.42 +59.01 | TYC3875-00232-1 | 5 | 467 |
| 20.20 +65.27 | TYC4240-00760-1 | 5 | 467 |
| 06.73 -00.85 | TYC4799-01149-1 | 5 | 467 |
| 06.82 -00.85 | TYC4800-01765-1 | 5 | 467 |
| 09.59 -12.20 | TYC5469-01173-1 | 5 | 467 |
| 04.32 +57.56 | TYC3727-00174-1 | 5 | 465 |
| 06.81 +00.53 | TYC0148-01931-1 | 5 | 465 |
| 14.59 +09.62 | TYC0910-00584-1 | 5 | 465 |
| 17.18 +63.49 | TYC4202-00819-1 | 5 | 465 |
| 02.63 +24.72 | TYC1771-01369-1 | 5 | 464 |
| 09.40 +50.64 | TYC3431-01072-1 | 5 | 464 |
| 19.77 +34.23 | TYC2664-00065-1 | 5 | 464 |
| 17.23 +63.41 | TYC4202-00729-1 | 5 | 464 |
| 19.55 +47.78 | TYC3560-01875-1 | 5 | 463 |
| 03.54 -23.63 | TYC6447-00508-1 | 5 | 462 |
| 15.57 +54.23 | TYC3869-00152-1 | 5 | 461 |
| 03.49 -23.91 | TYC6446-00367-1 | 5 | 460 |
| 18.03 +26.24 | TYC2095-00858-1 | 5 | 460 |
| 05.12 -14.20 | TYC5342-00950-1 | 5 | 460 |
| 08.29 +61.95 | TYC4126-00696-1 | 5 | 459 |
| 01.85 -19.39 | TYC5857-01759-1 | 5 | 458 |
| 18.69 +47.73 | TYC3531-01933-1 | 5 | 457 |
| 02.90 +71.91 | TYC4321-01626-1 | 5 | 457 |
| 07.76 +28.04 | TYC1920-02198-1 | 5 | 455 |
| 11.84 +57.36 | TYC3835-00456-1 | 5 | 455 |
| 19.82 +43.93 | TYC3148-00324-1 | 5 | 452 |
| 19.22 +50.00 | TYC3550-00977-1 | 5 | 452 |
| 07.37 +20.37 | TYC1355-00371-1 | 5 | 452 |
| 07.38 +20.11 | TYC1355-01250-1 | 5 | 452 |
| 07.44 +24.31 | TYC1910-00525-1 | 5 | 452 |
| 06.22 -29.87 | TYC6517-02214-1 | 5 | 451 |
| 06.78 +32.53 | TYC2440-00381-1 | 5 | 451 |
| 06.79 +33.01 | TYC2440-00726-1 | 5 | 451 |
| 18.74 +36.72 | TYC2649-00552-1 | 5 | 451 |
| 00.86 +35.26 | TYC2284-00203-1 | 5 | 449 |
| 22.02 +18.84 | TYC1688-01968-1 | 5 | 449 |
| 17.89 +37.33 | TYC2620-00271-1 | 5 | 449 |
| 07.52 +33.89 | TYC2461-00956-1 | 5 | 449 |
| 19.39 +55.59 | TYC3925-00586-1 | 5 | 447 |
| 03.68 +32.33 | TYC2359-01106-1 | 5 | 446 |
| 16.85 +12.57 | TYC0983-00877-1 | 5 | 446 |
| 18.77 +42.40 | TYC3126-00780-1 | 5 | 446 |
| 00.67 +34.56 | TYC2283-00394-1 | 5 | 445 |
| 07.44 +23.62 | TYC1910-00848-1 | 5 | 444 |
| 05.60 +35.23 | TYC2412-00726-1 | 5 | 444 |
| 04.70 +18.83 | TYC1275-01275-1 | 5 | 444 |
| 19.14 +48.70 | TYC3546-00181-1 | 5 | 444 |
| 05.36 +78.81 | TYC4532-01940-1 | 5 | 443 |
| 19.28 +49.57 | TYC3550-00100-1 | 5 | 443 |
| 13.92 -32.10 | TYC7283-01194-1 | 5 | 443 |
| 08.66 -23.42 | TYC6571-01792-1 | 5 | 442 |
| 04.51 -25.72 | TYC6467-02145-1 | 5 | 442 |
| 06.41 -31.86 | TYC7073-02399-1 | 5 | 442 |
| 16.19 +43.92 | TYC3067-00478-1 | 5 | 442 |
| 03.21 -00.37 | TYC4708-00772-1 | 5 | 442 |
| 06.31 +41.28 | TYC2934-01172-1 | 5 | 442 |
| 08.87 +28.26 | TYC1949-02013-1 | 5 | 441 |
| 02.81 +72.61 | TYC4321-01875-1 | 5 | 441 |
| 02.58 +23.72 | TYC1767-00336-1 | 5 | 441 |
| 20.00 +44.98 | TYC3149-00058-1 | 5 | 440 |
| 18.97 +50.43 | TYC3549-00093-1 | 5 | 440 |
| 16.85 +11.37 | TYC0983-01838-1 | 5 | 439 |
| 12.21 +10.11 | TYC0869-00765-1 | 5 | 439 |
| 07.71 +27.02 | TYC1920-01615-1 | 5 | 438 |
| 09.59 +30.43 | TYC2494-00231-1 | 5 | 438 |
| 12.52 +22.73 | TYC1989-02167-1 | 5 | 438 |
| 13.20 +17.58 | TYC1454-00146-1 | 5 | 437 |
| 22.03 +18.75 | TYC1684-00961-1 | 5 | 437 |
| 07.60 -13.30 | TYC5409-01376-1 | 5 | 437 |
| 12.60 +74.14 | TYC4400-00437-1 | 5 | 437 |
| 03.47 +34.21 | TYC2350-01402-1 | 5 | 437 |
| 03.28 +31.03 | TYC2340-00598-1 | 5 | 437 |
| 02.01 +46.91 | TYC3284-00477-1 | 5 | 436 |
| 04.06 +81.95 | TYC4522-00900-1 | 5 | 435 |
| 04.80 +21.61 | TYC1292-01242-1 | 5 | 435 |
| 07.46 +24.32 | TYC1910-00871-1 | 5 | 434 |
| 08.42 +61.63 | TYC4127-00371-1 | 5 | 433 |
| 12.02 +76.82 | TYC4550-00351-1 | 5 | 433 |
| 05.13 -13.90 | TYC5342-01118-1 | 5 | 433 |
| 09.39 +50.63 | TYC3431-00892-1 | 5 | 433 |
| 11.04 -23.61 | TYC6636-00499-1 | 5 | 433 |
| 11.29 -23.66 | TYC6649-00283-1 | 5 | 433 |
| 12.04 +76.82 | TYC4550-00261-1 | 5 | 432 |
| 03.24 +30.64 | TYC2340-01005-1 | 5 | 431 |
| 04.81 -24.38 | TYC6465-01583-1 | 5 | 430 |
| 01.87 -19.86 | TYC5858-01743-1 | 5 | 429 |
| 19.85 +43.03 | TYC3145-00027-1 | 5 | 429 |
| 00.74 +20.23 | TYC1194-00301-1 | 5 | 429 |
| 11.46 +43.88 | TYC3015-01273-1 | 5 | 429 |
| 19.25 +51.22 | TYC3554-00005-1 | 5 | 428 |
| 11.72 +30.80 | TYC2524-02201-1 | 5 | 428 |
| 02.62 +24.64 | TYC1771-00372-1 | 5 | 428 |
| 23.19 +57.17 | TYC4006-00944-1 | 5 | 428 |
| 02.29 +43.74 | TYC2842-00318-1 | 5 | 428 |
| 04.29 +57.86 | TYC3727-02006-1 | 5 | 428 |
| 04.84 -24.10 | TYC6466-01800-1 | 5 | 428 |
| 13.94 +43.95 | TYC3033-00808-1 | 5 | 428 |
| 06.74 +00.79 | TYC0147-00480-1 | 5 | 427 |
| 07.24 +13.61 | TYC0774-00206-1 | 5 | 426 |
| 07.41 +20.46 | TYC1355-00031-1 | 5 | 425 |
| 11.44 +02.81 | TYC0267-00018-1 | 5 | 424 |
| 01.78 -16.64 | TYC5855-00673-1 | 5 | 424 |
| 08.20 +04.65 | TYC0203-00356-1 | 5 | 424 |
| 06.53 -00.62 | TYC4798-00216-1 | 5 | 424 |
| 12.47 +74.18 | TYC4400-00562-1 | 5 | 423 |
| 18.57 +35.64 | TYC2636-00140-1 | 5 | 423 |
| 04.86 +06.59 | TYC0096-00339-1 | 5 | 421 |
| 19.94 +44.89 | TYC3149-00028-1 | 5 | 421 |
| 11.76 +02.54 | TYC0275-00014-1 | 5 | 421 |
| 08.63 +11.84 | TYC0805-01222-1 | 5 | 419 |
| 00.60 +34.64 | TYC2270-00769-1 | 5 | 419 |
| 05.30 +05.88 | TYC0112-02277-1 | 5 | 419 |
| 13.80 +17.25 | TYC1460-00130-1 | 5 | 419 |
| 02.07 +25.54 | TYC1761-00171-1 | 5 | 419 |
| 05.06 +25.79 | TYC1849-00846-1 | 5 | 419 |
| 01.62 +41.20 | TYC2818-00352-1 | 5 | 419 |
| 10.35 +41.21 | TYC3004-00140-1 | 5 | 419 |
| 12.55 +44.78 | TYC3020-01958-1 | 5 | 419 |
| 06.93 +53.27 | TYC3767-01234-1 | 5 | 419 |
| 17.20 +63.13 | TYC4202-00624-1 | 5 | 419 |
| 19.48 +48.52 | TYC3547-00623-1 | 5 | 418 |
| 20.19 +18.84 | TYC1626-00389-1 | 5 | 418 |
| 01.43 +28.59 | TYC1754-00146-1 | 5 | 418 |
| 19.18 +50.51 | TYC3550-00085-1 | 5 | 418 |
| 19.21 +51.16 | TYC3554-00034-1 | 5 | 418 |
| 09.24 +23.17 | TYC1951-00054-1 | 5 | 417 |
| 00.72 +20.44 | TYC1194-00024-1 | 5 | 417 |
| 22.97 +20.41 | TYC1717-00001-1 | 5 | 417 |
| 11.39 +40.97 | TYC3013-00132-1 | 5 | 417 |
| 23.45 -20.52 | TYC6402-00987-1 | 5 | 417 |
| 02.18 +32.43 | TYC2313-00104-1 | 5 | 416 |
| 08.60 -30.78 | TYC7136-02078-1 | 5 | 416 |
| 08.88 +33.18 | TYC2488-01568-1 | 5 | 416 |
| 18.74 +47.80 | TYC3531-01443-1 | 5 | 415 |
| 18.79 +48.71 | TYC3544-00026-1 | 5 | 414 |
| 22.95 +20.63 | TYC1717-00645-1 | 5 | 414 |
| 10.96 +01.80 | TYC0255-00939-1 | 5 | 413 |
| 02.73 +49.55 | TYC3304-00009-1 | 5 | 412 |
| 02.96 -20.89 | TYC5870-00508-1 | 5 | 411 |
| 12.25 +09.99 | TYC0866-00372-1 | 5 | 411 |
| 13.53 +53.91 | TYC3850-00028-1 | 5 | 411 |
| 05.26 +07.50 | TYC0112-02036-1 | 5 | 411 |
| 18.83 +48.12 | TYC3544-00811-1 | 5 | 411 |
| 02.09 +24.99 | TYC1758-00113-1 | 5 | 410 |
| 05.23 +07.18 | TYC0111-01411-1 | 5 | 410 |
| 08.13 +18.85 | TYC1384-02025-1 | 5 | 410 |
| 08.16 +19.51 | TYC1385-01161-1 | 5 | 410 |
| 07.76 +50.11 | TYC3413-00036-1 | 5 | 410 |
| 19.84 +48.20 | TYC3561-00461-1 | 5 | 410 |
| 02.45 +37.16 | TYC2335-00039-1 | 5 | 407 |
| 00.64 +21.62 | TYC1193-00510-1 | 5 | 407 |
| 04.79 +11.96 | TYC0691-00810-1 | 5 | 406 |
| 08.98 +09.64 | TYC0811-01653-1 | 5 | 406 |
| 04.00 +30.57 | TYC2357-01354-1 | 5 | 406 |
| 04.90 +50.49 | TYC3352-01495-1 | 5 | 406 |
| 09.47 +60.49 | TYC4135-01235-1 | 5 | 406 |
| 04.88 +75.91 | TYC4511-01193-1 | 5 | 406 |
| 10.04 -15.38 | TYC6046-00310-1 | 5 | 406 |
| 05.33 +06.88 | TYC0112-01576-1 | 5 | 406 |
| 05.02 +05.13 | TYC0106-00946-1 | 5 | 406 |
| 04.42 +25.92 | TYC1820-00142-1 | 5 | 406 |
| 07.51 +29.19 | TYC1922-01688-1 | 5 | 406 |
| 05.01 +35.69 | TYC2400-01664-1 | 5 | 406 |
| 19.12 +37.82 | TYC3120-00328-1 | 5 | 406 |
| 18.99 +41.44 | TYC3127-01146-1 | 5 | 406 |
| 19.70 +40.42 | TYC3140-01417-1 | 5 | 406 |
| 08.55 +50.96 | TYC3422-01513-1 | 5 | 406 |
| 18.93 +45.66 | TYC3540-01222-1 | 5 | 406 |
| 19.33 +46.18 | TYC3543-00234-1 | 5 | 406 |
| 19.08 +47.93 | TYC3545-01014-1 | 5 | 406 |
| 19.12 +48.43 | TYC3545-02767-1 | 5 | 406 |
| 19.32 +50.31 | TYC3551-00066-1 | 5 | 406 |
| 19.39 +49.95 | TYC3551-00566-1 | 5 | 406 |
| 19.26 +51.71 | TYC3554-00877-1 | 5 | 406 |
| 19.29 +51.79 | TYC3554-01087-1 | 5 | 406 |
| 19.62 +50.51 | TYC3564-00016-1 | 5 | 406 |
| 04.50 +54.87 | TYC3736-01003-1 | 5 | 406 |
| 05.71 +54.21 | TYC3749-00720-1 | 5 | 406 |
| 01.64 +41.22 | TYC2818-00058-1 | 5 | 406 |
| 00.31 -15.18 | TYC5839-00832-1 | 5 | 405 |
| 02.78 +19.87 | TYC1226-01131-1 | 5 | 405 |
| 19.14 +40.33 | TYC3124-00509-1 | 5 | 405 |
| 01.45 +34.21 | TYC2300-00163-1 | 5 | 405 |
| 16.16 -20.84 | TYC6213-01774-1 | 5 | 405 |
| 00.63 +42.67 | TYC2792-00359-1 | 4 | 404 |
| 06.42 -31.56 | TYC7073-02296-2 | 4 | 403 |
| 09.58 -11.80 | TYC5469-01149-1 | 4 | 403 |
| 08.66 -23.58 | TYC6571-01768-1 | 4 | 403 |
| 19.22 +40.40 | TYC3125-00317-1 | 4 | 402 |
| 03.13 +20.85 | TYC1231-00324-1 | 4 | 402 |
| 23.19 +57.12 | TYC4006-00596-1 | 4 | 402 |
| 08.81 +15.87 | TYC1393-01838-1 | 4 | 401 |
| 08.81 +15.68 | TYC1393-01926-1 | 4 | 401 |
| 15.90 +15.46 | TYC1496-00016-1 | 4 | 401 |
| 00.46 -16.02 | TYC5840-00688-1 | 4 | 400 |
| 00.22 -11.84 | TYC5264-00052-1 | 4 | 400 |
| 23.62 +77.65 | TYC4606-01598-1 | 4 | 398 |
| 09.52 -11.43 | TYC5468-00771-1 | 4 | 398 |
| 00.70 -26.55 | TYC6423-01861-1 | 4 | 398 |
| 03.23 +24.99 | TYC1788-00823-1 | 4 | 396 |
| 03.13 +30.34 | TYC2339-01426-1 | 4 | 396 |
| 06.18 -29.63 | TYC6517-01900-1 | 4 | 396 |
| 00.26 +01.02 | TYC0002-01240-1 | 4 | 393 |
| 02.97 +19.12 | TYC1227-00986-1 | 4 | 393 |
| 23.23 +56.81 | TYC4006-01337-1 | 4 | 393 |
| 20.96 +11.02 | TYC1107-02090-1 | 4 | 393 |
| 09.23 -27.31 | TYC6595-01851-1 | 4 | 393 |
| 05.60 -15.09 | TYC5917-00983-1 | 4 | 392 |
| 19.18 +51.28 | TYC3554-00045-1 | 4 | 392 |
| 12.48 +22.87 | TYC1989-01653-1 | 4 | 392 |
| 13.46 +13.91 | TYC0898-00036-1 | 4 | 392 |
| 06.21 -30.34 | TYC7072-00554-1 | 4 | 392 |
| 23.44 +08.57 | TYC1162-00639-1 | 4 | 391 |
| 10.63 +57.79 | TYC3822-01093-1 | 4 | 390 |
| 18.76 +42.50 | TYC3126-01702-1 | 4 | 389 |
| 19.74 +41.14 | TYC3140-00637-1 | 4 | 389 |
| 08.91 +32.13 | TYC2485-01445-1 | 4 | 388 |
| 08.21 +51.03 | TYC3414-00988-1 | 4 | 388 |
| 11.24 +18.07 | TYC1437-00962-1 | 4 | 388 |
| 05.40 +79.12 | TYC4532-01701-1 | 4 | 388 |
| 02.73 +49.80 | TYC3304-01341-1 | 4 | 388 |
| 22.97 +38.44 | TYC3215-00350-1 | 4 | 388 |
| 23.66 +42.34 | TYC3239-00836-1 | 4 | 388 |
| 11.30 -24.23 | TYC6649-00941-1 | 4 | 388 |
| 11.45 +43.99 | TYC3015-01571-1 | 4 | 387 |
| 14.95 +53.49 | TYC3861-00659-1 | 4 | 386 |
| 19.87 +49.49 | TYC3565-00342-1 | 4 | 386 |
| 11.13 -30.42 | TYC7201-00425-1 | 4 | 386 |
| 07.26 +37.05 | TYC2463-00033-1 | 4 | 384 |
| 18.28 +37.41 | TYC2634-00257-1 | 4 | 384 |
| 18.11 +53.91 | TYC3903-00039-1 | 4 | 384 |
| 08.68 -23.28 | TYC6571-00211-1 | 4 | 384 |
| 00.65 +34.75 | TYC2283-00696-1 | 4 | 382 |
| 02.05 +25.25 | TYC1761-00457-1 | 4 | 382 |
| 08.98 +25.13 | TYC1953-01537-1 | 4 | 382 |
| 07.05 +78.30 | TYC4530-00055-1 | 4 | 381 |
| 08.06 +33.72 | TYC2472-01122-1 | 4 | 381 |
| 12.99 -27.39 | TYC6706-01191-1 | 4 | 380 |
| 12.67 -30.63 | TYC7247-00620-1 | 4 | 380 |
| 09.60 -21.43 | TYC6055-01001-1 | 4 | 379 |
| 08.07 -29.15 | TYC6566-01936-1 | 4 | 379 |
| 07.80 +50.21 | TYC3413-00013-1 | 4 | 378 |
| 08.57 -30.14 | TYC7136-02211-1 | 4 | 378 |
| 07.76 +39.19 | TYC2959-00084-1 | 4 | 378 |
| 18.78 +42.14 | TYC3126-00787-1 | 4 | 376 |
| 18.78 +42.09 | TYC3126-00920-1 | 4 | 376 |
| 22.95 +20.86 | TYC1717-00742-1 | 4 | 375 |
| 19.07 +36.50 | TYC2652-01679-1 | 4 | 375 |
| 19.11 +49.80 | TYC3549-02604-1 | 4 | 374 |
| 06.04 +44.17 | TYC2937-00774-1 | 4 | 374 |
| 03.20 -00.97 | TYC4708-00501-1 | 4 | 374 |
| 06.46 +11.07 | TYC0736-01583-1 | 4 | 374 |
| 03.17 +21.27 | TYC1231-00203-1 | 4 | 374 |
| 03.21 +25.11 | TYC1787-00865-1 | 4 | 374 |
| 03.14 +30.49 | TYC2339-01082-1 | 4 | 374 |
| 02.64 +42.23 | TYC2840-01029-1 | 4 | 374 |
| 06.45 +38.85 | TYC2927-00556-1 | 4 | 374 |
| 06.45 +39.03 | TYC2927-00596-1 | 4 | 374 |
| 23.49 +39.17 | TYC3230-00183-1 | 4 | 374 |
| 05.09 +69.65 | TYC4346-00675-1 | 4 | 374 |
| 20.07 +45.78 | TYC3559-00536-1 | 4 | 374 |
| 22.05 +18.77 | TYC1688-01639-1 | 4 | 373 |
| 22.45 -17.42 | TYC6385-00622-1 | 4 | 373 |
| 03.20 -01.43 | TYC4708-00423-1 | 4 | 373 |
| 03.27 +30.87 | TYC2340-01060-1 | 4 | 373 |
| 02.65 +42.17 | TYC2840-01219-1 | 4 | 373 |
| 04.31 +57.80 | TYC3727-00639-1 | 4 | 373 |
| 05.37 +79.36 | TYC4532-00201-1 | 4 | 373 |
| 09.60 +34.82 | TYC2504-01508-1 | 4 | 373 |
| 18.81 +48.28 | TYC3544-00772-1 | 4 | 373 |
| 05.27 +07.21 | TYC0112-00755-1 | 4 | 372 |
| 05.64 +06.17 | TYC0127-01081-1 | 4 | 372 |
| 23.56 +39.30 | TYC3231-00265-1 | 4 | 372 |
| 06.47 +58.01 | TYC3773-00011-1 | 4 | 372 |
| 11.97 +76.95 | TYC4550-01021-1 | 4 | 372 |
| 19.44 +47.97 | TYC3547-00123-1 | 4 | 372 |
| 05.75 +00.97 | TYC0116-00779-1 | 4 | 371 |
| 04.31 +57.85 | TYC3727-00298-1 | 4 | 371 |
| 02.81 -23.04 | TYC6437-00155-1 | 4 | 371 |
| 06.39 -28.87 | TYC6518-01287-1 | 4 | 371 |
| 08.60 -31.09 | TYC7136-02784-1 | 4 | 370 |
| 06.45 +00.02 | TYC0133-00729-1 | 4 | 370 |
| 23.15 +57.45 | TYC4006-01837-1 | 4 | 370 |
| 14.25 -19.96 | TYC6146-00725-1 | 4 | 370 |
| 03.27 +31.33 | TYC2340-00847-1 | 4 | 370 |
| 18.98 +50.57 | TYC3549-00322-1 | 4 | 370 |
| 05.77 +01.56 | TYC0116-00406-1 | 4 | 370 |
| 20.96 +10.83 | TYC1107-00904-1 | 4 | 369 |
| 20.97 +10.69 | TYC1107-01186-1 | 4 | 369 |
| 17.04 +47.00 | TYC3501-01239-1 | 4 | 369 |
| 17.04 +47.18 | TYC3501-01309-1 | 4 | 369 |
| 00.65 +21.35 | TYC1193-01564-1 | 4 | 369 |
| 08.02 -26.33 | TYC6562-00753-1 | 4 | 369 |
| 08.25 +12.06 | TYC0802-01323-1 | 4 | 369 |
| 16.99 +25.63 | TYC2063-00292-1 | 4 | 369 |
| 04.71 -30.10 | TYC7039-01273-1 | 4 | 368 |
| 22.17 +16.08 | TYC1681-00776-1 | 4 | 367 |
| 22.06 +26.36 | TYC2212-01341-1 | 4 | 367 |
| 14.94 +53.09 | TYC3861-00042-1 | 4 | 367 |
| 07.91 -26.04 | TYC6557-00055-1 | 4 | 366 |
| 15.37 +59.07 | TYC3874-00928-1 | 4 | 366 |
| 04.70 +18.77 | TYC1275-00934-1 | 4 | 365 |
| 05.40 +22.14 | TYC1308-02242-1 | 4 | 365 |
| 05.38 +87.08 | TYC4625-00115-1 | 4 | 365 |
| 16.54 +02.15 | TYC0386-00167-1 | 4 | 365 |
| 07.09 +76.21 | TYC4526-02008-1 | 4 | 365 |
| 12.35 +76.96 | TYC4550-00003-1 | 4 | 365 |
| 23.27 +31.36 | TYC2752-00014-1 | 4 | 365 |
| 04.58 +00.25 | TYC0082-00409-1 | 4 | 365 |
| 14.86 +06.03 | TYC0333-00084-1 | 4 | 365 |
| 07.25 +28.63 | TYC1908-01017-1 | 4 | 365 |
| 05.12 +35.76 | TYC2401-01036-1 | 4 | 365 |
| 19.09 +39.20 | TYC3120-00233-1 | 4 | 365 |
| 03.93 +46.36 | TYC3326-02816-1 | 4 | 365 |
| 07.48 +52.34 | TYC3405-00344-1 | 4 | 365 |
| 09.65 +50.02 | TYC3432-00245-1 | 4 | 365 |
| 20.34 +59.38 | TYC3949-00093-1 | 4 | 365 |
| 05.24 +86.65 | TYC4625-00053-1 | 4 | 365 |
| 06.70 -01.29 | TYC4799-00912-1 | 4 | 365 |
| 06.05 -16.06 | TYC5932-01364-1 | 4 | 365 |
| 10.87 -18.75 | TYC6079-01778-1 | 4 | 365 |
| 15.86 -18.38 | TYC6190-00285-1 | 4 | 365 |
| 19.20 +48.22 | TYC3546-00462-1 | 4 | 364 |
| 20.36 +59.58 | TYC3949-00039-1 | 4 | 363 |
| 10.14 -19.92 | TYC6067-00355-1 | 4 | 363 |
| 02.12 +23.66 | TYC1758-01687-1 | 4 | 363 |
| 08.33 +27.29 | TYC1936-01059-1 | 4 | 363 |
| 08.15 +46.97 | TYC3408-00229-1 | 4 | 363 |
| 09.11 +76.33 | TYC4541-01782-1 | 4 | 363 |
| 08.65 +60.09 | TYC4127-01836-1 | 4 | 361 |
| 10.05 +81.52 | TYC4548-01124-1 | 4 | 360 |
| 00.62 +34.89 | TYC2283-00104-1 | 4 | 360 |
| 08.87 +28.55 | TYC1949-00292-1 | 4 | 360 |
| 08.63 +59.66 | TYC3804-00159-1 | 4 | 360 |
| 09.61 -21.84 | TYC6055-00937-1 | 4 | 360 |
| 10.07 +79.31 | TYC4545-01714-1 | 4 | 359 |
| 09.57 +34.83 | TYC2497-00404-1 | 4 | 359 |
| 19.08 +49.67 | TYC3549-00462-1 | 4 | 359 |
| 19.16 +44.01 | TYC3132-00928-1 | 4 | 358 |
| 10.61 +35.54 | TYC2518-01136-1 | 4 | 358 |
| 08.30 +68.83 | TYC4374-01198-1 | 4 | 357 |
| 03.69 +76.01 | TYC4509-03367-1 | 4 | 357 |
| 14.88 +16.30 | TYC1478-00845-1 | 4 | 357 |
| 08.56 +64.39 | TYC4130-00002-1 | 4 | 357 |
| 10.95 +01.52 | TYC0255-00987-1 | 4 | 357 |
| 03.22 +24.98 | TYC1788-00289-1 | 4 | 357 |
| 05.12 +69.53 | TYC4346-01189-1 | 4 | 357 |
| 00.07 +65.62 | TYC4022-00681-1 | 4 | 356 |
| 18.74 +47.73 | TYC3531-01541-1 | 4 | 356 |
| 02.77 -23.15 | TYC6434-00601-1 | 4 | 356 |
| 12.02 +76.61 | TYC4550-01015-1 | 4 | 356 |
| 06.59 -27.73 | TYC6516-01464-1 | 4 | 354 |
| 03.65 +32.04 | TYC2359-00884-1 | 4 | 354 |
| 03.76 +40.62 | TYC2867-01625-1 | 4 | 354 |
| 04.43 +19.18 | TYC1273-00106-1 | 4 | 354 |
| 06.80 -00.78 | TYC4800-00294-1 | 4 | 354 |
| 09.57 -12.14 | TYC5468-00464-1 | 4 | 354 |
| 22.85 +35.38 | TYC2757-00320-1 | 4 | 353 |
| 19.46 +44.76 | TYC3146-00064-1 | 4 | 353 |
| 06.27 +55.95 | TYC3768-00204-1 | 4 | 353 |
| 05.69 +06.10 | TYC0127-00124-1 | 4 | 353 |
| 06.60 -19.38 | TYC5956-02139-1 | 4 | 352 |
| 04.26 +27.06 | TYC1823-00624-1 | 4 | 352 |
| 06.34 +02.11 | TYC0136-01404-1 | 4 | 352 |
| 04.10 +11.07 | TYC0670-00684-1 | 4 | 352 |
| 04.38 +09.91 | TYC0672-00640-1 | 4 | 352 |
| 06.98 +22.85 | TYC1894-00522-1 | 4 | 352 |
| 04.79 +31.48 | TYC2374-00285-1 | 4 | 352 |
| 08.43 +31.07 | TYC2483-01176-2 | 4 | 352 |
| 00.64 +42.39 | TYC2792-01220-1 | 4 | 352 |
| 02.77 +49.48 | TYC3304-01552-1 | 4 | 352 |
| 05.41 +63.44 | TYC4084-00012-1 | 4 | 352 |
| 07.29 +70.44 | TYC4364-00657-1 | 4 | 352 |
| 07.81 +70.24 | TYC4365-01595-1 | 4 | 352 |
| 08.37 -26.25 | TYC6560-01909-1 | 4 | 352 |
| 09.86 -27.31 | TYC6611-01151-1 | 4 | 352 |
| 23.28 +31.35 | TYC2752-00013-1 | 4 | 352 |
| 10.03 +62.72 | TYC4140-01092-1 | 4 | 351 |
| 06.90 +28.29 | TYC1906-00801-1 | 4 | 351 |
| 19.39 +55.47 | TYC3925-00038-1 | 4 | 351 |
| 19.02 +48.75 | TYC3545-00035-1 | 4 | 351 |
| 00.59 +34.48 | TYC2270-01049-1 | 4 | 351 |
| 23.96 -22.17 | TYC6411-00179-1 | 4 | 351 |
| 03.17 +20.98 | TYC1231-00105-1 | 4 | 351 |
| 11.73 +30.84 | TYC2524-01673-1 | 4 | 351 |
| 05.13 -26.79 | TYC6481-00100-1 | 4 | 349 |
| 13.96 +43.10 | TYC3033-00924-1 | 4 | 349 |
| 06.45 +29.76 | TYC1891-00536-1 | 4 | 348 |
| 09.70 +66.43 | TYC4142-01111-1 | 4 | 348 |
| 19.21 +40.36 | TYC3125-00357-1 | 4 | 348 |
| 19.04 +37.78 | TYC3120-00330-1 | 4 | 347 |
| 07.40 +20.68 | TYC1359-02312-1 | 4 | 347 |
| 23.30 +58.12 | TYC4006-00001-1 | 4 | 347 |
| 07.85 +32.88 | TYC2471-01306-1 | 4 | 347 |
| 13.01 -27.53 | TYC6706-00815-1 | 4 | 347 |
| 09.04 +16.38 | TYC1394-00265-1 | 4 | 347 |
| 18.90 +45.90 | TYC3540-00927-1 | 4 | 347 |
| 03.49 -24.10 | TYC6446-00342-1 | 4 | 347 |
| 10.36 +50.19 | TYC3441-00370-1 | 4 | 347 |
| 19.80 +48.42 | TYC3561-00896-1 | 4 | 346 |
| 16.35 -20.73 | TYC6214-00200-1 | 4 | 346 |
| 19.45 +37.45 | TYC2666-00003-1 | 4 | 345 |
| 12.46 +74.81 | TYC4400-00189-1 | 4 | 344 |
| 07.85 +32.78 | TYC2471-01316-1 | 4 | 344 |
| 16.96 +25.67 | TYC2063-00731-1 | 4 | 343 |
| 19.42 +50.50 | TYC3551-00196-1 | 4 | 343 |
| 08.33 +61.33 | TYC4126-02143-1 | 4 | 342 |
| 23.41 -20.39 | TYC6402-00716-1 | 4 | 342 |
| 06.64 -19.46 | TYC5956-00637-1 | 4 | 342 |
| 03.25 +52.59 | TYC3702-00403-1 | 4 | 342 |
| 08.87 +27.93 | TYC1949-01288-1 | 4 | 342 |
| 09.14 +08.54 | TYC0812-00004-1 | 4 | 341 |
| 05.95 +51.55 | TYC3373-01064-1 | 4 | 341 |
| 09.78 +01.73 | TYC0236-00762-1 | 4 | 341 |
| 09.79 +01.41 | TYC0236-01039-1 | 4 | 340 |
| 08.72 +65.22 | TYC4133-01904-1 | 4 | 340 |
| 11.58 +20.29 | TYC1440-01568-1 | 4 | 339 |
| 10.42 +30.19 | TYC2511-00424-1 | 4 | 339 |
| 07.31 +58.22 | TYC3792-01719-1 | 4 | 338 |
| 01.71 +20.05 | TYC1211-00006-1 | 4 | 338 |
| 18.97 +45.64 | TYC3541-00732-1 | 4 | 338 |
| 04.85 +06.30 | TYC0096-00466-1 | 4 | 338 |
| 03.18 +21.53 | TYC1231-00151-1 | 4 | 337 |
| 07.46 +24.46 | TYC1914-00551-1 | 4 | 336 |
| 19.84 +44.08 | TYC3149-01008-1 | 4 | 336 |
| 19.25 +50.05 | TYC3550-01729-1 | 4 | 336 |
| 00.87 +34.51 | TYC2284-00058-1 | 4 | 336 |
| 04.37 +58.15 | TYC3731-01955-1 | 4 | 335 |
| 03.19 +20.90 | TYC1231-00079-1 | 4 | 335 |
| 22.07 +26.55 | TYC2212-00045-1 | 4 | 335 |
| 00.88 +34.74 | TYC2284-00419-1 | 4 | 335 |
| 19.89 +48.43 | TYC3562-01036-1 | 4 | 334 |
| 18.76 +42.03 | TYC3126-01040-2 | 4 | 334 |
| 05.76 -00.05 | TYC4768-00039-1 | 4 | 334 |
| 04.90 -23.39 | TYC6466-00954-1 | 4 | 334 |
| 07.27 +13.79 | TYC0774-01663-1 | 4 | 334 |
| 21.25 -20.62 | TYC6355-01006-1 | 4 | 333 |
| 21.25 -20.94 | TYC6359-01206-1 | 4 | 333 |
| 11.03 +23.95 | TYC1978-00936-1 | 4 | 333 |
| 07.55 -14.02 | TYC5409-02070-1 | 4 | 333 |
| 05.76 +00.46 | TYC0116-01018-1 | 4 | 333 |
| 06.29 +41.19 | TYC2930-00053-1 | 4 | 333 |
| 11.84 +57.44 | TYC3835-00382-1 | 4 | 333 |
| 05.27 +07.14 | TYC0112-00410-1 | 4 | 333 |
| 03.20 -01.91 | TYC4708-00091-1 | 4 | 333 |
| 08.02 -26.20 | TYC6558-03032-1 | 4 | 331 |
| 10.33 +19.89 | TYC1423-00165-1 | 4 | 331 |
| 18.20 +54.23 | TYC3903-01491-1 | 4 | 331 |
| 06.54 -00.43 | TYC4798-00917-1 | 4 | 331 |
| 06.19 -29.64 | TYC6517-01880-1 | 4 | 330 |
| 04.79 -24.50 | TYC6465-01876-1 | 4 | 330 |
| 08.53 -30.01 | TYC7135-00327-1 | 4 | 328 |
| 08.56 +16.75 | TYC1392-00344-1 | 4 | 328 |
| 08.58 +17.00 | TYC1392-00536-1 | 4 | 328 |
| 11.79 +03.69 | TYC0275-00283-1 | 4 | 328 |
| 20.33 +59.56 | TYC3949-01379-1 | 4 | 327 |
| 07.30 +36.25 | TYC2463-00134-1 | 4 | 327 |
| 07.84 +51.61 | TYC3413-01919-1 | 4 | 327 |
| 12.36 +74.43 | TYC4400-00346-1 | 4 | 327 |
| 07.79 +50.10 | TYC3413-00058-1 | 4 | 326 |
| 14.94 +52.31 | TYC3480-00055-1 | 4 | 326 |
| 23.11 +20.85 | TYC1717-00540-1 | 4 | 326 |
| 23.44 +08.89 | TYC1162-00617-1 | 4 | 326 |
| 07.30 +58.09 | TYC3788-00728-1 | 4 | 326 |
| 08.93 +60.27 | TYC4128-01732-1 | 4 | 324 |
| 10.27 +05.92 | TYC0251-00396-1 | 4 | 323 |
| 10.35 +00.58 | TYC0246-01018-1 | 4 | 322 |
| 11.71 +02.83 | TYC0275-00072-1 | 4 | 322 |
| 17.87 +56.04 | TYC3906-00036-1 | 4 | 322 |
| 13.47 +13.95 | TYC0898-00056-1 | 4 | 321 |
| 10.41 +75.52 | TYC4542-02395-1 | 4 | 321 |
| 15.61 +53.49 | TYC3869-00077-1 | 4 | 321 |
| 08.51 -29.94 | TYC6582-02080-1 | 4 | 320 |
| 05.01 +56.06 | TYC3738-00128-1 | 4 | 320 |
| 10.36 +00.07 | TYC0246-01091-1 | 4 | 319 |
| 08.69 -23.54 | TYC6571-00648-1 | 4 | 319 |
| 04.46 -24.86 | TYC6457-03124-1 | 4 | 319 |
| 12.54 +44.68 | TYC3020-02138-1 | 4 | 319 |
| 09.24 +23.77 | TYC1951-00244-1 | 4 | 319 |
| 09.61 +18.63 | TYC1413-00491-1 | 4 | 318 |
| 10.08 +67.05 | TYC4143-00556-1 | 4 | 318 |
| 22.44 -16.98 | TYC6385-00817-1 | 4 | 318 |
| 05.42 +63.19 | TYC4084-00004-1 | 4 | 318 |
| 16.88 +11.97 | TYC0983-00259-1 | 4 | 317 |
| 01.26 +76.43 | TYC4493-01325-1 | 4 | 317 |
| 02.97 -21.04 | TYC5870-00507-1 | 4 | 315 |
| 09.32 +33.81 | TYC2496-00721-1 | 4 | 315 |
| 18.70 +06.02 | TYC0459-00991-1 | 4 | 315 |
| 05.64 +20.91 | TYC1310-00113-1 | 4 | 315 |
| 11.93 +76.94 | TYC4550-00659-1 | 4 | 315 |
| 03.27 +31.37 | TYC2340-00770-1 | 4 | 315 |
| 06.78 -00.63 | TYC4800-00228-1 | 4 | 315 |
| 08.20 +61.39 | TYC4126-02017-1 | 4 | 315 |
| 04.38 +39.40 | TYC2883-01541-1 | 3 | 314 |
| 02.74 +49.32 | TYC3304-01064-1 | 3 | 314 |
| 08.38 +01.95 | TYC0201-00063-1 | 3 | 313 |
| 05.61 +21.00 | TYC1310-00150-1 | 3 | 313 |
| 10.05 +62.91 | TYC4140-00710-1 | 3 | 313 |
| 10.98 +01.59 | TYC0255-00760-1 | 3 | 313 |
| 18.98 +48.66 | TYC3545-00027-1 | 3 | 313 |
| 15.88 -18.63 | TYC6191-00122-1 | 3 | 313 |
| 17.42 +27.32 | TYC2082-01555-1 | 3 | 312 |
| 19.62 +50.60 | TYC3564-00110-1 | 3 | 312 |
| 18.94 +51.28 | TYC3552-00342-1 | 3 | 312 |
| 06.54 +05.45 | TYC0154-01023-1 | 3 | 311 |
| 08.70 +04.57 | TYC0219-00725-1 | 3 | 311 |
| 08.71 +04.53 | TYC0219-01501-1 | 3 | 311 |
| 14.78 +00.97 | TYC0326-00417-1 | 3 | 311 |
| 15.15 +02.47 | TYC0339-00470-1 | 3 | 311 |
| 07.38 +20.24 | TYC1355-00449-1 | 3 | 311 |
| 12.36 +17.72 | TYC1445-00841-1 | 3 | 311 |
| 13.20 +17.30 | TYC1451-00129-1 | 3 | 311 |
| 14.88 +18.41 | TYC1481-00201-1 | 3 | 311 |
| 07.45 +24.28 | TYC1910-00253-1 | 3 | 311 |
| 15.34 +36.04 | TYC2569-01073-1 | 3 | 311 |
| 17.07 +33.04 | TYC2594-00747-1 | 3 | 311 |
| 17.15 +33.53 | TYC2595-00605-1 | 3 | 311 |
| 19.17 +34.54 | TYC2648-00175-1 | 3 | 311 |
| 18.71 +36.39 | TYC2649-00983-1 | 3 | 311 |
| 16.49 +38.51 | TYC3063-01974-1 | 3 | 311 |
| 16.35 +41.28 | TYC3065-01165-1 | 3 | 311 |
| 17.35 +38.20 | TYC3086-01752-1 | 3 | 311 |
| 19.94 +44.02 | TYC3149-00524-1 | 3 | 311 |
| 02.19 +51.79 | TYC3293-00874-1 | 3 | 311 |
| 05.18 +69.45 | TYC4346-01071-1 | 3 | 311 |
| 06.54 -00.65 | TYC4798-01473-1 | 3 | 311 |
| 14.03 -27.66 | TYC6737-00237-1 | 3 | 311 |
| 15.21 -25.25 | TYC6765-01581-1 | 3 | 311 |
| 12.71 -30.45 | TYC7247-00841-1 | 3 | 311 |
| 09.70 +71.60 | TYC4386-01638-1 | 3 | 310 |
| 06.50 +00.35 | TYC0146-01038-1 | 3 | 310 |
| 13.22 -31.97 | TYC7267-00014-1 | 3 | 310 |
| 09.82 +07.87 | TYC0828-01415-1 | 3 | 310 |
| 18.57 +35.85 | TYC2636-00187-1 | 3 | 310 |
| 19.09 +49.31 | TYC3549-02560-1 | 3 | 310 |
| 16.26 +10.18 | TYC0950-00528-1 | 3 | 309 |
| 08.44 +28.97 | TYC1947-00312-1 | 3 | 307 |
| 08.46 +29.09 | TYC1947-00823-1 | 3 | 307 |
| 19.60 +48.75 | TYC3560-00094-1 | 3 | 307 |
| 10.00 +61.88 | TYC4137-00548-1 | 3 | 307 |
| 23.97 -22.16 | TYC6411-00301-1 | 3 | 307 |
| 09.27 +23.40 | TYC1951-01566-1 | 3 | 307 |
| 08.79 +55.82 | TYC3801-00979-1 | 3 | 306 |
| 00.23 -12.19 | TYC5264-00115-1 | 3 | 306 |
| 08.19 +04.74 | TYC0203-00460-1 | 3 | 306 |
| 03.63 +31.92 | TYC2359-00696-1 | 3 | 306 |
| 04.31 +58.31 | TYC3731-01347-1 | 3 | 306 |
| 16.31 +40.47 | TYC3065-00013-1 | 3 | 305 |
| 06.80 -00.68 | TYC4800-01382-1 | 3 | 304 |
| 09.59 +31.56 | TYC2494-00349-1 | 3 | 302 |
| 10.34 -00.95 | TYC4905-00061-1 | 3 | 301 |
| 11.03 -31.25 | TYC7200-00542-1 | 3 | 301 |
| 09.24 +00.31 | TYC0227-01917-1 | 3 | 301 |
| 09.24 +00.36 | TYC0227-02009-1 | 3 | 301 |
| 09.77 +03.62 | TYC0239-02322-1 | 3 | 301 |
| 19.09 +49.92 | TYC3549-02884-1 | 3 | 301 |
| 20.04 +46.53 | TYC3558-01811-1 | 3 | 301 |
| 16.97 +47.23 | TYC3500-00776-1 | 3 | 301 |
| 20.96 +10.79 | TYC1107-01100-1 | 3 | 300 |
| 19.81 +48.21 | TYC3561-01202-1 | 3 | 298 |
| 02.69 +38.81 | TYC2845-00320-1 | 3 | 298 |
| 18.71 +47.14 | TYC3531-00074-1 | 3 | 298 |
| 07.76 +27.94 | TYC1920-00657-1 | 3 | 296 |
| 07.25 +14.30 | TYC0774-00707-1 | 3 | 296 |
| 07.82 +50.16 | TYC3413-00104-1 | 3 | 295 |
| 22.19 +15.85 | TYC1681-00726-1 | 3 | 295 |
| 22.19 +15.79 | TYC1681-00780-1 | 3 | 295 |
| 23.15 +18.34 | TYC1715-01435-1 | 3 | 295 |
| 06.47 +10.12 | TYC0736-00711-1 | 3 | 295 |
| 02.78 +70.98 | TYC4316-00146-1 | 3 | 295 |
| 17.00 +26.09 | TYC2063-00061-1 | 3 | 295 |
| 03.76 +40.55 | TYC2867-01368-1 | 3 | 295 |
| 01.32 +77.00 | TYC4497-01880-1 | 3 | 295 |
| 12.20 +09.56 | TYC0866-00746-1 | 3 | 295 |
| 05.10 -26.87 | TYC6481-00656-1 | 3 | 294 |
| 18.34 -11.77 | TYC5685-03566-1 | 3 | 294 |
| 09.98 -24.27 | TYC6604-01285-1 | 3 | 294 |
| 08.97 +66.40 | TYC4134-01087-1 | 3 | 293 |
| 07.61 -13.75 | TYC5409-03872-1 | 3 | 293 |
| 23.27 +57.79 | TYC4006-00048-1 | 3 | 292 |
| 03.14 +11.26 | TYC0651-00078-1 | 3 | 292 |
| 05.11 -26.97 | TYC6481-00624-1 | 3 | 292 |
| 18.73 +48.00 | TYC3531-01783-1 | 3 | 292 |
| 20.04 +44.31 | TYC3162-00036-1 | 3 | 291 |
| 18.75 +47.41 | TYC3531-00050-1 | 3 | 291 |
| 00.00 -22.59 | TYC6412-00335-1 | 3 | 290 |
| 19.73 +41.18 | TYC3140-00747-1 | 3 | 288 |
| 11.81 +77.08 | TYC4550-00096-1 | 3 | 288 |
| 07.88 +30.38 | TYC2467-01382-1 | 3 | 288 |
| 08.59 -29.88 | TYC6582-01957-1 | 3 | 288 |
| 06.41 -31.37 | TYC7073-02146-1 | 3 | 288 |
| 09.13 +46.51 | TYC3424-00653-1 | 3 | 287 |
| 09.11 +46.66 | TYC3424-01380-1 | 3 | 287 |
| 00.70 -26.46 | TYC6423-01884-1 | 3 | 287 |
| 06.20 -29.44 | TYC6517-01852-1 | 3 | 287 |
| 14.89 +52.48 | TYC3480-00275-1 | 3 | 286 |
| 10.22 -15.64 | TYC6059-00469-1 | 3 | 286 |
| 07.27 +14.30 | TYC0774-00441-1 | 3 | 286 |
| 00.33 +31.86 | TYC2261-00881-1 | 3 | 286 |
| 09.41 +20.57 | TYC1408-01691-1 | 3 | 285 |
| 03.57 -24.21 | TYC6447-00516-1 | 3 | 284 |
| 21.09 +03.73 | TYC0530-01743-1 | 3 | 283 |
| 04.72 +18.98 | TYC1275-01098-1 | 3 | 283 |
| 19.46 +00.23 | TYC0465-00951-1 | 3 | 283 |
| 19.19 +50.59 | TYC3550-00215-1 | 3 | 282 |
| 23.05 -00.31 | TYC5242-00481-1 | 3 | 282 |
| 10.36 +50.29 | TYC3441-01256-1 | 3 | 282 |
| 01.65 -15.33 | TYC5855-00735-1 | 3 | 281 |
| 19.81 +50.61 | TYC3565-00263-1 | 3 | 280 |
| 18.72 +47.70 | TYC3531-01823-1 | 3 | 280 |
| 00.15 +65.57 | TYC4022-00793-1 | 3 | 280 |
| 01.10 +76.85 | TYC4493-01468-1 | 3 | 280 |
| 21.24 -20.60 | TYC6355-00965-1 | 3 | 280 |
| 11.23 -19.57 | TYC6087-01198-1 | 3 | 279 |
| 05.75 +01.21 | TYC0115-00744-1 | 3 | 279 |
| 14.83 +05.98 | TYC0332-01460-1 | 3 | 279 |
| 19.81 +43.11 | TYC3144-00070-1 | 3 | 279 |
| 08.16 +61.57 | TYC4126-01686-1 | 3 | 279 |
| 19.76 +49.88 | TYC3565-00245-1 | 3 | 279 |
| 21.52 -21.12 | TYC6373-01042-1 | 3 | 279 |
| 19.80 +43.08 | TYC3144-00073-1 | 3 | 278 |
| 20.02 +44.31 | TYC3162-00230-1 | 3 | 278 |
| 18.79 +48.61 | TYC3544-00050-1 | 3 | 278 |
| 23.49 +38.82 | TYC3230-00248-1 | 3 | 278 |
| 03.31 +31.04 | TYC2341-01181-1 | 3 | 278 |
| 23.04 -00.37 | TYC5242-00451-1 | 3 | 275 |
| 06.72 -01.16 | TYC4799-02024-1 | 3 | 275 |
| 01.06 -22.44 | TYC5850-02578-1 | 3 | 275 |
| 01.07 -22.19 | TYC5853-00524-1 | 3 | 275 |
| 00.58 +34.65 | TYC2270-00501-1 | 3 | 275 |
| 06.74 -00.18 | TYC4799-00410-1 | 3 | 275 |
| 22.21 +16.12 | TYC1681-00295-1 | 3 | 275 |
| 07.37 +20.54 | TYC1355-00325-1 | 3 | 274 |
| 06.47 +00.05 | TYC0133-00371-1 | 3 | 274 |
| 03.23 +31.00 | TYC2340-01417-1 | 3 | 274 |
| 04.85 +06.59 | TYC0096-00445-1 | 3 | 274 |
| 18.29 +36.45 | TYC2634-01127-1 | 3 | 274 |
| 21.64 +30.72 | TYC2717-01557-1 | 3 | 274 |
| 08.30 -12.74 | TYC5434-02543-1 | 3 | 273 |
| 04.55 +04.51 | TYC0090-00798-1 | 3 | 273 |
| 09.50 +09.75 | TYC0820-00525-1 | 3 | 272 |
| 02.81 +08.56 | TYC0640-00795-1 | 3 | 271 |
| 19.31 +41.24 | TYC3125-00107-1 | 3 | 271 |
| 21.46 -21.71 | TYC6372-00009-1 | 3 | 271 |
| 11.49 +07.05 | TYC0270-00949-1 | 3 | 271 |
| 11.34 -23.12 | TYC6650-00938-1 | 3 | 270 |
| 03.51 -18.71 | TYC5876-00252-1 | 3 | 270 |
| 11.25 +18.73 | TYC1437-00002-1 | 3 | 270 |
| 06.20 -29.84 | TYC6517-02091-1 | 3 | 270 |
| 06.49 +00.20 | TYC0133-00009-1 | 3 | 270 |
| 07.23 +14.25 | TYC0774-00243-1 | 3 | 270 |
| 08.88 +28.12 | TYC1949-00239-1 | 3 | 270 |
| 09.24 +23.71 | TYC1951-00663-1 | 3 | 270 |
| 09.31 +33.98 | TYC2496-01321-1 | 3 | 270 |
| 10.97 +40.67 | TYC3009-00062-1 | 3 | 270 |
| 23.64 +42.51 | TYC3239-00096-1 | 3 | 270 |
| 08.03 -01.12 | TYC4846-00354-1 | 3 | 270 |
| 02.59 -12.59 | TYC5291-00330-1 | 3 | 270 |
| 15.56 +53.60 | TYC3869-00232-1 | 3 | 269 |
| 07.46 +85.48 | TYC4622-00675-1 | 3 | 269 |
| 09.87 -24.18 | TYC6603-00083-1 | 3 | 269 |
| 08.28 +62.21 | TYC4126-00202-1 | 3 | 269 |
| 22.98 +38.88 | TYC3215-00021-1 | 3 | 269 |
| 08.03 -01.23 | TYC4846-01810-1 | 3 | 269 |
| 02.59 -12.35 | TYC5288-00442-1 | 3 | 269 |
| 12.47 +73.96 | TYC4400-00900-1 | 3 | 269 |
| 19.12 +48.80 | TYC3549-02851-1 | 3 | 268 |
| 04.87 -15.27 | TYC5899-00078-1 | 3 | 268 |
| 00.60 +34.69 | TYC2283-00642-1 | 3 | 268 |
| 19.09 +50.01 | TYC3549-00465-1 | 3 | 268 |
| 05.27 +06.47 | TYC0112-01496-1 | 3 | 267 |
| 06.52 -00.76 | TYC4798-00210-1 | 3 | 267 |
| 08.67 -30.17 | TYC7136-00178-1 | 3 | 265 |
| 19.31 +40.94 | TYC3125-00011-1 | 3 | 265 |
| 05.52 +05.66 | TYC0126-01635-1 | 3 | 265 |
| 02.75 +11.39 | TYC0643-00682-1 | 3 | 265 |
| 04.39 +10.21 | TYC0672-00288-1 | 3 | 265 |
| 05.95 +11.09 | TYC0720-00528-1 | 3 | 265 |
| 05.95 +09.82 | TYC0720-00695-1 | 3 | 265 |
| 02.32 +34.91 | TYC2318-01416-1 | 3 | 265 |
| 02.33 +36.09 | TYC2322-02040-1 | 3 | 265 |
| 08.43 +31.69 | TYC2483-01033-1 | 3 | 265 |
| 19.44 +36.89 | TYC2666-00208-1 | 3 | 265 |
| 06.79 +43.28 | TYC2953-00298-1 | 3 | 265 |
| 19.83 +42.35 | TYC3144-00118-1 | 3 | 265 |
| 05.38 +46.20 | TYC3358-00560-1 | 3 | 265 |
| 07.50 +85.13 | TYC4622-00634-1 | 3 | 265 |
| 03.51 -15.54 | TYC5873-00185-1 | 3 | 265 |
| 03.75 -21.30 | TYC5887-01010-1 | 3 | 265 |
| 02.84 -25.52 | TYC6440-00413-1 | 3 | 265 |
| 07.20 +10.99 | TYC0766-01247-1 | 3 | 264 |
| 03.75 -22.31 | TYC5887-00664-1 | 3 | 264 |
| 20.19 +18.17 | TYC1622-01948-1 | 3 | 264 |
| 01.74 +19.86 | TYC1208-00371-1 | 3 | 264 |
| 03.34 -28.73 | TYC6445-00925-1 | 3 | 264 |
| 07.20 +09.91 | TYC0766-01006-1 | 3 | 263 |
| 02.09 +18.36 | TYC1210-00054-1 | 3 | 263 |
| 04.90 +16.70 | TYC1280-00246-1 | 3 | 263 |
| 02.87 +27.71 | TYC1789-00494-1 | 3 | 263 |
| 00.68 +34.59 | TYC2283-00220-1 | 3 | 263 |
| 23.11 +21.29 | TYC1717-00504-1 | 3 | 263 |
| 04.67 +45.33 | TYC3342-00420-1 | 3 | 262 |
| 15.57 +53.91 | TYC3869-00528-1 | 3 | 261 |
| 04.66 +44.96 | TYC2892-00008-1 | 3 | 261 |
| 00.86 +34.64 | TYC2284-00096-1 | 3 | 261 |
| 10.31 +49.23 | TYC3438-00093-1 | 3 | 261 |
| 01.36 +28.34 | TYC1754-00599-1 | 3 | 261 |
| 19.14 +43.73 | TYC3132-00269-1 | 3 | 261 |
| 04.20 +50.08 | TYC3336-02067-1 | 3 | 261 |
| 22.02 +18.64 | TYC1684-00691-1 | 3 | 261 |
| 18.33 -11.91 | TYC5685-01746-1 | 3 | 261 |
| 22.41 -17.22 | TYC6385-00809-1 | 3 | 261 |
| 20.95 +10.75 | TYC1107-01590-1 | 3 | 259 |
| 19.14 +51.25 | TYC3554-00509-1 | 3 | 257 |
| 02.54 -12.14 | TYC5284-01047-1 | 3 | 257 |
| 08.25 +61.12 | TYC4126-01676-1 | 3 | 257 |
| 11.75 +77.01 | TYC4550-00146-1 | 3 | 257 |
| 01.21 +76.89 | TYC4497-02439-1 | 3 | 257 |
| 08.66 +12.87 | TYC0805-00112-1 | 3 | 256 |
| 19.06 +39.21 | TYC3120-01059-1 | 3 | 256 |
| 19.87 +46.71 | TYC3557-00046-1 | 3 | 256 |
| 19.01 +39.43 | TYC3124-00968-1 | 3 | 256 |
| 02.91 +71.58 | TYC4321-02040-1 | 3 | 256 |
| 05.10 -13.81 | TYC5329-01218-1 | 3 | 256 |
| 05.95 -14.44 | TYC5360-00379-1 | 3 | 256 |
| 00.32 -15.16 | TYC5839-00710-1 | 3 | 256 |
| 00.65 +21.77 | TYC1193-00180-1 | 3 | 256 |
| 10.99 -31.05 | TYC7200-01176-1 | 3 | 256 |
| 05.10 -14.06 | TYC5329-01864-1 | 3 | 256 |
| 20.66 +42.34 | TYC3161-00164-1 | 3 | 256 |
| 18.88 +45.40 | TYC3540-02352-1 | 3 | 255 |
| 06.69 -01.31 | TYC4799-00958-1 | 3 | 255 |
| 05.53 +05.48 | TYC0122-00048-1 | 3 | 255 |
| 19.79 +50.61 | TYC3565-00155-1 | 3 | 255 |
| 20.35 +59.66 | TYC3949-01631-1 | 3 | 255 |
| 04.86 -16.32 | TYC5899-00689-1 | 3 | 255 |
| 07.37 +20.44 | TYC1355-01288-1 | 3 | 255 |
| 07.30 +58.36 | TYC3792-01038-1 | 3 | 255 |
| 04.86 +06.61 | TYC0096-00117-1 | 3 | 254 |
| 14.55 +21.27 | TYC1482-00881-1 | 3 | 254 |
| 06.04 +45.04 | TYC3374-00181-1 | 3 | 254 |
| 15.34 +59.02 | TYC3874-00903-1 | 3 | 254 |
| 15.35 +58.52 | TYC3874-01020-1 | 3 | 254 |
| 05.02 +70.06 | TYC4346-01194-1 | 3 | 254 |
| 01.30 +75.80 | TYC4493-01401-1 | 3 | 254 |
| 05.28 +06.66 | TYC0112-00322-1 | 3 | 253 |
| 18.89 +39.08 | TYC3119-00302-1 | 3 | 253 |
| 01.39 +76.07 | TYC4494-01187-1 | 3 | 253 |
| 06.79 -00.82 | TYC4800-00947-1 | 3 | 252 |
| 03.49 -23.43 | TYC6446-00127-1 | 3 | 252 |
| 03.11 +30.73 | TYC2339-00790-1 | 3 | 252 |
| 06.88 +24.35 | TYC1894-02328-1 | 3 | 251 |
| 15.82 +35.83 | TYC2578-00723-1 | 3 | 250 |
| 08.55 -29.63 | TYC6582-01367-1 | 3 | 249 |
| 15.98 -28.04 | TYC6787-01938-1 | 3 | 249 |
| 19.97 +44.12 | TYC3149-00488-1 | 3 | 249 |
| 07.63 -21.92 | TYC5992-04160-1 | 3 | 248 |
| 04.24 +57.76 | TYC3727-00318-1 | 3 | 248 |
| 14.30 -20.37 | TYC6147-00030-1 | 3 | 248 |
| 17.29 +29.17 | TYC2073-00019-1 | 3 | 248 |
| 10.12 +34.28 | TYC2506-00309-1 | 3 | 248 |
| 00.27 -12.11 | TYC5264-00764-1 | 3 | 248 |
| 01.06 -22.29 | TYC5850-01505-1 | 3 | 248 |
| 19.66 +50.53 | TYC3564-01358-1 | 3 | 247 |
| 20.33 +59.80 | TYC3949-01031-1 | 3 | 247 |
| 05.72 +01.31 | TYC0115-00910-1 | 3 | 247 |
| 15.62 +03.80 | TYC0358-00948-1 | 3 | 247 |
| 07.20 +13.95 | TYC0774-01387-1 | 3 | 247 |
| 08.66 +11.99 | TYC0805-00326-1 | 3 | 247 |
| 20.91 +10.80 | TYC1094-01046-1 | 3 | 247 |
| 14.82 +16.25 | TYC1478-00790-1 | 3 | 247 |
| 19.03 +25.76 | TYC2126-00001-1 | 3 | 247 |
| 03.23 +30.73 | TYC2340-00554-1 | 3 | 247 |
| 14.30 +35.77 | TYC2552-00040-1 | 3 | 247 |
| 19.18 +33.69 | TYC2644-00133-1 | 3 | 247 |
| 00.18 +78.72 | TYC4496-00213-1 | 3 | 247 |
| 00.11 +83.31 | TYC4615-00395-1 | 3 | 247 |
| 07.54 -22.48 | TYC5991-00045-1 | 3 | 247 |
| 01.06 -22.78 | TYC6422-00036-1 | 3 | 247 |
| 12.09 +76.86 | TYC4550-01071-1 | 3 | 246 |
| 19.22 +40.38 | TYC3125-00542-1 | 3 | 246 |
| 03.53 +10.07 | TYC0653-00298-1 | 3 | 246 |
| 08.60 +12.66 | TYC0805-00494-1 | 3 | 246 |
| 15.49 +15.04 | TYC1494-00525-1 | 3 | 246 |
| 13.53 -17.13 | TYC6125-00669-1 | 3 | 246 |
| 03.55 +09.82 | TYC0650-00098-1 | 3 | 245 |
| 03.28 +30.74 | TYC2340-00030-1 | 3 | 245 |
| 14.76 +37.31 | TYC2560-00782-1 | 3 | 245 |
| 19.30 +47.99 | TYC3546-02241-1 | 3 | 244 |
| 02.99 -20.56 | TYC5870-00024-1 | 3 | 244 |
| 19.03 +36.88 | TYC2651-01918-1 | 3 | 243 |
| 08.79 +64.31 | TYC4131-01472-1 | 3 | 243 |
| 08.91 +64.02 | TYC4131-00133-1 | 3 | 243 |
| 19.45 +47.88 | TYC3547-00381-1 | 3 | 242 |
| 19.09 +49.45 | TYC3549-01062-1 | 3 | 242 |
| 03.52 -23.86 | TYC6446-00331-1 | 3 | 242 |
| 04.21 +50.75 | TYC3340-02134-1 | 3 | 242 |
| 04.67 +19.09 | TYC1275-01227-1 | 3 | 242 |
| 01.07 -22.50 | TYC5853-00935-1 | 3 | 242 |
| 01.66 -15.14 | TYC5855-00337-1 | 3 | 242 |
| 19.78 +50.72 | TYC3569-00364-1 | 3 | 241 |
| 00.61 +34.62 | TYC2283-00376-1 | 3 | 241 |
| 02.55 -12.43 | TYC5284-01154-1 | 3 | 241 |
| 02.03 +25.44 | TYC1760-01991-1 | 3 | 241 |
| 03.17 -01.16 | TYC4707-00004-1 | 3 | 241 |
| 05.07 -14.56 | TYC5329-01960-1 | 3 | 241 |
| 16.31 +41.32 | TYC3065-00547-1 | 3 | 241 |
| 22.05 +26.46 | TYC2212-01373-1 | 3 | 239 |
| 23.17 +56.85 | TYC4006-01213-1 | 3 | 239 |
| 07.91 +08.01 | TYC0779-00868-1 | 3 | 238 |
| 04.84 -16.31 | TYC5899-00085-1 | 3 | 238 |
| 11.21 -18.98 | TYC6087-01287-1 | 3 | 238 |
| 19.98 +44.26 | TYC3149-00159-1 | 3 | 238 |
| 02.19 +32.47 | TYC2313-01069-1 | 3 | 238 |
| 08.69 +64.31 | TYC4130-00707-1 | 3 | 237 |
| 05.61 +20.84 | TYC1310-02643-1 | 3 | 236 |
| 06.90 +24.12 | TYC1894-01765-1 | 3 | 236 |
| 18.87 +45.23 | TYC3540-03108-1 | 3 | 236 |
| 05.13 -14.03 | TYC5342-01150-1 | 3 | 236 |
| 19.14 +43.87 | TYC3132-01803-1 | 3 | 236 |
| 05.13 +68.95 | TYC4342-00081-1 | 3 | 236 |
| 12.24 +10.42 | TYC0869-01079-1 | 3 | 236 |
| 06.53 +29.71 | TYC1891-00176-1 | 3 | 236 |
| 07.45 +23.53 | TYC1910-01552-1 | 3 | 236 |
| 18.79 +45.62 | TYC3540-02216-1 | 3 | 236 |
| 02.67 +72.18 | TYC4320-00980-1 | 3 | 236 |
| 02.97 +74.71 | TYC4325-00207-1 | 3 | 236 |
| 10.36 -00.47 | TYC4905-00618-1 | 3 | 236 |
| 05.10 -13.93 | TYC5329-01183-1 | 3 | 236 |
| 02.98 -21.00 | TYC5870-00052-1 | 3 | 236 |
| 03.05 -20.93 | TYC5871-00056-1 | 3 | 236 |
| 02.31 +32.51 | TYC2314-00995-1 | 3 | 236 |
| 07.24 +14.19 | TYC0774-00117-1 | 3 | 236 |
| 22.04 +18.83 | TYC1688-01982-1 | 3 | 236 |
| 04.33 +57.93 | TYC3727-00192-1 | 3 | 236 |
| 19.72 +48.58 | TYC3561-00001-1 | 3 | 235 |
| 07.22 +30.18 | TYC2451-01857-1 | 3 | 235 |
| 15.82 +35.92 | TYC2578-00555-1 | 3 | 235 |
| 18.94 +38.91 | TYC3119-00376-1 | 3 | 235 |
| 18.70 +43.96 | TYC3130-01859-1 | 3 | 235 |
| 06.79 -00.59 | TYC4800-01536-1 | 3 | 235 |
| 02.48 +37.50 | TYC2335-00887-1 | 3 | 235 |
| 07.51 +16.97 | TYC1364-01604-1 | 3 | 235 |
| 20.94 +11.01 | TYC1107-01972-1 | 3 | 234 |
| 05.28 +06.36 | TYC0112-02195-1 | 3 | 234 |
| 05.33 +79.70 | TYC4519-00109-1 | 3 | 234 |
| 06.52 -00.50 | TYC4798-00156-1 | 3 | 234 |
| 14.36 -17.22 | TYC6143-01414-1 | 3 | 234 |
| 07.34 +58.33 | TYC3793-01758-1 | 3 | 233 |
| 08.55 -30.17 | TYC7135-00333-1 | 3 | 233 |
| 08.04 -16.77 | TYC5995-00062-1 | 3 | 233 |
| 08.55 -29.48 | TYC6582-00876-1 | 3 | 233 |
| 19.58 +47.58 | TYC3560-00365-1 | 3 | 233 |
| 08.65 +64.18 | TYC4130-01477-1 | 3 | 233 |
| 13.92 -32.27 | TYC7283-01575-1 | 3 | 233 |
| 19.89 +48.26 | TYC3562-00135-1 | 3 | 232 |
| 00.57 +34.86 | TYC2270-00920-1 | 3 | 232 |
| 11.66 +30.83 | TYC2523-01971-1 | 3 | 231 |
| 22.89 +35.30 | TYC2757-02126-1 | 3 | 230 |
| 07.72 +27.98 | TYC1920-00235-1 | 3 | 230 |
| 02.39 +32.34 | TYC2314-00539-1 | 3 | 230 |
| 23.59 +77.92 | TYC4606-02119-1 | 3 | 230 |
| 08.03 -17.12 | TYC5999-00054-1 | 3 | 230 |
| 10.39 -29.84 | TYC6631-01061-1 | 3 | 229 |
| 08.25 +61.29 | TYC4126-02196-1 | 3 | 229 |
| 23.03 -00.35 | TYC5242-00492-1 | 3 | 229 |
| 22.93 +20.89 | TYC1717-02061-1 | 3 | 229 |
| 21.43 -21.66 | TYC6372-00597-1 | 3 | 229 |
| 02.31 +32.17 | TYC2314-00479-1 | 3 | 229 |
| 19.29 +44.10 | TYC3133-00035-1 | 3 | 229 |
| 14.19 +04.13 | TYC0320-01468-1 | 3 | 228 |
| 04.68 +22.83 | TYC1830-00278-1 | 3 | 228 |
| 03.69 +31.45 | TYC2355-00133-1 | 3 | 228 |
| 04.34 +57.57 | TYC3727-00132-1 | 3 | 228 |
| 13.80 -13.31 | TYC5559-00546-1 | 3 | 228 |
| 03.00 -21.24 | TYC5870-00721-1 | 3 | 228 |
| 07.74 +27.20 | TYC1920-00569-1 | 3 | 227 |
| 09.40 +49.86 | TYC3428-00278-1 | 3 | 227 |
| 19.16 +34.78 | TYC2648-00273-1 | 3 | 227 |
| 07.80 +39.01 | TYC2959-01880-1 | 3 | 227 |
| 19.64 +49.17 | TYC3564-01422-1 | 3 | 227 |
| 13.48 +13.73 | TYC0898-00165-1 | 3 | 226 |
| 04.37 +39.47 | TYC2883-01371-1 | 3 | 226 |
| 00.72 -26.63 | TYC6423-01369-1 | 3 | 226 |
| 02.04 +25.13 | TYC1760-02270-1 | 3 | 225 |
| 18.66 +43.86 | TYC3117-01813-1 | 3 | 225 |
| 19.81 +48.72 | TYC3561-00046-1 | 3 | 225 |
| 04.48 +67.41 | TYC4077-01105-1 | 2 | 224 |
| 06.64 -19.28 | TYC5956-02467-1 | 2 | 224 |
| 04.44 +57.88 | TYC3727-00693-1 | 2 | 224 |
| 10.33 +00.20 | TYC0245-01029-1 | 2 | 224 |
| 17.42 +26.96 | TYC2082-01509-1 | 2 | 224 |
| 18.68 +47.98 | TYC3531-01007-1 | 2 | 224 |
| 01.86 -19.27 | TYC5857-01915-1 | 2 | 224 |
| 04.69 +18.56 | TYC1271-00273-1 | 2 | 223 |
| 09.95 +62.12 | TYC4137-01246-1 | 2 | 223 |
| 21.43 -21.52 | TYC6372-00859-1 | 2 | 223 |
| 14.38 -17.61 | TYC6143-00786-1 | 2 | 222 |
| 06.44 +58.09 | TYC3772-00034-1 | 2 | 222 |
| 07.22 +14.43 | TYC0774-00039-1 | 2 | 221 |
| 09.88 -16.13 | TYC6045-00052-1 | 2 | 221 |
| 08.56 -30.09 | TYC7136-02425-1 | 2 | 220 |
| 06.36 -29.47 | TYC6518-00366-1 | 2 | 220 |
| 09.80 -15.64 | TYC6044-01161-1 | 2 | 220 |
| 08.36 +61.83 | TYC4126-00940-1 | 2 | 220 |
| 19.11 +48.45 | TYC3545-02863-1 | 2 | 220 |
| 08.21 +61.16 | TYC4126-02045-1 | 2 | 219 |
| 09.34 +25.08 | TYC1955-01776-1 | 2 | 217 |
| 19.70 +43.31 | TYC3148-01448-1 | 2 | 217 |
| 04.57 -25.45 | TYC6467-00112-1 | 2 | 217 |
| 09.57 +31.02 | TYC2494-00348-1 | 2 | 217 |
| 19.74 +48.01 | TYC3561-00587-1 | 2 | 216 |
| 03.28 +25.54 | TYC1788-00784-1 | 2 | 216 |
| 06.32 +40.99 | TYC2930-00413-1 | 2 | 216 |
| 02.33 +32.51 | TYC2314-01147-1 | 2 | 216 |
| 18.98 +38.53 | TYC3119-00831-1 | 2 | 216 |
| 04.82 +60.26 | TYC4078-01220-1 | 2 | 216 |
| 06.46 +10.94 | TYC0736-00397-1 | 2 | 216 |
| 04.39 +58.02 | TYC3727-00029-1 | 2 | 216 |
| 11.82 +57.80 | TYC3838-00379-1 | 2 | 216 |
| 01.75 -15.90 | TYC5855-00466-1 | 2 | 216 |
| 01.75 -15.71 | TYC5855-00674-1 | 2 | 216 |
| 08.23 +05.87 | TYC0207-00128-1 | 2 | 216 |
| 14.75 +01.14 | TYC0326-00426-1 | 2 | 216 |
| 23.25 +57.03 | TYC4006-00662-1 | 2 | 216 |
| 07.34 +58.31 | TYC3793-01802-1 | 2 | 216 |
| 06.38 -28.87 | TYC6518-01295-1 | 2 | 215 |
| 06.53 +05.27 | TYC0154-00947-1 | 2 | 215 |
| 08.24 +05.74 | TYC0207-01267-1 | 2 | 215 |
| 02.39 +32.33 | TYC2314-00805-1 | 2 | 215 |
| 21.65 +30.35 | TYC2717-00093-1 | 2 | 215 |
| 04.50 +04.70 | TYC0078-00914-1 | 2 | 215 |
| 01.78 +20.10 | TYC1212-00741-1 | 2 | 215 |
| 04.74 +19.01 | TYC1275-01183-1 | 2 | 215 |
| 02.06 +25.03 | TYC1761-00542-1 | 2 | 215 |
| 02.65 +41.68 | TYC2840-00796-1 | 2 | 215 |
| 02.69 +39.06 | TYC2845-00312-1 | 2 | 215 |
| 03.80 +41.01 | TYC2867-00933-1 | 2 | 215 |
| 03.79 +41.02 | TYC2867-00993-1 | 2 | 215 |
| 09.47 +50.93 | TYC3431-01319-1 | 2 | 215 |
| 15.39 +59.54 | TYC3874-00065-1 | 2 | 215 |
| 02.74 +71.59 | TYC4320-00557-1 | 2 | 215 |
| 02.78 +71.97 | TYC4320-01538-1 | 2 | 215 |
| 05.15 +69.39 | TYC4346-00786-1 | 2 | 215 |
| 06.60 -32.07 | TYC7078-00594-1 | 2 | 215 |
| 06.61 -32.00 | TYC7091-00165-1 | 2 | 215 |
| 12.96 +12.09 | TYC0886-00980-1 | 2 | 215 |
| 16.87 +11.72 | TYC0983-01381-1 | 2 | 215 |
| 00.67 +20.47 | TYC1193-01883-1 | 2 | 215 |
| 19.47 +42.27 | TYC3142-00035-1 | 2 | 215 |
| 19.44 +48.71 | TYC3547-00478-1 | 2 | 215 |
| 06.63 -27.59 | TYC6516-00699-1 | 2 | 215 |
| 20.14 +44.53 | TYC3162-00229-1 | 2 | 215 |
| 10.39 -01.15 | TYC4905-01048-1 | 2 | 215 |
| 07.61 -13.77 | TYC5409-03577-1 | 2 | 215 |
| 04.28 +57.92 | TYC3727-00291-1 | 2 | 215 |
| 06.51 +29.23 | TYC1891-01222-1 | 2 | 214 |
| 11.71 +27.25 | TYC1984-02235-1 | 2 | 214 |
| 11.67 +31.37 | TYC2523-01948-1 | 2 | 214 |
| 23.44 +39.21 | TYC3230-00397-1 | 2 | 214 |
| 06.46 +38.84 | TYC2927-00465-1 | 2 | 214 |
| 07.22 +20.99 | TYC1358-00858-1 | 2 | 213 |
| 09.63 -11.91 | TYC5469-01076-1 | 2 | 213 |
| 19.99 +45.93 | TYC3558-00186-1 | 2 | 212 |
| 06.74 +34.30 | TYC2443-00596-1 | 2 | 212 |
| 06.53 +29.76 | TYC1891-00032-1 | 2 | 211 |
| 06.45 -00.29 | TYC4785-00455-1 | 2 | 211 |
| 04.02 +32.92 | TYC2361-01594-1 | 2 | 211 |
| 12.52 +21.70 | TYC1447-02424-1 | 2 | 211 |
| 11.07 +44.26 | TYC3012-00499-1 | 2 | 211 |
| 18.73 +43.47 | TYC3130-01901-1 | 2 | 211 |
| 06.74 +00.91 | TYC0147-01589-1 | 2 | 211 |
| 03.22 -01.33 | TYC4708-00214-1 | 2 | 211 |
| 07.56 -22.35 | TYC5991-00709-1 | 2 | 211 |
| 03.04 +14.94 | TYC0647-00876-1 | 2 | 211 |
| 08.48 +53.79 | TYC3797-02224-1 | 2 | 211 |
| 19.39 +51.06 | TYC3555-00386-1 | 2 | 210 |
| 00.91 +01.08 | TYC0012-00020-1 | 2 | 210 |
| 02.99 +06.21 | TYC0054-00476-1 | 2 | 210 |
| 03.03 +00.24 | TYC0055-01086-1 | 2 | 210 |
| 03.09 +00.94 | TYC0055-01139-1 | 2 | 210 |
| 03.00 +07.41 | TYC0061-00684-1 | 2 | 210 |
| 03.63 +03.48 | TYC0067-01558-1 | 2 | 210 |
| 04.02 +04.14 | TYC0076-00730-1 | 2 | 210 |
| 04.46 +05.28 | TYC0081-01965-1 | 2 | 210 |
| 04.54 +04.37 | TYC0090-01166-1 | 2 | 210 |
| 05.30 +06.31 | TYC0112-01332-1 | 2 | 210 |
| 06.20 +06.92 | TYC0143-00721-1 | 2 | 210 |
| 06.52 +06.25 | TYC0158-03174-1 | 2 | 210 |
| 06.92 +05.69 | TYC0161-00421-1 | 2 | 210 |
| 07.38 +06.58 | TYC0177-01868-1 | 2 | 210 |
| 09.64 +03.76 | TYC0238-01774-1 | 2 | 210 |
| 10.35 +04.68 | TYC0249-01152-1 | 2 | 210 |
| 03.03 +13.04 | TYC0647-00275-1 | 2 | 210 |
| 03.34 +08.79 | TYC0649-00795-1 | 2 | 210 |
| 03.54 +10.81 | TYC0653-00072-1 | 2 | 210 |
| 03.90 +08.18 | TYC0658-00860-1 | 2 | 210 |
| 03.73 +10.85 | TYC0660-00904-1 | 2 | 210 |
| 03.90 +13.15 | TYC0664-00658-1 | 2 | 210 |
| 04.51 +14.46 | TYC0681-00741-1 | 2 | 210 |
| 04.62 +11.80 | TYC0690-01499-1 | 2 | 210 |
| 05.07 +11.89 | TYC0693-02292-1 | 2 | 210 |
| 04.74 +14.72 | TYC0695-00595-1 | 2 | 210 |
| 05.65 +10.25 | TYC0718-00464-1 | 2 | 210 |
| 05.69 +14.98 | TYC0726-00159-1 | 2 | 210 |
| 06.42 +10.86 | TYC0736-00008-1 | 2 | 210 |
| 06.22 +13.67 | TYC0742-01780-1 | 2 | 210 |
| 06.87 +11.15 | TYC0751-01559-1 | 2 | 210 |
| 07.47 +09.60 | TYC0768-01704-1 | 2 | 210 |
| 08.63 +12.90 | TYC0805-00032-1 | 2 | 210 |
| 09.13 +07.58 | TYC0812-00841-1 | 2 | 210 |
| 09.50 +09.93 | TYC0820-00182-1 | 2 | 210 |
| 10.39 +07.69 | TYC0838-01228-1 | 2 | 210 |
| 11.33 +14.16 | TYC0861-00317-1 | 2 | 210 |
| 12.98 +12.92 | TYC0889-00083-1 | 2 | 210 |
| 01.01 +19.95 | TYC1192-00531-1 | 2 | 210 |
| 00.66 +21.65 | TYC1193-00306-1 | 2 | 210 |
| 02.92 +16.10 | TYC1224-01004-1 | 2 | 210 |
| 03.03 +16.32 | TYC1225-00673-1 | 2 | 210 |
| 02.97 +19.56 | TYC1227-00870-1 | 2 | 210 |
| 03.38 +17.46 | TYC1237-00064-1 | 2 | 210 |
| 03.38 +17.96 | TYC1237-00068-1 | 2 | 210 |
| 03.71 +21.19 | TYC1247-01176-1 | 2 | 210 |
| 04.11 +17.79 | TYC1254-00855-1 | 2 | 210 |
| 04.09 +20.17 | TYC1258-00611-1 | 2 | 210 |
| 03.96 +21.94 | TYC1261-00676-1 | 2 | 210 |
| 03.97 +21.06 | TYC1261-01468-1 | 2 | 210 |
| 04.14 +22.33 | TYC1263-00832-1 | 2 | 210 |
| 04.32 +20.98 | TYC1276-00152-1 | 2 | 210 |
| 04.65 +20.92 | TYC1278-01754-1 | 2 | 210 |
| 04.75 +21.20 | TYC1279-01732-1 | 2 | 210 |
| 05.03 +15.85 | TYC1281-00888-1 | 2 | 210 |
| 05.44 +16.37 | TYC1296-01221-1 | 2 | 210 |
| 05.39 +16.98 | TYC1300-00164-1 | 2 | 210 |
| 05.58 +17.24 | TYC1301-01760-1 | 2 | 210 |
| 05.82 +21.14 | TYC1311-02055-1 | 2 | 210 |
| 06.28 +17.28 | TYC1319-01292-1 | 2 | 210 |
| 06.59 +18.36 | TYC1333-00088-1 | 2 | 210 |
| 06.68 +17.48 | TYC1334-02248-1 | 2 | 210 |
| 07.72 +18.59 | TYC1365-00505-1 | 2 | 210 |
| 08.31 +18.82 | TYC1386-01245-1 | 2 | 210 |
| 08.39 +22.18 | TYC1390-00147-1 | 2 | 210 |
| 08.65 +17.19 | TYC1392-00787-1 | 2 | 210 |
| 09.09 +19.83 | TYC1404-01118-1 | 2 | 210 |
| 09.51 +17.85 | TYC1406-00728-1 | 2 | 210 |
| 09.74 +16.75 | TYC1410-00039-1 | 2 | 210 |
| 09.90 +15.09 | TYC1411-00872-1 | 2 | 210 |
| 09.86 +18.54 | TYC1414-00726-1 | 2 | 210 |
| 10.18 +15.63 | TYC1419-01299-1 | 2 | 210 |
| 10.28 +19.62 | TYC1422-00057-1 | 2 | 210 |
| 10.25 +19.98 | TYC1422-00091-1 | 2 | 210 |
| 10.66 +21.24 | TYC1427-00545-1 | 2 | 210 |
| 10.82 +21.26 | TYC1434-01181-1 | 2 | 210 |
| 11.28 +18.28 | TYC1437-00260-1 | 2 | 210 |
| 12.48 +22.02 | TYC1447-01239-1 | 2 | 210 |
| 22.04 +18.70 | TYC1684-00591-1 | 2 | 210 |
| 22.08 +18.70 | TYC1684-00603-1 | 2 | 210 |
| 02.14 +23.52 | TYC1758-02110-1 | 2 | 210 |
| 02.15 +23.59 | TYC1758-02276-1 | 2 | 210 |
| 04.82 +26.79 | TYC1839-00725-1 | 2 | 210 |
| 05.29 +22.58 | TYC1846-00836-1 | 2 | 210 |
| 05.42 +23.50 | TYC1847-01060-1 | 2 | 210 |
| 05.53 +24.67 | TYC1852-01611-1 | 2 | 210 |
| 05.82 +23.71 | TYC1862-01203-1 | 2 | 210 |
| 06.02 +26.95 | TYC1872-00232-1 | 2 | 210 |
| 06.88 +24.21 | TYC1894-02478-1 | 2 | 210 |
| 06.92 +29.83 | TYC1906-00570-1 | 2 | 210 |
| 07.59 +23.97 | TYC1911-01325-1 | 2 | 210 |
| 07.35 +27.36 | TYC1917-02086-1 | 2 | 210 |
| 07.73 +28.03 | TYC1920-00300-1 | 2 | 210 |
| 07.37 +29.26 | TYC1921-02173-1 | 2 | 210 |
| 09.12 +22.87 | TYC1951-01445-1 | 2 | 210 |
| 09.12 +22.72 | TYC1951-01486-1 | 2 | 210 |
| 09.88 +28.80 | TYC1967-00124-1 | 2 | 210 |
| 10.28 +25.11 | TYC1972-01235-1 | 2 | 210 |
| 10.60 +29.24 | TYC1976-01045-1 | 2 | 210 |
| 11.26 +25.75 | TYC1981-00138-1 | 2 | 210 |
| 11.52 +25.27 | TYC1982-00824-1 | 2 | 210 |
| 22.07 +26.12 | TYC2208-00023-1 | 2 | 210 |
| 00.88 +34.64 | TYC2284-00076-1 | 2 | 210 |
| 01.42 +34.26 | TYC2300-00141-1 | 2 | 210 |
| 03.19 +30.96 | TYC2340-01051-1 | 2 | 210 |
| 03.03 +35.27 | TYC2347-00077-1 | 2 | 210 |
| 03.74 +34.77 | TYC2363-00959-1 | 2 | 210 |
| 04.38 +30.01 | TYC2372-00918-1 | 2 | 210 |
| 04.69 +30.64 | TYC2374-00570-1 | 2 | 210 |
| 04.54 +33.90 | TYC2381-00207-1 | 2 | 210 |
| 05.09 +31.02 | TYC2388-00620-1 | 2 | 210 |
| 05.04 +30.22 | TYC2388-01708-1 | 2 | 210 |
| 05.17 +30.29 | TYC2389-00285-1 | 2 | 210 |
| 05.39 +30.62 | TYC2390-01514-1 | 2 | 210 |
| 05.62 +31.54 | TYC2404-00719-1 | 2 | 210 |
| 05.82 +32.83 | TYC2409-00707-1 | 2 | 210 |
| 05.69 +34.50 | TYC2412-01242-1 | 2 | 210 |
| 06.04 +32.77 | TYC2423-00053-1 | 2 | 210 |
| 06.09 +35.03 | TYC2427-00541-1 | 2 | 210 |
| 06.30 +35.12 | TYC2428-00065-1 | 2 | 210 |
| 07.08 +35.94 | TYC2450-01182-1 | 2 | 210 |
| 07.51 +33.94 | TYC2461-00546-1 | 2 | 210 |
| 08.01 +32.04 | TYC2472-01678-1 | 2 | 210 |
| 08.80 +32.16 | TYC2484-00182-1 | 2 | 210 |
| 08.88 +33.24 | TYC2488-00362-1 | 2 | 210 |
| 10.07 +32.45 | TYC2503-00331-1 | 2 | 210 |
| 11.74 +31.34 | TYC2524-01333-1 | 2 | 210 |
| 11.73 +30.77 | TYC2524-01336-1 | 2 | 210 |
| 15.99 +33.49 | TYC2575-01344-1 | 2 | 210 |
| 15.87 +35.98 | TYC2578-00475-1 | 2 | 210 |
| 18.57 +35.34 | TYC2632-01026-1 | 2 | 210 |
| 19.04 +36.74 | TYC2651-01906-1 | 2 | 210 |
| 23.24 +31.45 | TYC2751-00603-1 | 2 | 210 |
| 00.63 +42.78 | TYC2792-00039-1 | 2 | 210 |
| 12.52 +44.88 | TYC3020-01917-1 | 2 | 210 |
| 15.33 +42.02 | TYC3048-00040-1 | 2 | 210 |
| 14.98 +44.40 | TYC3050-00373-1 | 2 | 210 |
| 18.72 +42.62 | TYC3126-00099-1 | 2 | 210 |
| 02.06 +46.72 | TYC3280-01423-1 | 2 | 210 |
| 02.96 +48.42 | TYC3301-00160-1 | 2 | 210 |
| 04.90 +50.85 | TYC3356-01150-1 | 2 | 210 |
| 06.06 +47.11 | TYC3378-01183-1 | 2 | 210 |
| 06.46 +48.56 | TYC3380-00527-1 | 2 | 210 |
| 07.00 +48.53 | TYC3395-00553-1 | 2 | 210 |
| 07.43 +48.39 | TYC3397-00024-1 | 2 | 210 |
| 06.87 +49.65 | TYC3398-02053-1 | 2 | 210 |
| 08.17 +48.19 | TYC3411-01260-1 | 2 | 210 |
| 08.01 +49.32 | TYC3411-01433-1 | 2 | 210 |
| 07.77 +50.11 | TYC3413-00302-1 | 2 | 210 |
| 08.27 +45.98 | TYC3415-01262-1 | 2 | 210 |
| 08.61 +46.51 | TYC3416-01305-1 | 2 | 210 |
| 09.49 +45.81 | TYC3425-00058-1 | 2 | 210 |
| 09.47 +45.93 | TYC3425-00372-1 | 2 | 210 |
| 10.19 +49.54 | TYC3437-00871-1 | 2 | 210 |
| 09.90 +50.90 | TYC3439-01391-1 | 2 | 210 |
| 17.02 +47.33 | TYC3501-00437-1 | 2 | 210 |
| 03.27 +57.75 | TYC3710-00272-1 | 2 | 210 |
| 04.44 +52.69 | TYC3719-01114-1 | 2 | 210 |
| 04.13 +56.93 | TYC3726-00486-1 | 2 | 210 |
| 03.77 +58.67 | TYC3728-00332-1 | 2 | 210 |
| 03.78 +59.67 | TYC3729-00272-1 | 2 | 210 |
| 05.08 +58.48 | TYC3746-02035-1 | 2 | 210 |
| 05.71 +54.15 | TYC3749-00780-1 | 2 | 210 |
| 05.37 +59.00 | TYC3760-02087-1 | 2 | 210 |
| 05.81 +58.69 | TYC3762-01232-1 | 2 | 210 |
| 06.23 +52.64 | TYC3764-00125-1 | 2 | 210 |
| 06.93 +55.34 | TYC3771-00733-1 | 2 | 210 |
| 07.45 +53.09 | TYC3781-00158-1 | 2 | 210 |
| 07.82 +52.86 | TYC3783-00404-1 | 2 | 210 |
| 08.00 +53.48 | TYC3783-00816-1 | 2 | 210 |
| 07.92 +59.78 | TYC3795-00258-1 | 2 | 210 |
| 08.25 +52.55 | TYC3796-00083-1 | 2 | 210 |
| 08.68 +55.79 | TYC3801-00263-1 | 2 | 210 |
| 08.38 +59.78 | TYC3803-01705-1 | 2 | 210 |
| 09.05 +54.70 | TYC3805-00820-1 | 2 | 210 |
| 13.56 +53.38 | TYC3850-00275-1 | 2 | 210 |
| 15.38 +58.95 | TYC3874-00703-1 | 2 | 210 |
| 18.15 +54.34 | TYC3903-00043-1 | 2 | 210 |
| 04.54 +63.84 | TYC4073-01132-1 | 2 | 210 |
| 05.43 +61.39 | TYC4080-00553-1 | 2 | 210 |
| 05.41 +63.02 | TYC4084-00330-1 | 2 | 210 |
| 06.64 +64.57 | TYC4105-01421-1 | 2 | 210 |
| 07.44 +60.73 | TYC4112-01585-1 | 2 | 210 |
| 07.32 +67.12 | TYC4123-01105-1 | 2 | 210 |
| 08.93 +60.99 | TYC4128-00694-1 | 2 | 210 |
| 09.05 +61.69 | TYC4128-00802-1 | 2 | 210 |
| 10.55 +62.51 | TYC4147-01222-1 | 2 | 210 |
| 06.35 +68.09 | TYC4345-00254-1 | 2 | 210 |
| 06.12 +68.44 | TYC4345-00960-1 | 2 | 210 |
| 06.09 +70.96 | TYC4349-00122-1 | 2 | 210 |
| 06.48 +68.04 | TYC4358-01274-1 | 2 | 210 |
| 07.07 +68.76 | TYC4359-00223-1 | 2 | 210 |
| 07.09 +71.14 | TYC4363-00383-1 | 2 | 210 |
| 06.61 +71.82 | TYC4366-00020-1 | 2 | 210 |
| 04.90 +75.71 | TYC4511-01184-1 | 2 | 210 |
| 03.06 +79.83 | TYC4516-01695-1 | 2 | 210 |
| 07.55 +79.07 | TYC4535-00226-1 | 2 | 210 |
| 08.28 +75.87 | TYC4540-01529-1 | 2 | 210 |
| 08.24 +80.02 | TYC4546-00596-1 | 2 | 210 |
| 09.41 +80.48 | TYC4547-00239-1 | 2 | 210 |
| 03.01 +84.41 | TYC4620-02099-1 | 2 | 210 |
| 06.24 +84.65 | TYC4622-01153-1 | 2 | 210 |
| 07.78 -01.30 | TYC4832-02492-1 | 2 | 210 |
| 11.21 -00.79 | TYC4922-00350-1 | 2 | 210 |
| 23.04 -00.30 | TYC5242-00428-1 | 2 | 210 |
| 05.30 -13.76 | TYC5343-00461-1 | 2 | 210 |
| 08.30 -12.36 | TYC5434-01389-1 | 2 | 210 |
| 08.97 -14.28 | TYC5456-00524-1 | 2 | 210 |
| 09.57 -11.90 | TYC5468-00298-1 | 2 | 210 |
| 09.57 -11.80 | TYC5468-00448-1 | 2 | 210 |
| 00.49 -16.37 | TYC5840-00335-1 | 2 | 210 |
| 01.12 -22.13 | TYC5853-00486-1 | 2 | 210 |
| 03.02 -17.63 | TYC5867-00216-1 | 2 | 210 |
| 04.51 -19.75 | TYC5894-00213-1 | 2 | 210 |
| 05.30 -17.33 | TYC5906-00627-1 | 2 | 210 |
| 05.45 -16.16 | TYC5915-00574-1 | 2 | 210 |
| 05.74 -20.51 | TYC5926-01654-1 | 2 | 210 |
| 06.56 -15.62 | TYC5948-00649-1 | 2 | 210 |
| 07.02 -21.58 | TYC5975-00593-1 | 2 | 210 |
| 07.59 -20.44 | TYC5987-01944-1 | 2 | 210 |
| 08.83 -18.92 | TYC6021-01073-1 | 2 | 210 |
| 09.77 -21.54 | TYC6056-01077-1 | 2 | 210 |
| 09.87 -20.96 | TYC6057-00538-1 | 2 | 210 |
| 11.31 -18.67 | TYC6087-00647-1 | 2 | 210 |
| 11.99 -19.99 | TYC6097-01392-1 | 2 | 210 |
| 13.30 -18.24 | TYC6116-00230-1 | 2 | 210 |
| 03.25 -26.10 | TYC6442-01256-1 | 2 | 210 |
| 04.05 -22.53 | TYC6455-00030-1 | 2 | 210 |
| 04.30 -24.43 | TYC6457-02533-1 | 2 | 210 |
| 05.50 -25.21 | TYC6480-01432-1 | 2 | 210 |
| 06.81 -23.41 | TYC6521-02296-1 | 2 | 210 |
| 09.23 -27.89 | TYC6595-00701-1 | 2 | 210 |
| 10.92 -26.69 | TYC6643-00908-1 | 2 | 210 |
| 11.27 -24.08 | TYC6649-00714-1 | 2 | 210 |
| 12.69 -30.58 | TYC7247-00471-1 | 2 | 210 |
| 14.85 +06.25 | TYC0333-01207-1 | 2 | 210 |
| 16.52 +02.32 | TYC0386-00209-1 | 2 | 210 |
| 13.17 +17.44 | TYC1451-00013-1 | 2 | 210 |
| 13.18 +17.57 | TYC1454-00291-1 | 2 | 210 |
| 14.53 +21.63 | TYC1482-00192-1 | 2 | 210 |
| 19.93 +41.54 | TYC3145-02372-1 | 2 | 210 |
| 17.01 +47.10 | TYC3501-01276-1 | 2 | 210 |
| 18.92 +48.74 | TYC3544-00692-1 | 2 | 210 |
| 14.00 -27.56 | TYC6737-00131-1 | 2 | 210 |
| 15.24 -25.52 | TYC6766-00965-1 | 2 | 210 |
| 15.25 -25.13 | TYC6766-01205-1 | 2 | 210 |
| 13.20 -32.10 | TYC7267-01593-1 | 2 | 210 |
| 21.20 +14.54 | TYC1117-00629-1 | 2 | 209 |
| 23.29 +18.30 | TYC1716-01366-1 | 2 | 209 |
| 22.94 +21.03 | TYC1717-00736-1 | 2 | 209 |
| 01.39 +28.64 | TYC1754-00007-1 | 2 | 209 |
| 00.24 -12.32 | TYC5264-00118-1 | 2 | 209 |
| 22.42 -17.03 | TYC6385-00624-1 | 2 | 209 |
| 22.42 -17.19 | TYC6385-00794-1 | 2 | 209 |
| 00.43 +78.53 | TYC4496-00017-1 | 2 | 209 |
| 08.72 +04.26 | TYC0219-00002-1 | 2 | 209 |
| 08.64 +12.77 | TYC0805-00058-1 | 2 | 209 |
| 01.00 +19.97 | TYC1192-00877-1 | 2 | 209 |
| 05.03 +16.58 | TYC1281-00501-1 | 2 | 209 |
| 09.38 +19.99 | TYC1405-00100-1 | 2 | 209 |
| 21.62 +30.20 | TYC2717-00055-1 | 2 | 209 |
| 14.96 +44.41 | TYC3050-00368-1 | 2 | 209 |
| 07.14 +52.52 | TYC3780-00693-1 | 2 | 209 |
| 15.60 +53.57 | TYC3869-00188-1 | 2 | 209 |
| 04.12 +69.15 | TYC4328-00128-1 | 2 | 209 |
| 07.78 -00.93 | TYC4832-02399-1 | 2 | 209 |
| 08.89 -17.31 | TYC6017-00360-1 | 2 | 209 |
| 00.76 +07.50 | TYC0017-01006-1 | 2 | 209 |
| 03.41 +00.33 | TYC0057-00256-1 | 2 | 209 |
| 03.10 +04.06 | TYC0058-00502-1 | 2 | 209 |
| 03.10 +06.29 | TYC0061-00637-1 | 2 | 209 |
| 03.84 +01.11 | TYC0066-00288-1 | 2 | 209 |
| 03.84 +03.00 | TYC0069-00062-1 | 2 | 209 |
| 04.46 +05.40 | TYC0081-00866-1 | 2 | 209 |
| 05.02 +00.85 | TYC0098-00941-1 | 2 | 209 |
| 06.80 +00.61 | TYC0148-01689-1 | 2 | 209 |
| 07.74 +03.36 | TYC0183-01837-1 | 2 | 209 |
| 08.42 +00.05 | TYC0197-01913-1 | 2 | 209 |
| 08.04 +03.94 | TYC0202-00120-1 | 2 | 209 |
| 09.55 +05.82 | TYC0241-01284-1 | 2 | 209 |
| 11.44 +02.78 | TYC0267-00035-1 | 2 | 209 |
| 03.26 +14.13 | TYC0655-00579-1 | 2 | 209 |
| 03.80 +10.63 | TYC0661-00722-1 | 2 | 209 |
| 04.47 +13.65 | TYC0680-00695-1 | 2 | 209 |
| 05.13 +14.23 | TYC0710-00114-1 | 2 | 209 |
| 05.65 +09.89 | TYC0718-00908-1 | 2 | 209 |
| 08.60 +14.21 | TYC0809-00577-1 | 2 | 209 |
| 11.12 +09.73 | TYC0848-00411-1 | 2 | 209 |
| 11.19 +11.13 | TYC0851-01127-1 | 2 | 209 |
| 13.01 +12.59 | TYC0889-00754-1 | 2 | 209 |
| 03.68 +18.21 | TYC1239-00743-1 | 2 | 209 |
| 04.13 +21.38 | TYC1262-00788-1 | 2 | 209 |
| 05.61 +20.21 | TYC1306-00158-1 | 2 | 209 |
| 05.59 +21.43 | TYC1309-01306-1 | 2 | 209 |
| 07.47 +15.30 | TYC1360-01313-1 | 2 | 209 |
| 07.97 +18.74 | TYC1367-01801-1 | 2 | 209 |
| 07.50 +19.98 | TYC1368-01429-1 | 2 | 209 |
| 08.41 +17.42 | TYC1383-00203-1 | 2 | 209 |
| 08.64 +16.98 | TYC1392-00604-1 | 2 | 209 |
| 09.21 +20.28 | TYC1407-00051-1 | 2 | 209 |
| 09.75 +16.52 | TYC1410-00702-1 | 2 | 209 |
| 12.34 +17.70 | TYC1445-00658-1 | 2 | 209 |
| 23.30 +18.70 | TYC1716-00025-1 | 2 | 209 |
| 03.30 +25.56 | TYC1788-00856-1 | 2 | 209 |
| 03.87 +26.69 | TYC1808-01041-1 | 2 | 209 |
| 04.74 +24.10 | TYC1830-01464-1 | 2 | 209 |
| 05.18 +25.74 | TYC1850-00737-1 | 2 | 209 |
| 05.64 +24.90 | TYC1865-01596-1 | 2 | 209 |
| 06.48 +27.02 | TYC1887-01649-1 | 2 | 209 |
| 06.90 +24.41 | TYC1898-00786-1 | 2 | 209 |
| 07.39 +29.53 | TYC1921-02230-1 | 2 | 209 |
| 07.96 +28.07 | TYC1934-00116-1 | 2 | 209 |
| 10.94 +29.19 | TYC1980-01059-1 | 2 | 209 |
| 04.78 +32.86 | TYC2378-00043-1 | 2 | 209 |
| 05.63 +31.15 | TYC2404-00997-1 | 2 | 209 |
| 06.25 +30.55 | TYC2420-00530-1 | 2 | 209 |
| 06.79 +31.60 | TYC2436-00429-1 | 2 | 209 |
| 06.95 +30.82 | TYC2437-00163-1 | 2 | 209 |
| 06.83 +35.10 | TYC2444-00142-1 | 2 | 209 |
| 07.08 +35.39 | TYC2446-00816-1 | 2 | 209 |
| 08.10 +34.21 | TYC2476-00956-1 | 2 | 209 |
| 07.96 +33.84 | TYC2476-01305-1 | 2 | 209 |
| 08.88 +33.20 | TYC2488-00875-1 | 2 | 209 |
| 09.12 +30.39 | TYC2492-00552-1 | 2 | 209 |
| 09.60 +34.93 | TYC2504-01498-1 | 2 | 209 |
| 10.49 +32.08 | TYC2511-00203-1 | 2 | 209 |
| 18.59 +35.62 | TYC2632-01282-1 | 2 | 209 |
| 19.42 +36.52 | TYC2666-00044-1 | 2 | 209 |
| 02.73 +38.59 | TYC2845-01017-1 | 2 | 209 |
| 18.88 +41.22 | TYC3123-00558-1 | 2 | 209 |
| 18.99 +40.95 | TYC3123-01264-1 | 2 | 209 |
| 19.01 +39.84 | TYC3124-00722-1 | 2 | 209 |
| 19.02 +39.41 | TYC3124-01262-1 | 2 | 209 |
| 18.78 +42.87 | TYC3126-00375-1 | 2 | 209 |
| 18.93 +42.21 | TYC3127-00824-1 | 2 | 209 |
| 19.09 +42.91 | TYC3128-00301-1 | 2 | 209 |
| 19.11 +41.86 | TYC3128-00490-1 | 2 | 209 |
| 19.23 +43.04 | TYC3129-00585-1 | 2 | 209 |
| 19.25 +42.82 | TYC3129-00601-1 | 2 | 209 |
| 19.25 +43.11 | TYC3129-00609-1 | 2 | 209 |
| 19.00 +44.56 | TYC3131-01027-1 | 2 | 209 |
| 19.22 +44.11 | TYC3133-00792-1 | 2 | 209 |
| 19.67 +39.34 | TYC3136-00348-1 | 2 | 209 |
| 19.36 +39.94 | TYC3138-00054-1 | 2 | 209 |
| 19.38 +39.84 | TYC3138-00108-1 | 2 | 209 |
| 19.69 +41.13 | TYC3140-00831-1 | 2 | 209 |
| 19.67 +41.11 | TYC3140-00915-1 | 2 | 209 |
| 19.68 +41.08 | TYC3140-00917-1 | 2 | 209 |
| 19.71 +40.42 | TYC3140-01473-1 | 2 | 209 |
| 19.97 +40.89 | TYC3141-00266-1 | 2 | 209 |
| 20.00 +40.61 | TYC3141-01372-1 | 2 | 209 |
| 19.83 +41.99 | TYC3144-00017-1 | 2 | 209 |
| 19.80 +42.89 | TYC3144-00058-1 | 2 | 209 |
| 19.83 +41.90 | TYC3144-00282-1 | 2 | 209 |
| 19.76 +41.63 | TYC3144-00787-1 | 2 | 209 |
| 20.02 +44.67 | TYC3162-00631-1 | 2 | 209 |
| 20.12 +44.22 | TYC3162-01633-1 | 2 | 209 |
| 02.69 +49.56 | TYC3304-01096-1 | 2 | 209 |
| 03.51 +51.54 | TYC3324-00245-1 | 2 | 209 |
| 06.21 +50.45 | TYC3383-00105-1 | 2 | 209 |
| 07.38 +52.24 | TYC3405-00098-1 | 2 | 209 |
| 07.31 +51.76 | TYC3405-00163-1 | 2 | 209 |
| 07.69 +47.17 | TYC3406-02150-1 | 2 | 209 |
| 07.69 +49.05 | TYC3409-00241-1 | 2 | 209 |
| 09.03 +51.11 | TYC3430-01324-1 | 2 | 209 |
| 09.96 +46.97 | TYC3433-01205-1 | 2 | 209 |
| 10.36 +50.23 | TYC3441-00569-1 | 2 | 209 |
| 18.74 +48.20 | TYC3531-00309-1 | 2 | 209 |
| 18.90 +46.83 | TYC3540-01169-1 | 2 | 209 |
| 18.94 +45.66 | TYC3540-01328-1 | 2 | 209 |
| 18.79 +47.20 | TYC3544-00069-1 | 2 | 209 |
| 18.86 +47.72 | TYC3544-01695-1 | 2 | 209 |
| 18.97 +48.44 | TYC3545-00741-1 | 2 | 209 |
| 18.96 +47.03 | TYC3545-02241-1 | 2 | 209 |
| 19.12 +48.50 | TYC3545-02679-1 | 2 | 209 |
| 18.89 +48.93 | TYC3548-00841-1 | 2 | 209 |
| 19.10 +50.42 | TYC3549-00657-1 | 2 | 209 |
| 19.09 +49.94 | TYC3549-01586-1 | 2 | 209 |
| 19.29 +50.46 | TYC3550-00001-1 | 2 | 209 |
| 19.26 +51.69 | TYC3554-01053-1 | 2 | 209 |
| 19.52 +46.08 | TYC3556-00249-1 | 2 | 209 |
| 19.97 +45.39 | TYC3558-00762-1 | 2 | 209 |
| 20.05 +45.29 | TYC3558-01052-1 | 2 | 209 |
| 19.96 +45.77 | TYC3558-01248-1 | 2 | 209 |
| 19.55 +47.82 | TYC3560-00021-1 | 2 | 209 |
| 19.54 +47.71 | TYC3560-00079-1 | 2 | 209 |
| 19.53 +47.85 | TYC3560-00131-1 | 2 | 209 |
| 19.52 +47.83 | TYC3560-00151-1 | 2 | 209 |
| 20.01 +47.30 | TYC3562-01742-1 | 2 | 209 |
| 03.76 +59.59 | TYC3728-00939-1 | 2 | 209 |
| 03.83 +58.54 | TYC3729-00813-1 | 2 | 209 |
| 06.71 +55.81 | TYC3770-01315-1 | 2 | 209 |
| 07.60 +54.71 | TYC3786-00858-1 | 2 | 209 |
| 11.24 +56.20 | TYC3827-00170-1 | 2 | 209 |
| 15.57 +54.02 | TYC3869-00364-1 | 2 | 209 |
| 04.28 +66.45 | TYC4076-00331-1 | 2 | 209 |
| 07.63 +61.39 | TYC4112-00735-1 | 2 | 209 |
| 08.30 +63.73 | TYC4129-02199-1 | 2 | 209 |
| 09.41 +66.17 | TYC4141-01224-1 | 2 | 209 |
| 02.80 +71.97 | TYC4321-01904-1 | 2 | 209 |
| 07.86 +69.78 | TYC4365-01605-1 | 2 | 209 |
| 08.94 +71.19 | TYC4378-01665-1 | 2 | 209 |
| 02.77 +75.08 | TYC4508-00867-1 | 2 | 209 |
| 07.05 +79.03 | TYC4534-00476-1 | 2 | 209 |
| 06.13 +82.18 | TYC4537-00591-1 | 2 | 209 |
| 06.97 +81.16 | TYC4538-00504-1 | 2 | 209 |
| 09.61 +76.36 | TYC4541-01665-1 | 2 | 209 |
| 06.30 +84.39 | TYC4622-00697-1 | 2 | 209 |
| 06.26 +84.90 | TYC4622-00837-1 | 2 | 209 |
| 11.78 +88.75 | TYC4644-00117-1 | 2 | 209 |
| 03.23 -01.10 | TYC4708-00243-1 | 2 | 209 |
| 06.42 -00.87 | TYC4785-01229-1 | 2 | 209 |
| 08.98 -14.17 | TYC5456-00854-1 | 2 | 209 |
| 10.04 -13.46 | TYC5488-00430-1 | 2 | 209 |
| 01.90 -19.73 | TYC5858-01966-1 | 2 | 209 |
| 02.73 -16.88 | TYC5863-00314-1 | 2 | 209 |
| 05.06 -17.26 | TYC5904-00118-1 | 2 | 209 |
| 05.59 -15.57 | TYC5916-01853-1 | 2 | 209 |
| 05.74 -20.15 | TYC5926-01564-1 | 2 | 209 |
| 06.23 -17.63 | TYC5937-01603-1 | 2 | 209 |
| 06.17 -21.75 | TYC5945-00834-1 | 2 | 209 |
| 07.38 -20.05 | TYC5974-00560-1 | 2 | 209 |
| 07.56 -16.02 | TYC5979-01867-1 | 2 | 209 |
| 09.89 -21.03 | TYC6057-00559-1 | 2 | 209 |
| 10.28 -18.07 | TYC6064-00773-1 | 2 | 209 |
| 02.77 -23.17 | TYC6434-01568-1 | 2 | 209 |
| 03.23 -23.66 | TYC6439-00412-1 | 2 | 209 |
| 03.39 -29.52 | TYC6452-00909-1 | 2 | 209 |
| 04.46 -24.58 | TYC6457-02810-1 | 2 | 209 |
| 04.41 -29.25 | TYC6463-01294-1 | 2 | 209 |
| 05.03 -26.65 | TYC6481-00038-1 | 2 | 209 |
| 06.45 -25.93 | TYC6511-00686-1 | 2 | 209 |
| 06.40 -28.86 | TYC6518-01288-1 | 2 | 209 |
| 06.80 -23.03 | TYC6521-01255-1 | 2 | 209 |
| 10.29 -27.01 | TYC6626-00830-1 | 2 | 209 |
| 10.65 -23.42 | TYC6633-00905-1 | 2 | 209 |
| 06.61 -32.39 | TYC7091-01316-1 | 2 | 209 |
| 17.89 -30.45 | TYC7378-00388-1 | 2 | 209 |
| 07.56 -22.19 | TYC5991-00262-1 | 2 | 208 |
| 23.96 -22.72 | TYC6982-00448-1 | 2 | 208 |
| 09.03 +06.84 | TYC0232-00717-1 | 2 | 208 |
| 10.53 +00.79 | TYC0253-01007-1 | 2 | 208 |
| 09.64 +19.01 | TYC1413-00634-1 | 2 | 208 |
| 11.31 +18.18 | TYC1437-00022-1 | 2 | 208 |
| 18.26 +36.42 | TYC2634-00166-1 | 2 | 208 |
| 00.64 +42.86 | TYC2792-00033-1 | 2 | 208 |
| 17.20 +63.64 | TYC4202-00619-1 | 2 | 208 |
| 07.51 -15.84 | TYC5979-01008-1 | 2 | 208 |
| 05.98 +19.07 | TYC1320-00657-1 | 2 | 208 |
| 18.95 +39.77 | TYC3123-00623-1 | 2 | 208 |
| 19.10 +42.75 | TYC3128-00237-1 | 2 | 208 |
| 18.75 +43.25 | TYC3130-00387-1 | 2 | 208 |
| 19.70 +41.23 | TYC3140-00627-1 | 2 | 208 |
| 19.99 +40.50 | TYC3141-01364-1 | 2 | 208 |
| 19.73 +44.83 | TYC3148-00815-1 | 2 | 208 |
| 19.97 +43.86 | TYC3149-00788-1 | 2 | 208 |
| 18.79 +46.10 | TYC3540-02649-1 | 2 | 208 |
| 18.78 +47.02 | TYC3544-00216-1 | 2 | 208 |
| 19.09 +47.41 | TYC3545-02752-1 | 2 | 208 |
| 18.80 +48.77 | TYC3548-01129-1 | 2 | 208 |
| 19.28 +51.92 | TYC3554-01214-1 | 2 | 208 |
| 19.99 +46.84 | TYC3558-00901-1 | 2 | 208 |
| 19.87 +47.32 | TYC3561-00822-1 | 2 | 208 |
| 08.22 +55.98 | TYC3799-00767-1 | 2 | 208 |
| 20.34 +59.75 | TYC3949-00089-1 | 2 | 208 |
| 07.59 -13.95 | TYC5409-02888-1 | 2 | 208 |
| 19.12 +44.84 | TYC3132-00232-1 | 2 | 208 |
| 18.80 +47.23 | TYC3544-00126-1 | 2 | 208 |
| 06.26 +40.88 | TYC2930-02097-1 | 2 | 207 |
| 18.57 +65.33 | TYC4222-00009-1 | 2 | 207 |
| 03.00 +00.47 | TYC0048-01139-1 | 2 | 207 |
| 05.06 +03.08 | TYC0102-00103-1 | 2 | 207 |
| 03.55 +10.66 | TYC0653-00330-1 | 2 | 207 |
| 08.91 +33.25 | TYC2488-00461-1 | 2 | 207 |
| 03.04 -20.10 | TYC5871-00120-1 | 2 | 207 |
| 18.10 +26.20 | TYC2095-00163-1 | 2 | 207 |
| 07.75 +38.90 | TYC2959-01832-1 | 2 | 207 |
| 19.98 +43.96 | TYC3149-00840-1 | 2 | 207 |
| 07.55 -22.26 | TYC5991-00281-1 | 2 | 206 |
| 19.72 +41.16 | TYC3140-00955-1 | 2 | 206 |
| 18.72 +47.14 | TYC3531-01176-1 | 2 | 206 |
| 18.72 +47.22 | TYC3531-01608-1 | 2 | 206 |
| 18.74 +47.68 | TYC3531-01974-1 | 2 | 206 |
| 18.73 +47.62 | TYC3531-02205-1 | 2 | 206 |
| 15.29 +71.21 | TYC4414-00043-1 | 2 | 206 |
| 03.53 -24.05 | TYC6446-00283-1 | 2 | 206 |
| 04.55 -25.49 | TYC6467-01875-1 | 2 | 206 |
| 03.35 +01.09 | TYC0057-00854-1 | 2 | 206 |
| 05.60 +35.46 | TYC2412-00290-1 | 2 | 206 |
| 18.72 +47.56 | TYC3531-00190-1 | 2 | 206 |
| 19.26 +45.34 | TYC3542-00955-1 | 2 | 206 |
| 05.08 -13.84 | TYC5329-01033-1 | 2 | 206 |
| 03.05 +01.37 | TYC0055-00561-1 | 2 | 206 |
| 18.74 +43.55 | TYC3130-02162-1 | 2 | 206 |
| 00.29 -15.00 | TYC5267-01080-1 | 2 | 206 |
| 19.93 +47.72 | TYC3562-00485-1 | 2 | 206 |
| 06.74 +01.17 | TYC0147-01917-1 | 2 | 205 |
| 07.22 +14.62 | TYC0774-00045-1 | 2 | 205 |
| 04.68 +19.09 | TYC1275-01229-1 | 2 | 205 |
| 22.04 +26.20 | TYC2207-00093-1 | 2 | 205 |
| 09.32 +50.73 | TYC3431-00611-1 | 2 | 205 |
| 04.32 +58.23 | TYC3731-01344-1 | 2 | 205 |
| 05.12 -27.33 | TYC6481-00969-1 | 2 | 205 |
| 06.58 -27.15 | TYC6516-01636-1 | 2 | 205 |
| 21.26 +14.74 | TYC1117-01793-1 | 2 | 205 |
| 00.50 -16.36 | TYC5840-00430-1 | 2 | 205 |
| 00.25 +01.21 | TYC0002-00403-1 | 2 | 205 |
| 07.62 -14.19 | TYC5409-00312-1 | 2 | 205 |
| 00.68 +34.82 | TYC2283-00362-1 | 2 | 204 |
| 04.78 -24.27 | TYC6465-01477-1 | 2 | 204 |
| 19.49 +48.74 | TYC3547-00028-1 | 2 | 203 |
| 09.27 +50.68 | TYC3431-00667-1 | 2 | 203 |
| 09.40 +19.96 | TYC1405-00082-1 | 2 | 203 |
| 01.45 +28.78 | TYC1754-00161-1 | 2 | 203 |
| 03.18 -01.03 | TYC4708-00428-1 | 2 | 203 |
| 11.00 -23.76 | TYC6635-00682-1 | 2 | 203 |
| 18.34 -11.81 | TYC5685-03152-1 | 2 | 202 |
| 09.56 +30.18 | TYC2494-00077-1 | 2 | 202 |
| 19.48 +41.25 | TYC3138-00205-1 | 2 | 202 |
| 05.12 -27.05 | TYC6481-01220-1 | 2 | 202 |
| 23.24 +31.32 | TYC2751-00241-1 | 2 | 202 |
| 19.58 +41.94 | TYC3143-00016-1 | 2 | 202 |
| 19.22 +38.95 | TYC3121-00085-1 | 2 | 202 |
| 19.93 +48.35 | TYC3562-00660-1 | 2 | 202 |
| 19.50 +47.40 | TYC3547-02062-1 | 2 | 202 |
| 21.09 +03.72 | TYC0530-02118-1 | 2 | 202 |
| 21.09 +03.99 | TYC0534-00770-1 | 2 | 202 |
| 22.18 +16.20 | TYC1681-00515-1 | 2 | 202 |
| 22.03 +18.94 | TYC1688-01697-1 | 2 | 202 |
| 01.54 +29.33 | TYC1755-01726-1 | 2 | 202 |
| 00.61 +34.73 | TYC2283-00345-1 | 2 | 202 |
| 02.00 +46.42 | TYC3280-00418-1 | 2 | 202 |
| 20.35 +59.32 | TYC3949-01869-1 | 2 | 202 |
| 05.26 +79.26 | TYC4519-00808-1 | 2 | 202 |
| 23.61 +77.84 | TYC4606-01114-1 | 2 | 202 |
| 02.55 -12.62 | TYC5284-01283-1 | 2 | 202 |
| 01.08 -22.45 | TYC5853-00909-1 | 2 | 202 |
| 21.24 -21.08 | TYC6359-01252-1 | 2 | 202 |
| 21.52 -21.06 | TYC6373-00896-1 | 2 | 202 |
| 23.23 -22.46 | TYC6401-01120-1 | 2 | 202 |
| 03.36 -28.86 | TYC6452-01125-1 | 2 | 202 |
| 23.21 -23.01 | TYC6972-00470-1 | 2 | 202 |
| 00.99 +20.27 | TYC1195-01632-1 | 2 | 202 |
| 01.75 +19.89 | TYC1208-00649-1 | 2 | 202 |
| 03.23 +25.55 | TYC1788-00578-1 | 2 | 202 |
| 12.53 +22.97 | TYC1989-01906-1 | 2 | 202 |
| 02.36 +32.57 | TYC2314-01115-1 | 2 | 202 |
| 02.64 +42.20 | TYC2840-01491-1 | 2 | 202 |
| 02.68 +38.76 | TYC2845-01479-1 | 2 | 202 |
| 22.97 +38.32 | TYC3215-00834-1 | 2 | 202 |
| 02.87 +71.56 | TYC4321-02122-1 | 2 | 202 |
| 05.28 +79.39 | TYC4519-00896-1 | 2 | 202 |
| 12.13 +77.25 | TYC4550-00081-1 | 2 | 202 |
| 02.54 -12.46 | TYC5284-01163-1 | 2 | 202 |
| 02.58 -12.75 | TYC5291-00474-1 | 2 | 202 |
| 03.02 -21.13 | TYC5870-00532-1 | 2 | 202 |
| 00.73 -26.83 | TYC6423-01626-1 | 2 | 202 |
| 03.33 -29.14 | TYC6445-00237-1 | 2 | 202 |
| 05.11 -26.57 | TYC6481-00076-1 | 2 | 202 |
| 16.86 +11.76 | TYC0983-01201-1 | 2 | 202 |
| 17.88 +56.83 | TYC3910-00224-1 | 2 | 202 |
| 10.22 +08.38 | TYC0837-00528-1 | 2 | 201 |
| 10.23 +08.49 | TYC0837-00603-1 | 2 | 201 |
| 19.62 +38.86 | TYC3135-00290-1 | 2 | 201 |
| 10.95 +48.61 | TYC3446-00513-1 | 2 | 201 |
| 10.96 +48.80 | TYC3446-00888-1 | 2 | 201 |
| 18.70 +47.80 | TYC3531-01691-1 | 2 | 201 |
| 22.95 +20.91 | TYC1717-00021-1 | 2 | 200 |
| 19.70 +42.49 | TYC3144-01102-1 | 2 | 200 |
| 19.67 +42.45 | TYC3144-02551-1 | 2 | 200 |
| 19.62 +48.50 | TYC3560-00602-1 | 2 | 200 |
| 04.85 +06.02 | TYC0096-00942-1 | 2 | 200 |
| 03.15 +29.88 | TYC1795-00191-1 | 2 | 200 |
| 02.62 +38.84 | TYC2832-01754-1 | 2 | 200 |
| 18.91 +41.29 | TYC3127-01062-1 | 2 | 200 |
| 19.43 +38.57 | TYC3134-00161-1 | 2 | 200 |
| 18.95 +45.22 | TYC3541-02639-1 | 2 | 200 |
| 18.95 +45.11 | TYC3541-02654-1 | 2 | 200 |
| 19.46 +46.48 | TYC3543-00129-1 | 2 | 200 |
| 19.35 +45.75 | TYC3543-01474-1 | 2 | 200 |
| 09.75 -27.19 | TYC6610-00287-1 | 2 | 200 |
| 04.41 +25.41 | TYC1820-00805-1 | 2 | 200 |
| 00.64 +34.75 | TYC2283-00664-1 | 2 | 200 |
| 19.58 +38.93 | TYC3135-00343-1 | 2 | 200 |
| 19.57 +38.87 | TYC3135-00739-1 | 2 | 200 |
| 08.27 +61.35 | TYC4126-02333-1 | 2 | 200 |
| 19.16 +43.44 | TYC3132-01389-1 | 2 | 199 |
| 19.35 +40.20 | TYC3138-00262-1 | 2 | 199 |
| 19.21 +50.54 | TYC3550-00375-1 | 2 | 199 |
| 08.83 -19.24 | TYC6021-00720-1 | 2 | 199 |
| 19.36 +45.84 | TYC3543-02874-1 | 2 | 199 |
| 06.05 +44.35 | TYC2937-01313-1 | 2 | 199 |
| 19.16 +43.56 | TYC3132-01175-1 | 2 | 198 |
| 03.01 -27.43 | TYC6441-00258-1 | 2 | 198 |
| 09.52 -11.31 | TYC5468-00974-1 | 2 | 197 |
| 03.73 +00.29 | TYC0065-01111-1 | 2 | 197 |
| 03.66 +04.84 | TYC0067-01546-1 | 2 | 197 |
| 04.16 +04.98 | TYC0076-00439-1 | 2 | 197 |
| 04.45 +07.09 | TYC0081-00701-1 | 2 | 197 |
| 05.88 +02.85 | TYC0121-00062-1 | 2 | 197 |
| 05.51 +06.37 | TYC0126-01457-1 | 2 | 197 |
| 06.98 +05.50 | TYC0157-00313-1 | 2 | 197 |
| 06.52 +05.80 | TYC0158-03110-1 | 2 | 197 |
| 07.17 +01.73 | TYC0163-00342-1 | 2 | 197 |
| 07.77 +00.76 | TYC0180-01581-1 | 2 | 197 |
| 07.95 +01.52 | TYC0181-01473-1 | 2 | 197 |
| 07.61 +05.99 | TYC0190-01068-1 | 2 | 197 |
| 08.42 +01.08 | TYC0197-00617-1 | 2 | 197 |
| 08.04 +04.21 | TYC0202-00060-1 | 2 | 197 |
| 08.47 +06.93 | TYC0209-00295-1 | 2 | 197 |
| 08.66 +06.87 | TYC0223-01232-1 | 2 | 197 |
| 02.58 +12.06 | TYC0642-00833-1 | 2 | 197 |
| 03.49 +09.74 | TYC0650-01264-1 | 2 | 197 |
| 03.52 +13.68 | TYC0656-00523-1 | 2 | 197 |
| 03.90 +08.62 | TYC0658-00940-1 | 2 | 197 |
| 04.02 +14.26 | TYC0665-00272-1 | 2 | 197 |
| 03.95 +13.09 | TYC0665-00425-1 | 2 | 197 |
| 04.35 +08.37 | TYC0668-01176-1 | 2 | 197 |
| 04.97 +09.64 | TYC0688-01256-1 | 2 | 197 |
| 05.51 +08.59 | TYC0701-01546-1 | 2 | 197 |
| 05.49 +10.33 | TYC0704-01310-1 | 2 | 197 |
| 05.27 +13.03 | TYC0707-00654-1 | 2 | 197 |
| 05.94 +10.44 | TYC0720-00004-1 | 2 | 197 |
| 05.89 +11.99 | TYC0724-00720-1 | 2 | 197 |
| 06.14 +09.58 | TYC0734-01073-1 | 2 | 197 |
| 06.49 +10.98 | TYC0736-00997-1 | 2 | 197 |
| 07.05 +09.14 | TYC0749-02886-1 | 2 | 197 |
| 08.24 +13.05 | TYC0802-01776-1 | 2 | 197 |
| 08.59 +14.10 | TYC0809-01217-1 | 2 | 197 |
| 08.78 +07.89 | TYC0810-00022-1 | 2 | 197 |
| 09.02 +10.58 | TYC0814-00822-1 | 2 | 197 |
| 09.17 +10.38 | TYC0815-00562-1 | 2 | 197 |
| 09.59 +09.89 | TYC0821-00003-1 | 2 | 197 |
| 09.28 +13.69 | TYC0825-01072-1 | 2 | 197 |
| 09.97 +10.03 | TYC0832-00872-1 | 2 | 197 |
| 10.25 +09.14 | TYC0837-00198-1 | 2 | 197 |
| 20.66 +11.20 | TYC1092-00263-1 | 2 | 197 |
| 20.65 +11.23 | TYC1092-00429-1 | 2 | 197 |
| 21.22 +14.78 | TYC1117-01119-1 | 2 | 197 |
| 02.98 +15.27 | TYC1224-01407-1 | 2 | 197 |
| 03.02 +16.46 | TYC1225-01182-1 | 2 | 197 |
| 02.79 +19.80 | TYC1226-01359-1 | 2 | 197 |
| 03.56 +15.54 | TYC1234-00493-1 | 2 | 197 |
| 03.31 +18.08 | TYC1236-00742-1 | 2 | 197 |
| 03.42 +17.83 | TYC1237-00220-1 | 2 | 197 |
| 03.50 +18.59 | TYC1238-00341-1 | 2 | 197 |
| 04.01 +20.38 | TYC1258-00341-1 | 2 | 197 |
| 04.44 +16.75 | TYC1265-00241-1 | 2 | 197 |
| 04.52 +19.88 | TYC1273-01001-1 | 2 | 197 |
| 05.27 +16.87 | TYC1283-01297-1 | 2 | 197 |
| 05.07 +18.98 | TYC1290-00981-1 | 2 | 197 |
| 05.24 +20.72 | TYC1295-00130-1 | 2 | 197 |
| 05.46 +18.74 | TYC1300-01006-1 | 2 | 197 |
| 06.33 +16.18 | TYC1315-00870-1 | 2 | 197 |
| 06.28 +17.52 | TYC1319-01382-1 | 2 | 197 |
| 07.15 +15.60 | TYC1345-00419-1 | 2 | 197 |
| 07.33 +21.09 | TYC1358-02172-1 | 2 | 197 |
| 08.10 +15.53 | TYC1376-01480-1 | 2 | 197 |
| 08.39 +21.91 | TYC1390-00319-1 | 2 | 197 |
| 09.09 +19.28 | TYC1404-00868-1 | 2 | 197 |
| 09.30 +19.38 | TYC1405-00879-1 | 2 | 197 |
| 13.79 +17.31 | TYC1460-00578-1 | 2 | 197 |
| 02.98 +26.86 | TYC1790-00782-1 | 2 | 197 |
| 04.86 +26.77 | TYC1839-00506-1 | 2 | 197 |
| 05.22 +24.36 | TYC1846-01555-1 | 2 | 197 |
| 05.91 +26.75 | TYC1871-01467-1 | 2 | 197 |
| 05.93 +27.98 | TYC1871-01541-1 | 2 | 197 |
| 06.60 +26.54 | TYC1888-00163-1 | 2 | 197 |
| 06.53 +29.51 | TYC1891-00612-1 | 2 | 197 |
| 07.37 +27.03 | TYC1917-01984-1 | 2 | 197 |
| 07.89 +22.68 | TYC1925-00497-1 | 2 | 197 |
| 09.42 +24.69 | TYC1952-01065-1 | 2 | 197 |
| 09.04 +25.86 | TYC1953-01293-1 | 2 | 197 |
| 09.81 +25.01 | TYC1963-01549-1 | 2 | 197 |
| 10.24 +22.57 | TYC1969-00825-1 | 2 | 197 |
| 11.37 +24.76 | TYC1981-00606-1 | 2 | 197 |
| 19.61 +28.51 | TYC2150-04889-1 | 2 | 197 |
| 22.06 +26.48 | TYC2212-01395-1 | 2 | 197 |
| 00.35 +31.82 | TYC2261-00848-1 | 2 | 197 |
| 02.98 +35.67 | TYC2338-00881-1 | 2 | 197 |
| 03.29 +31.03 | TYC2340-00960-1 | 2 | 197 |
| 03.31 +32.68 | TYC2345-01922-1 | 2 | 197 |
| 04.60 +32.46 | TYC2377-01340-1 | 2 | 197 |
| 05.04 +30.04 | TYC2388-01460-1 | 2 | 197 |
| 05.66 +31.08 | TYC2404-00070-1 | 2 | 197 |
| 05.68 +32.94 | TYC2408-01118-1 | 2 | 197 |
| 06.01 +32.12 | TYC2423-00097-1 | 2 | 197 |
| 06.43 +33.06 | TYC2425-00641-1 | 2 | 197 |
| 06.37 +34.59 | TYC2429-00504-1 | 2 | 197 |
| 06.72 +34.12 | TYC2443-00203-1 | 2 | 197 |
| 07.27 +35.14 | TYC2459-00003-1 | 2 | 197 |
| 07.90 +32.33 | TYC2471-00002-1 | 2 | 197 |
| 08.15 +34.71 | TYC2477-00080-1 | 2 | 197 |
| 08.71 +31.86 | TYC2484-00592-1 | 2 | 197 |
| 09.54 +34.40 | TYC2497-00735-1 | 2 | 197 |
| 09.73 +33.49 | TYC2504-00068-1 | 2 | 197 |
| 09.80 +37.40 | TYC2508-00361-1 | 2 | 197 |
| 16.01 +33.32 | TYC2576-00693-1 | 2 | 197 |
| 19.08 +37.48 | TYC2652-00223-1 | 2 | 197 |
| 19.49 +37.49 | TYC2666-00023-1 | 2 | 197 |
| 21.63 +30.67 | TYC2717-01833-1 | 2 | 197 |
| 02.44 +37.58 | TYC2831-01098-1 | 2 | 197 |
| 02.44 +37.70 | TYC2831-01320-1 | 2 | 197 |
| 03.86 +44.99 | TYC2876-00288-1 | 2 | 197 |
| 08.75 +44.35 | TYC2988-00198-1 | 2 | 197 |
| 18.98 +39.11 | TYC3119-00144-1 | 2 | 197 |
| 18.97 +39.18 | TYC3119-00347-1 | 2 | 197 |
| 18.96 +39.18 | TYC3119-00397-1 | 2 | 197 |
| 19.02 +39.05 | TYC3120-00020-1 | 2 | 197 |
| 19.01 +39.07 | TYC3120-00038-1 | 2 | 197 |
| 19.12 +38.96 | TYC3120-00285-1 | 2 | 197 |
| 19.12 +39.25 | TYC3120-00377-1 | 2 | 197 |
| 19.17 +38.98 | TYC3120-00531-1 | 2 | 197 |
| 19.15 +38.73 | TYC3120-00677-1 | 2 | 197 |
| 19.03 +39.37 | TYC3120-01722-1 | 2 | 197 |
| 19.28 +38.96 | TYC3121-00077-1 | 2 | 197 |
| 18.93 +40.02 | TYC3123-00515-1 | 2 | 197 |
| 18.93 +39.83 | TYC3123-00675-1 | 2 | 197 |
| 18.97 +39.96 | TYC3123-00789-1 | 2 | 197 |
| 18.89 +40.11 | TYC3123-01075-1 | 2 | 197 |
| 18.89 +40.50 | TYC3123-01120-1 | 2 | 197 |
| 18.87 +41.11 | TYC3123-01582-1 | 2 | 197 |
| 18.95 +41.12 | TYC3123-01598-1 | 2 | 197 |
| 19.07 +39.38 | TYC3124-00602-1 | 2 | 197 |
| 19.13 +40.31 | TYC3124-01125-1 | 2 | 197 |
| 19.31 +39.93 | TYC3125-00287-1 | 2 | 197 |
| 19.22 +40.98 | TYC3125-00360-1 | 2 | 197 |
| 19.20 +40.45 | TYC3125-01096-1 | 2 | 197 |
| 18.75 +42.64 | TYC3126-00537-1 | 2 | 197 |
| 18.98 +42.63 | TYC3127-00794-1 | 2 | 197 |
| 18.91 +41.88 | TYC3127-01759-1 | 2 | 197 |
| 19.14 +41.58 | TYC3128-00194-1 | 2 | 197 |
| 19.16 +43.05 | TYC3128-00729-1 | 2 | 197 |
| 19.13 +41.96 | TYC3128-01388-1 | 2 | 197 |
| 19.12 +41.95 | TYC3128-01782-1 | 2 | 197 |
| 18.77 +44.27 | TYC3130-00467-1 | 2 | 197 |
| 18.77 +44.15 | TYC3130-01141-1 | 2 | 197 |
| 18.75 +44.20 | TYC3130-01172-1 | 2 | 197 |
| 18.83 +44.32 | TYC3130-01475-1 | 2 | 197 |
| 18.87 +45.00 | TYC3131-00006-1 | 2 | 197 |
| 18.86 +44.16 | TYC3131-00306-1 | 2 | 197 |
| 18.86 +44.33 | TYC3131-00386-1 | 2 | 197 |
| 18.86 +43.59 | TYC3131-00641-1 | 2 | 197 |
| 18.99 +44.99 | TYC3131-00760-1 | 2 | 197 |
| 18.88 +43.87 | TYC3131-00790-1 | 2 | 197 |
| 19.05 +44.97 | TYC3132-00004-1 | 2 | 197 |
| 19.07 +44.36 | TYC3132-00096-1 | 2 | 197 |
| 19.07 +44.29 | TYC3132-00348-1 | 2 | 197 |
| 19.10 +43.76 | TYC3132-00997-1 | 2 | 197 |
| 19.06 +44.58 | TYC3132-01102-1 | 2 | 197 |
| 19.03 +44.26 | TYC3132-01372-1 | 2 | 197 |
| 19.32 +44.96 | TYC3133-00001-1 | 2 | 197 |
| 19.31 +44.96 | TYC3133-00010-1 | 2 | 197 |
| 19.32 +43.99 | TYC3133-00068-1 | 2 | 197 |
| 19.33 +44.81 | TYC3133-00121-1 | 2 | 197 |
| 19.45 +37.92 | TYC3134-00002-1 | 2 | 197 |
| 19.42 +39.11 | TYC3134-00011-1 | 2 | 197 |
| 19.43 +37.95 | TYC3134-00020-1 | 2 | 197 |
| 19.44 +37.70 | TYC3134-00036-1 | 2 | 197 |
| 19.50 +37.73 | TYC3134-00074-1 | 2 | 197 |
| 19.42 +38.72 | TYC3134-00175-1 | 2 | 197 |
| 19.60 +38.72 | TYC3135-00087-1 | 2 | 197 |
| 19.66 +39.08 | TYC3135-00233-1 | 2 | 197 |
| 19.62 +38.76 | TYC3135-00271-1 | 2 | 197 |
| 19.59 +39.12 | TYC3135-00286-1 | 2 | 197 |
| 19.71 +38.89 | TYC3136-00087-1 | 2 | 197 |
| 19.71 +39.10 | TYC3136-01030-1 | 2 | 197 |
| 19.35 +40.95 | TYC3138-00025-1 | 2 | 197 |
| 19.40 +40.40 | TYC3138-00064-1 | 2 | 197 |
| 19.37 +40.15 | TYC3138-00164-1 | 2 | 197 |
| 19.64 +40.19 | TYC3139-00079-1 | 2 | 197 |
| 19.66 +40.03 | TYC3139-00101-1 | 2 | 197 |
| 19.83 +41.15 | TYC3140-00075-1 | 2 | 197 |
| 19.81 +40.74 | TYC3140-00409-1 | 2 | 197 |
| 19.79 +40.73 | TYC3140-01229-1 | 2 | 197 |
| 19.70 +40.32 | TYC3140-01505-1 | 2 | 197 |
| 19.75 +40.29 | TYC3140-01566-1 | 2 | 197 |
| 19.87 +39.88 | TYC3141-00591-1 | 2 | 197 |
| 19.99 +40.96 | TYC3141-00800-1 | 2 | 197 |
| 19.97 +40.32 | TYC3141-03186-1 | 2 | 197 |
| 19.86 +40.54 | TYC3141-03443-1 | 2 | 197 |
| 19.35 +43.06 | TYC3142-00004-1 | 2 | 197 |
| 19.45 +42.73 | TYC3142-00010-1 | 2 | 197 |
| 19.36 +42.92 | TYC3142-00012-1 | 2 | 197 |
| 19.39 +42.27 | TYC3142-00017-1 | 2 | 197 |
| 19.48 +41.60 | TYC3142-00025-1 | 2 | 197 |
| 19.46 +43.06 | TYC3142-00032-1 | 2 | 197 |
| 19.45 +42.68 | TYC3142-00074-1 | 2 | 197 |
| 19.50 +42.73 | TYC3142-00082-1 | 2 | 197 |
| 19.48 +43.06 | TYC3142-00092-1 | 2 | 197 |
| 19.36 +43.10 | TYC3142-00158-1 | 2 | 197 |
| 19.47 +43.00 | TYC3142-00168-1 | 2 | 197 |
| 19.46 +43.08 | TYC3142-00242-1 | 2 | 197 |
| 19.44 +42.25 | TYC3142-00255-1 | 2 | 197 |
| 19.44 +42.48 | TYC3142-00722-1 | 2 | 197 |
| 19.61 +42.47 | TYC3143-00277-1 | 2 | 197 |
| 19.52 +42.35 | TYC3143-01021-1 | 2 | 197 |
| 19.66 +42.76 | TYC3143-01113-1 | 2 | 197 |
| 19.55 +41.47 | TYC3143-01438-1 | 2 | 197 |
| 19.55 +41.51 | TYC3143-01514-1 | 2 | 197 |
| 19.80 +41.90 | TYC3144-00088-1 | 2 | 197 |
| 19.81 +43.10 | TYC3144-00153-1 | 2 | 197 |
| 19.82 +41.94 | TYC3144-00238-1 | 2 | 197 |
| 19.80 +42.00 | TYC3144-00254-1 | 2 | 197 |
| 19.80 +41.71 | TYC3144-00262-1 | 2 | 197 |
| 19.81 +43.08 | TYC3144-00439-1 | 2 | 197 |
| 19.72 +43.05 | TYC3144-00770-1 | 2 | 197 |
| 19.74 +42.12 | TYC3144-02313-1 | 2 | 197 |
| 19.87 +41.72 | TYC3145-00562-1 | 2 | 197 |
| 19.86 +43.07 | TYC3145-00633-1 | 2 | 197 |
| 19.49 +44.48 | TYC3146-00011-1 | 2 | 197 |
| 19.42 +44.50 | TYC3146-00035-1 | 2 | 197 |
| 19.44 +44.51 | TYC3146-00077-1 | 2 | 197 |
| 19.38 +44.91 | TYC3146-00082-1 | 2 | 197 |
| 19.38 +44.77 | TYC3146-00136-1 | 2 | 197 |
| 19.48 +43.88 | TYC3146-00401-1 | 2 | 197 |
| 19.48 +43.76 | TYC3146-00518-1 | 2 | 197 |
| 19.66 +43.80 | TYC3147-00226-1 | 2 | 197 |
| 19.65 +43.84 | TYC3147-00272-1 | 2 | 197 |
| 19.67 +44.02 | TYC3147-00534-1 | 2 | 197 |
| 19.79 +43.82 | TYC3148-00248-1 | 2 | 197 |
| 19.74 +44.59 | TYC3148-00819-1 | 2 | 197 |
| 19.74 +44.67 | TYC3148-01037-1 | 2 | 197 |
| 19.92 +44.97 | TYC3149-00067-1 | 2 | 197 |
| 19.89 +44.72 | TYC3149-00094-1 | 2 | 197 |
| 19.98 +43.77 | TYC3149-00249-1 | 2 | 197 |
| 19.99 +43.77 | TYC3149-00264-1 | 2 | 197 |
| 19.87 +44.86 | TYC3149-00423-1 | 2 | 197 |
| 19.89 +44.70 | TYC3149-00447-1 | 2 | 197 |
| 19.96 +44.62 | TYC3149-00913-1 | 2 | 197 |
| 19.94 +43.61 | TYC3149-00952-1 | 2 | 197 |
| 19.99 +43.51 | TYC3149-01148-1 | 2 | 197 |
| 19.92 +44.14 | TYC3149-01229-1 | 2 | 197 |
| 20.06 +44.24 | TYC3162-00072-1 | 2 | 197 |
| 20.07 +44.16 | TYC3162-00094-1 | 2 | 197 |
| 20.09 +44.12 | TYC3162-00162-1 | 2 | 197 |
| 20.08 +44.72 | TYC3162-00281-1 | 2 | 197 |
| 20.04 +44.53 | TYC3162-00517-1 | 2 | 197 |
| 20.03 +44.31 | TYC3162-00552-1 | 2 | 197 |
| 20.05 +44.37 | TYC3162-01075-1 | 2 | 197 |
| 20.09 +44.03 | TYC3162-01747-1 | 2 | 197 |
| 02.77 +49.72 | TYC3304-00381-1 | 2 | 197 |
| 02.78 +50.04 | TYC3304-01166-1 | 2 | 197 |
| 03.43 +45.49 | TYC3312-00114-1 | 2 | 197 |
| 03.16 +48.56 | TYC3314-00886-1 | 2 | 197 |
| 03.07 +50.83 | TYC3322-00708-1 | 2 | 197 |
| 05.38 +47.00 | TYC3362-01743-1 | 2 | 197 |
| 06.66 +46.59 | TYC3377-00366-1 | 2 | 197 |
| 06.33 +52.30 | TYC3387-00656-1 | 2 | 197 |
| 06.88 +49.84 | TYC3398-00198-1 | 2 | 197 |
| 07.94 +49.28 | TYC3410-02866-1 | 2 | 197 |
| 07.88 +50.35 | TYC3413-00201-1 | 2 | 197 |
| 08.51 +52.19 | TYC3422-00138-1 | 2 | 197 |
| 09.77 +45.74 | TYC3433-00511-1 | 2 | 197 |
| 10.00 +48.02 | TYC3437-00196-1 | 2 | 197 |
| 18.75 +46.76 | TYC3527-01639-1 | 2 | 197 |
| 18.73 +47.43 | TYC3531-01336-1 | 2 | 197 |
| 18.91 +46.84 | TYC3540-00254-1 | 2 | 197 |
| 18.92 +45.16 | TYC3540-00517-1 | 2 | 197 |
| 18.89 +45.44 | TYC3540-02280-1 | 2 | 197 |
| 18.90 +45.98 | TYC3540-02454-1 | 2 | 197 |
| 18.89 +45.34 | TYC3540-02596-1 | 2 | 197 |
| 18.89 +45.75 | TYC3540-03070-1 | 2 | 197 |
| 18.88 +45.52 | TYC3540-03078-1 | 2 | 197 |
| 18.96 +46.03 | TYC3541-00094-1 | 2 | 197 |
| 19.02 +46.49 | TYC3541-00113-1 | 2 | 197 |
| 19.02 +45.64 | TYC3541-00932-1 | 2 | 197 |
| 19.01 +46.42 | TYC3541-00973-1 | 2 | 197 |
| 18.95 +45.67 | TYC3541-01666-1 | 2 | 197 |
| 19.02 +46.37 | TYC3541-02171-1 | 2 | 197 |
| 19.10 +46.76 | TYC3541-02930-1 | 2 | 197 |
| 19.15 +46.70 | TYC3542-00038-1 | 2 | 197 |
| 19.18 +46.86 | TYC3542-00046-1 | 2 | 197 |
| 19.21 +46.87 | TYC3542-00073-1 | 2 | 197 |
| 19.20 +46.40 | TYC3542-00098-1 | 2 | 197 |
| 19.18 +46.83 | TYC3542-00157-1 | 2 | 197 |
| 19.18 +46.39 | TYC3542-00176-1 | 2 | 197 |
| 19.19 +46.80 | TYC3542-00400-1 | 2 | 197 |
| 19.29 +46.10 | TYC3542-00964-1 | 2 | 197 |
| 19.28 +45.29 | TYC3542-01207-1 | 2 | 197 |
| 19.27 +46.11 | TYC3542-01224-1 | 2 | 197 |
| 19.25 +46.05 | TYC3542-01372-1 | 2 | 197 |
| 19.27 +45.98 | TYC3542-01644-1 | 2 | 197 |
| 19.42 +46.86 | TYC3543-00154-1 | 2 | 197 |
| 19.42 +45.93 | TYC3543-01026-1 | 2 | 197 |
| 18.90 +47.07 | TYC3544-00138-1 | 2 | 197 |
| 18.84 +48.38 | TYC3544-01066-1 | 2 | 197 |
| 18.82 +48.18 | TYC3544-01204-1 | 2 | 197 |
| 18.77 +47.12 | TYC3544-01245-1 | 2 | 197 |
| 18.90 +47.76 | TYC3544-02201-1 | 2 | 197 |
| 18.75 +47.30 | TYC3544-02521-1 | 2 | 197 |
| 18.76 +47.43 | TYC3544-02531-1 | 2 | 197 |
| 18.91 +47.50 | TYC3544-02548-1 | 2 | 197 |
| 18.85 +47.59 | TYC3544-02584-1 | 2 | 197 |
| 19.02 +48.04 | TYC3545-00040-1 | 2 | 197 |
| 19.06 +47.77 | TYC3545-00132-1 | 2 | 197 |
| 19.06 +48.51 | TYC3545-00135-1 | 2 | 197 |
| 18.98 +48.54 | TYC3545-00273-1 | 2 | 197 |
| 19.06 +47.76 | TYC3545-00558-1 | 2 | 197 |
| 19.08 +48.36 | TYC3545-01033-1 | 2 | 197 |
| 19.17 +48.74 | TYC3546-00028-1 | 2 | 197 |
| 19.21 +47.33 | TYC3546-00042-1 | 2 | 197 |
| 19.19 +47.05 | TYC3546-00141-1 | 2 | 197 |
| 19.23 +47.26 | TYC3546-00221-1 | 2 | 197 |
| 19.13 +48.26 | TYC3546-00808-1 | 2 | 197 |
| 19.21 +47.61 | TYC3546-02321-1 | 2 | 197 |
| 19.38 +48.00 | TYC3547-00432-1 | 2 | 197 |
| 19.44 +48.49 | TYC3547-00683-1 | 2 | 197 |
| 19.46 +48.18 | TYC3547-00869-1 | 2 | 197 |
| 19.45 +47.80 | TYC3547-00971-1 | 2 | 197 |
| 19.47 +47.57 | TYC3547-01761-1 | 2 | 197 |
| 19.48 +47.57 | TYC3547-01762-1 | 2 | 197 |
| 19.42 +47.61 | TYC3547-02857-1 | 2 | 197 |
| 18.96 +49.49 | TYC3549-01573-1 | 2 | 197 |
| 18.97 +49.42 | TYC3549-01627-1 | 2 | 197 |
| 19.02 +49.34 | TYC3549-02380-1 | 2 | 197 |
| 19.05 +49.60 | TYC3549-02531-1 | 2 | 197 |
| 19.18 +49.36 | TYC3550-00074-1 | 2 | 197 |
| 19.28 +50.53 | TYC3550-00087-1 | 2 | 197 |
| 19.28 +50.22 | TYC3550-00277-1 | 2 | 197 |
| 19.31 +49.94 | TYC3550-00295-1 | 2 | 197 |
| 19.30 +50.56 | TYC3550-00313-1 | 2 | 197 |
| 19.48 +49.16 | TYC3551-00281-1 | 2 | 197 |
| 19.46 +49.53 | TYC3551-00970-1 | 2 | 197 |
| 19.21 +51.05 | TYC3554-00090-1 | 2 | 197 |
| 19.19 +51.05 | TYC3554-00253-1 | 2 | 197 |
| 19.33 +51.16 | TYC3555-00277-1 | 2 | 197 |
| 19.63 +46.79 | TYC3556-00054-1 | 2 | 197 |
| 19.54 +45.74 | TYC3556-00397-1 | 2 | 197 |
| 19.60 +46.82 | TYC3556-01482-1 | 2 | 197 |
| 19.69 +46.35 | TYC3556-03273-1 | 2 | 197 |
| 19.68 +46.27 | TYC3556-03496-1 | 2 | 197 |
| 19.78 +46.73 | TYC3557-00013-1 | 2 | 197 |
| 19.86 +45.26 | TYC3557-00162-1 | 2 | 197 |
| 19.76 +46.66 | TYC3557-00274-1 | 2 | 197 |
| 19.77 +46.77 | TYC3557-00385-1 | 2 | 197 |
| 19.70 +46.35 | TYC3557-00716-1 | 2 | 197 |
| 19.87 +46.56 | TYC3557-00727-1 | 2 | 197 |
| 19.87 +46.57 | TYC3557-01001-1 | 2 | 197 |
| 19.79 +46.52 | TYC3557-01525-1 | 2 | 197 |
| 19.88 +46.12 | TYC3558-00778-1 | 2 | 197 |
| 19.88 +46.13 | TYC3558-00926-1 | 2 | 197 |
| 19.91 +46.30 | TYC3558-01357-1 | 2 | 197 |
| 20.03 +45.03 | TYC3558-02425-1 | 2 | 197 |
| 20.07 +45.42 | TYC3559-00732-1 | 2 | 197 |
| 19.53 +48.59 | TYC3560-00028-1 | 2 | 197 |
| 19.52 +48.62 | TYC3560-00120-1 | 2 | 197 |
| 19.60 +47.37 | TYC3560-00743-1 | 2 | 197 |
| 19.54 +47.56 | TYC3560-01361-1 | 2 | 197 |
| 19.71 +47.71 | TYC3561-00086-1 | 2 | 197 |
| 19.80 +47.30 | TYC3561-00132-1 | 2 | 197 |
| 19.72 +48.65 | TYC3561-00262-1 | 2 | 197 |
| 19.72 +48.57 | TYC3561-00313-1 | 2 | 197 |
| 19.84 +47.29 | TYC3561-00399-1 | 2 | 197 |
| 19.82 +47.00 | TYC3561-00894-1 | 2 | 197 |
| 19.86 +47.85 | TYC3561-01148-1 | 2 | 197 |
| 19.85 +46.92 | TYC3561-01415-1 | 2 | 197 |
| 19.90 +48.50 | TYC3562-00426-1 | 2 | 197 |
| 19.90 +47.23 | TYC3562-01583-1 | 2 | 197 |
| 19.88 +47.28 | TYC3562-01866-1 | 2 | 197 |
| 19.96 +47.55 | TYC3562-01953-1 | 2 | 197 |
| 19.62 +49.61 | TYC3564-00075-1 | 2 | 197 |
| 19.62 +50.58 | TYC3564-00086-1 | 2 | 197 |
| 19.60 +50.32 | TYC3564-00126-1 | 2 | 197 |
| 19.63 +49.71 | TYC3564-00343-1 | 2 | 197 |
| 19.63 +49.44 | TYC3564-00415-1 | 2 | 197 |
| 19.63 +50.25 | TYC3564-00750-1 | 2 | 197 |
| 19.64 +49.03 | TYC3564-02286-1 | 2 | 197 |
| 19.67 +49.57 | TYC3564-02914-1 | 2 | 197 |
| 19.69 +49.11 | TYC3564-03085-1 | 2 | 197 |
| 19.84 +49.51 | TYC3565-00426-1 | 2 | 197 |
| 19.84 +49.64 | TYC3565-00977-1 | 2 | 197 |
| 19.74 +51.36 | TYC3569-00042-1 | 2 | 197 |
| 19.70 +51.25 | TYC3569-00317-1 | 2 | 197 |
| 03.42 +53.21 | TYC3703-00658-1 | 2 | 197 |
| 03.90 +52.71 | TYC3717-00077-1 | 2 | 197 |
| 04.30 +58.91 | TYC3731-01519-1 | 2 | 197 |
| 05.32 +58.04 | TYC3743-01638-1 | 2 | 197 |
| 05.39 +55.00 | TYC3752-00656-1 | 2 | 197 |
| 05.37 +59.11 | TYC3760-02140-1 | 2 | 197 |
| 05.80 +58.35 | TYC3762-01572-1 | 2 | 197 |
| 06.79 +57.66 | TYC3774-01081-1 | 2 | 197 |
| 07.44 +53.10 | TYC3781-00121-1 | 2 | 197 |
| 07.41 +57.47 | TYC3789-01496-1 | 2 | 197 |
| 07.62 +57.44 | TYC3790-01254-1 | 2 | 197 |
| 07.92 +59.74 | TYC3795-00092-1 | 2 | 197 |
| 09.23 +52.55 | TYC3806-01277-1 | 2 | 197 |
| 09.95 +54.58 | TYC3814-00299-1 | 2 | 197 |
| 04.39 +60.75 | TYC4065-01095-1 | 2 | 197 |
| 03.93 +63.29 | TYC4067-00045-1 | 2 | 197 |
| 06.85 +62.39 | TYC4101-01228-1 | 2 | 197 |
| 06.12 +63.98 | TYC4103-01054-1 | 2 | 197 |
| 06.86 +61.55 | TYC4110-00828-1 | 2 | 197 |
| 07.37 +60.15 | TYC4111-01025-1 | 2 | 197 |
| 07.12 +64.26 | TYC4118-00305-1 | 2 | 197 |
| 09.36 +60.55 | TYC4135-01165-1 | 2 | 197 |
| 06.10 +67.80 | TYC4345-00448-1 | 2 | 197 |
| 06.12 +68.62 | TYC4345-00999-1 | 2 | 197 |
| 06.20 +69.79 | TYC4349-00751-1 | 2 | 197 |
| 06.78 +73.79 | TYC4370-00653-1 | 2 | 197 |
| 08.24 +72.74 | TYC4380-01888-1 | 2 | 197 |
| 08.84 +72.58 | TYC4381-01453-1 | 2 | 197 |
| 06.57 +77.01 | TYC4529-01687-1 | 2 | 197 |
| 06.17 +82.28 | TYC4537-01009-1 | 2 | 197 |
| 04.35 +83.97 | TYC4617-00631-1 | 2 | 197 |
| 06.27 +84.12 | TYC4618-00354-1 | 2 | 197 |
| 03.05 +84.48 | TYC4620-02166-1 | 2 | 197 |
| 02.84 -00.68 | TYC4700-00516-1 | 2 | 197 |
| 05.99 -00.00 | TYC4769-00812-1 | 2 | 197 |
| 14.79 -00.36 | TYC4986-00393-1 | 2 | 197 |
| 02.49 -13.31 | TYC5284-00881-1 | 2 | 197 |
| 02.50 -13.32 | TYC5284-01228-1 | 2 | 197 |
| 04.16 -11.97 | TYC5315-00213-1 | 2 | 197 |
| 04.83 -14.86 | TYC5328-00609-1 | 2 | 197 |
| 08.22 -15.00 | TYC5438-01132-1 | 2 | 197 |
| 09.67 -13.05 | TYC5469-00536-1 | 2 | 197 |
| 09.73 -13.27 | TYC5486-00356-1 | 2 | 197 |
| 13.83 -13.05 | TYC5559-00171-1 | 2 | 197 |
| 22.89 -14.45 | TYC5819-00419-1 | 2 | 197 |
| 22.88 -14.33 | TYC5819-01046-1 | 2 | 197 |
| 03.56 -15.30 | TYC5874-00343-1 | 2 | 197 |
| 03.42 -21.14 | TYC5879-00536-1 | 2 | 197 |
| 04.35 -21.01 | TYC5896-01490-1 | 2 | 197 |
| 04.64 -21.01 | TYC5898-01364-1 | 2 | 197 |
| 05.46 -16.57 | TYC5915-01552-1 | 2 | 197 |
| 05.70 -15.56 | TYC5917-00547-1 | 2 | 197 |
| 05.74 -20.22 | TYC5926-01616-1 | 2 | 197 |
| 06.58 -15.47 | TYC5948-01244-1 | 2 | 197 |
| 06.70 -15.05 | TYC5949-01297-1 | 2 | 197 |
| 07.44 -15.49 | TYC5966-01025-1 | 2 | 197 |
| 07.49 -17.49 | TYC5983-01310-1 | 2 | 197 |
| 07.60 -19.97 | TYC5987-00948-1 | 2 | 197 |
| 07.57 -22.16 | TYC5991-00245-1 | 2 | 197 |
| 08.03 -17.37 | TYC5999-00525-1 | 2 | 197 |
| 08.93 -18.93 | TYC6021-00082-1 | 2 | 197 |
| 03.05 -22.72 | TYC6438-00591-1 | 2 | 197 |
| 03.48 -26.43 | TYC6449-00142-1 | 2 | 197 |
| 03.47 -27.24 | TYC6449-00844-1 | 2 | 197 |
| 03.87 -28.96 | TYC6454-00600-1 | 2 | 197 |
| 03.92 -26.54 | TYC6458-00004-1 | 2 | 197 |
| 04.78 -22.68 | TYC6465-00028-1 | 2 | 197 |
| 05.02 -23.85 | TYC6466-01145-1 | 2 | 197 |
| 04.89 -26.06 | TYC6469-00849-1 | 2 | 197 |
| 04.52 -29.84 | TYC6470-01302-1 | 2 | 197 |
| 05.14 -26.96 | TYC6481-00997-1 | 2 | 197 |
| 05.30 -28.32 | TYC6486-00532-1 | 2 | 197 |
| 05.88 -27.31 | TYC6499-00666-1 | 2 | 197 |
| 06.88 -27.77 | TYC6530-00707-1 | 2 | 197 |
| 06.90 -27.66 | TYC6530-00917-1 | 2 | 197 |
| 07.77 -23.26 | TYC6540-01870-1 | 2 | 197 |
| 08.30 -25.32 | TYC6560-01067-1 | 2 | 197 |
| 08.87 -29.59 | TYC6584-01962-1 | 2 | 197 |
| 09.31 -24.33 | TYC6587-01705-1 | 2 | 197 |
| 09.38 -22.77 | TYC6588-00244-1 | 2 | 197 |
| 22.80 +34.91 | TYC2744-02491-1 | 2 | 197 |
| 22.89 +35.53 | TYC2757-01104-1 | 2 | 197 |
| 19.72 +50.54 | TYC3565-00025-1 | 2 | 197 |
| 02.84 +73.32 | TYC4325-01185-1 | 2 | 197 |
| 06.71 -01.29 | TYC4799-00626-1 | 2 | 197 |
| 06.45 -26.37 | TYC6515-01896-1 | 2 | 197 |
| 08.53 -29.94 | TYC6582-02089-1 | 2 | 197 |
| 12.56 +21.84 | TYC1448-01608-1 | 2 | 197 |
| 03.18 +30.46 | TYC2340-01064-1 | 2 | 197 |
| 09.57 +34.56 | TYC2497-00496-1 | 2 | 197 |
| 02.40 +38.05 | TYC2831-00280-1 | 2 | 197 |
| 06.30 +40.86 | TYC2930-00373-1 | 2 | 197 |
| 02.82 +49.68 | TYC3305-00026-1 | 2 | 197 |
| 08.67 -29.80 | TYC6583-02255-1 | 2 | 197 |
| 09.34 +34.11 | TYC2496-01064-1 | 2 | 197 |
| 00.63 +42.72 | TYC2792-00191-1 | 2 | 197 |
| 18.98 +40.06 | TYC3123-01531-1 | 2 | 197 |
| 08.36 -26.39 | TYC6564-03465-1 | 2 | 197 |
| 08.12 -29.36 | TYC6567-01460-1 | 2 | 197 |
| 10.99 -31.23 | TYC7200-00821-1 | 2 | 197 |
| 10.49 +00.55 | TYC0246-00939-1 | 2 | 197 |
| 21.10 +03.67 | TYC0530-01674-1 | 2 | 196 |
| 21.10 +03.80 | TYC0534-00988-1 | 2 | 196 |
| 06.49 +11.03 | TYC0736-00219-1 | 2 | 196 |
| 13.01 +12.73 | TYC0889-00859-1 | 2 | 196 |
| 03.31 +18.03 | TYC1236-00749-1 | 2 | 196 |
| 19.61 +28.54 | TYC2150-03321-1 | 2 | 196 |
| 19.10 +37.33 | TYC2652-00166-1 | 2 | 196 |
| 06.06 +44.27 | TYC2937-01719-1 | 2 | 196 |
| 19.33 +40.02 | TYC3125-00110-1 | 2 | 196 |
| 19.33 +40.13 | TYC3125-00220-1 | 2 | 196 |
| 19.27 +40.16 | TYC3125-01796-1 | 2 | 196 |
| 19.16 +41.56 | TYC3128-01286-1 | 2 | 196 |
| 19.13 +44.91 | TYC3132-00658-1 | 2 | 196 |
| 19.13 +44.76 | TYC3132-00670-1 | 2 | 196 |
| 19.23 +43.21 | TYC3133-00577-1 | 2 | 196 |
| 19.46 +37.67 | TYC3134-00040-1 | 2 | 196 |
| 19.73 +39.32 | TYC3136-00478-1 | 2 | 196 |
| 19.52 +42.43 | TYC3143-00407-1 | 2 | 196 |
| 19.64 +42.33 | TYC3143-01077-1 | 2 | 196 |
| 19.94 +43.66 | TYC3149-00766-1 | 2 | 196 |
| 20.08 +44.67 | TYC3162-00057-1 | 2 | 196 |
| 20.06 +44.26 | TYC3162-00304-1 | 2 | 196 |
| 03.43 +45.30 | TYC3312-00885-1 | 2 | 196 |
| 19.10 +46.88 | TYC3545-02930-1 | 2 | 196 |
| 19.34 +50.45 | TYC3551-00732-1 | 2 | 196 |
| 19.63 +46.13 | TYC3556-00103-1 | 2 | 196 |
| 19.83 +45.15 | TYC3557-00108-1 | 2 | 196 |
| 19.61 +48.29 | TYC3560-00406-1 | 2 | 196 |
| 19.71 +48.40 | TYC3561-00101-1 | 2 | 196 |
| 06.72 -01.44 | TYC4799-01186-1 | 2 | 196 |
| 20.06 +29.98 | TYC2153-01956-1 | 2 | 196 |
| 21.63 +30.65 | TYC2717-01709-1 | 2 | 196 |
| 19.19 +40.49 | TYC3125-01346-1 | 2 | 196 |
| 19.95 +41.98 | TYC3145-00220-1 | 2 | 196 |
| 19.88 +44.66 | TYC3149-00199-1 | 2 | 196 |
| 09.27 +51.38 | TYC3431-00348-1 | 2 | 196 |
| 18.76 +47.03 | TYC3544-00024-1 | 2 | 196 |
| 18.75 +47.56 | TYC3544-01705-1 | 2 | 196 |
| 06.45 -25.36 | TYC6511-00638-1 | 2 | 196 |
| 14.77 +01.12 | TYC0326-01238-1 | 2 | 196 |
| 15.17 +02.52 | TYC0339-00343-1 | 2 | 196 |
| 08.62 +13.06 | TYC0805-01310-1 | 2 | 196 |
| 09.29 +13.14 | TYC0825-01144-1 | 2 | 196 |
| 09.78 +13.10 | TYC0834-00283-1 | 2 | 196 |
| 15.88 +12.60 | TYC0951-00555-1 | 2 | 196 |
| 00.74 +20.70 | TYC1194-00841-1 | 2 | 196 |
| 08.39 +21.74 | TYC1390-00413-1 | 2 | 196 |
| 08.84 +17.13 | TYC1393-00685-1 | 2 | 196 |
| 09.30 +16.15 | TYC1402-00111-1 | 2 | 196 |
| 13.76 +17.24 | TYC1460-00133-1 | 2 | 196 |
| 01.40 +28.98 | TYC1754-01006-1 | 2 | 196 |
| 07.35 +27.61 | TYC1917-01296-1 | 2 | 196 |
| 07.38 +29.24 | TYC1921-00931-1 | 2 | 196 |
| 08.67 +24.82 | TYC1942-01425-1 | 2 | 196 |
| 09.32 +23.14 | TYC1952-01764-1 | 2 | 196 |
| 03.34 +31.09 | TYC2341-01205-1 | 2 | 196 |
| 15.97 +33.38 | TYC2575-01294-1 | 2 | 196 |
| 02.30 +44.29 | TYC2842-01685-1 | 2 | 196 |
| 06.03 +44.19 | TYC2937-00466-1 | 2 | 196 |
| 07.81 +39.43 | TYC2963-01614-1 | 2 | 196 |
| 16.48 +38.12 | TYC3063-01904-1 | 2 | 196 |
| 19.12 +39.89 | TYC3124-00016-1 | 2 | 196 |
| 19.82 +44.80 | TYC3148-00064-1 | 2 | 196 |
| 07.85 +52.48 | TYC3413-02516-1 | 2 | 196 |
| 08.84 +47.74 | TYC3420-01226-1 | 2 | 196 |
| 19.08 +49.49 | TYC3549-00906-1 | 2 | 196 |
| 20.02 +46.85 | TYC3558-00047-1 | 2 | 196 |
| 18.12 +54.13 | TYC3903-00840-1 | 2 | 196 |
| 08.67 +64.87 | TYC4130-00673-1 | 2 | 196 |
| 08.81 +66.78 | TYC4134-01209-1 | 2 | 196 |
| 06.88 +69.27 | TYC4359-01157-1 | 2 | 196 |
| 13.84 -12.69 | TYC5559-00037-1 | 2 | 196 |
| 01.86 -19.91 | TYC5857-01045-1 | 2 | 196 |
| 09.34 -20.29 | TYC6037-01162-1 | 2 | 196 |
| 03.31 -29.04 | TYC6445-00157-1 | 2 | 196 |
| 09.80 -27.84 | TYC6611-00239-1 | 2 | 196 |
| 03.99 -00.82 | TYC4718-00063-1 | 2 | 196 |
| 09.94 -24.24 | TYC6604-01263-1 | 2 | 196 |
| 10.99 -31.06 | TYC7200-00957-1 | 2 | 196 |
| 11.38 +24.74 | TYC1981-01601-1 | 2 | 195 |
| 19.21 +43.21 | TYC3133-00557-1 | 2 | 195 |
| 19.95 +41.88 | TYC3145-00162-1 | 2 | 195 |
| 20.04 +44.33 | TYC3162-01140-1 | 2 | 195 |
| 18.88 +48.98 | TYC3548-01729-1 | 2 | 195 |
| 20.03 +45.07 | TYC3558-02493-1 | 2 | 195 |
| 19.51 +48.53 | TYC3560-01650-1 | 2 | 195 |
| 16.88 +12.43 | TYC0983-00467-1 | 2 | 195 |
| 19.19 +40.69 | TYC3125-00499-1 | 2 | 195 |
| 03.07 -20.72 | TYC5871-00586-1 | 2 | 195 |
| 04.82 +06.60 | TYC0096-00325-1 | 2 | 195 |
| 19.97 +22.30 | TYC1628-01920-1 | 2 | 195 |
| 19.82 +45.00 | TYC3148-00019-1 | 2 | 195 |
| 19.38 +46.68 | TYC3543-01062-1 | 2 | 195 |
| 19.06 +48.64 | TYC3545-00339-1 | 2 | 195 |
| 09.55 -12.16 | TYC5468-00880-1 | 2 | 195 |
| 20.14 +44.42 | TYC3162-00935-1 | 2 | 195 |
| 12.52 +74.28 | TYC4400-00425-1 | 2 | 195 |
| 11.99 -20.50 | TYC6100-00310-1 | 2 | 195 |
| 09.48 +45.32 | TYC3425-01513-1 | 2 | 195 |
| 04.87 +11.39 | TYC0692-00548-1 | 2 | 194 |
| 05.80 +01.21 | TYC0116-00180-1 | 2 | 194 |
| 03.28 +25.18 | TYC1788-01190-1 | 2 | 194 |
| 08.25 +60.57 | TYC4126-00137-1 | 2 | 194 |
| 11.27 +17.82 | TYC1437-00562-1 | 2 | 194 |
| 10.88 -31.29 | TYC7199-00387-1 | 2 | 194 |
| 16.18 +44.27 | TYC3067-00387-1 | 2 | 194 |
| 19.42 +38.49 | TYC3134-00034-1 | 2 | 194 |
| 10.32 +40.88 | TYC3004-00198-1 | 2 | 193 |
| 06.37 -31.37 | TYC7073-00594-1 | 2 | 193 |
| 08.23 +04.83 | TYC0203-00700-1 | 2 | 193 |
| 15.88 +15.23 | TYC1496-01882-1 | 2 | 193 |
| 05.13 +69.79 | TYC4346-01141-1 | 2 | 193 |
| 19.40 +00.69 | TYC0465-00348-1 | 2 | 193 |
| 19.40 +00.61 | TYC0465-00382-1 | 2 | 193 |
| 17.15 +33.30 | TYC2595-01465-1 | 2 | 193 |
| 17.15 +33.46 | TYC2595-01470-1 | 2 | 193 |
| 18.72 +36.41 | TYC2649-00912-1 | 2 | 193 |
| 18.71 +36.51 | TYC2649-00996-1 | 2 | 193 |
| 19.68 +50.50 | TYC3564-02969-1 | 2 | 193 |
| 18.19 +54.30 | TYC3903-01542-1 | 2 | 193 |
| 20.24 -00.86 | TYC5161-00879-1 | 2 | 193 |
| 22.19 +16.40 | TYC1681-00325-1 | 2 | 193 |
| 23.30 +18.73 | TYC1716-01473-1 | 2 | 193 |
| 22.88 -14.61 | TYC5819-00485-1 | 2 | 193 |
| 20.23 -00.67 | TYC5161-00504-1 | 2 | 192 |
| 01.89 -19.58 | TYC5858-01848-1 | 2 | 192 |
| 20.95 +10.93 | TYC1107-01718-1 | 2 | 192 |
| 21.00 +10.80 | TYC1107-01903-1 | 2 | 192 |
| 19.20 +40.65 | TYC3125-00535-1 | 2 | 191 |
| 11.81 +76.55 | TYC4550-00738-1 | 2 | 191 |
| 06.48 +29.77 | TYC1891-00264-1 | 2 | 191 |
| 11.98 -20.03 | TYC6100-00509-1 | 2 | 190 |
| 07.71 +29.33 | TYC1924-01476-1 | 2 | 190 |
| 19.83 +48.39 | TYC3561-00044-1 | 2 | 190 |
| 05.93 -23.44 | TYC6491-01631-1 | 2 | 190 |
| 11.76 +14.29 | TYC0870-00023-1 | 2 | 190 |
| 04.60 +79.99 | TYC4518-00488-1 | 2 | 190 |
| 19.22 +41.11 | TYC3125-00466-1 | 2 | 190 |
| 00.35 +83.03 | TYC4615-00566-1 | 2 | 189 |
| 12.25 +73.90 | TYC4399-01073-1 | 2 | 189 |
| 03.41 -12.13 | TYC5299-01490-1 | 2 | 189 |
| 06.69 -15.36 | TYC5949-00023-1 | 2 | 189 |
| 08.48 +21.99 | TYC1391-00331-1 | 2 | 189 |
| 07.05 -11.90 | TYC5389-01691-1 | 2 | 188 |
| 19.81 +48.10 | TYC3561-00002-1 | 2 | 188 |
| 03.11 -21.02 | TYC5871-00701-1 | 2 | 188 |
| 09.59 +35.05 | TYC2500-00486-1 | 2 | 188 |
| 05.94 +00.16 | TYC0117-00515-1 | 2 | 188 |
| 09.44 +02.33 | TYC0228-00046-1 | 2 | 188 |
| 05.95 -00.38 | TYC4769-01167-1 | 2 | 188 |
| 03.61 -13.53 | TYC5309-01068-1 | 2 | 188 |
| 07.19 -12.96 | TYC5402-01775-1 | 2 | 188 |
| 04.76 -20.68 | TYC5898-00081-1 | 2 | 188 |
| 05.32 -16.97 | TYC5906-00353-1 | 2 | 188 |
| 05.31 -18.03 | TYC5906-01072-1 | 2 | 188 |
| 08.97 -15.33 | TYC6014-01677-1 | 2 | 188 |
| 03.09 -24.59 | TYC6438-00680-1 | 2 | 188 |
| 04.82 -28.20 | TYC6471-00759-1 | 2 | 188 |
| 06.30 -24.73 | TYC6510-00353-1 | 2 | 188 |
| 06.40 -26.64 | TYC6514-01748-1 | 2 | 188 |
| 07.89 -26.85 | TYC6561-02216-1 | 2 | 188 |
| 09.02 -24.61 | TYC6589-00522-1 | 2 | 188 |
| 09.02 -25.34 | TYC6589-00954-1 | 2 | 188 |
| 09.23 -28.26 | TYC6599-00170-1 | 2 | 188 |
| 06.94 +31.72 | TYC2437-00178-1 | 2 | 188 |
| 09.37 +49.98 | TYC3428-00387-1 | 2 | 188 |
| 19.66 +46.74 | TYC3556-03169-1 | 2 | 188 |
| 18.75 +43.56 | TYC3130-01591-1 | 2 | 187 |
| 03.63 +72.38 | TYC4335-00040-1 | 2 | 187 |
| 01.81 -15.53 | TYC5855-01134-1 | 2 | 187 |
| 03.86 -29.06 | TYC6454-00895-1 | 2 | 187 |
| 05.49 -24.48 | TYC6480-00023-1 | 2 | 187 |
| 07.58 -23.31 | TYC6539-00066-1 | 2 | 187 |
| 08.47 -25.88 | TYC6573-05093-1 | 2 | 187 |
| 09.15 -26.42 | TYC6594-02204-1 | 2 | 187 |
| 19.60 +28.04 | TYC2146-00144-1 | 2 | 187 |
| 19.17 +43.85 | TYC3132-02040-1 | 2 | 187 |
| 07.52 -22.18 | TYC5991-00121-1 | 2 | 187 |
| 13.01 +12.32 | TYC0886-00790-1 | 2 | 187 |
| 03.52 -23.62 | TYC6446-00249-1 | 2 | 187 |
| 06.00 +00.71 | TYC0117-01059-1 | 2 | 186 |
| 07.04 +05.53 | TYC0170-00446-1 | 2 | 186 |
| 05.79 +29.10 | TYC1874-01573-1 | 2 | 186 |
| 04.94 +63.77 | TYC4087-01545-1 | 2 | 186 |
| 04.83 -29.05 | TYC6471-01114-1 | 2 | 186 |
| 09.96 -23.39 | TYC6604-00937-1 | 2 | 186 |
| 07.14 -30.18 | TYC7090-00790-1 | 2 | 186 |
| 09.57 -31.71 | TYC7154-01498-1 | 2 | 186 |
| 05.39 +17.53 | TYC1300-00327-1 | 2 | 185 |
| 05.79 +28.52 | TYC1874-00469-1 | 2 | 185 |
| 04.20 +30.42 | TYC2371-02386-1 | 2 | 185 |
| 09.84 +50.95 | TYC3439-00783-1 | 2 | 185 |
| 18.68 +48.25 | TYC3531-00025-1 | 2 | 185 |
| 06.97 +59.58 | TYC3779-00463-1 | 2 | 185 |
| 01.56 +29.29 | TYC1755-01708-1 | 2 | 185 |
| 07.50 +17.08 | TYC1364-01482-1 | 2 | 185 |
| 07.16 +30.15 | TYC2438-00546-1 | 2 | 185 |
| 00.01 -22.07 | TYC5844-00122-1 | 2 | 185 |
| 08.57 +16.55 | TYC1392-00389-1 | 2 | 184 |
| 11.13 +17.40 | TYC1430-00874-1 | 2 | 184 |
| 04.83 +29.45 | TYC1843-00791-1 | 2 | 184 |
| 07.88 +29.61 | TYC1937-00618-1 | 2 | 184 |
| 03.53 +34.61 | TYC2350-00012-1 | 2 | 184 |
| 09.45 +47.62 | TYC3428-01137-1 | 2 | 184 |
| 09.14 +65.29 | TYC4134-00096-1 | 2 | 184 |
| 00.47 -15.96 | TYC5840-00646-1 | 2 | 184 |
| 09.78 -16.76 | TYC6044-01469-1 | 2 | 184 |
| 09.80 -16.38 | TYC6044-01697-1 | 2 | 184 |
| 08.57 -22.68 | TYC6570-00512-1 | 2 | 184 |
| 05.30 +06.82 | TYC0112-01127-1 | 2 | 184 |
| 06.52 +03.77 | TYC0154-01696-1 | 2 | 184 |
| 07.38 +05.34 | TYC0173-01290-1 | 2 | 184 |
| 07.38 +05.95 | TYC0177-02974-1 | 2 | 184 |
| 07.75 +01.65 | TYC0180-01539-1 | 2 | 184 |
| 08.99 +02.62 | TYC0217-00291-1 | 2 | 184 |
| 09.00 +02.86 | TYC0217-01058-1 | 2 | 184 |
| 09.07 +03.88 | TYC0229-00904-1 | 2 | 184 |
| 09.95 +01.35 | TYC0237-01073-1 | 2 | 184 |
| 10.77 +04.35 | TYC0257-00720-1 | 2 | 184 |
| 10.77 +05.29 | TYC0260-00186-1 | 2 | 184 |
| 10.68 +05.84 | TYC0260-00703-1 | 2 | 184 |
| 05.81 +08.37 | TYC0715-00766-1 | 2 | 184 |
| 08.44 +09.04 | TYC0796-00017-1 | 2 | 184 |
| 09.74 +09.67 | TYC0828-00503-1 | 2 | 184 |
| 11.53 +19.71 | TYC1440-00157-1 | 2 | 184 |
| 05.83 +26.25 | TYC1866-00857-1 | 2 | 184 |
| 07.41 -02.10 | TYC4821-00873-1 | 2 | 184 |
| 10.92 -00.12 | TYC4914-00037-1 | 2 | 184 |
| 07.05 -12.37 | TYC5389-01730-1 | 2 | 184 |
| 07.05 -13.78 | TYC5393-00586-1 | 2 | 184 |
| 09.01 -11.71 | TYC5452-01937-1 | 2 | 184 |
| 10.04 -12.24 | TYC5484-00727-1 | 2 | 184 |
| 03.38 -21.85 | TYC5879-00396-1 | 2 | 184 |
| 05.27 -15.34 | TYC5902-00825-1 | 2 | 184 |
| 06.56 -16.31 | TYC5948-01413-1 | 2 | 184 |
| 06.86 -18.69 | TYC5954-00470-1 | 2 | 184 |
| 06.83 -19.18 | TYC5958-01524-1 | 2 | 184 |
| 07.13 -20.40 | TYC5972-02676-1 | 2 | 184 |
| 07.13 -22.32 | TYC5976-00550-1 | 2 | 184 |
| 08.90 -17.89 | TYC6017-00982-1 | 2 | 184 |
| 05.21 -23.85 | TYC6474-00995-1 | 2 | 184 |
| 05.54 -28.07 | TYC6484-00037-1 | 2 | 184 |
| 06.65 -23.18 | TYC6508-00528-1 | 2 | 184 |
| 06.65 -24.65 | TYC6512-00208-1 | 2 | 184 |
| 07.21 -23.42 | TYC6524-02963-1 | 2 | 184 |
| 07.91 -24.67 | TYC6557-01003-1 | 2 | 184 |
| 08.75 -22.82 | TYC6571-00902-1 | 2 | 184 |
| 08.69 -24.56 | TYC6575-01835-1 | 2 | 184 |
| 09.37 -23.32 | TYC6588-00700-1 | 2 | 184 |
| 09.37 -23.75 | TYC6588-00965-1 | 2 | 184 |
| 09.23 -27.88 | TYC6595-00629-1 | 2 | 184 |
| 10.32 -29.79 | TYC6630-01617-1 | 2 | 184 |
| 08.61 -30.58 | TYC7136-01284-1 | 2 | 184 |
| 03.47 -24.04 | TYC6446-00451-1 | 2 | 184 |
| 19.09 +43.70 | TYC3132-00173-1 | 2 | 184 |
| 05.81 +07.50 | TYC0715-01314-1 | 2 | 184 |
| 08.56 +07.90 | TYC0797-01097-1 | 2 | 184 |
| 09.35 +19.80 | TYC1405-00076-1 | 2 | 184 |
| 08.08 +28.52 | TYC1938-01045-1 | 2 | 184 |
| 05.59 +35.79 | TYC2416-00619-1 | 2 | 184 |
| 09.45 +49.28 | TYC3428-01048-1 | 2 | 184 |
| 04.63 +60.98 | TYC4078-00493-1 | 2 | 184 |
| 06.25 -00.70 | TYC4784-01502-1 | 2 | 184 |
| 08.32 -14.86 | TYC5439-01584-1 | 2 | 184 |
| 02.72 -22.71 | TYC6434-00217-1 | 2 | 184 |
| 02.74 -23.81 | TYC6434-00417-1 | 2 | 184 |
| 06.52 -27.72 | TYC6515-01528-1 | 2 | 184 |
| 09.21 +16.37 | TYC1401-02085-1 | 2 | 184 |
| 11.65 +25.91 | TYC1982-01729-1 | 2 | 184 |
| 11.45 +03.41 | TYC0267-00186-1 | 2 | 184 |
| 23.46 +08.99 | TYC1162-00506-1 | 2 | 184 |
| 08.54 +71.62 | TYC4378-00549-1 | 2 | 184 |
| 19.81 +40.93 | TYC3140-00351-1 | 2 | 183 |
| 06.43 +00.25 | TYC0133-00745-1 | 2 | 183 |
| 06.56 +06.05 | TYC0158-03278-1 | 2 | 183 |
| 07.77 +01.76 | TYC0180-00183-1 | 2 | 183 |
| 08.56 +07.29 | TYC0222-00202-1 | 2 | 183 |
| 10.11 +04.49 | TYC0247-00520-1 | 2 | 183 |
| 10.36 +03.96 | TYC0249-01271-1 | 2 | 183 |
| 08.99 +08.02 | TYC0811-00777-1 | 2 | 183 |
| 09.32 +20.14 | TYC1408-00051-1 | 2 | 183 |
| 18.71 +48.46 | TYC3531-00211-1 | 2 | 183 |
| 19.26 +50.00 | TYC3550-00139-1 | 2 | 183 |
| 10.51 +60.28 | TYC4144-01447-1 | 2 | 183 |
| 07.69 +71.50 | TYC4369-00367-1 | 2 | 183 |
| 05.00 +79.61 | TYC4519-00968-1 | 2 | 183 |
| 06.26 -01.48 | TYC4784-01965-1 | 2 | 183 |
| 07.10 -00.58 | TYC4814-02359-1 | 2 | 183 |
| 09.41 -00.48 | TYC4881-00285-1 | 2 | 183 |
| 06.27 -13.73 | TYC5375-01208-1 | 2 | 183 |
| 06.74 -13.96 | TYC5390-00245-1 | 2 | 183 |
| 10.89 -11.13 | TYC5502-00329-1 | 2 | 183 |
| 05.04 -21.56 | TYC5912-00102-1 | 2 | 183 |
| 08.56 -19.74 | TYC6019-03291-1 | 2 | 183 |
| 10.73 -19.32 | TYC6078-00907-1 | 2 | 183 |
| 05.50 -25.86 | TYC6480-01556-1 | 2 | 183 |
| 05.20 -27.71 | TYC6482-00574-1 | 2 | 183 |
| 05.79 -26.26 | TYC6498-01349-1 | 2 | 183 |
| 06.28 -23.22 | TYC6506-02471-1 | 2 | 183 |
| 07.77 -24.29 | TYC6540-04390-1 | 2 | 183 |
| 08.62 -30.99 | TYC7136-01992-1 | 2 | 183 |
| 07.12 +03.96 | TYC0170-02071-1 | 2 | 183 |
| 06.14 +12.07 | TYC0738-00806-1 | 2 | 183 |
| 05.65 +26.12 | TYC1865-02054-1 | 2 | 183 |
| 07.58 +29.87 | TYC1923-00692-1 | 2 | 183 |
| 10.35 -01.60 | TYC4905-01077-1 | 2 | 183 |
| 03.23 -29.72 | TYC6445-00705-1 | 2 | 183 |
| 22.42 -17.41 | TYC6385-00604-1 | 2 | 183 |
| 08.20 +02.79 | TYC0199-01735-1 | 2 | 182 |
| 04.86 +08.84 | TYC0684-00559-1 | 2 | 182 |
| 09.22 +16.60 | TYC1401-00534-1 | 2 | 182 |
| 05.85 +24.48 | TYC1866-02444-1 | 2 | 182 |
| 08.45 +32.72 | TYC2486-00678-1 | 2 | 182 |
| 08.00 +52.44 | TYC3414-02480-1 | 2 | 182 |
| 04.39 -00.11 | TYC4727-00341-1 | 2 | 182 |
| 04.82 -00.60 | TYC4736-00910-1 | 2 | 182 |
| 03.55 -17.48 | TYC5873-00005-1 | 2 | 182 |
| 06.61 -18.11 | TYC5952-01517-1 | 2 | 182 |
| 07.60 -20.41 | TYC5987-01917-1 | 2 | 182 |
| 03.61 -23.37 | TYC6447-00890-1 | 2 | 182 |
| 04.32 -27.83 | TYC6463-00441-1 | 2 | 182 |
| 07.57 -23.35 | TYC6539-01225-1 | 2 | 182 |
| 07.18 +01.97 | TYC0167-00054-1 | 2 | 182 |
| 08.69 +05.12 | TYC0219-00751-1 | 2 | 182 |
| 11.39 +03.16 | TYC0267-00969-1 | 2 | 182 |
| 20.94 +10.84 | TYC1094-02025-1 | 2 | 182 |
| 10.27 +16.31 | TYC1419-00766-1 | 2 | 182 |
| 02.03 +25.24 | TYC1760-01255-1 | 2 | 182 |
| 04.99 +31.01 | TYC2388-01000-1 | 2 | 182 |
| 02.19 +43.85 | TYC2842-00972-1 | 2 | 182 |
| 09.29 +50.71 | TYC3431-00765-1 | 2 | 182 |
| 04.83 +79.48 | TYC4519-00481-1 | 2 | 182 |
| 10.05 -13.81 | TYC5488-00739-1 | 2 | 182 |
| 09.88 -16.38 | TYC6045-00872-1 | 2 | 182 |
| 06.42 -29.83 | TYC6518-02420-1 | 2 | 182 |
| 03.11 +30.63 | TYC2339-01146-1 | 2 | 182 |
| 01.83 -19.53 | TYC5857-01140-1 | 2 | 182 |
| 04.39 -27.41 | TYC6460-02445-1 | 2 | 182 |
| 08.59 +46.73 | TYC3416-01671-1 | 2 | 181 |
| 06.20 +05.88 | TYC0143-00250-1 | 2 | 181 |
| 09.15 +04.82 | TYC0229-02533-1 | 2 | 181 |
| 10.96 +02.65 | TYC0258-00738-1 | 2 | 181 |
| 03.05 +17.01 | TYC1225-00641-1 | 2 | 181 |
| 06.23 +15.08 | TYC1314-01554-1 | 2 | 181 |
| 03.09 +28.89 | TYC1795-01065-1 | 2 | 181 |
| 03.87 +25.61 | TYC1804-00633-1 | 2 | 181 |
| 10.23 +23.23 | TYC1969-00720-1 | 2 | 181 |
| 07.79 +32.18 | TYC2458-00209-1 | 2 | 181 |
| 04.94 +46.13 | TYC3344-01728-1 | 2 | 181 |
| 10.47 +46.60 | TYC3435-00966-1 | 2 | 181 |
| 07.30 +58.98 | TYC3792-01387-1 | 2 | 181 |
| 10.14 -20.03 | TYC6067-00044-1 | 2 | 181 |
| 08.69 +04.14 | TYC0219-01293-1 | 2 | 181 |
| 10.27 +11.98 | TYC0840-00775-1 | 2 | 181 |
| 04.88 +67.39 | TYC4091-01463-1 | 2 | 181 |
| 07.47 +04.80 | TYC0173-01280-1 | 2 | 180 |
| 10.80 +13.38 | TYC0852-01335-1 | 2 | 180 |
| 09.67 +17.72 | TYC1413-00553-1 | 2 | 180 |
| 11.03 +43.97 | TYC3012-00089-1 | 2 | 180 |
| 06.78 +45.55 | TYC3390-00831-1 | 2 | 180 |
| 08.53 +52.42 | TYC3422-00358-1 | 2 | 180 |
| 19.09 +50.14 | TYC3549-01640-1 | 2 | 180 |
| 07.91 +59.31 | TYC3795-00648-1 | 2 | 180 |
| 10.33 +52.85 | TYC3815-01507-1 | 2 | 180 |
| 10.61 +58.58 | TYC3822-01000-1 | 2 | 180 |
| 04.90 -18.82 | TYC5907-00121-1 | 2 | 180 |
| 07.57 -15.11 | TYC5979-02350-1 | 2 | 180 |
| 05.33 -23.44 | TYC6475-01384-1 | 2 | 180 |
| 07.22 -31.62 | TYC7103-00774-1 | 2 | 180 |
| 09.62 +01.52 | TYC0235-01971-1 | 2 | 180 |
| 08.16 +21.43 | TYC1389-00316-1 | 2 | 180 |
| 05.81 +57.26 | TYC3758-00662-1 | 2 | 180 |
| 08.32 +62.34 | TYC4126-00438-1 | 2 | 180 |
| 08.33 +61.16 | TYC4126-01766-1 | 2 | 180 |
| 07.70 -12.29 | TYC5418-00771-1 | 2 | 180 |
| 07.18 +01.36 | TYC0163-00152-1 | 2 | 179 |
| 06.91 +29.09 | TYC1906-01038-1 | 2 | 179 |
| 10.68 +29.64 | TYC1979-00285-1 | 2 | 179 |
| 11.01 +44.60 | TYC3012-02135-1 | 2 | 179 |
| 05.97 +52.11 | TYC3373-00830-1 | 2 | 179 |
| 06.47 +49.16 | TYC3384-00570-1 | 2 | 179 |
| 06.46 +49.47 | TYC3384-00986-1 | 2 | 179 |
| 06.22 +51.75 | TYC3387-00436-1 | 2 | 179 |
| 08.15 +50.55 | TYC3414-01526-1 | 2 | 179 |
| 09.46 +50.74 | TYC3431-01401-1 | 2 | 179 |
| 04.59 +59.70 | TYC3744-00129-1 | 2 | 179 |
| 10.85 +55.59 | TYC3826-00278-1 | 2 | 179 |
| 04.91 +65.38 | TYC4087-01436-1 | 2 | 179 |
| 06.92 -29.39 | TYC6534-01589-1 | 2 | 179 |
| 07.45 +23.78 | TYC1910-00196-1 | 2 | 179 |
| 09.84 +13.42 | TYC0834-00236-1 | 2 | 179 |
| 09.80 -27.68 | TYC6611-00515-1 | 2 | 179 |
| 03.05 +17.45 | TYC1225-00857-1 | 2 | 179 |
| 07.85 +21.02 | TYC1374-01891-1 | 2 | 179 |
| 04.44 +35.39 | TYC2380-01266-1 | 2 | 179 |
| 07.99 +30.99 | TYC2468-01679-1 | 2 | 179 |
| 11.13 +35.35 | TYC2522-00220-1 | 2 | 179 |
| 10.45 +45.91 | TYC3435-00219-1 | 2 | 179 |
| 05.67 -12.77 | TYC5354-00564-1 | 2 | 179 |
| 23.41 -20.48 | TYC6402-00741-1 | 2 | 179 |
| 06.03 -26.18 | TYC6496-01586-1 | 2 | 179 |
| 08.72 +12.42 | TYC0813-00239-1 | 2 | 179 |
| 08.84 +28.72 | TYC1949-00624-1 | 2 | 179 |
| 03.81 +30.43 | TYC2356-01036-1 | 2 | 179 |
| 07.33 +47.87 | TYC3397-00899-1 | 2 | 179 |
| 03.74 +74.84 | TYC4339-01492-1 | 2 | 179 |
| 04.97 +08.35 | TYC0684-00885-1 | 2 | 178 |
| 03.17 +17.54 | TYC1228-00368-1 | 2 | 178 |
| 08.58 +17.45 | TYC1392-00830-1 | 2 | 178 |
| 06.38 +25.91 | TYC1882-00850-1 | 2 | 178 |
| 04.03 +33.65 | TYC2361-00148-1 | 2 | 178 |
| 03.71 +36.59 | TYC2367-00786-1 | 2 | 178 |
| 11.60 +42.35 | TYC3015-02310-1 | 2 | 178 |
| 04.47 +51.63 | TYC3341-00102-1 | 2 | 178 |
| 05.97 +51.51 | TYC3373-00715-1 | 2 | 178 |
| 05.81 +59.17 | TYC3762-01450-1 | 2 | 178 |
| 06.76 +67.94 | TYC4358-00008-1 | 2 | 178 |
| 04.73 +75.12 | TYC4511-00040-1 | 2 | 178 |
| 23.57 +76.81 | TYC4602-01242-1 | 2 | 178 |
| 06.03 -25.94 | TYC6496-01225-1 | 2 | 178 |
| 11.41 +02.57 | TYC0267-00121-1 | 2 | 178 |
| 07.86 +10.10 | TYC0783-00594-1 | 2 | 178 |
| 05.71 +15.24 | TYC1298-00163-1 | 2 | 178 |
| 05.60 +36.11 | TYC2416-00139-1 | 2 | 178 |
| 05.17 +69.11 | TYC4342-01082-1 | 2 | 178 |
| 06.82 -01.53 | TYC4800-02069-1 | 2 | 178 |
| 08.02 -00.17 | TYC4846-02083-1 | 2 | 178 |
| 09.85 +14.27 | TYC0834-00001-1 | 2 | 178 |
| 04.67 +18.87 | TYC1275-01225-1 | 2 | 178 |
| 19.55 +40.94 | TYC3139-00106-1 | 2 | 177 |
| 09.38 +01.72 | TYC0228-00819-1 | 2 | 177 |
| 08.34 +27.20 | TYC1936-01149-1 | 2 | 177 |
| 19.06 +39.37 | TYC3120-01712-1 | 2 | 177 |
| 06.47 +68.01 | TYC4358-00881-1 | 2 | 177 |
| 18.73 +06.04 | TYC0459-00837-1 | 2 | 177 |
| 05.25 +04.08 | TYC0107-01845-1 | 2 | 177 |
| 08.03 +13.96 | TYC0792-01670-1 | 2 | 177 |
| 21.22 +14.69 | TYC1117-00971-1 | 2 | 177 |
| 05.83 +21.34 | TYC1311-01180-1 | 2 | 177 |
| 05.82 +21.62 | TYC1311-01404-1 | 2 | 177 |
| 05.90 +22.11 | TYC1324-02547-1 | 2 | 177 |
| 07.18 +19.76 | TYC1353-01493-1 | 2 | 177 |
| 12.50 +22.37 | TYC1447-02095-1 | 2 | 177 |
| 02.03 +24.43 | TYC1757-00610-1 | 2 | 177 |
| 03.14 +25.24 | TYC1787-00630-1 | 2 | 177 |
| 03.26 +25.31 | TYC1788-00442-1 | 2 | 177 |
| 07.26 +27.17 | TYC1917-02482-1 | 2 | 177 |
| 05.12 +36.63 | TYC2401-00144-1 | 2 | 177 |
| 06.64 +30.78 | TYC2435-00944-1 | 2 | 177 |
| 19.08 +42.30 | TYC3128-00596-1 | 2 | 177 |
| 18.75 +44.24 | TYC3130-02035-1 | 2 | 177 |
| 19.64 +42.37 | TYC3143-01079-1 | 2 | 177 |
| 19.34 +43.41 | TYC3146-00048-1 | 2 | 177 |
| 19.01 +48.60 | TYC3545-00461-1 | 2 | 177 |
| 19.15 +48.33 | TYC3546-00636-1 | 2 | 177 |
| 19.71 +46.34 | TYC3557-00899-1 | 2 | 177 |
| 19.61 +48.33 | TYC3560-00644-1 | 2 | 177 |
| 19.85 +47.19 | TYC3561-00657-1 | 2 | 177 |
| 19.63 +49.48 | TYC3564-00663-1 | 2 | 177 |
| 08.18 +61.06 | TYC4126-01858-1 | 2 | 177 |
| 04.39 -24.95 | TYC6457-03010-1 | 2 | 177 |
| 06.41 -27.79 | TYC6514-01559-1 | 2 | 177 |
| 05.69 +05.82 | TYC0127-00840-1 | 2 | 177 |
| 10.25 +13.18 | TYC0843-00716-1 | 2 | 177 |
| 04.45 +19.46 | TYC1273-00668-1 | 2 | 177 |
| 05.61 +20.34 | TYC1306-00381-1 | 2 | 177 |
| 02.65 +24.70 | TYC1771-01395-1 | 2 | 177 |
| 03.21 +25.71 | TYC1787-00218-1 | 2 | 177 |
| 22.05 +26.58 | TYC2212-01603-1 | 2 | 177 |
| 02.41 +32.30 | TYC2327-00922-1 | 2 | 177 |
| 05.79 +35.83 | TYC2417-00101-1 | 2 | 177 |
| 06.48 +30.08 | TYC2422-00942-1 | 2 | 177 |
| 09.52 +35.11 | TYC2500-00056-1 | 2 | 177 |
| 06.29 +41.25 | TYC2930-00235-1 | 2 | 177 |
| 06.29 +40.94 | TYC2930-00617-1 | 2 | 177 |
| 11.37 +40.83 | TYC3013-00165-1 | 2 | 177 |
| 00.24 -11.74 | TYC5264-00910-1 | 2 | 177 |
| 07.58 -13.66 | TYC5409-03245-1 | 2 | 177 |
| 07.30 -31.06 | TYC7103-00980-1 | 2 | 177 |
| 11.47 +07.81 | TYC0856-00161-1 | 2 | 177 |
| 18.94 +33.05 | TYC2643-00568-1 | 2 | 177 |
| 19.14 +43.85 | TYC3132-01846-1 | 2 | 177 |
| 00.62 +34.69 | TYC2283-00301-1 | 2 | 177 |
| 11.82 +57.68 | TYC3838-00365-1 | 2 | 177 |
| 10.28 +19.67 | TYC1422-00468-1 | 2 | 177 |
| 20.03 +45.75 | TYC3558-01674-1 | 2 | 177 |
| 00.62 +34.62 | TYC2283-00426-1 | 2 | 176 |
| 00.26 -11.99 | TYC5264-00619-1 | 2 | 176 |
| 00.27 -12.08 | TYC5264-00632-1 | 2 | 176 |
| 00.31 -15.38 | TYC5839-00896-1 | 2 | 176 |
| 00.00 -22.41 | TYC5844-00119-1 | 2 | 176 |
| 00.27 +01.16 | TYC0002-01108-1 | 2 | 176 |
| 00.26 +01.24 | TYC0002-01245-1 | 2 | 176 |
| 17.43 +27.22 | TYC2082-01191-1 | 2 | 176 |
| 15.85 +35.58 | TYC2578-00938-1 | 2 | 176 |
| 18.80 +48.62 | TYC3544-00488-1 | 2 | 176 |
| 18.79 +48.49 | TYC3544-00593-1 | 2 | 176 |
| 19.34 +50.34 | TYC3551-00394-1 | 2 | 176 |
| 14.95 +53.55 | TYC3861-01446-1 | 2 | 176 |
| 15.57 +53.88 | TYC3869-00565-1 | 2 | 176 |
| 20.34 +59.57 | TYC3949-01099-1 | 2 | 176 |
| 23.25 +57.99 | TYC4006-00271-1 | 2 | 176 |
| 23.26 +57.96 | TYC4006-00323-1 | 2 | 176 |
| 15.26 +71.85 | TYC4414-00287-1 | 2 | 176 |
| 23.64 +77.59 | TYC4606-00148-1 | 2 | 176 |
| 23.64 +77.53 | TYC4606-00236-1 | 2 | 176 |
| 06.72 -01.31 | TYC4799-00039-1 | 2 | 176 |
| 17.84 -29.74 | TYC6840-01029-1 | 2 | 176 |
| 18.08 -29.02 | TYC6854-02150-1 | 2 | 176 |
| 18.08 -28.87 | TYC6854-03453-1 | 2 | 176 |
| 08.60 -30.02 | TYC7136-02257-1 | 2 | 176 |
| 05.52 +08.66 | TYC0701-00427-1 | 2 | 176 |
| 03.12 +25.86 | TYC1787-00446-1 | 2 | 176 |
| 08.81 +32.66 | TYC2488-01061-1 | 2 | 176 |
| 19.40 +37.81 | TYC3134-01183-1 | 2 | 176 |
| 22.88 +38.36 | TYC3215-00160-1 | 2 | 176 |
| 06.55 +49.82 | TYC3384-00852-1 | 2 | 176 |
| 05.72 +54.84 | TYC3753-01335-1 | 2 | 176 |
| 03.20 -00.16 | TYC4708-00625-1 | 2 | 176 |
| 00.27 -15.35 | TYC5839-00850-1 | 2 | 176 |
| 08.60 -23.51 | TYC6570-03132-1 | 2 | 176 |
| 08.51 -29.67 | TYC6582-02277-1 | 2 | 176 |
| 08.78 -31.92 | TYC7141-02662-1 | 2 | 176 |
| 04.88 +12.18 | TYC0692-01433-1 | 2 | 176 |
| 09.12 +21.94 | TYC1407-01864-1 | 2 | 176 |
| 21.98 +19.02 | TYC1687-00051-1 | 2 | 176 |
| 06.53 +29.49 | TYC1891-00132-1 | 2 | 176 |
| 08.84 +27.55 | TYC1949-01616-1 | 2 | 176 |
| 07.08 +38.36 | TYC2943-00824-1 | 2 | 176 |
| 04.99 +47.86 | TYC3348-00142-1 | 2 | 176 |
| 03.13 +57.52 | TYC3710-00110-1 | 2 | 176 |
| 09.26 +53.55 | TYC3806-00328-1 | 2 | 176 |
| 06.79 -00.78 | TYC4800-00720-1 | 2 | 176 |
| 05.67 -14.23 | TYC5358-00614-1 | 2 | 176 |
| 11.66 +42.65 | TYC3015-01763-1 | 2 | 176 |
| 06.30 +25.97 | TYC1882-00172-1 | 2 | 176 |
| 18.64 +04.13 | TYC0455-00157-1 | 2 | 175 |
| 18.64 +04.56 | TYC0455-00524-1 | 2 | 175 |
| 21.07 +03.66 | TYC0530-02029-1 | 2 | 175 |
| 21.08 +03.89 | TYC0534-00856-1 | 2 | 175 |
| 20.63 +11.16 | TYC1092-00285-1 | 2 | 175 |
| 21.19 +14.54 | TYC1116-01143-1 | 2 | 175 |
| 00.97 +20.44 | TYC1195-01635-1 | 2 | 175 |
| 03.88 +19.48 | TYC1257-00799-1 | 2 | 175 |
| 20.39 +17.20 | TYC1635-01877-1 | 2 | 175 |
| 20.39 +17.02 | TYC1635-02106-1 | 2 | 175 |
| 22.16 +16.13 | TYC1681-00315-1 | 2 | 175 |
| 22.02 +19.19 | TYC1688-01747-1 | 2 | 175 |
| 22.95 +20.49 | TYC1717-00119-1 | 2 | 175 |
| 16.93 +25.98 | TYC2063-00409-1 | 2 | 175 |
| 16.93 +25.87 | TYC2063-00633-1 | 2 | 175 |
| 19.87 +27.64 | TYC2148-02996-1 | 2 | 175 |
| 19.87 +27.97 | TYC2148-03017-1 | 2 | 175 |
| 00.34 +31.55 | TYC2261-01216-1 | 2 | 175 |
| 00.34 +32.16 | TYC2265-01348-1 | 2 | 175 |
| 21.60 +30.48 | TYC2717-00543-1 | 2 | 175 |
| 21.60 +30.30 | TYC2717-00853-1 | 2 | 175 |
| 05.37 +58.82 | TYC3760-00786-1 | 2 | 175 |
| 17.84 +56.53 | TYC3910-00474-1 | 2 | 175 |
| 09.71 +67.04 | TYC4142-01025-1 | 2 | 175 |
| 18.35 +65.45 | TYC4222-00825-1 | 2 | 175 |
| 20.22 +64.92 | TYC4240-00148-1 | 2 | 175 |
| 20.22 +65.68 | TYC4244-00518-1 | 2 | 175 |
| 00.27 -14.95 | TYC5267-00827-1 | 2 | 175 |
| 04.83 -16.09 | TYC5899-00895-1 | 2 | 175 |
| 17.32 -19.67 | TYC6241-00369-1 | 2 | 175 |
| 21.54 -21.41 | TYC6373-00207-1 | 2 | 175 |
| 22.41 -17.08 | TYC6385-00819-1 | 2 | 175 |
| 15.99 -27.63 | TYC6787-00883-1 | 2 | 175 |
| 16.36 -26.60 | TYC6802-00688-1 | 2 | 175 |
| 17.89 -28.18 | TYC6853-00039-1 | 2 | 175 |
| 17.97 -29.10 | TYC6853-02575-1 | 2 | 175 |
| 17.97 -29.40 | TYC6853-03991-1 | 2 | 175 |
| 18.10 -30.22 | TYC7391-01135-1 | 2 | 175 |
| 18.10 -31.29 | TYC7391-01500-1 | 2 | 175 |
| 21.59 -31.60 | TYC7474-00823-1 | 2 | 175 |
| 06.74 +00.41 | TYC0147-01358-1 | 2 | 175 |
| 06.68 +18.20 | TYC1334-01851-1 | 2 | 175 |
| 05.94 +29.12 | TYC1875-02410-1 | 2 | 175 |
| 07.71 +22.51 | TYC1912-00808-1 | 2 | 175 |
| 03.73 +39.93 | TYC2867-00245-1 | 2 | 175 |
| 08.01 -01.53 | TYC4846-02208-1 | 2 | 175 |
| 06.48 +10.89 | TYC0736-00110-1 | 2 | 175 |
| 03.22 +25.14 | TYC1788-01093-1 | 2 | 175 |
| 07.36 +58.01 | TYC3789-01576-1 | 2 | 175 |
| 03.02 -20.73 | TYC5870-00133-1 | 2 | 175 |
| 06.61 -19.37 | TYC5956-01339-1 | 2 | 175 |
| 10.04 -15.61 | TYC6046-01383-1 | 2 | 175 |
| 08.68 -23.23 | TYC6571-00253-1 | 2 | 175 |
| 08.69 -23.42 | TYC6571-02094-1 | 2 | 175 |
| 08.66 -23.47 | TYC6571-02482-1 | 2 | 175 |
| 11.35 -23.03 | TYC6650-00323-1 | 2 | 175 |
| 10.67 +09.68 | TYC0839-01005-1 | 2 | 174 |
| 04.46 +36.10 | TYC2384-00186-1 | 2 | 174 |
| 03.85 +40.58 | TYC2868-00027-1 | 2 | 174 |
| 06.54 +51.23 | TYC3388-00636-1 | 2 | 174 |
| 08.58 +68.60 | TYC4375-01663-1 | 2 | 174 |
| 01.38 +28.36 | TYC1754-00562-1 | 2 | 174 |
| 07.30 +29.36 | TYC1921-00741-1 | 2 | 174 |
| 22.03 +26.38 | TYC2211-01194-1 | 2 | 174 |
| 14.29 -19.96 | TYC6147-00656-1 | 2 | 174 |
| 07.32 +37.29 | TYC2463-00995-1 | 2 | 174 |
| 06.60 -29.20 | TYC6520-01904-1 | 2 | 174 |
| 04.64 +01.46 | TYC0083-00721-1 | 2 | 174 |
| 09.03 +14.83 | TYC0818-00686-1 | 2 | 174 |
| 05.35 +33.99 | TYC2398-01269-1 | 2 | 174 |
| 06.86 +06.02 | TYC0160-00724-1 | 2 | 174 |
| 10.70 +15.83 | TYC1428-01322-1 | 2 | 174 |
| 05.46 +27.89 | TYC1856-00813-1 | 2 | 174 |
| 08.00 +25.02 | TYC1930-00066-1 | 2 | 174 |
| 02.25 +51.80 | TYC3306-00275-1 | 2 | 174 |
| 19.68 +50.61 | TYC3564-02949-1 | 2 | 174 |
| 04.36 +57.89 | TYC3727-00317-1 | 2 | 174 |
| 10.42 +25.56 | TYC1973-00905-1 | 2 | 173 |
| 06.68 +16.86 | TYC1330-01814-1 | 2 | 173 |
| 07.01 +27.63 | TYC1903-01154-1 | 2 | 173 |
| 07.97 +33.40 | TYC2472-01109-1 | 2 | 173 |
| 03.94 +45.64 | TYC3327-02903-1 | 2 | 173 |
| 09.28 +67.34 | TYC4141-00667-1 | 2 | 173 |
| 05.79 +14.30 | TYC0727-00913-1 | 2 | 172 |
| 04.69 +22.49 | TYC1279-00373-1 | 2 | 172 |
| 07.77 +16.46 | TYC1362-00575-1 | 2 | 172 |
| 05.15 +32.70 | TYC2393-00326-1 | 2 | 172 |
| 08.29 +64.11 | TYC4129-02482-1 | 2 | 172 |
| 08.96 +71.59 | TYC4378-02015-1 | 2 | 172 |
| 03.08 -22.40 | TYC5871-00316-1 | 2 | 172 |
| 06.15 +08.02 | TYC0730-02367-1 | 2 | 172 |
| 06.29 +08.53 | TYC0731-02566-1 | 2 | 172 |
| 06.46 +11.31 | TYC0740-01163-1 | 2 | 172 |
| 05.14 +79.00 | TYC4519-00982-1 | 2 | 172 |
| 12.55 -27.75 | TYC6690-00516-1 | 2 | 172 |
| 02.25 +51.66 | TYC3293-00972-1 | 2 | 172 |
| 00.65 +34.85 | TYC2283-00112-1 | 2 | 172 |
| 08.59 +14.64 | TYC0809-00088-1 | 2 | 171 |
| 06.97 +21.73 | TYC1356-01333-1 | 2 | 171 |
| 04.53 +24.93 | TYC1833-00138-1 | 2 | 171 |
| 05.71 +30.76 | TYC2405-00804-1 | 2 | 171 |
| 10.43 +30.58 | TYC2511-00875-1 | 2 | 171 |
| 06.87 +48.96 | TYC3398-00845-1 | 2 | 171 |
| 04.48 +54.60 | TYC3736-00975-1 | 2 | 171 |
| 04.39 +61.50 | TYC4065-01321-1 | 2 | 171 |
| 08.96 +71.23 | TYC4378-02085-1 | 2 | 171 |
| 03.23 +31.08 | TYC2340-00898-1 | 2 | 171 |
| 00.29 -14.93 | TYC5267-00603-1 | 2 | 171 |
| 08.25 -28.17 | TYC6568-00101-1 | 2 | 171 |
| 17.24 +04.96 | TYC0407-00274-1 | 2 | 171 |
| 06.18 -29.88 | TYC6517-02271-1 | 2 | 171 |
| 05.95 +09.43 | TYC0720-00747-1 | 2 | 170 |
| 09.27 +15.14 | TYC1402-01494-1 | 2 | 170 |
| 06.84 +29.56 | TYC1906-00392-1 | 2 | 170 |
| 06.85 +30.01 | TYC2436-00594-1 | 2 | 170 |
| 06.75 +34.65 | TYC2444-00367-1 | 2 | 170 |
| 07.73 +53.35 | TYC3782-01435-1 | 2 | 170 |
| 03.63 +74.24 | TYC4339-00792-1 | 2 | 170 |
| 08.32 +17.26 | TYC1382-00487-1 | 2 | 170 |
| 08.36 +54.69 | TYC3797-00740-1 | 2 | 170 |
| 04.16 +05.70 | TYC0079-00308-1 | 2 | 170 |
| 10.66 +34.99 | TYC2515-00364-1 | 2 | 170 |
| 04.89 +06.70 | TYC0097-00342-1 | 2 | 170 |
| 04.88 +06.67 | TYC0097-00529-1 | 2 | 170 |
| 05.27 +06.92 | TYC0112-01568-1 | 2 | 170 |
| 22.06 +25.96 | TYC2208-01207-1 | 2 | 170 |
| 15.86 +35.20 | TYC2578-01408-1 | 2 | 170 |
| 13.93 +43.98 | TYC3033-00783-1 | 2 | 170 |
| 06.56 +57.67 | TYC3773-01650-1 | 2 | 170 |
| 07.36 +57.95 | TYC3789-00670-1 | 2 | 170 |
| 08.63 +64.72 | TYC4130-01594-1 | 2 | 170 |
| 06.63 -19.73 | TYC5956-01605-1 | 2 | 170 |
| 06.63 -19.01 | TYC5956-02131-1 | 2 | 170 |
| 06.38 -29.14 | TYC6518-01282-1 | 2 | 170 |
| 04.74 +13.83 | TYC0695-01074-1 | 2 | 170 |
| 08.62 +19.56 | TYC1395-02132-1 | 2 | 170 |
| 03.28 +25.47 | TYC1788-00964-1 | 2 | 170 |
| 11.12 +24.83 | TYC1978-01868-1 | 2 | 170 |
| 10.77 +32.68 | TYC2515-01002-1 | 2 | 170 |
| 08.50 -29.73 | TYC6581-00143-1 | 2 | 170 |
| 00.28 -15.47 | TYC5839-00842-1 | 2 | 170 |
| 01.86 -19.49 | TYC5857-01409-1 | 2 | 170 |
| 23.17 +57.31 | TYC4006-00448-1 | 2 | 170 |
| 06.50 +29.82 | TYC1891-00653-1 | 2 | 170 |
| 08.65 -23.47 | TYC6571-01937-1 | 2 | 170 |
| 05.81 +21.23 | TYC1311-02205-1 | 2 | 169 |
| 07.42 +22.81 | TYC1910-01420-1 | 2 | 169 |
| 20.93 +10.65 | TYC1094-01024-1 | 2 | 169 |
| 03.11 +30.53 | TYC2339-01328-1 | 2 | 169 |
| 03.65 +31.40 | TYC2355-00392-1 | 2 | 169 |
| 00.22 -12.09 | TYC5264-00160-1 | 2 | 169 |
| 02.96 -20.80 | TYC5870-00314-1 | 2 | 169 |
| 03.17 +21.49 | TYC1231-00093-1 | 2 | 168 |
| 03.38 +27.37 | TYC1805-00282-1 | 2 | 168 |
| 04.58 +25.02 | TYC1833-00934-1 | 2 | 168 |
| 17.39 +27.52 | TYC2082-00216-1 | 2 | 168 |
| 17.39 +27.34 | TYC2082-01113-1 | 2 | 168 |
| 00.69 +34.90 | TYC2283-00659-1 | 2 | 168 |
| 05.09 +52.27 | TYC3357-00347-1 | 2 | 168 |
| 09.99 +47.17 | TYC3433-01223-1 | 2 | 168 |
| 01.28 +76.52 | TYC4493-01690-1 | 2 | 168 |
| 06.27 +41.05 | TYC2930-01569-1 | 2 | 168 |
| 19.48 +38.43 | TYC3134-00030-1 | 2 | 168 |
| 07.39 +20.56 | TYC1355-00329-1 | 2 | 167 |
| 09.59 +34.68 | TYC2497-00300-1 | 2 | 167 |
| 07.36 +58.31 | TYC3793-01856-1 | 2 | 167 |
| 12.55 +74.46 | TYC4400-00414-1 | 2 | 167 |
| 06.60 -19.40 | TYC5956-01721-1 | 2 | 167 |
| 11.30 -23.86 | TYC6649-00730-1 | 2 | 167 |
| 11.29 -23.94 | TYC6649-00793-1 | 2 | 167 |
| 11.35 -23.05 | TYC6650-00322-1 | 2 | 167 |
| 11.31 -24.08 | TYC6650-00477-1 | 2 | 167 |
| 13.02 -27.63 | TYC6706-01027-1 | 2 | 167 |
| 06.62 -32.23 | TYC7091-00665-1 | 2 | 167 |
| 06.62 -32.39 | TYC7091-01180-1 | 2 | 167 |
| 20.13 +44.42 | TYC3162-00869-1 | 2 | 167 |
| 07.08 +22.55 | TYC1895-01167-1 | 2 | 167 |
| 05.73 +01.56 | TYC0115-00560-1 | 2 | 167 |
| 08.25 -29.38 | TYC6568-02053-1 | 2 | 167 |
| 10.86 -30.95 | TYC7199-01641-1 | 2 | 167 |
| 05.08 -14.11 | TYC5329-01545-1 | 2 | 167 |
| 18.91 +04.21 | TYC0457-00275-1 | 2 | 167 |
| 18.53 +06.76 | TYC0458-01462-1 | 2 | 167 |
| 06.61 -19.05 | TYC5956-02094-1 | 2 | 167 |
| 17.88 -30.27 | TYC7378-00350-1 | 2 | 167 |
| 03.13 +25.93 | TYC1787-00554-1 | 2 | 166 |
| 08.49 +31.46 | TYC2483-00802-1 | 2 | 166 |
| 04.47 +45.61 | TYC3329-02320-1 | 2 | 166 |
| 06.33 -28.84 | TYC6518-01101-1 | 2 | 166 |
| 19.98 +43.02 | TYC3145-00163-1 | 2 | 166 |
| 03.12 +30.90 | TYC2339-00646-1 | 2 | 166 |
| 14.33 -17.44 | TYC6143-01018-1 | 2 | 166 |
| 11.35 -23.16 | TYC6650-00447-1 | 2 | 165 |
| 11.77 +02.87 | TYC0275-00039-1 | 2 | 165 |
| 04.22 +09.32 | TYC0667-00034-1 | 2 | 165 |
| 06.48 -28.62 | TYC6519-00664-1 | 2 | 165 |
| 09.82 +34.08 | TYC2505-00437-1 | 2 | 165 |
| 09.82 +34.19 | TYC2505-00465-1 | 2 | 165 |
| 09.58 +13.40 | TYC0827-00456-1 | 2 | 165 |
| 05.02 +56.48 | TYC3742-00795-1 | 2 | 165 |
| 08.39 +01.91 | TYC0201-00123-1 | 2 | 164 |
| 12.50 +21.97 | TYC1447-02157-1 | 2 | 164 |
| 10.16 +25.01 | TYC1971-00727-1 | 2 | 164 |
| 12.51 +22.89 | TYC1989-02120-1 | 2 | 164 |
| 09.59 +34.96 | TYC2497-00222-1 | 2 | 164 |
| 07.42 +79.49 | TYC4535-00588-1 | 2 | 164 |
| 08.32 -12.55 | TYC5435-02464-1 | 2 | 164 |
| 06.88 +24.64 | TYC1898-01428-1 | 2 | 164 |
| 03.21 -02.20 | TYC4708-01134-1 | 2 | 164 |
| 05.79 -19.87 | TYC5926-00870-1 | 2 | 164 |
| 10.20 -20.01 | TYC6067-00670-1 | 2 | 164 |
| 09.38 +20.16 | TYC1408-01762-1 | 2 | 164 |
| 15.44 +59.12 | TYC3875-00543-1 | 2 | 164 |
| 08.34 +61.41 | TYC4126-00898-1 | 2 | 164 |
| 08.70 -23.37 | TYC6571-00508-1 | 2 | 164 |
| 05.82 +33.38 | TYC2409-01975-1 | 2 | 164 |
| 16.53 +38.71 | TYC3063-00267-1 | 2 | 164 |
| 03.85 -28.85 | TYC6454-00875-1 | 2 | 164 |
| 03.95 -25.22 | TYC6458-00713-1 | 2 | 164 |
| 04.36 +39.51 | TYC2883-01548-1 | 2 | 164 |
| 08.60 +30.64 | TYC2483-00071-1 | 2 | 163 |
| 05.13 +70.34 | TYC4346-00847-1 | 2 | 163 |
| 04.93 +08.40 | TYC0684-00427-1 | 2 | 163 |
| 06.21 +11.84 | TYC0738-00258-1 | 2 | 163 |
| 07.76 +61.50 | TYC4113-01090-1 | 2 | 163 |
| 05.02 +79.35 | TYC4519-00937-1 | 2 | 163 |
| 02.06 +46.66 | TYC3280-00781-1 | 2 | 163 |
| 11.07 -00.20 | TYC4921-00239-1 | 2 | 163 |
| 19.49 +48.67 | TYC3547-00394-1 | 2 | 162 |
| 03.02 +45.73 | TYC3310-00221-1 | 2 | 162 |
| 02.20 +43.45 | TYC2842-01340-1 | 2 | 162 |
| 09.10 +51.19 | TYC3430-00704-1 | 2 | 162 |
| 16.34 -20.65 | TYC6214-00255-1 | 2 | 162 |
| 06.14 -29.93 | TYC6517-02157-1 | 2 | 162 |
| 02.13 +31.97 | TYC2313-00565-1 | 2 | 162 |
| 02.68 +49.29 | TYC3304-00669-1 | 2 | 162 |
| 01.84 -19.72 | TYC5857-02010-1 | 2 | 162 |
| 08.85 +64.65 | TYC4131-00984-1 | 2 | 161 |
| 02.54 +25.01 | TYC1771-01005-1 | 2 | 161 |
| 02.08 +31.86 | TYC2308-00247-1 | 2 | 161 |
| 19.29 +44.03 | TYC3133-00214-1 | 2 | 161 |
| 07.87 +31.74 | TYC2467-00620-1 | 2 | 161 |
| 03.14 +20.86 | TYC1231-00322-1 | 2 | 161 |
| 04.32 +58.08 | TYC3727-00120-1 | 2 | 160 |
| 08.85 +63.94 | TYC4131-00751-1 | 2 | 160 |
| 00.55 +34.73 | TYC2270-01100-1 | 2 | 160 |
| 10.04 +63.12 | TYC4140-00884-1 | 2 | 160 |
| 10.31 -30.10 | TYC7183-00830-1 | 2 | 160 |
| 11.41 +44.25 | TYC3015-00999-1 | 2 | 160 |
| 06.32 +53.51 | TYC3764-02358-1 | 2 | 160 |
| 05.22 -27.09 | TYC6482-00932-1 | 2 | 160 |
| 08.57 -29.85 | TYC6582-01662-1 | 2 | 160 |
| 11.08 -30.24 | TYC7200-00483-1 | 2 | 160 |
| 10.92 +01.18 | TYC0255-00682-1 | 2 | 159 |
| 08.11 +61.76 | TYC4126-00014-1 | 2 | 159 |
| 18.25 +65.68 | TYC4213-01512-1 | 2 | 159 |
| 02.06 +25.73 | TYC1761-00185-1 | 2 | 159 |
| 11.08 -29.97 | TYC6648-00514-1 | 2 | 159 |
| 10.34 +54.00 | TYC3815-00495-1 | 2 | 159 |
| 17.28 +29.32 | TYC2073-01865-1 | 2 | 159 |
| 03.57 +10.64 | TYC0653-00184-1 | 2 | 158 |
| 11.09 -24.28 | TYC6636-01151-1 | 2 | 158 |
| 07.92 +09.33 | TYC0780-00031-1 | 2 | 158 |
| 02.43 +07.19 | TYC0045-01050-1 | 2 | 157 |
| 03.22 +00.05 | TYC0056-01152-1 | 2 | 157 |
| 14.86 +05.68 | TYC0333-01170-1 | 2 | 157 |
| 00.31 +14.96 | TYC0601-00067-1 | 2 | 157 |
| 21.16 +14.00 | TYC1116-00544-1 | 2 | 157 |
| 02.99 +19.29 | TYC1227-00494-1 | 2 | 157 |
| 03.96 +20.65 | TYC1261-01452-1 | 2 | 157 |
| 20.37 +17.61 | TYC1635-01985-1 | 2 | 157 |
| 22.87 +20.75 | TYC1710-00511-1 | 2 | 157 |
| 11.68 +26.77 | TYC1984-01964-1 | 2 | 157 |
| 16.90 +25.80 | TYC2063-00517-1 | 2 | 157 |
| 19.07 +25.94 | TYC2126-00433-1 | 2 | 157 |
| 19.98 +22.56 | TYC2141-01070-1 | 2 | 157 |
| 00.68 +34.75 | TYC2283-00115-1 | 2 | 157 |
| 19.74 +33.33 | TYC2660-00520-1 | 2 | 157 |
| 19.74 +33.53 | TYC2660-00549-1 | 2 | 157 |
| 19.42 +37.14 | TYC2666-01015-1 | 2 | 157 |
| 02.63 +41.21 | TYC2836-00015-1 | 2 | 157 |
| 14.97 +44.08 | TYC3050-00601-1 | 2 | 157 |
| 14.96 +44.00 | TYC3050-00680-1 | 2 | 157 |
| 19.69 +39.35 | TYC3136-01028-1 | 2 | 157 |
| 19.88 +44.87 | TYC3149-00473-1 | 2 | 157 |
| 19.86 +43.58 | TYC3149-01009-1 | 2 | 157 |
| 20.02 +44.43 | TYC3162-00785-1 | 2 | 157 |
| 07.01 +48.89 | TYC3399-01903-1 | 2 | 157 |
| 19.18 +45.60 | TYC3542-00549-1 | 2 | 157 |
| 19.21 +50.79 | TYC3554-00266-1 | 2 | 157 |
| 19.87 +46.84 | TYC3557-01133-1 | 2 | 157 |
| 19.88 +46.58 | TYC3558-00527-1 | 2 | 157 |
| 19.87 +47.29 | TYC3561-00330-1 | 2 | 157 |
| 19.90 +47.22 | TYC3562-01590-1 | 2 | 157 |
| 19.74 +48.98 | TYC3565-00002-1 | 2 | 157 |
| 07.09 +58.36 | TYC3779-01240-1 | 2 | 157 |
| 15.32 +59.46 | TYC3874-00543-1 | 2 | 157 |
| 18.38 +65.62 | TYC4222-00855-1 | 2 | 157 |
| 02.66 +70.92 | TYC4316-00028-1 | 2 | 157 |
| 15.19 +70.68 | TYC4414-01316-1 | 2 | 157 |
| 05.01 +78.57 | TYC4515-00093-1 | 2 | 157 |
| 06.82 -00.55 | TYC4800-01141-1 | 2 | 157 |
| 05.10 -14.67 | TYC5329-00016-1 | 2 | 157 |
| 05.21 -13.69 | TYC5342-00398-1 | 2 | 157 |
| 16.48 -13.45 | TYC5639-00436-1 | 2 | 157 |
| 16.35 -20.69 | TYC6214-00224-1 | 2 | 157 |
| 06.61 -27.52 | TYC6516-00838-1 | 2 | 157 |
| 18.10 -29.34 | TYC6854-01860-1 | 2 | 157 |
| 13.93 -32.40 | TYC7283-00732-1 | 2 | 157 |
| 06.46 +10.74 | TYC0736-01411-1 | 2 | 157 |
| 22.12 +18.94 | TYC1688-01663-1 | 2 | 157 |
| 06.98 +23.78 | TYC1895-00988-1 | 2 | 157 |
| 03.79 +39.59 | TYC2867-00058-1 | 2 | 157 |
| 04.07 +37.86 | TYC2877-00946-1 | 2 | 157 |
| 19.30 +47.85 | TYC3546-02186-1 | 2 | 157 |
| 09.94 -24.61 | TYC6608-00122-1 | 2 | 157 |
| 00.58 +34.83 | TYC2270-00894-1 | 2 | 157 |
| 06.38 -28.93 | TYC6518-01380-1 | 2 | 157 |
| 03.98 +20.49 | TYC1257-00103-1 | 2 | 157 |
| 13.17 +17.53 | TYC1454-01134-1 | 2 | 157 |
| 08.89 +32.84 | TYC2488-01013-1 | 2 | 157 |
| 19.99 +46.96 | TYC3562-02318-1 | 2 | 157 |
| 08.89 +85.54 | TYC4635-01160-1 | 2 | 157 |
| 08.66 -23.38 | TYC6571-03192-1 | 2 | 157 |
| 16.96 +26.56 | TYC2067-00421-1 | 2 | 156 |
| 04.08 +37.33 | TYC2370-02401-1 | 2 | 156 |
| 07.11 +37.99 | TYC2943-01164-1 | 2 | 156 |
| 19.86 +44.69 | TYC3149-00481-1 | 2 | 156 |
| 19.30 +50.48 | TYC3550-00115-1 | 2 | 156 |
| 19.22 +50.98 | TYC3554-00289-1 | 2 | 156 |
| 19.65 +50.40 | TYC3564-00274-1 | 2 | 156 |
| 19.39 +56.16 | TYC3925-00079-1 | 2 | 156 |
| 15.19 +71.23 | TYC4414-00630-1 | 2 | 156 |
| 13.72 +17.76 | TYC1463-00324-1 | 2 | 156 |
| 02.64 +24.89 | TYC1771-00743-1 | 2 | 156 |
| 07.31 +37.14 | TYC2463-00825-1 | 2 | 156 |
| 02.32 +43.65 | TYC2842-00214-1 | 2 | 156 |
| 02.32 +43.95 | TYC2842-00308-1 | 2 | 156 |
| 11.75 +14.33 | TYC0870-00011-1 | 2 | 156 |
| 04.16 +22.14 | TYC1263-00610-1 | 2 | 156 |
| 15.32 +36.60 | TYC2569-00636-1 | 2 | 156 |
| 16.35 +41.38 | TYC3065-00618-1 | 2 | 156 |
| 20.08 +45.44 | TYC3559-01884-1 | 2 | 156 |
| 23.53 +77.82 | TYC4606-00838-1 | 2 | 156 |
| 05.97 +51.31 | TYC3373-00507-1 | 2 | 156 |
| 18.93 +48.70 | TYC3544-00653-1 | 2 | 156 |
| 20.12 +44.65 | TYC3162-00005-1 | 2 | 156 |
| 10.39 +00.55 | TYC0246-00151-1 | 2 | 156 |
| 02.52 +24.53 | TYC1771-00075-1 | 2 | 156 |
| 03.73 +40.92 | TYC2867-01101-1 | 2 | 156 |
| 11.62 +42.93 | TYC3015-01656-1 | 2 | 156 |
| 03.07 -21.09 | TYC5871-00157-1 | 2 | 156 |
| 15.86 -19.26 | TYC6194-01311-1 | 2 | 156 |
| 00.91 +01.02 | TYC0012-00199-1 | 2 | 156 |
| 03.38 +01.86 | TYC0057-00350-1 | 2 | 156 |
| 03.38 +01.65 | TYC0057-00677-1 | 2 | 156 |
| 04.57 +00.34 | TYC0082-00484-1 | 2 | 156 |
| 04.93 +07.27 | TYC0097-00184-1 | 2 | 156 |
| 04.93 +07.43 | TYC0097-00236-1 | 2 | 156 |
| 04.88 +06.47 | TYC0097-01039-1 | 2 | 156 |
| 06.34 +02.44 | TYC0136-00910-1 | 2 | 156 |
| 06.52 +00.17 | TYC0146-00966-1 | 2 | 156 |
| 06.52 +00.14 | TYC0146-01142-1 | 2 | 156 |
| 06.50 +00.19 | TYC0146-01254-1 | 2 | 156 |
| 06.75 +00.87 | TYC0147-01751-1 | 2 | 156 |
| 07.52 +02.04 | TYC0182-00590-1 | 2 | 156 |
| 08.06 +07.23 | TYC0206-00223-1 | 2 | 156 |
| 08.05 +07.33 | TYC0206-00259-1 | 2 | 156 |
| 10.61 +05.15 | TYC0259-00082-1 | 2 | 156 |
| 10.59 +05.09 | TYC0259-01200-1 | 2 | 156 |
| 10.83 +06.81 | TYC0260-00109-1 | 2 | 156 |
| 10.81 +06.70 | TYC0260-00433-1 | 2 | 156 |
| 11.75 +03.57 | TYC0275-00757-1 | 2 | 156 |
| 14.80 +00.43 | TYC0326-00945-1 | 2 | 156 |
| 14.85 +05.77 | TYC0333-00249-1 | 2 | 156 |
| 14.84 +05.90 | TYC0333-00841-1 | 2 | 156 |
| 16.56 +02.03 | TYC0386-00130-1 | 2 | 156 |
| 00.32 +14.31 | TYC0601-00613-1 | 2 | 156 |
| 04.10 +11.20 | TYC0670-00663-1 | 2 | 156 |
| 04.37 +09.92 | TYC0672-01005-1 | 2 | 156 |
| 04.85 +10.09 | TYC0687-01102-1 | 2 | 156 |
| 05.49 +09.48 | TYC0704-01971-1 | 2 | 156 |
| 06.68 +14.71 | TYC0758-01045-1 | 2 | 156 |
| 07.86 +09.35 | TYC0779-00195-1 | 2 | 156 |
| 07.86 +09.46 | TYC0783-00883-1 | 2 | 156 |
| 08.65 +12.70 | TYC0805-00894-1 | 2 | 156 |
| 09.45 +14.77 | TYC0826-00218-1 | 2 | 156 |
| 11.74 +14.77 | TYC0863-00055-1 | 2 | 156 |
| 12.03 +08.61 | TYC0865-00999-1 | 2 | 156 |
| 12.02 +09.02 | TYC0865-01096-1 | 2 | 156 |
| 11.75 +14.83 | TYC0870-00936-1 | 2 | 156 |
| 14.60 +09.26 | TYC0910-00745-1 | 2 | 156 |
| 16.27 +10.08 | TYC0950-00875-1 | 2 | 156 |
| 15.89 +12.96 | TYC0951-00037-1 | 2 | 156 |
| 04.50 +16.85 | TYC1265-00193-1 | 2 | 156 |
| 04.51 +16.56 | TYC1265-00233-1 | 2 | 156 |
| 05.41 +22.10 | TYC1308-01951-1 | 2 | 156 |
| 06.34 +16.21 | TYC1315-00008-1 | 2 | 156 |
| 06.65 +15.13 | TYC1329-00397-1 | 2 | 156 |
| 06.65 +15.14 | TYC1329-01534-1 | 2 | 156 |
| 06.62 +19.96 | TYC1337-01042-1 | 2 | 156 |
| 06.61 +19.83 | TYC1337-01454-1 | 2 | 156 |
| 07.46 +19.73 | TYC1355-00497-1 | 2 | 156 |
| 07.46 +19.58 | TYC1355-01208-1 | 2 | 156 |
| 07.55 +17.13 | TYC1364-00235-1 | 2 | 156 |
| 09.29 +16.71 | TYC1402-00527-1 | 2 | 156 |
| 09.29 +16.63 | TYC1402-01504-1 | 2 | 156 |
| 09.36 +21.70 | TYC1408-00411-1 | 2 | 156 |
| 09.36 +21.80 | TYC1408-00603-1 | 2 | 156 |
| 10.30 +17.45 | TYC1419-00396-1 | 2 | 156 |
| 10.31 +17.50 | TYC1420-00045-1 | 2 | 156 |
| 11.72 +18.61 | TYC1438-00003-1 | 2 | 156 |
| 11.73 +18.47 | TYC1438-00508-1 | 2 | 156 |
| 12.48 +21.74 | TYC1447-01638-1 | 2 | 156 |
| 23.17 +18.30 | TYC1715-01077-1 | 2 | 156 |
| 23.31 +18.62 | TYC1716-01017-1 | 2 | 156 |
| 23.11 +21.30 | TYC1718-00090-1 | 2 | 156 |
| 02.11 +23.22 | TYC1758-00818-1 | 2 | 156 |
| 03.22 +25.41 | TYC1788-00311-1 | 2 | 156 |
| 03.22 +25.22 | TYC1788-01081-1 | 2 | 156 |
| 03.65 +24.37 | TYC1799-00491-1 | 2 | 156 |
| 04.51 +27.43 | TYC1837-00101-1 | 2 | 156 |
| 04.49 +27.40 | TYC1837-00510-1 | 2 | 156 |
| 05.08 +29.08 | TYC1857-00705-1 | 2 | 156 |
| 05.07 +28.74 | TYC1857-01621-1 | 2 | 156 |
| 05.75 +25.52 | TYC1866-00136-1 | 2 | 156 |
| 05.77 +25.54 | TYC1866-00271-1 | 2 | 156 |
| 06.96 +23.12 | TYC1894-01750-1 | 2 | 156 |
| 06.90 +28.46 | TYC1906-00803-1 | 2 | 156 |
| 07.18 +29.89 | TYC1908-00003-1 | 2 | 156 |
| 07.24 +28.27 | TYC1908-01027-1 | 2 | 156 |
| 08.07 +24.49 | TYC1930-01719-1 | 2 | 156 |
| 08.09 +24.46 | TYC1931-00470-1 | 2 | 156 |
| 08.32 +27.13 | TYC1936-00545-1 | 2 | 156 |
| 08.60 +23.26 | TYC1942-00741-1 | 2 | 156 |
| 08.62 +23.03 | TYC1942-02430-1 | 2 | 156 |
| 09.22 +23.15 | TYC1951-01667-1 | 2 | 156 |
| 16.16 +26.97 | TYC2038-01417-1 | 2 | 156 |
| 17.44 +27.09 | TYC2082-00927-1 | 2 | 156 |
| 18.04 +26.48 | TYC2099-02539-1 | 2 | 156 |
| 03.46 +35.02 | TYC2350-00477-1 | 2 | 156 |
| 03.46 +34.90 | TYC2350-01511-1 | 2 | 156 |
| 04.16 +33.68 | TYC2362-00016-1 | 2 | 156 |
| 04.77 +31.54 | TYC2374-00229-1 | 2 | 156 |
| 05.11 +35.39 | TYC2397-00319-1 | 2 | 156 |
| 05.76 +32.87 | TYC2409-00853-1 | 2 | 156 |
| 05.78 +32.97 | TYC2409-01326-1 | 2 | 156 |
| 05.76 +35.43 | TYC2413-00435-1 | 2 | 156 |
| 06.40 +32.59 | TYC2425-00300-1 | 2 | 156 |
| 06.39 +32.11 | TYC2425-00323-1 | 2 | 156 |
| 06.31 +35.26 | TYC2429-01085-1 | 2 | 156 |
| 07.19 +30.22 | TYC2438-00250-1 | 2 | 156 |
| 06.62 +34.54 | TYC2443-00566-1 | 2 | 156 |
| 07.25 +32.80 | TYC2455-00374-1 | 2 | 156 |
| 07.24 +32.59 | TYC2455-01836-1 | 2 | 156 |
| 08.43 +31.25 | TYC2483-00393-1 | 2 | 156 |
| 08.72 +31.51 | TYC2484-01047-1 | 2 | 156 |
| 08.72 +31.77 | TYC2484-01314-1 | 2 | 156 |
| 08.77 +35.89 | TYC2490-00092-1 | 2 | 156 |
| 08.77 +35.81 | TYC2490-00366-1 | 2 | 156 |
| 09.56 +30.55 | TYC2494-00215-1 | 2 | 156 |
| 09.21 +33.65 | TYC2496-00410-1 | 2 | 156 |
| 09.21 +33.51 | TYC2496-00486-1 | 2 | 156 |
| 10.54 +31.46 | TYC2511-00613-1 | 2 | 156 |
| 10.55 +31.51 | TYC2511-00622-1 | 2 | 156 |
| 10.75 +32.17 | TYC2512-00192-1 | 2 | 156 |
| 10.76 +32.08 | TYC2512-00300-1 | 2 | 156 |
| 18.96 +33.05 | TYC2643-00588-1 | 2 | 156 |
| 21.64 +30.60 | TYC2717-01677-1 | 2 | 156 |
| 03.82 +40.67 | TYC2867-01126-1 | 2 | 156 |
| 04.43 +39.35 | TYC2879-00068-1 | 2 | 156 |
| 04.43 +39.38 | TYC2883-01470-1 | 2 | 156 |
| 06.87 +40.80 | TYC2946-00632-1 | 2 | 156 |
| 13.95 +43.41 | TYC3033-00978-1 | 2 | 156 |
| 14.98 +43.98 | TYC3050-01159-1 | 2 | 156 |
| 16.50 +38.49 | TYC3063-02040-1 | 2 | 156 |
| 19.04 +38.22 | TYC3120-00338-1 | 2 | 156 |
| 18.75 +42.51 | TYC3126-01578-1 | 2 | 156 |
| 18.75 +42.55 | TYC3126-01579-1 | 2 | 156 |
| 19.74 +42.19 | TYC3144-01758-1 | 2 | 156 |
| 19.95 +43.93 | TYC3149-00534-1 | 2 | 156 |
| 20.66 +42.18 | TYC3161-00192-1 | 2 | 156 |
| 03.92 +46.74 | TYC3326-02895-1 | 2 | 156 |
| 06.05 +46.76 | TYC3374-01468-1 | 2 | 156 |
| 08.15 +47.29 | TYC3408-00307-1 | 2 | 156 |
| 08.63 +46.18 | TYC3416-00884-1 | 2 | 156 |
| 08.62 +46.13 | TYC3416-01488-1 | 2 | 156 |
| 09.64 +50.16 | TYC3432-00246-1 | 2 | 156 |
| 09.97 +51.33 | TYC3439-00520-1 | 2 | 156 |
| 11.41 +49.36 | TYC3453-01756-1 | 2 | 156 |
| 13.26 +46.00 | TYC3460-01651-1 | 2 | 156 |
| 13.26 +46.19 | TYC3460-02449-1 | 2 | 156 |
| 17.05 +47.22 | TYC3501-00626-1 | 2 | 156 |
| 18.97 +51.18 | TYC3553-00109-1 | 2 | 156 |
| 19.80 +47.74 | TYC3561-00185-1 | 2 | 156 |
| 19.79 +47.33 | TYC3561-00318-1 | 2 | 156 |
| 19.85 +48.17 | TYC3561-01538-1 | 2 | 156 |
| 19.88 +48.31 | TYC3562-00914-1 | 2 | 156 |
| 03.81 +58.02 | TYC3725-01432-1 | 2 | 156 |
| 03.81 +57.60 | TYC3725-01748-1 | 2 | 156 |
| 06.94 +55.41 | TYC3771-00552-1 | 2 | 156 |
| 06.94 +55.55 | TYC3771-01120-1 | 2 | 156 |
| 07.46 +52.56 | TYC3781-00702-1 | 2 | 156 |
| 09.18 +59.18 | TYC3811-01066-1 | 2 | 156 |
| 09.19 +59.10 | TYC3812-01489-1 | 2 | 156 |
| 11.83 +57.75 | TYC3838-00707-1 | 2 | 156 |
| 13.54 +53.72 | TYC3850-00654-1 | 2 | 156 |
| 19.40 +55.69 | TYC3925-00173-1 | 2 | 156 |
| 04.78 +64.44 | TYC4086-00230-1 | 2 | 156 |
| 04.77 +64.81 | TYC4086-01633-1 | 2 | 156 |
| 06.92 +61.03 | TYC4110-01622-1 | 2 | 156 |
| 07.53 +60.72 | TYC4112-00763-1 | 2 | 156 |
| 07.53 +60.62 | TYC4112-01712-1 | 2 | 156 |
| 09.01 +66.31 | TYC4134-01236-1 | 2 | 156 |
| 20.21 +65.24 | TYC4240-00236-1 | 2 | 156 |
| 07.23 +70.35 | TYC4364-00712-1 | 2 | 156 |
| 07.78 +70.50 | TYC4365-01142-1 | 2 | 156 |
| 06.63 +72.86 | TYC4366-00067-1 | 2 | 156 |
| 06.63 +72.95 | TYC4366-00843-1 | 2 | 156 |
| 07.74 +74.34 | TYC4373-00490-1 | 2 | 156 |
| 07.78 +74.51 | TYC4373-00571-1 | 2 | 156 |
| 11.05 +73.75 | TYC4391-00045-1 | 2 | 156 |
| 11.05 +73.41 | TYC4391-00411-1 | 2 | 156 |
| 07.12 +76.80 | TYC4526-00228-1 | 2 | 156 |
| 07.07 +76.91 | TYC4530-01116-1 | 2 | 156 |
| 09.03 +76.46 | TYC4541-02307-1 | 2 | 156 |
| 10.46 +77.41 | TYC4542-01207-1 | 2 | 156 |
| 10.46 +77.59 | TYC4545-00430-1 | 2 | 156 |
| 03.73 -00.79 | TYC4717-00126-1 | 2 | 156 |
| 03.74 -00.69 | TYC4717-00502-1 | 2 | 156 |
| 06.80 -00.74 | TYC4800-01154-1 | 2 | 156 |
| 07.78 -00.92 | TYC4832-01104-1 | 2 | 156 |
| 07.78 -01.13 | TYC4832-01837-1 | 2 | 156 |
| 05.12 -12.36 | TYC5338-00770-1 | 2 | 156 |
| 05.13 -12.59 | TYC5338-01446-1 | 2 | 156 |
| 07.62 -13.82 | TYC5409-02582-1 | 2 | 156 |
| 07.62 -13.99 | TYC5409-02718-1 | 2 | 156 |
| 08.63 -13.38 | TYC5441-00585-1 | 2 | 156 |
| 08.64 -13.46 | TYC5441-01186-1 | 2 | 156 |
| 13.84 -13.05 | TYC5559-00166-1 | 2 | 156 |
| 13.83 -13.17 | TYC5559-00371-1 | 2 | 156 |
| 03.00 -20.72 | TYC5870-00671-1 | 2 | 156 |
| 03.81 -17.75 | TYC5884-00438-1 | 2 | 156 |
| 03.82 -17.94 | TYC5884-01513-1 | 2 | 156 |
| 04.67 -15.17 | TYC5892-00984-1 | 2 | 156 |
| 04.65 -15.04 | TYC5892-01098-1 | 2 | 156 |
| 06.07 -16.03 | TYC5932-00973-1 | 2 | 156 |
| 06.62 -19.17 | TYC5956-01347-1 | 2 | 156 |
| 06.60 -19.29 | TYC5956-01948-1 | 2 | 156 |
| 06.61 -19.09 | TYC5956-02415-1 | 2 | 156 |
| 08.48 -15.96 | TYC5998-01944-1 | 2 | 156 |
| 08.15 -20.55 | TYC6004-01064-1 | 2 | 156 |
| 08.15 -20.33 | TYC6004-01358-1 | 2 | 156 |
| 08.89 -17.29 | TYC6017-00433-1 | 2 | 156 |
| 08.89 -17.23 | TYC6017-00467-1 | 2 | 156 |
| 08.93 -22.48 | TYC6025-02106-1 | 2 | 156 |
| 08.95 -22.48 | TYC6026-00029-1 | 2 | 156 |
| 10.01 -15.50 | TYC6046-00892-1 | 2 | 156 |
| 10.14 -19.56 | TYC6067-00041-1 | 2 | 156 |
| 10.87 -18.65 | TYC6079-01235-1 | 2 | 156 |
| 11.99 -20.15 | TYC6100-00270-1 | 2 | 156 |
| 14.32 -20.31 | TYC6147-00265-1 | 2 | 156 |
| 16.11 -18.45 | TYC6204-00086-1 | 2 | 156 |
| 16.11 -18.39 | TYC6204-00216-1 | 2 | 156 |
| 01.11 -22.63 | TYC6422-00160-1 | 2 | 156 |
| 04.94 -23.29 | TYC6466-00651-1 | 2 | 156 |
| 07.80 -27.28 | TYC6548-00113-1 | 2 | 156 |
| 07.81 -27.50 | TYC6548-01046-1 | 2 | 156 |
| 08.37 -26.30 | TYC6564-02663-1 | 2 | 156 |
| 09.95 -24.00 | TYC6604-00699-1 | 2 | 156 |
| 09.75 -27.05 | TYC6610-00185-1 | 2 | 156 |
| 09.86 -27.55 | TYC6611-00359-1 | 2 | 156 |
| 13.92 -28.98 | TYC6728-00884-1 | 2 | 156 |
| 13.92 -28.87 | TYC6728-01110-1 | 2 | 156 |
| 04.04 -30.51 | TYC7030-00092-1 | 2 | 156 |
| 04.03 -30.60 | TYC7030-00128-1 | 2 | 156 |
| 05.22 -31.47 | TYC7048-00616-1 | 2 | 156 |
| 05.22 -31.21 | TYC7048-00658-1 | 2 | 156 |
| 06.05 -30.61 | TYC7071-01549-1 | 2 | 156 |
| 06.03 -30.37 | TYC7071-01617-1 | 2 | 156 |
| 08.97 -31.66 | TYC7138-00633-1 | 2 | 156 |
| 08.96 -31.57 | TYC7138-00687-1 | 2 | 156 |
| 00.01 +77.56 | TYC4496-02063-1 | 2 | 155 |
| 06.20 -29.95 | TYC6517-02179-1 | 2 | 155 |
| 15.92 +36.20 | TYC2578-00334-1 | 2 | 155 |
| 09.86 -16.18 | TYC6044-01004-1 | 2 | 155 |
| 06.49 +00.37 | TYC0133-02031-1 | 2 | 155 |
| 09.74 +15.82 | TYC1410-00470-1 | 2 | 155 |
| 08.64 +12.54 | TYC0805-00632-1 | 2 | 154 |
| 10.23 +67.86 | TYC4384-00080-1 | 2 | 154 |
| 03.24 +31.60 | TYC2340-00401-1 | 2 | 154 |
| 09.00 +65.22 | TYC4134-00360-1 | 2 | 154 |
| 05.76 -13.35 | TYC5359-01219-1 | 2 | 154 |
| 05.49 +09.31 | TYC0700-00141-1 | 2 | 154 |
| 06.70 +14.80 | TYC0758-01146-1 | 2 | 154 |
| 16.26 +10.00 | TYC0950-00948-1 | 2 | 154 |
| 09.45 +15.15 | TYC1403-00296-1 | 2 | 154 |
| 15.88 +15.66 | TYC1496-02104-1 | 2 | 154 |
| 03.66 +24.24 | TYC1799-00623-1 | 2 | 154 |
| 06.29 +35.09 | TYC2428-00933-1 | 2 | 154 |
| 06.62 +34.62 | TYC2443-00064-1 | 2 | 154 |
| 06.07 +46.69 | TYC3374-01346-1 | 2 | 154 |
| 09.97 +51.46 | TYC3439-00401-1 | 2 | 154 |
| 06.93 +60.77 | TYC4110-01603-1 | 2 | 154 |
| 08.92 +61.16 | TYC4128-00276-1 | 2 | 154 |
| 08.96 +61.30 | TYC4128-00766-1 | 2 | 154 |
| 05.10 -13.69 | TYC5329-01865-1 | 2 | 154 |
| 05.14 -14.16 | TYC5342-01141-1 | 2 | 154 |
| 07.63 -13.89 | TYC5409-03239-1 | 2 | 154 |
| 04.31 -24.05 | TYC6457-02564-1 | 2 | 154 |
| 07.52 +02.14 | TYC0182-00432-1 | 2 | 154 |
| 04.85 +09.82 | TYC0687-01374-1 | 2 | 154 |
| 06.33 +15.97 | TYC1315-01154-1 | 2 | 154 |
| 04.31 -23.92 | TYC6457-03268-1 | 2 | 154 |
| 03.06 +21.21 | TYC1231-01405-1 | 2 | 153 |
| 09.17 +17.88 | TYC1404-00262-1 | 2 | 153 |
| 11.31 +25.30 | TYC1981-00752-1 | 2 | 153 |
| 05.94 +74.37 | TYC4356-00177-1 | 2 | 153 |
| 02.80 +07.56 | TYC0640-00131-1 | 2 | 153 |
| 15.23 +71.34 | TYC4414-00045-1 | 2 | 153 |
| 05.19 -13.75 | TYC5342-00610-1 | 2 | 153 |
| 04.88 +06.56 | TYC0097-00302-1 | 2 | 152 |
| 06.73 -01.32 | TYC4799-01434-1 | 2 | 152 |
| 07.73 +27.75 | TYC1920-00135-1 | 2 | 152 |
| 08.35 +01.91 | TYC0200-01440-1 | 2 | 152 |
| 02.48 +37.48 | TYC2335-00305-1 | 2 | 152 |
| 04.67 +19.21 | TYC1275-00510-1 | 2 | 152 |
| 23.10 +21.10 | TYC1717-01028-1 | 2 | 152 |
| 17.42 +27.64 | TYC2082-00424-1 | 2 | 152 |
| 04.79 -24.78 | TYC6465-00482-1 | 2 | 152 |
| 09.52 -12.22 | TYC5468-00684-1 | 2 | 151 |
| 06.54 -30.24 | TYC7074-01128-1 | 2 | 151 |
| 20.18 +17.96 | TYC1622-00464-1 | 2 | 151 |
| 09.60 +35.11 | TYC2500-00718-1 | 2 | 151 |
| 07.19 +13.80 | TYC0774-01705-1 | 2 | 150 |
| 20.94 +10.82 | TYC1107-01881-1 | 2 | 150 |
| 20.94 +10.93 | TYC1107-02140-1 | 2 | 150 |
| 01.04 +20.09 | TYC1195-01136-1 | 2 | 150 |
| 19.79 +33.35 | TYC2660-00114-1 | 2 | 150 |
| 22.83 +35.54 | TYC2757-00226-1 | 2 | 150 |
| 19.20 +48.97 | TYC3550-01222-1 | 2 | 150 |
| 08.54 -30.50 | TYC7135-01867-1 | 2 | 150 |
| 21.63 -31.28 | TYC7487-00738-1 | 2 | 150 |
| 15.55 +15.56 | TYC1494-00922-1 | 2 | 150 |
| 12.51 +22.04 | TYC1447-02403-1 | 2 | 150 |
| 16.94 +25.94 | TYC2063-00705-1 | 2 | 150 |
| 19.49 +37.73 | TYC3134-00056-1 | 2 | 149 |
| 09.73 +54.14 | TYC3807-00276-1 | 2 | 149 |
| 07.99 -01.31 | TYC4833-02216-1 | 2 | 149 |
| 17.96 +26.02 | TYC2094-00385-1 | 2 | 149 |
| 08.47 -11.34 | TYC5436-02018-1 | 2 | 149 |
| 09.59 -11.14 | TYC5465-00157-1 | 2 | 149 |
| 08.89 -26.92 | TYC6580-02351-1 | 2 | 149 |
| 08.46 +61.34 | TYC4127-01639-1 | 2 | 149 |
| 05.50 +27.11 | TYC1856-01060-1 | 2 | 148 |
| 19.81 +27.91 | TYC2147-00655-1 | 2 | 148 |
| 12.36 +74.90 | TYC4400-00269-1 | 2 | 148 |
| 04.87 -16.02 | TYC5899-00437-1 | 2 | 147 |
| 20.00 +44.89 | TYC3149-00001-1 | 2 | 147 |
| 09.23 +48.40 | TYC3427-01285-1 | 2 | 147 |
| 12.37 +74.17 | TYC4400-01085-1 | 2 | 147 |
| 06.19 +11.98 | TYC0738-00038-1 | 2 | 147 |
| 08.05 +21.14 | TYC1388-00504-1 | 2 | 147 |
| 06.08 +75.49 | TYC4525-00453-1 | 2 | 147 |
| 05.37 +07.14 | TYC0112-00154-1 | 2 | 146 |
| 03.13 +32.85 | TYC2343-01845-1 | 2 | 146 |
| 04.20 +47.30 | TYC3332-01264-1 | 2 | 146 |
| 10.36 -00.86 | TYC4905-00353-1 | 2 | 146 |
| 06.62 -14.39 | TYC5377-02915-1 | 2 | 145 |
| 04.51 -18.85 | TYC5894-01910-1 | 2 | 145 |
| 10.99 +07.38 | TYC0261-00612-1 | 2 | 145 |
| 08.42 +31.48 | TYC2483-00792-1 | 2 | 145 |
| 17.33 +38.63 | TYC3073-00550-1 | 2 | 145 |
| 07.10 +64.85 | TYC4118-00835-1 | 2 | 145 |
| 04.79 -24.21 | TYC6465-01593-1 | 2 | 145 |
| 07.28 +83.62 | TYC4618-00692-1 | 2 | 144 |
| 10.33 +12.32 | TYC0840-00043-1 | 2 | 144 |
| 06.78 -18.06 | TYC5953-00867-1 | 2 | 144 |
| 10.97 +02.13 | TYC0255-00462-1 | 2 | 144 |
| 01.00 +20.82 | TYC1195-01060-1 | 2 | 144 |
| 01.38 +28.84 | TYC1754-00642-1 | 2 | 144 |
| 10.05 +61.52 | TYC4137-00733-1 | 2 | 143 |
| 09.00 -13.64 | TYC5456-01910-1 | 2 | 143 |
| 08.75 -23.69 | TYC6571-02308-1 | 2 | 143 |
| 08.54 -30.36 | TYC7135-01129-1 | 2 | 143 |
| 08.55 -31.33 | TYC7136-02462-1 | 2 | 143 |
| 04.45 +19.51 | TYC1273-00832-1 | 2 | 143 |
| 04.66 +22.98 | TYC1830-01261-1 | 2 | 143 |
| 09.37 +50.75 | TYC3431-00818-1 | 2 | 143 |
| 02.05 +46.57 | TYC3280-00645-1 | 2 | 143 |
| 07.47 +81.11 | TYC4539-01144-1 | 2 | 143 |
| 21.56 -21.03 | TYC6373-00256-1 | 2 | 143 |
| 13.75 +47.49 | TYC3463-00150-1 | 2 | 143 |
| 04.97 -26.90 | TYC6469-01437-1 | 2 | 143 |
| 23.96 -22.59 | TYC6982-00383-1 | 2 | 143 |
| 02.53 -12.17 | TYC5284-00134-1 | 2 | 143 |
| 18.79 +42.72 | TYC3126-00552-1 | 2 | 142 |
| 10.78 +05.01 | TYC0260-00044-1 | 2 | 142 |
| 07.98 +24.84 | TYC1930-01701-1 | 2 | 142 |
| 05.15 +56.14 | TYC3739-00330-1 | 2 | 142 |
| 10.21 +20.48 | TYC1425-01342-1 | 2 | 142 |
| 07.80 +61.57 | TYC4113-01100-1 | 2 | 142 |
| 03.18 +21.12 | TYC1231-00400-1 | 2 | 141 |
| 02.63 +71.91 | TYC4320-01599-1 | 2 | 141 |
| 08.73 +13.31 | TYC0816-01097-1 | 2 | 141 |
| 01.38 +28.03 | TYC1754-02324-1 | 2 | 141 |
| 10.80 +27.55 | TYC1979-02475-1 | 2 | 141 |
| 23.05 +57.02 | TYC3993-01252-1 | 2 | 141 |
| 05.49 +73.65 | TYC4355-00135-1 | 2 | 141 |
| 08.51 +07.90 | TYC0796-01753-1 | 2 | 140 |
| 05.86 -29.59 | TYC6502-01425-1 | 2 | 140 |
| 08.58 +08.88 | TYC0797-00215-1 | 2 | 140 |
| 05.35 +07.32 | TYC0112-00300-1 | 2 | 139 |
| 09.56 -29.47 | TYC6613-01073-1 | 2 | 139 |
| 09.24 +23.32 | TYC1951-01164-1 | 2 | 139 |
| 06.25 +70.60 | TYC4349-00654-1 | 2 | 138 |
| 04.33 +15.01 | TYC1264-00016-1 | 2 | 138 |
| 12.29 +74.48 | TYC4400-00383-1 | 2 | 138 |
| 12.51 -27.66 | TYC6690-00181-1 | 2 | 138 |
| 11.09 +44.06 | TYC3012-00005-1 | 2 | 138 |
| 04.41 +57.79 | TYC3727-00377-1 | 2 | 138 |
| 19.84 +08.40 | TYC1058-02357-1 | 2 | 138 |
| 13.85 +16.57 | TYC1460-00014-1 | 2 | 138 |
| 07.68 +28.39 | TYC1924-01520-1 | 2 | 138 |
| 19.08 +34.48 | TYC2648-01029-1 | 2 | 138 |
| 19.01 +39.34 | TYC3120-01690-1 | 2 | 138 |
| 19.31 +38.60 | TYC3121-01106-1 | 2 | 138 |
| 18.26 +65.03 | TYC4209-00643-1 | 2 | 138 |
| 06.02 -13.83 | TYC5361-01241-1 | 2 | 138 |
| 13.87 -13.46 | TYC5559-00336-1 | 2 | 138 |
| 06.62 -18.62 | TYC5952-00001-1 | 2 | 138 |
| 08.61 -23.76 | TYC6570-01391-1 | 2 | 138 |
| 08.54 -30.33 | TYC7135-00809-1 | 2 | 138 |
| 05.74 +01.46 | TYC0115-00764-1 | 2 | 138 |
| 08.26 +01.75 | TYC0196-02467-1 | 2 | 138 |
| 08.16 +04.76 | TYC0203-01622-1 | 2 | 138 |
| 10.28 +07.36 | TYC0251-00565-1 | 2 | 138 |
| 18.63 +05.24 | TYC0455-00364-1 | 2 | 138 |
| 18.63 +04.53 | TYC0455-00923-1 | 2 | 138 |
| 20.27 +05.20 | TYC0504-01823-1 | 2 | 138 |
| 20.27 +05.04 | TYC0504-02238-1 | 2 | 138 |
| 03.13 +11.48 | TYC0651-00140-1 | 2 | 138 |
| 07.27 +13.32 | TYC0774-01216-1 | 2 | 138 |
| 18.49 +12.34 | TYC1031-00127-1 | 2 | 138 |
| 18.49 +11.30 | TYC1031-00420-1 | 2 | 138 |
| 14.90 +17.31 | TYC1478-00126-1 | 2 | 138 |
| 20.39 +16.81 | TYC1631-01657-1 | 2 | 138 |
| 20.44 +19.24 | TYC1640-02044-1 | 2 | 138 |
| 21.17 +15.03 | TYC1649-01790-1 | 2 | 138 |
| 09.29 +23.32 | TYC1951-01364-1 | 2 | 138 |
| 10.34 +29.26 | TYC1975-00621-1 | 2 | 138 |
| 16.95 +24.94 | TYC2063-00214-1 | 2 | 138 |
| 19.94 +22.93 | TYC2140-00910-1 | 2 | 138 |
| 09.53 +33.97 | TYC2497-00595-1 | 2 | 138 |
| 09.81 +34.78 | TYC2505-00133-1 | 2 | 138 |
| 15.82 +34.74 | TYC2575-00506-1 | 2 | 138 |
| 15.77 +35.16 | TYC2577-00533-1 | 2 | 138 |
| 19.14 +35.51 | TYC2648-00508-1 | 2 | 138 |
| 19.23 +32.24 | TYC2657-01355-1 | 2 | 138 |
| 05.99 +44.04 | TYC2924-00833-1 | 2 | 138 |
| 06.27 +40.92 | TYC2930-00109-1 | 2 | 138 |
| 18.75 +41.47 | TYC3126-00215-1 | 2 | 138 |
| 18.86 +44.98 | TYC3131-00022-1 | 2 | 138 |
| 19.42 +38.76 | TYC3134-00292-1 | 2 | 138 |
| 06.90 +50.81 | TYC3402-00089-1 | 2 | 138 |
| 08.73 +51.89 | TYC3422-00244-1 | 2 | 138 |
| 08.74 +50.42 | TYC3422-01472-1 | 2 | 138 |
| 09.47 +50.22 | TYC3431-00012-1 | 2 | 138 |
| 12.47 +45.25 | TYC3456-00388-1 | 2 | 138 |
| 13.73 +47.46 | TYC3463-00472-1 | 2 | 138 |
| 19.15 +51.45 | TYC3554-01324-1 | 2 | 138 |
| 19.86 +46.64 | TYC3557-00955-1 | 2 | 138 |
| 20.06 +45.73 | TYC3559-00216-1 | 2 | 138 |
| 08.92 +53.57 | TYC3805-01775-1 | 2 | 138 |
| 10.28 +56.00 | TYC3818-01078-1 | 2 | 138 |
| 15.33 +58.71 | TYC3874-01142-1 | 2 | 138 |
| 19.36 +55.47 | TYC3925-00795-1 | 2 | 138 |
| 20.44 +59.97 | TYC3949-00013-1 | 2 | 138 |
| 23.04 +56.99 | TYC3993-01123-1 | 2 | 138 |
| 23.23 +57.95 | TYC4006-00223-1 | 2 | 138 |
| 10.53 +60.47 | TYC4144-00979-1 | 2 | 138 |
| 20.33 +60.43 | TYC4233-01018-1 | 2 | 138 |
| 02.83 +68.04 | TYC4313-01102-1 | 2 | 138 |
| 11.01 +78.63 | TYC4552-01505-1 | 2 | 138 |
| 10.39 -00.50 | TYC4905-00383-1 | 2 | 138 |
| 03.02 -21.10 | TYC5870-00859-1 | 2 | 138 |
| 10.17 -20.93 | TYC6071-00147-1 | 2 | 138 |
| 11.92 -20.50 | TYC6100-00652-1 | 2 | 138 |
| 14.40 -16.77 | TYC6139-00166-1 | 2 | 138 |
| 14.25 -20.11 | TYC6146-00588-1 | 2 | 138 |
| 07.64 -27.57 | TYC6547-00503-1 | 2 | 138 |
| 09.65 -23.71 | TYC6602-01523-1 | 2 | 138 |
| 10.36 -29.32 | TYC6631-01179-1 | 2 | 138 |
| 05.64 +06.08 | TYC0127-00040-1 | 2 | 138 |
| 02.76 +49.68 | TYC3304-00729-1 | 2 | 138 |
| 12.47 +74.25 | TYC4400-00065-1 | 2 | 138 |
| 03.46 -23.87 | TYC6446-01139-1 | 2 | 138 |
| 04.81 -24.23 | TYC6465-01551-1 | 2 | 138 |
| 04.90 -23.50 | TYC6466-01563-1 | 2 | 138 |
| 11.24 +25.47 | TYC1981-01158-1 | 2 | 138 |
| 18.46 +65.57 | TYC4222-00675-1 | 2 | 137 |
| 00.75 -26.54 | TYC6423-01908-1 | 2 | 137 |
| 11.03 +02.15 | TYC0262-00155-1 | 2 | 137 |
| 10.53 +26.85 | TYC1973-00671-1 | 2 | 137 |
| 19.01 +37.00 | TYC2651-00303-1 | 2 | 137 |
| 02.53 +41.20 | TYC2836-00037-1 | 2 | 137 |
| 10.44 -01.55 | TYC4905-00571-1 | 2 | 137 |
| 04.88 -22.46 | TYC5911-00222-1 | 2 | 137 |
| 08.20 +05.49 | TYC0203-00252-1 | 2 | 137 |
| 10.79 +04.68 | TYC0257-00023-1 | 2 | 137 |
| 09.52 +12.65 | TYC0826-00410-1 | 2 | 137 |
| 11.43 +14.55 | TYC0862-00479-1 | 2 | 137 |
| 15.84 +13.03 | TYC0951-00245-1 | 2 | 137 |
| 15.92 +12.37 | TYC0951-00419-1 | 2 | 137 |
| 19.89 +07.90 | TYC1058-01324-1 | 2 | 137 |
| 19.89 +09.28 | TYC1058-01329-1 | 2 | 137 |
| 09.64 +32.40 | TYC2501-00125-1 | 2 | 137 |
| 18.71 +42.35 | TYC3126-00156-1 | 2 | 137 |
| 09.15 +46.78 | TYC3424-00705-1 | 2 | 137 |
| 15.33 +58.50 | TYC3874-01306-1 | 2 | 137 |
| 19.30 +55.65 | TYC3924-00132-1 | 2 | 137 |
| 23.23 +58.54 | TYC4010-00974-1 | 2 | 137 |
| 04.66 +66.87 | TYC4090-00911-1 | 2 | 137 |
| 08.16 +61.02 | TYC4126-02289-1 | 2 | 137 |
| 20.14 +64.93 | TYC4240-00065-1 | 2 | 137 |
| 01.30 +75.72 | TYC4493-01265-1 | 2 | 137 |
| 12.06 +76.07 | TYC4550-01081-1 | 2 | 137 |
| 11.11 -14.16 | TYC5507-01592-1 | 2 | 137 |
| 13.81 -13.03 | TYC5559-00434-1 | 2 | 137 |
| 09.88 -21.04 | TYC6057-01185-1 | 2 | 137 |
| 15.86 -18.94 | TYC6194-01445-1 | 2 | 137 |
| 06.16 -25.62 | TYC6509-01315-1 | 2 | 137 |
| 09.91 -23.31 | TYC6603-01375-1 | 2 | 137 |
| 11.04 -23.41 | TYC6636-00694-1 | 2 | 137 |
| 10.53 +12.62 | TYC0844-00947-1 | 2 | 136 |
| 20.39 +17.55 | TYC1635-01626-1 | 2 | 136 |
| 11.38 +32.02 | TYC2520-00285-1 | 2 | 136 |
| 19.36 +56.26 | TYC3929-01300-1 | 2 | 136 |
| 20.25 +59.64 | TYC3949-00029-1 | 2 | 136 |
| 07.64 +62.40 | TYC4116-00623-1 | 2 | 136 |
| 06.93 +72.34 | TYC4367-01170-1 | 2 | 136 |
| 09.73 +72.04 | TYC4386-01130-1 | 2 | 136 |
| 11.11 -13.98 | TYC5507-01406-1 | 2 | 136 |
| 06.79 +00.24 | TYC0148-02516-1 | 2 | 136 |
| 15.81 +12.72 | TYC0938-00010-1 | 2 | 136 |
| 10.09 +17.86 | TYC1415-00168-1 | 2 | 136 |
| 13.15 +17.12 | TYC1450-01061-1 | 2 | 136 |
| 06.98 +24.48 | TYC1898-00334-1 | 2 | 136 |
| 11.23 +25.90 | TYC1981-00208-1 | 2 | 136 |
| 16.91 +26.33 | TYC2067-00084-1 | 2 | 136 |
| 17.99 +26.04 | TYC2094-00783-1 | 2 | 136 |
| 07.47 +33.59 | TYC2456-02084-1 | 2 | 136 |
| 15.91 +36.18 | TYC2578-00346-1 | 2 | 136 |
| 15.02 +52.93 | TYC3861-00092-1 | 2 | 136 |
| 20.32 +65.01 | TYC4241-01628-1 | 2 | 136 |
| 23.02 -01.22 | TYC5242-01173-1 | 2 | 136 |
| 07.68 -14.26 | TYC5422-02727-1 | 2 | 136 |
| 15.83 +35.55 | TYC2578-00984-1 | 2 | 136 |
| 04.32 +58.06 | TYC3727-00425-1 | 2 | 136 |
| 10.16 +17.58 | TYC1422-01328-1 | 2 | 135 |
| 14.87 +17.41 | TYC1478-00199-1 | 2 | 135 |
| 15.32 +58.99 | TYC3874-00477-1 | 2 | 135 |
| 11.87 +76.93 | TYC4550-00344-1 | 2 | 135 |
| 07.58 -13.25 | TYC5409-01280-1 | 2 | 135 |
| 19.94 +22.40 | TYC1628-02167-1 | 2 | 135 |
| 00.70 +34.49 | TYC2283-00953-1 | 2 | 135 |
| 07.49 +33.37 | TYC2456-01857-1 | 2 | 135 |
| 11.33 +41.31 | TYC3015-00036-1 | 2 | 135 |
| 19.08 +46.83 | TYC3541-02869-1 | 2 | 135 |
| 10.53 +59.38 | TYC3822-00248-1 | 2 | 135 |
| 14.37 -20.50 | TYC6147-00233-1 | 2 | 135 |
| 15.86 -19.17 | TYC6194-01033-1 | 2 | 135 |
| 21.40 -22.06 | TYC6372-00472-1 | 2 | 135 |
| 04.09 +80.48 | TYC4518-01498-1 | 1 | 134 |
| 11.02 +76.85 | TYC4549-00445-1 | 1 | 134 |
| 10.97 -24.47 | TYC6639-00165-1 | 1 | 134 |
| 04.63 +22.99 | TYC1830-01158-1 | 1 | 134 |
| 15.11 +72.49 | TYC4414-00198-1 | 1 | 134 |
| 12.63 -30.37 | TYC7247-00632-1 | 1 | 134 |
| 20.07 +45.85 | TYC3559-00298-1 | 1 | 134 |
| 09.38 +45.91 | TYC3425-00583-1 | 1 | 134 |
| 04.75 -24.03 | TYC6465-01075-1 | 1 | 134 |
| 08.33 -28.66 | TYC6568-00631-1 | 1 | 134 |
| 22.22 +16.31 | TYC1681-00401-1 | 1 | 134 |
| 07.28 +14.36 | TYC0775-00439-1 | 1 | 133 |
| 22.99 +38.95 | TYC3215-00083-1 | 1 | 133 |
| 07.39 +47.51 | TYC3397-01350-1 | 1 | 133 |
| 06.82 -00.44 | TYC4800-00123-1 | 1 | 133 |
| 06.81 -00.82 | TYC4800-00126-1 | 1 | 133 |
| 03.14 +03.46 | TYC0058-00239-1 | 1 | 132 |
| 09.88 +71.41 | TYC4386-00933-1 | 1 | 132 |
| 08.85 +28.10 | TYC1949-00889-1 | 1 | 132 |
| 22.93 +20.58 | TYC1706-00149-1 | 1 | 132 |
| 07.94 +23.03 | TYC1925-00795-1 | 1 | 131 |
| 07.31 +55.93 | TYC3784-00785-1 | 1 | 131 |
| 05.33 +12.68 | TYC0707-01407-1 | 1 | 131 |
| 03.23 +31.69 | TYC2340-00163-1 | 1 | 131 |
| 05.18 +79.09 | TYC4519-01190-1 | 1 | 131 |
| 14.87 +05.80 | TYC0333-00607-1 | 1 | 131 |
| 19.77 +15.03 | TYC1615-01133-1 | 1 | 130 |
| 06.44 +29.16 | TYC1891-00934-1 | 1 | 130 |
| 09.87 +24.42 | TYC1961-00176-1 | 1 | 130 |
| 18.76 +35.58 | TYC2646-01870-1 | 1 | 130 |
| 20.54 +34.06 | TYC2693-01003-1 | 1 | 130 |
| 06.66 +55.41 | TYC3769-02214-1 | 1 | 130 |
| 06.50 +06.21 | TYC0145-02313-1 | 1 | 130 |
| 00.31 +31.99 | TYC2265-01258-1 | 1 | 130 |
| 19.49 +38.73 | TYC3134-00139-1 | 1 | 130 |
| 19.47 +41.04 | TYC3138-00035-1 | 1 | 130 |
| 06.56 -18.61 | TYC5952-00033-1 | 1 | 130 |
| 06.88 +40.87 | TYC2946-00405-1 | 1 | 130 |
| 04.55 +68.00 | TYC4329-01779-1 | 1 | 130 |
| 10.33 +20.40 | TYC1426-01208-1 | 1 | 129 |
| 19.43 +48.51 | TYC3547-00351-1 | 1 | 129 |
| 19.26 +51.11 | TYC3554-00518-1 | 1 | 129 |
| 05.64 +17.97 | TYC1302-00387-1 | 1 | 129 |
| 23.20 +57.51 | TYC4006-00050-1 | 1 | 129 |
| 03.56 -23.60 | TYC6447-00465-1 | 1 | 129 |
| 21.22 -20.70 | TYC6359-00946-1 | 1 | 129 |
| 15.97 -28.01 | TYC6787-01673-1 | 1 | 129 |
| 06.78 +00.25 | TYC0148-02360-1 | 1 | 128 |
| 11.94 +76.96 | TYC4550-00390-1 | 1 | 128 |
| 04.92 -23.20 | TYC6466-00642-1 | 1 | 128 |
| 06.60 -27.78 | TYC6516-01014-1 | 1 | 128 |
| 06.82 +40.87 | TYC2945-01760-1 | 1 | 127 |
| 12.69 +74.25 | TYC4400-00008-1 | 1 | 127 |
| 02.14 +18.86 | TYC1217-00436-1 | 1 | 126 |
| 05.21 -26.91 | TYC6482-00867-1 | 1 | 126 |
| 05.99 +28.24 | TYC1875-02032-1 | 1 | 126 |
| 07.51 +33.77 | TYC2461-00696-1 | 1 | 126 |
| 19.04 +37.27 | TYC2651-01977-1 | 1 | 126 |
| 08.04 -00.90 | TYC4846-00974-1 | 1 | 126 |
| 05.59 +20.70 | TYC1309-01843-1 | 1 | 126 |
| 01.00 +20.26 | TYC1195-00998-1 | 1 | 126 |
| 22.06 +18.67 | TYC1684-00635-1 | 1 | 126 |
| 02.63 +24.58 | TYC1771-00746-1 | 1 | 126 |
| 22.06 +26.63 | TYC2212-00797-1 | 1 | 126 |
| 02.35 +32.43 | TYC2314-00261-1 | 1 | 126 |
| 02.57 -12.26 | TYC5288-00387-1 | 1 | 126 |
| 00.48 -16.01 | TYC5840-00953-1 | 1 | 126 |
| 09.36 +05.98 | TYC0234-01040-1 | 1 | 125 |
| 06.98 +05.11 | TYC0157-00425-1 | 1 | 125 |
| 10.99 +02.13 | TYC0255-00385-1 | 1 | 125 |
| 00.31 +14.47 | TYC0601-00135-1 | 1 | 125 |
| 00.66 +21.67 | TYC1193-00374-1 | 1 | 125 |
| 11.90 +16.47 | TYC1441-01140-1 | 1 | 125 |
| 22.22 +16.36 | TYC1681-00226-1 | 1 | 125 |
| 02.74 +38.67 | TYC2845-01210-1 | 1 | 125 |
| 07.76 +38.76 | TYC2959-00628-1 | 1 | 125 |
| 20.65 +42.68 | TYC3161-00035-1 | 1 | 125 |
| 10.34 +50.33 | TYC3441-00552-1 | 1 | 125 |
| 04.94 +64.83 | TYC4087-01011-1 | 1 | 125 |
| 06.19 +70.27 | TYC4349-00735-1 | 1 | 125 |
| 05.29 +82.68 | TYC4617-01919-1 | 1 | 125 |
| 09.92 -15.90 | TYC6045-00488-1 | 1 | 125 |
| 09.92 -15.70 | TYC6045-00885-1 | 1 | 125 |
| 08.25 -28.98 | TYC6568-00405-1 | 1 | 125 |
| 08.82 +51.11 | TYC3423-01293-1 | 1 | 125 |
| 08.19 +04.83 | TYC0203-00786-1 | 1 | 125 |
| 02.19 +32.08 | TYC2313-00457-1 | 1 | 125 |
| 19.79 +46.35 | TYC3557-01106-1 | 1 | 125 |
| 05.74 +01.26 | TYC0115-01234-1 | 1 | 125 |
| 23.26 +57.97 | TYC4006-00029-1 | 1 | 125 |
| 03.57 -23.95 | TYC6447-00361-1 | 1 | 124 |
| 10.42 -00.86 | TYC4905-00016-1 | 1 | 124 |
| 08.86 -29.74 | TYC6584-01784-1 | 1 | 124 |
| 19.81 +50.67 | TYC3569-00071-1 | 1 | 123 |
| 09.48 +50.73 | TYC3431-01393-1 | 1 | 123 |
| 10.27 +84.44 | TYC4636-01222-1 | 1 | 123 |
| 14.47 -15.24 | TYC6152-00516-1 | 1 | 123 |
| 10.34 -29.13 | TYC6631-00926-1 | 1 | 123 |
| 10.57 +08.75 | TYC0839-00739-1 | 1 | 123 |
| 11.50 +20.12 | TYC1440-00727-1 | 1 | 123 |
| 02.54 +42.45 | TYC2840-01641-1 | 1 | 123 |
| 04.51 -13.82 | TYC5320-01641-1 | 1 | 123 |
| 06.33 -28.42 | TYC6518-01819-1 | 1 | 123 |
| 16.51 +38.62 | TYC3063-01721-1 | 1 | 123 |
| 19.03 +41.55 | TYC3128-01653-1 | 1 | 122 |
| 10.95 +40.51 | TYC3009-00007-1 | 1 | 122 |
| 23.42 +39.29 | TYC3230-00187-1 | 1 | 122 |
| 03.42 -21.98 | TYC5879-00285-1 | 1 | 122 |
| 08.29 +04.70 | TYC0204-00884-1 | 1 | 121 |
| 08.19 +05.98 | TYC0207-00917-1 | 1 | 121 |
| 08.20 +05.70 | TYC0207-01289-1 | 1 | 121 |
| 07.44 +47.16 | TYC3397-01174-1 | 1 | 121 |
| 07.28 -30.50 | TYC7103-01579-1 | 1 | 121 |
| 19.05 +37.46 | TYC2651-01908-1 | 1 | 121 |
| 08.13 +30.68 | TYC2469-01274-1 | 1 | 121 |
| 09.96 -23.97 | TYC6604-00746-1 | 1 | 121 |
| 09.86 +27.13 | TYC1964-01457-1 | 1 | 120 |
| 09.57 -13.24 | TYC5472-00060-1 | 1 | 120 |
| 10.28 +41.80 | TYC3004-00212-1 | 1 | 120 |
| 11.28 +17.76 | TYC1437-00994-1 | 1 | 120 |
| 14.47 -16.31 | TYC6152-00692-1 | 1 | 119 |
| 19.85 +46.68 | TYC3557-00064-1 | 1 | 118 |
| 20.06 +29.86 | TYC2153-02004-1 | 1 | 118 |
| 19.31 +44.28 | TYC3133-00038-1 | 1 | 118 |
| 19.31 +44.21 | TYC3133-00104-1 | 1 | 118 |
| 05.20 +07.48 | TYC0111-00315-1 | 1 | 118 |
| 05.70 +06.12 | TYC0127-00992-1 | 1 | 118 |
| 03.29 +09.88 | TYC0649-01431-1 | 1 | 118 |
| 16.80 +12.05 | TYC0970-01615-1 | 1 | 118 |
| 18.48 +10.98 | TYC1027-00164-1 | 1 | 118 |
| 04.50 +19.44 | TYC1273-00925-1 | 1 | 118 |
| 01.38 +28.37 | TYC1754-00523-1 | 1 | 118 |
| 03.29 +29.93 | TYC1796-00246-1 | 1 | 118 |
| 03.28 +29.95 | TYC1796-00330-1 | 1 | 118 |
| 22.03 +26.18 | TYC2207-00427-1 | 1 | 118 |
| 00.60 +34.47 | TYC2270-00317-1 | 1 | 118 |
| 00.55 +34.54 | TYC2270-00399-1 | 1 | 118 |
| 00.65 +35.13 | TYC2283-00029-1 | 1 | 118 |
| 06.43 +39.03 | TYC2927-00605-1 | 1 | 118 |
| 19.48 +41.01 | TYC3138-00509-1 | 1 | 118 |
| 16.99 +46.90 | TYC3500-01138-1 | 1 | 118 |
| 16.33 -19.63 | TYC6210-00005-1 | 1 | 118 |
| 02.79 -24.27 | TYC6434-01383-1 | 1 | 118 |
| 17.03 -28.63 | TYC6823-02292-1 | 1 | 118 |
| 17.88 -28.56 | TYC6853-01401-1 | 1 | 118 |
| 03.08 +05.08 | TYC0061-00711-1 | 1 | 118 |
| 04.80 +06.16 | TYC0096-00886-1 | 1 | 118 |
| 05.27 +07.37 | TYC0112-00310-1 | 1 | 118 |
| 06.74 +01.67 | TYC0147-00289-1 | 1 | 118 |
| 08.15 +05.05 | TYC0203-00494-1 | 1 | 118 |
| 09.58 +06.38 | TYC0241-00578-1 | 1 | 118 |
| 18.88 +03.68 | TYC0453-00375-1 | 1 | 118 |
| 10.36 +09.24 | TYC0837-01031-1 | 1 | 118 |
| 12.17 +09.45 | TYC0866-00850-1 | 1 | 118 |
| 16.21 +09.36 | TYC0945-01009-1 | 1 | 118 |
| 07.73 +22.21 | TYC1373-01500-1 | 1 | 118 |
| 12.46 +22.00 | TYC1447-00820-1 | 1 | 118 |
| 06.48 +29.69 | TYC1891-00871-1 | 1 | 118 |
| 08.89 +27.91 | TYC1949-00157-1 | 1 | 118 |
| 14.53 +22.86 | TYC2015-00421-1 | 1 | 118 |
| 15.98 +27.47 | TYC2037-01484-1 | 1 | 118 |
| 16.07 +27.48 | TYC2038-01477-1 | 1 | 118 |
| 22.00 +26.16 | TYC2207-00329-1 | 1 | 118 |
| 00.55 +35.21 | TYC2270-00184-1 | 1 | 118 |
| 02.48 +37.40 | TYC2335-01139-1 | 1 | 118 |
| 19.72 +34.67 | TYC2664-00396-1 | 1 | 118 |
| 02.75 +38.39 | TYC2845-00095-1 | 1 | 118 |
| 06.27 +41.03 | TYC2930-00811-1 | 1 | 118 |
| 14.88 +44.19 | TYC3049-00624-1 | 1 | 118 |
| 18.78 +42.91 | TYC3126-00500-1 | 1 | 118 |
| 19.62 +38.59 | TYC3135-00039-1 | 1 | 118 |
| 19.66 +42.91 | TYC3143-00397-1 | 1 | 118 |
| 19.86 +42.14 | TYC3145-01084-1 | 1 | 118 |
| 18.79 +45.07 | TYC3540-00808-1 | 1 | 118 |
| 18.97 +49.57 | TYC3549-00369-1 | 1 | 118 |
| 19.06 +49.66 | TYC3549-00808-1 | 1 | 118 |
| 19.24 +50.98 | TYC3554-00050-1 | 1 | 118 |
| 20.08 +45.34 | TYC3559-01144-1 | 1 | 118 |
| 15.01 +53.26 | TYC3861-00152-1 | 1 | 118 |
| 18.17 +55.62 | TYC3907-00811-1 | 1 | 118 |
| 18.34 +65.80 | TYC4226-01114-1 | 1 | 118 |
| 05.04 +69.09 | TYC4342-01013-1 | 1 | 118 |
| 04.81 +68.18 | TYC4342-02396-1 | 1 | 118 |
| 12.35 +74.97 | TYC4400-00146-1 | 1 | 118 |
| 10.31 -01.45 | TYC4904-01820-1 | 1 | 118 |
| 00.19 -11.93 | TYC5264-00070-1 | 1 | 118 |
| 09.55 -12.45 | TYC5468-00432-1 | 1 | 118 |
| 13.77 -13.62 | TYC5552-00086-1 | 1 | 118 |
| 18.31 -11.72 | TYC5685-03906-1 | 1 | 118 |
| 10.23 -21.80 | TYC6071-00117-1 | 1 | 118 |
| 11.95 -19.54 | TYC6097-01498-1 | 1 | 118 |
| 21.46 -21.89 | TYC6372-00034-1 | 1 | 118 |
| 01.08 -22.69 | TYC6422-00117-1 | 1 | 118 |
| 16.37 -26.82 | TYC6802-00832-1 | 1 | 118 |
| 03.17 +25.13 | TYC1787-01137-1 | 1 | 118 |
| 03.25 +25.03 | TYC1788-00623-1 | 1 | 118 |
| 04.32 +57.75 | TYC3727-00835-1 | 1 | 118 |
| 04.91 -23.29 | TYC6466-00589-1 | 1 | 118 |
| 05.58 +21.03 | TYC1309-00893-1 | 1 | 118 |
| 09.32 +19.18 | TYC1405-01424-1 | 1 | 118 |
| 22.21 +16.15 | TYC1681-00383-1 | 1 | 118 |
| 17.43 +27.05 | TYC2082-01268-1 | 1 | 118 |
| 03.14 +30.27 | TYC2339-01565-1 | 1 | 118 |
| 19.79 +34.46 | TYC2664-01334-1 | 1 | 118 |
| 05.73 +62.10 | TYC4098-00897-1 | 1 | 118 |
| 08.67 +64.40 | TYC4130-01719-1 | 1 | 118 |
| 00.25 -11.86 | TYC5264-00137-1 | 1 | 118 |
| 09.75 -27.14 | TYC6610-00457-1 | 1 | 118 |
| 04.26 -32.10 | TYC7037-00038-1 | 1 | 118 |
| 04.27 -31.92 | TYC7037-00045-1 | 1 | 118 |
| 05.25 +06.57 | TYC0112-02262-1 | 1 | 117 |
| 14.73 +00.89 | TYC0326-01160-1 | 1 | 117 |
| 23.10 +18.52 | TYC1714-00471-1 | 1 | 117 |
| 22.00 +26.56 | TYC2211-01336-1 | 1 | 117 |
| 18.27 +36.37 | TYC2634-00264-1 | 1 | 117 |
| 19.31 +56.06 | TYC3924-00496-1 | 1 | 117 |
| 23.11 +57.71 | TYC3993-00517-1 | 1 | 117 |
| 20.45 +60.14 | TYC4233-02214-1 | 1 | 117 |
| 01.07 +77.25 | TYC4497-00996-1 | 1 | 117 |
| 11.95 -21.02 | TYC6100-01121-1 | 1 | 117 |
| 06.63 -27.71 | TYC6516-01694-1 | 1 | 117 |
| 12.96 -26.88 | TYC6705-01141-1 | 1 | 117 |
| 06.52 +05.55 | TYC0154-00369-1 | 1 | 117 |
| 04.84 +12.21 | TYC0691-00020-1 | 1 | 117 |
| 16.86 +11.49 | TYC0983-02037-1 | 1 | 117 |
| 23.00 +38.86 | TYC3215-00065-1 | 1 | 117 |
| 05.22 +78.32 | TYC4515-00097-1 | 1 | 117 |
| 02.60 -12.60 | TYC5291-00390-1 | 1 | 117 |
| 18.31 -12.28 | TYC5685-03609-1 | 1 | 117 |
| 03.59 -24.24 | TYC6447-00585-1 | 1 | 117 |
| 06.48 -28.57 | TYC6519-01630-1 | 1 | 117 |
| 08.24 +05.94 | TYC0207-01003-1 | 1 | 117 |
| 11.37 +06.91 | TYC0270-00209-1 | 1 | 117 |
| 02.19 +32.32 | TYC2313-00409-1 | 1 | 117 |
| 19.30 +42.06 | TYC3129-00053-1 | 1 | 117 |
| 07.96 +49.53 | TYC3410-01968-1 | 1 | 117 |
| 02.81 +71.43 | TYC4321-01211-1 | 1 | 117 |
| 02.80 +71.77 | TYC4321-02063-1 | 1 | 117 |
| 08.38 +19.90 | TYC1386-01605-1 | 1 | 116 |
| 04.96 -23.10 | TYC6466-00498-1 | 1 | 116 |
| 05.74 +00.50 | TYC0115-00473-1 | 1 | 116 |
| 08.58 +09.27 | TYC0797-00072-1 | 1 | 116 |
| 01.02 +20.81 | TYC1195-01417-1 | 1 | 116 |
| 20.23 +18.10 | TYC1622-00428-1 | 1 | 116 |
| 23.34 +19.04 | TYC1716-02212-1 | 1 | 116 |
| 01.41 +29.07 | TYC1754-01191-1 | 1 | 116 |
| 17.16 +33.07 | TYC2595-01492-1 | 1 | 116 |
| 22.83 +35.39 | TYC2757-01898-1 | 1 | 116 |
| 01.97 +46.49 | TYC3280-01631-1 | 1 | 116 |
| 10.09 +68.44 | TYC4383-01140-1 | 1 | 116 |
| 22.86 -14.10 | TYC5819-00752-1 | 1 | 116 |
| 06.32 -17.57 | TYC5938-00582-1 | 1 | 116 |
| 09.89 -16.25 | TYC6045-00842-1 | 1 | 116 |
| 10.04 -16.12 | TYC6046-00969-1 | 1 | 116 |
| 16.33 -20.78 | TYC6214-00141-1 | 1 | 116 |
| 21.44 -21.83 | TYC6372-00311-1 | 1 | 116 |
| 21.43 -21.63 | TYC6372-00336-1 | 1 | 116 |
| 21.57 -21.05 | TYC6373-00264-1 | 1 | 116 |
| 17.86 -30.27 | TYC7378-00459-1 | 1 | 116 |
| 00.24 +01.30 | TYC0002-00960-1 | 1 | 116 |
| 08.62 +12.56 | TYC0805-01092-1 | 1 | 116 |
| 23.42 +08.57 | TYC1162-01134-1 | 1 | 116 |
| 00.81 +20.36 | TYC1194-00659-1 | 1 | 116 |
| 03.25 +21.52 | TYC1244-00377-1 | 1 | 116 |
| 07.33 +20.22 | TYC1354-00637-1 | 1 | 116 |
| 20.46 +19.09 | TYC1640-01578-1 | 1 | 116 |
| 23.11 +21.39 | TYC1718-00030-1 | 1 | 116 |
| 15.26 +36.34 | TYC2569-00149-1 | 1 | 116 |
| 17.30 +37.44 | TYC2604-01196-1 | 1 | 116 |
| 18.73 +35.71 | TYC2649-00048-1 | 1 | 116 |
| 19.79 +33.64 | TYC2660-00286-1 | 1 | 116 |
| 19.72 +34.37 | TYC2664-01092-1 | 1 | 116 |
| 19.07 +44.96 | TYC3132-00015-1 | 1 | 116 |
| 19.05 +44.84 | TYC3132-00156-1 | 1 | 116 |
| 09.44 +45.53 | TYC3425-01563-1 | 1 | 116 |
| 02.94 +71.57 | TYC4321-02129-1 | 1 | 116 |
| 02.49 +74.24 | TYC4324-01069-1 | 1 | 116 |
| 23.95 +77.39 | TYC4606-01142-1 | 1 | 116 |
| 06.87 -00.91 | TYC4800-01186-1 | 1 | 116 |
| 03.19 -20.67 | TYC5871-00569-1 | 1 | 116 |
| 20.25 -27.14 | TYC6914-00741-1 | 1 | 116 |
| 00.77 +07.69 | TYC0604-01274-1 | 1 | 116 |
| 04.13 +17.84 | TYC1254-00785-1 | 1 | 116 |
| 17.28 +29.34 | TYC2073-00598-1 | 1 | 116 |
| 11.43 +40.94 | TYC3013-00780-1 | 1 | 116 |
| 19.23 +41.06 | TYC3125-00862-1 | 1 | 116 |
| 07.63 -13.92 | TYC5409-02903-1 | 1 | 116 |
| 01.97 +46.85 | TYC3280-00691-1 | 1 | 116 |
| 04.50 +05.26 | TYC0081-01866-1 | 1 | 116 |
| 04.52 +03.47 | TYC0086-00797-1 | 1 | 116 |
| 19.25 +50.52 | TYC3550-00747-1 | 1 | 116 |
| 22.88 -29.41 | TYC6977-00753-1 | 1 | 116 |
| 02.52 +24.79 | TYC1771-00135-1 | 1 | 115 |
| 08.40 -11.08 | TYC5431-02201-1 | 1 | 115 |
| 13.30 -17.68 | TYC6116-00153-1 | 1 | 115 |
| 10.96 -23.94 | TYC6635-00655-1 | 1 | 115 |
| 02.23 +32.34 | TYC2313-00721-1 | 1 | 115 |
| 19.20 +46.66 | TYC3542-01498-1 | 1 | 115 |
| 03.39 +30.98 | TYC2341-00449-1 | 1 | 114 |
| 02.19 +33.10 | TYC2313-01449-1 | 1 | 114 |
| 11.32 +40.93 | TYC3010-01919-1 | 1 | 114 |
| 04.86 -16.22 | TYC5899-00852-1 | 1 | 114 |
| 23.10 +21.43 | TYC1717-01096-1 | 1 | 114 |
| 02.11 +23.26 | TYC1758-00894-1 | 1 | 114 |
| 02.10 +23.17 | TYC1758-01120-1 | 1 | 114 |
| 23.51 +77.64 | TYC4606-00426-1 | 1 | 114 |
| 01.07 -22.34 | TYC5853-00898-1 | 1 | 114 |
| 24.00 -22.82 | TYC6982-00970-1 | 1 | 114 |
| 05.94 +28.16 | TYC1875-01861-1 | 1 | 113 |
| 22.20 +16.36 | TYC1681-00234-1 | 1 | 113 |
| 03.78 +31.64 | TYC2356-00630-1 | 1 | 113 |
| 00.89 +00.83 | TYC0012-00171-1 | 1 | 113 |
| 06.54 +00.20 | TYC0146-01162-1 | 1 | 113 |
| 01.55 +29.61 | TYC1755-01436-1 | 1 | 113 |
| 02.06 +25.20 | TYC1761-00889-1 | 1 | 113 |
| 23.41 -21.04 | TYC6402-01081-1 | 1 | 113 |
| 23.55 +77.55 | TYC4606-01992-1 | 1 | 113 |
| 08.13 +61.13 | TYC4126-02025-1 | 1 | 112 |
| 03.22 +30.73 | TYC2340-01031-1 | 1 | 112 |
| 06.30 -28.75 | TYC6518-01619-1 | 1 | 112 |
| 00.23 +01.38 | TYC0002-01558-1 | 1 | 111 |
| 01.25 +76.30 | TYC4493-01800-1 | 1 | 111 |
| 15.62 +05.61 | TYC0358-00781-1 | 1 | 111 |
| 19.95 +21.98 | TYC1628-01403-1 | 1 | 111 |
| 18.69 +35.61 | TYC2645-00572-1 | 1 | 111 |
| 19.72 +33.63 | TYC2660-00007-1 | 1 | 111 |
| 07.55 -14.36 | TYC5409-00798-1 | 1 | 111 |
| 09.87 -23.93 | TYC6603-01779-1 | 1 | 111 |
| 20.33 +64.39 | TYC4241-01640-1 | 1 | 111 |
| 06.38 -31.80 | TYC7073-00066-1 | 1 | 111 |
| 07.56 +33.99 | TYC2461-00710-1 | 1 | 111 |
| 11.36 -23.29 | TYC6650-00462-1 | 1 | 111 |
| 09.62 +04.05 | TYC0238-01228-1 | 1 | 110 |
| 14.31 +34.95 | TYC2549-00395-1 | 1 | 110 |
| 22.11 +18.96 | TYC1688-01776-1 | 1 | 110 |
| 01.91 -19.40 | TYC5858-01744-1 | 1 | 110 |
| 19.92 +41.29 | TYC3145-02530-1 | 1 | 110 |
| 03.14 +31.01 | TYC2339-00955-1 | 1 | 110 |
| 20.12 +44.23 | TYC3162-01207-1 | 1 | 110 |
| 18.73 +47.94 | TYC3531-00451-1 | 1 | 110 |
| 18.86 +47.69 | TYC3544-01763-1 | 1 | 110 |
| 19.38 +51.47 | TYC3555-00444-1 | 1 | 110 |
| 19.38 +51.97 | TYC3555-01386-1 | 1 | 110 |
| 19.54 +46.47 | TYC3556-00276-1 | 1 | 110 |
| 20.09 +45.49 | TYC3559-01998-1 | 1 | 110 |
| 19.87 +49.40 | TYC3565-00916-1 | 1 | 110 |
| 19.87 +49.57 | TYC3565-01349-1 | 1 | 110 |
| 13.44 +13.49 | TYC0897-00400-1 | 1 | 109 |
| 07.24 +15.29 | TYC1346-00407-1 | 1 | 109 |
| 03.65 +32.23 | TYC2359-00652-1 | 1 | 109 |
| 18.70 +37.22 | TYC2649-00207-1 | 1 | 109 |
| 18.47 +65.33 | TYC4222-00053-1 | 1 | 109 |
| 05.75 +00.82 | TYC0116-00571-1 | 1 | 108 |
| 05.75 +02.01 | TYC0119-01143-1 | 1 | 108 |
| 00.86 +35.00 | TYC2284-00115-1 | 1 | 108 |
| 00.87 +35.16 | TYC2284-00569-1 | 1 | 108 |
| 03.70 +31.74 | TYC2355-00178-1 | 1 | 108 |
| 08.77 +64.79 | TYC4131-00548-1 | 1 | 108 |
| 08.39 +61.68 | TYC4127-00748-1 | 1 | 108 |
| 17.25 +05.24 | TYC0407-00606-1 | 1 | 107 |
| 17.24 +05.09 | TYC0407-00744-1 | 1 | 107 |
| 01.71 +20.13 | TYC1211-00915-1 | 1 | 107 |
| 02.61 +24.29 | TYC1767-00521-1 | 1 | 107 |
| 02.61 +25.02 | TYC1771-00537-1 | 1 | 107 |
| 17.42 +26.81 | TYC2082-00023-1 | 1 | 107 |
| 00.86 +35.04 | TYC2284-00443-1 | 1 | 107 |
| 02.32 +32.28 | TYC2314-00383-1 | 1 | 107 |
| 02.32 +32.43 | TYC2314-01393-1 | 1 | 107 |
| 03.28 +31.64 | TYC2340-00531-1 | 1 | 107 |
| 19.66 +40.89 | TYC3139-00228-1 | 1 | 107 |
| 03.05 +48.92 | TYC3318-01346-1 | 1 | 107 |
| 08.67 +47.21 | TYC3416-00171-1 | 1 | 107 |
| 17.00 +46.88 | TYC3501-00235-1 | 1 | 107 |
| 02.80 +72.14 | TYC4320-00854-1 | 1 | 107 |
| 02.79 +71.72 | TYC4320-01019-1 | 1 | 107 |
| 01.87 -19.23 | TYC5858-01704-1 | 1 | 107 |
| 03.57 -15.61 | TYC5874-00102-1 | 1 | 107 |
| 12.56 +74.40 | TYC4400-00290-1 | 1 | 106 |
| 05.75 +00.56 | TYC0115-00667-1 | 1 | 106 |
| 05.75 +01.55 | TYC0115-01054-1 | 1 | 106 |
| 03.97 -21.51 | TYC5888-00234-1 | 1 | 106 |
| 07.61 +05.43 | TYC0186-00315-1 | 1 | 106 |
| 03.07 +10.71 | TYC0651-00214-1 | 1 | 106 |
| 00.75 +20.02 | TYC1194-00080-1 | 1 | 106 |
| 23.14 +18.65 | TYC1715-00238-1 | 1 | 106 |
| 02.08 +24.99 | TYC1758-00324-1 | 1 | 106 |
| 11.62 +24.99 | TYC1982-01632-1 | 1 | 106 |
| 17.06 +32.96 | TYC2594-00277-1 | 1 | 106 |
| 18.71 +36.93 | TYC2649-00434-1 | 1 | 106 |
| 04.43 +57.72 | TYC3727-00046-1 | 1 | 106 |
| 05.81 +58.86 | TYC3762-02038-1 | 1 | 106 |
| 08.65 +64.21 | TYC4130-00261-1 | 1 | 106 |
| 20.17 +65.42 | TYC4240-00028-1 | 1 | 106 |
| 06.19 +71.57 | TYC4353-01311-1 | 1 | 106 |
| 04.49 +81.69 | TYC4522-01008-1 | 1 | 106 |
| 06.42 -01.37 | TYC4785-02066-1 | 1 | 106 |
| 00.24 -12.24 | TYC5264-00011-1 | 1 | 106 |
| 08.28 -12.41 | TYC5434-00585-1 | 1 | 106 |
| 10.79 -21.85 | TYC6081-01938-1 | 1 | 106 |
| 03.03 -28.56 | TYC6444-00131-1 | 1 | 106 |
| 11.29 -29.13 | TYC6661-00616-1 | 1 | 106 |
| 11.16 -24.29 | TYC6636-01100-1 | 1 | 106 |
| 07.96 -27.70 | TYC6562-01816-1 | 1 | 106 |
| 10.30 +19.72 | TYC1422-00492-1 | 1 | 106 |
| 01.56 +29.70 | TYC1755-01138-1 | 1 | 106 |
| 00.38 +31.85 | TYC2261-00225-1 | 1 | 106 |
| 07.14 -29.95 | TYC6536-00302-1 | 1 | 106 |
| 20.09 +45.53 | TYC3559-00682-1 | 1 | 106 |
| 07.29 -31.05 | TYC7103-00656-1 | 1 | 105 |
| 06.44 +53.59 | TYC3764-01892-1 | 1 | 105 |
| 15.37 +58.96 | TYC3874-01478-1 | 1 | 105 |
| 21.21 +14.53 | TYC1117-00101-1 | 1 | 105 |
| 19.05 +39.20 | TYC3120-00825-1 | 1 | 105 |
| 19.60 +44.11 | TYC3147-01206-1 | 1 | 105 |
| 18.75 +48.12 | TYC3544-00733-1 | 1 | 105 |
| 20.05 +46.41 | TYC3558-00757-1 | 1 | 105 |
| 06.20 -29.62 | TYC6517-01969-1 | 1 | 105 |
| 00.69 +34.52 | TYC2283-00792-1 | 1 | 104 |
| 08.93 +04.35 | TYC0221-00609-1 | 1 | 104 |
| 15.95 +28.70 | TYC2040-01706-1 | 1 | 104 |
| 18.63 +04.30 | TYC0455-01354-1 | 1 | 103 |
| 04.82 -23.97 | TYC6465-01448-1 | 1 | 103 |
| 10.05 +61.75 | TYC4137-00322-1 | 1 | 103 |
| 03.38 -29.95 | TYC6452-01014-1 | 1 | 103 |
| 06.55 +05.99 | TYC0158-02732-1 | 1 | 103 |
| 05.49 +10.67 | TYC0704-03003-1 | 1 | 103 |
| 04.66 +23.26 | TYC1830-00522-1 | 1 | 103 |
| 07.73 +27.90 | TYC1920-00110-1 | 1 | 103 |
| 11.65 +25.57 | TYC1982-01958-1 | 1 | 103 |
| 22.06 +26.15 | TYC2208-00137-1 | 1 | 103 |
| 00.37 +31.83 | TYC2261-00845-1 | 1 | 103 |
| 06.82 +33.71 | TYC2440-00884-1 | 1 | 103 |
| 07.52 +33.98 | TYC2461-00894-1 | 1 | 103 |
| 18.73 +36.95 | TYC2649-00476-1 | 1 | 103 |
| 05.85 +58.39 | TYC3762-02358-1 | 1 | 103 |
| 20.20 +65.05 | TYC4240-00088-1 | 1 | 103 |
| 07.59 -13.82 | TYC5409-03244-1 | 1 | 103 |
| 08.29 -12.54 | TYC5434-01439-1 | 1 | 103 |
| 13.00 -27.73 | TYC6706-01283-1 | 1 | 103 |
| 10.63 +26.46 | TYC1979-00937-1 | 1 | 102 |
| 09.50 +31.20 | TYC2494-00515-1 | 1 | 102 |
| 09.52 +31.14 | TYC2494-00578-1 | 1 | 102 |
| 06.91 -19.39 | TYC5958-00012-1 | 1 | 102 |
| 06.15 -29.66 | TYC6517-01731-1 | 1 | 102 |
| 10.89 -31.22 | TYC7199-00498-1 | 1 | 102 |
| 10.56 +08.83 | TYC0839-00380-1 | 1 | 102 |
| 09.04 +16.11 | TYC1394-01102-1 | 1 | 102 |
| 05.62 +24.97 | TYC1865-01335-1 | 1 | 102 |
| 09.35 +49.82 | TYC3428-00237-1 | 1 | 102 |
| 07.71 -10.64 | TYC5414-02427-1 | 1 | 102 |
| 10.94 -23.54 | TYC6635-00957-1 | 1 | 102 |
| 11.21 -24.20 | TYC6649-00884-1 | 1 | 102 |
| 03.09 +10.87 | TYC0651-00410-1 | 1 | 102 |
| 16.48 +38.35 | TYC3063-01615-1 | 1 | 102 |
| 08.58 -30.13 | TYC7136-02093-1 | 1 | 102 |
| 08.66 -23.37 | TYC6571-01894-1 | 1 | 102 |
| 10.89 +40.94 | TYC3009-00196-1 | 1 | 102 |
| 19.91 +48.78 | TYC3566-00325-1 | 1 | 102 |
| 07.48 +83.51 | TYC4618-00140-1 | 1 | 102 |
| 22.40 -17.15 | TYC6385-00620-1 | 1 | 102 |
| 11.54 +20.07 | TYC1440-00759-1 | 1 | 101 |
| 05.51 -19.60 | TYC5924-01014-1 | 1 | 101 |
| 06.32 -28.70 | TYC6518-01766-1 | 1 | 101 |
| 07.28 -31.61 | TYC7103-01014-1 | 1 | 101 |
| 19.13 +51.28 | TYC3554-00234-1 | 1 | 101 |
| 19.78 +50.76 | TYC3569-00363-1 | 1 | 101 |
| 04.51 +21.00 | TYC1277-01274-1 | 1 | 100 |
| 10.08 +54.18 | TYC3815-00232-1 | 1 | 100 |
| 09.01 +53.89 | TYC3805-00597-1 | 1 | 100 |
| 11.39 +43.12 | TYC3015-00403-1 | 1 | 100 |
| 07.48 -22.55 | TYC6538-00248-1 | 1 | 100 |
| 08.14 +05.21 | TYC0203-00798-1 | 1 | 100 |
| 10.94 +13.96 | TYC0853-00761-1 | 1 | 100 |
| 09.34 +24.86 | TYC1952-00843-1 | 1 | 100 |
| 10.38 -23.77 | TYC6619-01154-1 | 1 | 100 |
| 10.38 -24.12 | TYC6619-01294-1 | 1 | 100 |
| 08.68 -30.35 | TYC7136-00490-1 | 1 | 100 |
| 08.64 +29.57 | TYC1948-00227-1 | 1 | 100 |
| 08.90 +33.14 | TYC2488-00523-1 | 1 | 100 |